\documentclass[hdr, twoside, final]{unswthesis}
\usepackage{mystyle} 

\makeatletter
\fancypagestyle{noHeading}{
        \fancyhead{}
        \renewcommand{\headrulewidth}{0pt}
}
\thesistitle{Orthogonal Time Frequency Space (OTFS) Modulation for Wireless Communications}
\thesisschool{School of Electrical Engineering and Telecommunications}
\thesisauthor{Shuangyang Li}
\thesisZid{ }
\thesisdegree{Doctor of Philosophy}
\thesisdate{August 2022}
\thesissupervisor{Jinhong Yuan}

\DeclareGraphicsExtensions{.pdf,.jpeg,.png}
\graphicspath{{2-intro/}{3-literature/}{4-techchapter/}{5-techchapter/}{6-techchapter/}{8-appendices/}}

\usepackage{amsmath,bm}
\begin{document}

    \renewcommand{\bibname}{References}
    \frontmatter
    \maketitle


    \chapter*{Abbreviations} \label{abbreviations}
\addcontentsline{toc}{chapter}{\protect\numberline{}{Abbreviations}}
\markboth{ABBREVIATIONS}{}

\begin{longtable}[t]{ll}
\textbf{AMP} \quad\quad&\mbox{Approximate message passing} \vspace{0.1in}\\
\textbf{AoA} \quad\quad&\mbox{Angle-of-arrival} \vspace{0.1in}\\
\textbf{AoD} \quad\quad&\mbox{Angle-of-departure} \vspace{0.1in}\\
\textbf{AWGN} \quad\quad&\mbox{Additive white Gaussian noise} \vspace{0.1in}\\
\textbf{BCH} \quad\quad&\mbox{Bose-Chaudhuri-Hocquenghem} \vspace{0.1in}\\
\textbf{BER} \quad\quad&\mbox{Bit error rate} \vspace{0.1in}\\
\textbf{BPSK} \quad\quad&\mbox{Binary phase shift keying} \vspace{0.1in}\\
\textbf{BS} \quad\quad&\mbox{Base station} \vspace{0.1in}\\
\textbf{B5G} \quad\quad&\mbox{Beyond the fifth-generation} \vspace{0.1in}\\
\textbf{CDMA} \quad\quad&\mbox{Code-division multiple access} \vspace{0.1in}\\
\textbf{CP} \quad\quad&\mbox{Cyclic prefix} \vspace{0.1in}\\
\textbf{CSI} \quad\quad&\mbox{Channel state information} \vspace{0.1in}\\
\textbf{CSIT} \quad\quad&\mbox{Channel state information at transmitter} \vspace{0.1in}\\
\textbf{DD} \quad\quad&\mbox{Delay-Doppler } \vspace{0.1in}\\
\textbf{DDA} \quad\quad&\mbox{Delay-Doppler-angular} \vspace{0.1in}\\
\textbf{DFT} \quad\quad&\mbox{Discrete Fourier transform } \vspace{0.1in}\\
\textbf{DoF} \quad\quad&\mbox{Degrees of freedom } \vspace{0.1in}\\
\textbf{DTFT} \quad\quad&\mbox{Discrete-time Fourier transform } \vspace{0.1in}\\
\textbf{DZT} \quad\quad&\mbox{Discrete Zak transform} \vspace{0.1in}\\
\textbf{E-UTRA} \quad\quad&\mbox{Evolved Universal Terrestrial Radio Access } \vspace{0.1in}\\
\textbf{EXIT} \quad\quad&\mbox{Extrinsic information transfer} \vspace{0.1in}\\
\textbf{FER} \quad\quad&\mbox{Frame error rate} \vspace{0.1in}\\
\textbf{FFT} \quad\quad&\mbox{Fast Fourier transform} \vspace{0.1in}\\
\textbf{GFDM} \quad\quad&\mbox{Generalized frequency-division multiplexing} \vspace{0.1in}\\
\textbf{IDZT} \quad\quad&\mbox{Inverse Discrete Zak transform} \vspace{0.1in}\\
\textbf{IFFT} \quad\quad&\mbox{Inverse fast Fourier transform} \vspace{0.1in}\\
\textbf{ICI} \quad\quad&\mbox{Intercarrier interference} \vspace{0.1in}\\
\textbf{ISAC} \quad\quad&\mbox{Integrated sensing and communications} \vspace{0.1in}\\
\textbf{ISFFT}  \quad\quad&\mbox{Inverse symplectic finite Fourier transform} \vspace{0.1in}\\
\textbf{ISI} \quad\quad&\mbox{Intersymbol interference} \vspace{0.1in}\\
\textbf{IZT} \quad\quad&\mbox{Inverse Zak transform} \vspace{0.1in}\\
\textbf{KL} \quad\quad&\mbox{Kullback-Leibler} \vspace{0.1in}\\
\textbf{LoS} \quad\quad&\mbox{Line-of-sight} \vspace{0.1in}\\
\textbf{LTV} \quad\quad&\mbox{Linear time-varying} \vspace{0.1in}\\
\textbf{MAP} \quad\quad&\mbox{Maximum \emph{a posteriori}} \vspace{0.1in}\\
\textbf{MCA} \quad\quad&\mbox{Mobile communications on board aircraft} \vspace{0.1in}\\
\textbf{MIMO} \quad\quad&\mbox{Multiple-input multiple-output} \vspace{0.1in}\\
\textbf{MISO} \quad\quad&\mbox{Multiple-input single-output} \vspace{0.1in}\\
\textbf{ML} \quad\quad&\mbox{Maximum-likelihood} \vspace{0.1in}\\
\textbf{MLSE} \quad\quad&\mbox{Maximum-likelihood sequence estimation} \vspace{0.1in}\\
\textbf{mmWave} \quad\quad&\mbox{Millimeter wave} \vspace{0.1in}\\
\textbf{MRC} \quad\quad&\mbox{Maximum ratio combining} \vspace{0.1in}\\
\textbf{MSE} \quad\quad&\mbox{Mean square error} \vspace{0.1in}\\
\textbf{LEO} \quad\quad&\mbox{Low-earth-orbit} \vspace{0.1in}\\
\textbf{LDPC} \quad\quad&\mbox{Low-density parity-check} \vspace{0.1in}\\
\textbf{L-MMSE} \quad\quad&\mbox{Linear minimum mean square error} \vspace{0.1in}\\
\textbf{LTE} \quad\quad&\mbox{Long term evolution} \vspace{0.1in}\\
\textbf{OAMP} \quad\quad&\mbox{Orthogonal approximate message passing} \vspace{0.1in}\\
\textbf{ODDM} \quad\quad&\mbox{Orthogonal delay-doppler division multiplexing} \vspace{0.1in}\\
\textbf{OFDM} \quad\quad&\mbox{Orthogonal frequency-division multiplexing} \vspace{0.1in}\\
\textbf{OSTF} \quad\quad&\mbox{Orthogonal short time Fourier} \vspace{0.1in}\\
\textbf{PAPR} \quad\quad&\mbox{Peak-to-average power ratio} \vspace{0.1in}\\
\textbf{PEP} \quad\quad&\mbox{Pairwise-error probability} \vspace{0.1in}\\
\textbf{PIC} \quad\quad&\mbox{Parallel interference cancellation} \vspace{0.1in}\\
\textbf{PSD} \quad\quad&\mbox{Power spectral density} \vspace{0.1in}\\
\textbf{QPSK} \quad\quad&\mbox{quadrature phase shift keying} \vspace{0.1in}\\
\textbf{RCS} \quad\quad&\mbox{Radar cross section} \vspace{0.1in}\\
\textbf{RMS} \quad\quad&\mbox{Root-mean-square} \vspace{0.1in}\\
\textbf{SFFT}  \quad\quad&\mbox{Symplectic finite Fourier transform} \vspace{0.1in}\\
\textbf{SIC}  \quad\quad&\mbox{Successive interference cancellation} \vspace{0.1in}\\
\textbf{SINR}  \quad\quad&\mbox{Signal-to-interference-plus-noise ratio} \vspace{0.1in}\\
\textbf{SNR}  \quad\quad&\mbox{Signal-to-noise ratio} \vspace{0.1in}\\
\textbf{SOMP}  \quad\quad&\mbox{Structured orthogonal matching pursuit} \vspace{0.1in}\\
\textbf{SPA}  \quad\quad&\mbox{Sum-product algorithm} \vspace{0.1in}\\
\textbf{TD} \quad\quad&\mbox{Time-delay} \vspace{0.1in}\\
\textbf{TDMA} \quad\quad&\mbox{Time division multiple access} \vspace{0.1in}\\
\textbf{TDS} \quad\quad&\mbox{Time-delay-spatial} \vspace{0.1in}\\
\textbf{TF} \quad\quad&\mbox{Time-frequency} \vspace{0.1in}\\
\textbf{UAMP} \quad\quad&\mbox{AMP with a unitary transformation} \vspace{0.1in}\\
\textbf{UAV} \quad\quad&\mbox{Unmanned aerial vehicle} \vspace{0.1in}\\
\textbf{UE} \quad\quad&\mbox{User equipment} \vspace{0.1in}\\
\textbf{US} \quad\quad&\mbox{Uncorrelated scattering} \vspace{0.1in}\\
\textbf{VOFDM} \quad\quad&\mbox{Vector OFDM} \vspace{0.1in}\\
\textbf{WSS} \quad\quad&\mbox{Wide-sense stationary} \vspace{0.1in}\\
\textbf{WSSUS} \quad\quad&\mbox{Wide-sense stationary uncorrelated scattering} \vspace{0.1in}\\
\textbf{ZF} \quad\quad&\mbox{Zero forcing} \vspace{0.1in}\\
\textbf{ZT} \quad\quad&\mbox{Zak transform} \vspace{0.1in}\\
\textbf{2D} \quad\quad&\mbox{Two-dimensional} \vspace{0.1in}\\
\textbf{3D} \quad\quad&\mbox{Three-dimensional} \vspace{0.1in}\\
\textbf{3GPP} \quad\quad&\mbox{3rd Generation Partnership Project} \vspace{0.1in}\\
\textbf{5G} \quad\quad&\mbox{5-th generation} \vspace{0.1in}\\
\textbf{16-QAM} \quad\quad&\mbox{16-quadrature amplitude modulation} \vspace{0.1in}\\
\end{longtable}

    \chapter*{List of Notations} \label{listofnotations}
\addcontentsline{toc}{chapter}{\protect\numberline{}{List of Notations}}
\markboth{LIST OF NOTATIONS}{}
Scalars, vectors, and matrices are written in italic, boldface lower-case and upper-case letters, respectively, e.g., $x$, $\mathbf{x}$, and $\mathbf{X}$.
\begin{longtable}[t]{ll}
$x^\mathrm{*}$ \quad\quad&\mbox{Conjugate of $x$} \vspace{0.1in}\\
$\mathbf{X}^\mathrm{T}$ \quad\quad&\mbox{Transpose of $\mathbf{X}$} \vspace{0.1in}\\
$\mathbf{X}^\mathrm{H}$ \quad\quad&\mbox{Hermitian transpose of $\mathbf{X}$ } \vspace{0.1in}\\
$\left\lceil \cdot \right\rceil $\quad\quad&\mbox{Round up operation } \vspace{0.1in}\\
$\textrm{vec}\left( {\bf{X}} \right)$\quad\quad&\mbox{Vectorization of $\mathbf{X}$} \vspace{0.1in}\\
${\left[ {\cdot} \right]_x}$\quad\quad&\mbox{Modulo operation with $x$ } \vspace{0.1in}\\
$\delta \left( x \right)$ and $\delta \left[ x \right]$ \quad\quad&\mbox{Dirichlet function with a continuous/discerte variable  $x$ } \vspace{0.1in}\\
$\otimes$ \quad\quad&\mbox{Kronecker product operator } \vspace{0.1in}\\
$\Pr\{\cdot\}$ \quad\quad&\mbox{Probability of an event } \vspace{0.1in}\\
${\cal X}\{\cdot\}$ \quad\quad&\mbox{Returns the cardinality of a set} \vspace{0.1in}\\
$\propto$ \quad\quad&\mbox{Both sides of the equation are multiplicatively connected to a constant}\vspace{0.1in}\\
$\mathbf{I}_{N}$ \quad\quad&\mbox{$N$ dimension identity matrix} \vspace{0.1in}\\
${\mathbb{E}}\left\{ \cdot \right\}$ \quad\quad&\mbox{Statistical expectation} \vspace{0.1in}\\
${\mathbb{Z}}$ \quad\quad&\mbox{Set of all integers} \vspace{0.1in}\\
$\mathcal{CN}(\mu,\sigma^2)$ \quad\quad&\mbox{Circularly symmetric complex Gaussian random variable:}\vspace{0.1in} \\
\quad\quad&\mbox{the real and imaginary parts are i.i.d. $\mathcal{N}(\mu/2,\sigma^2/2)$} \vspace{0.1in}\\
$\ln(\cdot)$ \quad\quad&\mbox{Natural logarithm} \vspace{0.1in}\\
$\max\left\{\cdot\right\}$ \quad\quad&\mbox{Return the maximum element of its input} \vspace{0.1in}\\
$\min\left\{\cdot\right\}$ \quad\quad&\mbox{Return the minimum element of its input} \vspace{0.1in}\\
${\cal O}\left( \cdot \right)$ \quad\quad&\mbox{Asymptotical Description for the order of computational complexity} \vspace{0.1in}\\

\end{longtable}

    \chapter{Abstract}

The orthogonal time frequency space (OTFS) modulation is a recently proposed multi-carrier transmission scheme, which innovatively multiplexes the information symbols in the delay-Doppler (DD) domain  instead of the conventional time-frequency (TF) domain. The DD domain symbol multiplexing gives rise to a direct interaction between the DD domain information symbols and DD domain channel responses, which are usually quasi-static, compact, separable, and potentially sparse. Therefore, OTFS modulation enjoys appealing advantages over the conventional orthogonal frequency-division multiplexing (OFDM) modulation for wireless communications.

In this thesis, we investigate the related subjects of OTFS modulation for wireless communications, specifically focusing on its signal detection, performance analysis, and applications. In specific, we first offer a literature review on the OTFS modulation in Chapter~1. Furthermore, a summary of wireless channels is given in Chapter 2. In particular, we discuss the characteristics of wireless channels in different domains and compare their properties.
In Chapter 3, we present a detailed derivation of the OTFS concept based on the theory of Zak transform (ZT) and discrete Zak transform (DZT). We unveil the connections between OTFS modulation and DZT, where the DD domain interpretations of key components for modulation, such as pulse shaping, and matched-filtering, are highlighted.

The main research contributions of this thesis appear in Chapter 4 to Chapter 7. In Chapter 4, we introduce the hybrid maximum a posteriori (MAP) and parallel interference cancellation (PIC) detection. This detection approach exploits the power discrepancy among different resolvable paths and can obtain near-optimal error performance with a reduced complexity.

In Chapter 5, we propose the cross domain iterative detection for OTFS modulation by leveraging the unitary transformations among different domains. After presenting the key concepts of the cross domain iterative detection, we study its performance via state evolution. We show that the cross domain iterative detection can approach the optimal error performance theoretically. Our numerical results agree with our theoretical analysis and demonstrate a significant performance improvement compared to conventional OTFS detection methods.

In Chapter 6, we investigate the error performance for coded OTFS systems based on the pairwise-error probability (PEP) analysis. We show that there exists a fundamental trade-off between the coding gain and the diversity gain for coded OTFS systems. According to this trade-off, we further provide some rule-of-thumb guidelines for code design in OTFS systems.

In Chapter 7, we study the potential of OTFS modulation in integrated sensing and communication (ISAC) transmissions. We propose the concept of spatial-spreading to facilitate the ISAC design, which is able to discretize the angular domain, resulting in simple and insightful input-output relationships for both radar sensing and communication. Based on spatial-spreading, we verify the effectiveness of OTFS modulation in ISAC transmissions and demonstrate the performance improvements in comparison to the OFDM counterpart.

A summary of this thesis is presented in Chapter 8, where we also discuss some potential research directions on OTFS modulation. The concept of OTFS modulation and the elegant theory of DD domain communication may have opened a new gate for the development of wireless communications, which is worthy to be further explored.

\chapter{Publications and Presentations}

\section*{List of Publications}

\subsection*{Journal Papers:}
\begin{enumerate}[J-1.]
\itemsep=1.0em

\item   \textbf{S. Li}, W. Yuan, Z. Wei, and J. Yuan, ``Cross Domain Iterative Detection for Orthogonal Time Frequency Space Modulation,'' to appear in \emph{\textbf{IEEE Transactions on Wireless Communications},} vol. 21, no. 4, pp. 2227-2242, Apr. 2022.

\item   \textbf{S. Li}, W. Yuan, Z. Wei, J. Yuan, B. Bai, D. W. K. Ng, and Y. Xie, ``Hybrid MAP and PIC detection for OTFS modulation,'' \emph{\textbf{IEEE Transactions on Vehicular Technology}}, vol. 70, no. 7, pp. 7193-7198, Jul. 2021.

\item   \textbf{S. Li}, J. Yuan, W. Yuan, Z. Wei, B. Bai, and D. W. K. Ng, ``Performance Analysis of Coded OTFS Systems over High-Mobility Channels,'' \emph{\textbf{IEEE Transactions on Wireless Communications}}, vol. 20, no. 9, pp. 6033-6048, Apr. 2021.

\item   \textbf{S. Li}, W. Yuan, C. Liu, Z. Wei,  J. Yuan, B. Bai, and D. W. K. Ng, ``A Novel ISAC Transmission Framework based on Spatially-Spread Orthogonal Time Frequency Space Modulation,'' to appear in \emph{\textbf{IEEE Journal on Selected Areas in Communications},} 2022.

\end{enumerate}
\subsection*{Conference Papers:}

\begin{enumerate}[C-1.]
\itemsep=1.0em

\item   \textbf{S. Li}, W. Yuan, Z. Wei, R. He, B. Ai, B. Bai, and J. Yuan, ``A Tutorial to Orthogonal Time Frequency Space Modulation for Future Wireless Communications,'' in \emph{\textbf{IEEE International Communication Conference in China (ICCC) Workshop},} Xiamen, China, 2021. pp. 439-443.

\item   \textbf{S. Li}, W. Yuan, J. Yuan, and Giuseppe Caire, ``On the Potential of Spatially-Spread Orthogonal Time Frequency Space Modulation for ISAC Transmissions,'' to appear in \emph{\textbf{IEEE International Conference on Acoustics, Speech and Signal Processing (ICASSP)},} Singapore, 2022.

\item   \textbf{S. Li}, J. Yuan, W. Yuan, Z. Wei, B. Bai, and D. W. K. Ng, ``On the Performance of Coded OTFS Modulation over High-Mobility Channels," in \emph{\textbf{IEEE International Communication Conference (ICC) Workshop},} Montreal, Canada, 2021. pp. 1-6.

\end{enumerate}

The following papers, which are not included in the thesis, have also been published
by the author.

\subsection*{Journal Papers:}
\begin{enumerate}[J-1.]
\itemsep=1.0em

\item   \textbf{S. Li}, Z. Wei, W. Yuan, J. Yuan, B. Bai, D. W. K. Ng, and L. Hanzo, ``Faster-than-Nyquist Asynchronous NOMA Outperforms Synchronous NOMA,'' to appear in \emph{\textbf{IEEE Journal on Selected Areas in Communications},} 2022.

\item \textbf{S. Li}, W. Yuan, J. Yuan, B. Bai, D. W. K. Ng, and L. Hanzo, ``Time-Domain vs. Frequency-Domain Equalization for FTN Signaling,'' \emph{\textbf{IEEE Transactions on Vehicular Technology}}, vol. 69, no. 8, pp. 9174-9179, Aug. 2020.

\item   \textbf{S. Li}, J. Yuan, B. Bai, and N. Benvenuto, ``Code based Channel Shortening for Faster-than-Nyquist Signaling: Reduced-Complexity Detection and Code Design,'' \emph{\textbf{IEEE Transactions on Communications}}, vol. 68, no. 7, pp. 3996-4011, Jul. 2020.

\item   \textbf{S. Li}, B. Bai, J. Zhou, P. Chen, and Z. Yu, ``Reduced-Complexity Equalization for Faster-than-Nyquist Signaling: New methods based on Ungerboeck Observation Model,'' \emph{\textbf{IEEE Transactions on Communications},} vol. 66, no. 3, pp. 1190-1204, Mar. 2018.

\item   \textbf{S. Li}, B. Bai, J. Zhou, Q. He, and Q. Li, ``Superposition Coded Modulation based Faster-than-Nyquist Signaling,'' \emph{\textbf{Wireless Communications and Mobile Computing},} vol. 2018, Article ID 4181626, 10 pages, 2018.

\item   Z. Wei, W. Yuan, \textbf{S. Li}, J. Yuan, and D. W. K. Ng, ``Off-grid Channel Estimation with Sparse Bayesian Learning for OTFS Systems,'' to appear in \emph{\textbf{IEEE Transactions on Wireless Communications}.} 2022.

\item   W. Yuan, \textbf{S. Li}, Z. Wei, J. Yuan, and D. W. K. Ng, ``Data-Aided Channel Estimation for OTFS Systems with A Superimposed Pilot and Data Transmission Scheme,'' \emph{\textbf{IEEE Wireless Communications Letters},} vol. 10, no. 9, pp. 1954-1958, Sept. 2021.

\item   B. Liu, \textbf{S. Li}, Y. Xie, and J. Yuan, ``A Novel Sum-Product Detection Algorithm for Faster-than-Nyquist Signaling: A Deep Learning Approach,''  \emph{\textbf{IEEE Transactions on Communications},}  vol. 69, no. 9, pp. 5975-5987, Sept. 2021.

\item   R. Chong, \textbf{S. Li}, J. Yuan, and D. W. K. Ng, ``Achievable Rate Upper-Bounds of Uplink Multiuser OTFS Transmissions,'' to appear in \emph{\textbf{IEEE Wireless Communications Letters}}, vol. 11, no. 4, pp. 791-795, Apr. 2022.

\item   W. Yuan, \textbf{S. Li}, L. Xiang, and D. W. K. Ng, ``Distributed Estimation Framework for Beyond 5G Intelligent Vehicular Networks,'' \emph{\textbf{IEEE Open Journal of Vehicular Technology}.} vol. 1, pp. 190-214, Nov. 2020.



\item   P. Kang, K. Cai, X. He, \textbf{S. Li}, and J. Yuan, ``Generalized Mutual Information-Maximizing Quantized Decoding of LDPC Codes with Layered Scheduling,'' to appear in \emph{\textbf{IEEE Transactions on Vehicular Technology},} 2022.

\item   M. Liu, \textbf{S. Li}, C. Zhang, B. Wang, B. Bai, ``Coded Orthogonal Time Frequency Space Modulation,'' \emph{\textbf{ZTE Communications}}, vol. 19, no. 4, pp. 54-62, Dec. 2021.

\item   Y. K. Enku, B. Bai, F. Wan, C. U. Guyo, I. N. Tiba, C. Zhang, and \textbf{S. Li}, ``Two-dimensional Convolutional Neural Network based Signal Detection for OTFS Systems,'' \emph{\textbf{IEEE Wireless Communications Letters}}, vol. 10, no. 11, pp. 2514-2518, Nov. 2021.

\item   W. Yuan, Z. Wei, \textbf{S. Li}, J. Yuan, and D. W. K. Ng, ``Integrated Radar Sensing and Communication-assisted Orthogonal Time Frequency Space Transmission for Vehicular Networks,'' \emph{\textbf{IEEE Journal of Selected Topics in Signal Processing},} vol. 15, no. 6, pp. 1515-1528, Nov. 2021.

\item   Z. Wei, W. Yuan, \textbf{S. Li}, J. Yuan, B. Ganesh, R. Hadani, and L. Hanzo, ``Orthogonal Time-Frequency Space Modulation: A Promising Next Generation Waveform,'' \emph{\textbf{IEEE Wireless Communications},} vol. 28, no. 4, pp. 136-144, August 2021.

\item   Z. Wei, W. Yuan, \textbf{S. Li}, J. Yuan, and D. W. K. Ng, ``Transmitter and Receiver Window Designs for Orthogonal Time Frequency Space Modulation,'' \emph{\textbf{IEEE Transactions on Communications}}, vol. 69, no. 4, pp. 2207-2223, Apr. 2021.
\item   D. Shi, W. Yuan, \textbf{S. Li}, N. Wu, and D. W. K. Ng, ``Cycle-Slip Detection and Correction for Carrier Phase Synchronization in Coded Systems,'' \emph{\textbf{IEEE Communications Letters}.} vol. 25, no. 1, pp. 113-116, Jan. 2021.
\item   W. Yuan, C. Liu, F. Liu, \textbf{S. Li}, and D. W. K. Ng, ``Learning-based Predictive Beamforming for UAV Communications with Jittering,'' \emph{\textbf{IEEE Wireless Communications Letters}.} vol. 9, no. 11, pp. 1970-1974, Nov. 2020.


\item   J. Zhang, B. Bai, \textbf{S. Li}, M. Zhu, and H. Li, ``Tail-Biting Globally-Coupled LDPC Codes,'' \emph{\textbf{IEEE Transactions on Communications},} vol. 67, no. 12, pp. 8206-8219, Dec. 2019.

\item J. Zhang, B. Bai, M. Zhu, \textbf{S. Li}, and H. Li, ``Protograph-based Globally-Coupled LDPC Codes over the Gaussian Channel with Burst Erasures,'' \emph{\textbf{IEEE Access},} vol. 7, pp. 153853-153868, 2019.

\end{enumerate}

\subsection*{Conference Papers:}
\begin{enumerate}[C-1.]
\itemsep=1.0em

\item   \textbf{S. Li}, Z. Wei, W. Yuan, J. Yuan, B. Bai, and D. W. K. Ng, ``On the Achievable Rates of Uplink NOMA with Asynchronized Transmission,'' in \emph{\textbf{ IEEE Wireless Communications and Networking Conference (WCNC)},} Nanjing, China, 2021, pp. 1-7.

\item   \textbf{S. Li}, J. Yuan, and B. Bai, ``Code based Channel Shortening for Faster-than-Nyquist Signaling,'' in \emph{\textbf{IEEE International Communication Conference (ICC)}. }Dublin, Ireland, 2020, pp. 1-6.

\item \textbf{S. Li}, J. Zhang, B. Bai, X. Ma, and J. Yuan, ``Self-Superposition Transmission: A Novel Method for Enhancing Performance of Convolutional Codes,'' in \emph{\textbf{IEEE International Symposium on Turbo Codes \emph{\&} Iterative Information Processing (ISTC)}}, Hong Kong, 2018, pp. 1-5.

\item   H. Wen, W. Yuan, and \textbf{S. Li}, ``Downlink OTFS Non-Orthogonal Multiple Access Receiver Design based on Cross-Domain Detection" to appear in \emph{\textbf{IEEE International Conference on Communications (ICC) Workshop},} Seoul, 2022.

\item   R. Chong, \textbf{S. Li}, W. Yuan, and J. Yuan, ``Outage Analysis for OTFS-based Single User and Multi-User Transmissions" to appear in \emph{\textbf{IEEE International Conference on Communications (ICC) Workshop},} Seoul, 2022.

\item   Y. K. Enku, B. Bai, \textbf{S. Li}, M. Liu, and I. N. Tiba, ``Deep-Learning based Signal Detection for MIMO-OTFS Systems" to appear in \emph{\textbf{IEEE International Conference on Communications (ICC) Workshop},} Seoul, 2022.


\item  Z. Wei, W. Yuan, \textbf{S. Li}, J. Yuan, and D. W. K. Ng, ``Performance Analysis and Window Design for Channel Estimation of OTFS Modulation,'' in \emph{\textbf{IEEE International Conference on Communications (ICC)},} Montreal, Canada, 2021, pp. 1-5.

\item   Z. Wei, W. Yuan, \textbf{S. Li}, J. Yuan, and D. W. K. Ng, ``A New Off-grid Channel Estimation Method with Sparse Bayesian Learning for OTFS Systems,'' in \emph{\textbf{IEEE Global Communication Conference (GlobeCom)}}, Madrid, Spain, 2021, pp. 1-7.

\item   W. Yuan, \textbf{S. Li}, Z. Wei, J. Yuan, and D. W. K. Ng, ``Bypassing Channel Estimation for OTFS Transmission: An Integrated Sensing and Communication Solution,'' in \emph{\textbf{ IEEE Wireless Communications and Networking Conference (WCNC) Workshop},} Nanjing, China, 2021, pp. 1-6.

\item B. Liu, \textbf{S. Li}, Y. Xie, and J. Yuan, ``Deep Learning Assisted Sum-Product Detection Algorithm for Faster-than-Nyquist Signaling,'' in \emph{\textbf{IEEE Information Theory Workshop (ITW)}}, Visby, Sweden, 2019, pp. 1-5.

\item M. Liu, \textbf{S. Li}, Q. Li, and B. Bai, ``Faster-than-Nyquist Signaling based Adaptive Modulation and Coding,'' in \emph{\textbf{IEEE International Conference on Wireless Communications and Signal Processing (WCSP)}}, Hangzhou, 2018, pp. 1-5.

\end{enumerate}
\section*{List of Presentations}

%

\subsection*{Poster presentations:}

\begin{itemize}
\item ``Self-superposition transmission: A novel method for enhancing performance of convolutional codes," in \emph{\textbf{Australia Information Theory School (AusITS)}}, Sydney, 2019.
\end{itemize}

    \tableofcontents
    \listoffigures  
    \listoftables  

    \mainmatter
    \pagestyle{fancy}
        \fancyhf{}
        \fancyhead[LE]{\leftmark}
        \fancyhead[RO]{\rightmark}
        \fancyfoot[C]{\thepage}
        \renewcommand{\headrulewidth}{1pt}
        \setcounter{secnumdepth}{3}

    \chapter{Introduction}
Beyond the fifth-generation (B5G) wireless communication systems are required to accommodate various emerging applications in high-mobility environments, such as mobile communications on board aircraft (MCA), low-earth-orbit (LEO) satellites, high-speed trains, and unmanned aerial vehicles (UAVs) \cite{Meyer2019road,cai2020joint,Weijie2020learning,Fotouhi2019survey}.
On top of that, ultra-higher data rate requirements push mobile providers to utilize higher frequency bands, such as millimeter wave (mmWave) bands, where a huge chunk of the spectrum is available.
However, wireless communications in high-mobility scenarios at high carrier frequencies suffer from severe Doppler spread effect, caused by the relative motion between transceivers.
Consequently, the currently deployed orthogonal frequency division multiplexing (OFDM) modulation may not be able to support efficient and reliable communications in such scenarios \cite{hwang2008ofdm}.
Therefore, as a potential solution to supporting heterogeneous requirements of B5G wireless systems, the recently proposed orthogonal time frequency space (OTFS) modulation has attracted substantial attention \cite{Hadani2017orthogonal}.

In high-mobility scenarios, wireless channels are usually doubly-dispersive in the time-frequency (TF) domain \cite{hlawatsch2011wireless,tse2005fundamentals}. In specific, the time dispersion is caused by the effect of multi-path, while the frequency dispersion is caused by the Doppler shifts.
Conventionally, OFDM modulation can efficiently mitigate the intersymbol interference (ISI) induced by the time dispersion by introducing a cyclic prefix (CP). However, the success of OFDM modulation relies deeply on maintaining the orthogonality among all the sub-carriers. Note that perfect orthogonality is highly impractical at the receiver, especially in high-mobility environments, due to the exceedingly large frequency dispersion, and
consequently, the performance of conventional OFDM systems is unsatisfactory in such scenarios \cite{hwang2008ofdm}. On the other hand, by invoking the two-dimensional (2D) symplectic finite Fourier
transform (SFFT), OTFS modulates the information symbols in the delay-Doppler (DD) domain, where the channel parameters are relatively stable compared to those in the TF domain \cite{hlawatsch2011wireless}. More importantly, it can be shown that by modulating the information symbols in the DD domain rather than the TF domain, each symbol principally experiences the whole fluctuations of the TF channel over an OTFS frame, and thus OTFS modulation offers the potential to exploiting the full channel diversity, achieving a better error performance compared to that of the conventional OFDM modulation in a high-mobility environment~\cite{Hadani2017orthogonal}.

Although OTFS modulation has shown great potentials for future wireless communications, there are still many theoretical and practical issues of OTFS modulation that need to be solved.
From a theoretical point of view, the intrinsic connections between OTFS modulation and DD domain signal processing have not been fully understood. OTFS modulation is built upon the elegant mathematical theory of Zak transform (ZT)~\cite{Hadani2017orthogonal}. But the interpretation of OTFS modulation and demodulation in the DD domain by using ZT has long been missing in the literature. Furthermore,  the theoretical error performance advantages of OTFS systems over the conventional OFDM systems
have not been thoroughly studied yet, especially for coded cases. We note that there are several previous works \cite{Raviteja2019effective,Biglieri2019error,Surabhi2019on} on the error performance analysis of OTFS systems. However, these works mainly
considered the uncoded case and their analysis may not be directly extended to the coded cases.
Consequently, the error performance of coded OTFS systems has not been fully studied.
As commonly recognized, channel coding is an efficient tool to combat fading and channel impairments and thus is a key enabler for reliable communications between users with high-mobility \cite{wu2016survey}. For OTFS modulation, the transformation from the TF domain to the DD domain provides the potential of exploiting the full TF diversity. In this case, a good channel code needs to couple the coded symbols to the 2D OTFS modulation, in order to exploit the full diversity and in the meantime maximize the coding gain.
However, it is still unknown what is the key coding parameter determining the coding gain for OTFS modulation.

From a practical point of view, OTFS transmission usually requires advanced and complex detection methods. The rationale behind this observation is that the DD domain channel usually has few non-zero responses as the result of the separability of DD domain channels. Consequently, a sequence-wise equalization is needed for OTFS to achieve a good error performance, whereas the OFDM counterpart only requires a symbol-wise detection (single-tap equalization) as the channel response in the TF domain can be characterized by a point-wise multiplication.
Furthermore, the combination between OTFS modulation and emerging wireless technologies is also of great practical importance. For example, radar sensing is expected to be an important service for future wireless networks. Therefore, a joint design for integrated sensing and communication (ISAC) transmission using OTFS waveforms is worth to be investigated.

In this thesis, we will study the related aspects of OTFS modulation. It is worthwhile to point out that OTFS modulation is a relatively new concept and its development is still in infancy. There are some concepts and interpretations of OTFS modulation that are still unclear in the literature.
However, the authors would like to explain the OTFS modulation in a systematic way to the best of their capability.

\section{Main Contributions and Organization of the Thesis}
Throughout this thesis, we will discuss several related subjects of OTFS modulation, and the main contributions of this thesis can be summarized as follows.
\begin{itemize}
\item We provide a literature review of the recent advances of OTFS modulation. The review includes the related works in OTFS concepts, channel estimation, signal detection, performance analysis, and applications.
\item We present the derivation on OTFS modulation by using the concepts of ZT and discrete Zak transform (DZT). We show the intrinsic connections between DD domain signal processing and conventional modulation components, such as pulse shaping, and derive the corresponding input-output relationships for OTFS modulation. Several examples are given to show the properties of the OTFS waveform and a summary of the OTFS system model in the vector form is presented with various channel conditions.

\item We study the hybrid maximum \emph{a posteriori} (MAP) and parallel interference cancellation (PIC) (Hybrid-MAP-PIC algorithm) detection for OTFS modulation. The Hybrid-MAP-PIC algorithm is motivated by the fact that different paths may have highly diverse channel gains, such that it is possible to select the paths with large gains for MAP detection, while performing PIC detection for the remaining paths, without scarifying too much on the error performance. We give detailed descriptions of the Hybrid-MAP-PIC algorithm and also provide the error performance analysis and parameter selection. Our numerical results agree with our analysis and show a marginal performance loss (less than 1 dB) to the optimal detection.
    The related contents of the hybrid MAP and PIC detection have been presented in
    \begin{itemize}
    \item  S. Li, W. Yuan, Z. Wei, J. Yuan, B. Bai, D. W. K. Ng, and Y. Xie, ``Hybrid MAP and PIC detection for OTFS modulation,'' \emph{IEEE Trans. Veh. Technol.}, vol. 70, no. 7, pp. 7193-7198, Jul. 2021.
    \end{itemize}

\item The cross domain iterative detection for OTFS modulation is investigated. The core idea of the cross domain iterative detection is to perform the message passing detection in different domains and pass the extrinsic information iteratively via the unitary transformation. In particular, we apply two basic estimation/detection algorithms in different domains and show that the cross domain iterative detection enjoys promising error performance even in very severe and complex fractional Doppler cases. State evolution is derived to characterize the convergence performance of the algorithm. Our numerical results agree with our theoretical analysis and demonstrate a significant performance improvement compared to conventional OTFS detection methods.
    The related contents of the cross domain iterative detection have been presented in
    \begin{itemize}
    \item  S. Li, W. Yuan, Z. Wei, and J. Yuan, ``Cross domain iterative detection for orthogonal time frequency space modulation,'' \emph{IEEE Trans. Wireless Commun.}, vol. 21, no. 4, pp. 2227-2242, Apr. 2022.
    \end{itemize}

\item We provide the analysis of the error performance of coded OTFS systems based on the study of pairwise error probability (PEP). By leveraging advanced bounding techniques, we show that the OTFS transmission has a fundamental trade-off between the diversity gain and coding gain. In particular, we find that the diversity gain of OTFS systems improves with the number of resolvable paths, while the coding gain declines. Rule-of-thumb guidelines for code design in OTFS systems are also given. Those analytical results are explicitly verified by our numerical results.
    The related contents of the performance analysis for coded OTFS systems have been presented in
    \begin{itemize}
    \item  S. Li, J. Yuan, W. Yuan, Z. Wei, B. Bai, and D. W. K. Ng, ``Performance analysis of coded OTFS systems over high-mobility channels," \emph{IEEE Trans. Wireless Commun.}, vol. 20, no. 9, pp. 6033-6048, Apr. 2021.
    \item  S. Li, J. Yuan, W. Yuan, Z. Wei, B. Bai, and D. W. K. Ng, ``On the performance of coded OTFS modulation over high-mobility channels," \emph{IEEE Int. Commun. Conf. Workshop}, pp. 1-6, 2021.
    \end{itemize}

\item We consider the application of OTFS modulation in ISAC transmissions. To facilitate the system design, we extend the OTFS idea to the spatial domain by introducing the concept of spatial-spreading. The key novelty of this concept is the discretization of the angular domain, which results in simple and insightful input-output relation for both radar sensing and communication functionalities. We derive the system model based on the spatially-spread OTFS (SS-OTFS) modulation for both radar sensing and communication and propose simple and direct estimation, power allocation, and precoding schemes. Our numerical results verify the effectiveness of the considered SS-OTFS framework and demonstrate the performance improvements in comparison to the OFDM counterpart.
    The related contents of the ISAC designs using OTFS waveforms have been presented in
    \begin{itemize}
    \item  S. Li, W. Yuan, C. Liu, Z. Wei, J. Yuan, B. Bai, and D. W. K. Ng, ``A novel ISAC transmission framework based on spatially-spread orthogonal time frequency space modulation,'' to appear in \emph{IEEE J. Sel. Areas Commun.}, 2022.
     \item  S. Li, W. Yuan, J. Yuan, and G. Caire, ``On the potential of spatially-spread orthogonal time frequency space modulation for ISAC transmissions,'' to appear in \emph{Proc. - ICASSP IEEE Int. Conf. Acoust. Speech Signal Process.}, pp. 1-5, 2022.
    \end{itemize}

\end{itemize}

\section{Literature Review}\label{c:literature}

The technology of OTFS modulation has been developed rapidly ever since the pioneering paper of Hadani, \emph{et. al.}~\cite{Hadani2017orthogonal}, and several summaries about
the developments of OTFS modulation have also appeared in the literature~\cite{hadani2018whitebook,Zhiqiang_magzine,Shuangyang2021tutorial}.
In this section, we aim to provide a literature review of the recent advances of OTFS modulation. In particular, we will summarize the related works of OTFS modulation based on different subjects of the topic, including OTFS concept and implementation, channel estimation, signal detection, performance analysis, and applications.

\subsection{OTFS Concept and Implementation}
The concept of OTFS modulation was built upon the elegant mathematical theory of ZT~\cite{zak1967finite,janssen1988zak}, but the intrinsic connections between OTFS modulation and ZT were not discussed in detail in the first paper of OTFS~\cite{Hadani2017orthogonal}. In~\cite{mohammed2021derivation}, the author provided an interesting interpretation of OTFS from the ZT point of view, where a rigorous derivation of OTFS modulation from the first principles is given. In particular, some important and interesting properties of OTFS modulation have been discussed and explained, such as the twisted convolution, and DD domain localization. A recent work~\cite{lampel2021orthogonal} also gives a direct implementation of OTFS modulation based on the DZT~\cite{Bolcskei1994Gabor,bolcskei1997discrete}, where the authors have shown that the OTFS modulation can be simply implemented by a DZT transformation. More importantly, this paper also considered the pulse shaping issue of OTFS modulation from the DD domain interpretations, where insightful conclusions are also obtained. A recent book chapter also explains the OTFS modulation using ZT~\cite{Viterbo2022DDcommunications}, where the authors have shown that bandlimiting filters and finite-time windows can be used to formulate OTFS modulation with DD domain Nyquist pulse shaping.

Although the OTFS modulation can be directly implemented in the DD domain via ZT/DZT, the implementation of OTFS modulation via the TF domain has also been widely considered in the literature~\cite{Hadani2017orthogonal,Hadani2018OTFS_long}. The idea of the TF domain implementation is first to transform the DD domain information symbols into the TF domain by using the inverse symplectic finite Fourier transform (ISFFT) and then modulates the TF domain symbols based on conventional TF domain multi-carrier modulators, such as OFDM modulator~\cite{hadani2018whitebook}.
Furthermore, depending on how to insert the cyclic prefix (CP), there are two types OTFS that appeared in the literature, namely ZT/DZT-based OTFS~\cite{mohammed2021derivation,lampel2021orthogonal} and CP/OFDM-based OTFS~\cite{rezazadehreyhani2018analysis}. In specific, the ZT/DZT-based OTFS only requires one CP in the time-delay (TD) domain and consequently, there will be interference in both the TF domain and DD domain at the receiver side. Some derivations on the ZT/DZT-based OTFS, including the input-output relationship among different domains, could be found in~\cite{Raviteja2018interference} and~\cite{Raviteja2019practical}{\footnote{Strictly speaking, the work of~\cite{Raviteja2018interference} considered the zero-padding technique instead of inserting CP. However, it has been shown that the resultant input-output relationship is similar to the ZT/DZT-based OTFS~\cite{pandey2021low}. }}.
On the other hand, the CP/OFDM-based OTFS appends the CP similar to the conventional CP insertion for OFDM modulation, e.g., each OFDM symbol contains one CP. Therefore, after the CP removal, the resultant TF domain symbols will have no intersymbol interference (ISI)~\cite{rezazadehreyhani2018analysis}. A concise system model for CP/OFDM-based OTFS has been demonstrated in~\cite{rezazadehreyhani2018analysis}. It can be shown that these two implementations have distinct properties, and we will provide more detailed discussions on DZT-based OTFS in Chapter~3.

The comparisons and connections between OTFS modulation and conventional multi-carrier modulation schemes have also been discussed in the literature. For example, the connection between OTFS modulation and the generalized frequency-division multiplexing (GFDM) modulation was discussed in~\cite{nimr2018extended}, where the authors have shown that OTFS modulation could be implemented based on the GFDM framework with a simple permutation. In particular, the authors have shown that the permutation is the key ingredient for achieving the promising performance over practical wireless channels.
A recent DD domain communication scheme called orthogonal delay-Doppler division multiplexing (ODDM) has been studied~\cite{Hailin2022ICC,Shencheng2022WCL}. The most interesting feature is that the information symbols can be modulated with respect to practical Nyquist pulses with full separation without violating the uncertainty principle~\cite{Hailin2022ICC}. Furthermore, it has been shown that ODDM can enjoy a better out-of-band emission performance than the conventional OTFS modulation. The connections and comparisons between OTFS modulation and other schemes, such as vector OFDM (VOFDM) modulation, and orthogonal short time Fourier (OSTF) modulation, have also been discussed before, and we refer the interested readers to~\cite{Yao2021OTFS} and~\cite{sayeed2021time} for more details.

\subsection{Channel Estimation}
The channel estimation is an important issue for OTFS systems.
In comparison to the conventional channel estimation in the TF domain, OTFS modulation enables DD domain channel estimation approaches that could capture the DD domain channel response with a reduced signaling overhead. Perhaps, the most famous channel estimation scheme is the one proposed in~\cite{Raviteja2019embedded}. In~\cite{Raviteja2019embedded}, an embedded pilot scheme for OTFS channel estimation has been proposed, where a sufficiently large guard interval is applied around the pilot to improve the acquisition of delay and Doppler responses. As the DD domain interaction between the transmitted signal and the channel response can be characterized by the twisted convolution, such a scheme can result in direct channel estimation by simply checking the received signal's value around the DD grid of the embedded pilot.
This approach has soon been extended to the multiple-input multiple-output (MIMO) case in~\cite{Ramachandran2018mimo}, where several pilot symbols are inserted
for different transmit antennas with sufficient guard spacing in the DD domain.
Although~\cite{Raviteja2019embedded} has offered a very simple and direct channel estimation framework, it may require a large guard space to perform well, especially when the DD domain channel sparsity does not hold, e.g., fractional Doppler case.
To overcome this issue, a superimposed pilot scheme has been considered in~\cite{Weijie2021superimposed}, where there is no guard interval around the pilot, such that the signaling overhead is minimized. Consequently, an iterative receiver is adopted to recover the superposition between pilot and information symbols, and the simulation results showed a similar error performance to that of~\cite{Raviteja2019embedded} at an expense of higher computation complexity.
Meanwhile. several compressed sensing-based approaches have been studied in the literature. For instance, the
authors in~\cite{Wenqian2019channel} proposed a three-dimensional (3D) structured orthogonal matching pursuit (SOMP) algorithm to estimate the delay-Doppler-angle domain channel via exploiting the 3D structured sparsity.
A 3D Newtonized orthogonal matching pursuit (NOMP) algorithm was proposed in~\cite{Muye2021new_path}, which exploits the fractional component in the Doppler and angle domains via Newton's method.
Furthermore, some recent works on channel estimation based on sparse Bayesian learning approaches have also appeared in~\cite{Lei2020sparse,Wei2021off,Zhiqiang2021new_offgrid}, which have demonstrated better error performances compared to the orthogonal matching pursuit-based approaches.

\subsection{Signal Detection}
The signal equalization for OTFS has also been a widely studied subject.
Many existing studies focused on the low-complexity detection for OTFS modulation.
In \cite{Raviteja2018interference}, the authors developed an iterative receiver based on the classic message passing algorithm (MPA), where the interference from other information symbols is treated as a Gaussian variable to reduce the detection complexity. However, due to the short cycles on the probabilistic graphical model, MPA may fail to converge and results in performance degradation.
To solve this issue, the authors of \cite{Yuan2019simple} proposed a convergence guaranteed receiver based on the variational Bayes framework.
The basic idea of this detector is to approximate the corresponding \emph{a posteriori} distribution of the optimal detection by exploiting the Kullback-Leibler (KL) divergence such that the message passing algorithm can be implemented based on a simpler graphical model. 
A hybrid detection scheme is proposed in~\cite{li2021hybrid}, where both \emph{maximum a posteriori} (MAP) detection and PIC detection are considered according to the channel coefficients. Simulation results show that this hybrid detection can approach the error performance of near-optimal symbol-wise MAP detection, especially for coded OTFS systems, but only requires a reduced complexity.
An OTFS detection approach based on approximate message passing (AMP) with a unitary transformation (UAMP) developed in~\cite{yuan2020iterative}, which not only enjoys a strong resilience to the error propagation due to the short cycles on the probabilistic graphical model but also enables an efficient implementation.
Two DD domain low-complexity linear equalizers for OTFS modulations were studied in~\cite{Surabhi2020low}, where the authors proposed simplified implementation of linear minimum mean squared error (L-MMSE) and zero-forcing (ZF) detections with only a logarithmic complexity by exploiting the DD domain channel properties.
A Rake-receiver-based OTFS detection was proposed in~\cite{Tharaj2020low_complexity}, where the zero-padding technique was also adopted to shape the corresponding channel matrix to assist the application of DD domain maximum ratio combining. An iterative MMSE detection zero-padded OTFS modulation was proposed in~\cite{Oliver2022ICC}, where the MMSE detection was applied in a window-by-window manner with the help of successive interference cancellation (SIC).

It should be noticed that most of the existing works on the OTFS detection take advantage of the DD domain channel sparsity to reduce the detection complexity.
However, when an OTFS frame duration is not sufficiently long, the resultant DD domain effective channel can be dense due to the insufficient resolution of the Doppler frequency, i.e., fractional Doppler~\cite{Raviteja2018interference}. In such a case, conventional detection methods may experience a very high detection complexity since the channel sparsity no longer holds.
In light of this, a cross domain iterative detection was proposed in~\cite{li2021cross}, where the extrinsic information is passed between the TD domain and DD domain via the corresponding unitary transformations in order to improve the error performance. This method has shown a robust error performance in the case of fractional Doppler without introducing additional complexity.
\subsection{Performance Analysis}
The unique properties of OTFS modulation has given rise to various advantages for the practical system design. For example, it has been reported that OTFS modulation enjoys the potential of achieving full channel diversity~\cite{Hadani2017orthogonal,Hadani2018OTFS_long}. In specific, the diversity performance of uncoded OTFS signals has been studied in~\cite{Surabhi2019on}, where the authors have shown that the diversity gain of uncoded OTFS systems can be one but the full diversity can be obtained by suitable precoding schemes. The \emph{effective} diversity was discussed in~\cite{Raviteja2019effective}, where the authors have shown that the full diversity can be achieved almost surely for the case of two resolvable paths when the frame size is sufficiently large, even for uncoded OTFS modulation systems.

The error performance of coded OTFS systems has been investigated in~\cite{li2021performance_analysis,Li2020on}, where the authors have shown that there is a trade-off between the diversity gain and the coding gain of OTFS systems. In particular, it has been shown in~\cite{li2021performance_analysis,Li2020on} that the diversity gain of OTFS systems improves with the number of resolvable paths, while the coding gain declines, which provided some insightful code design guidelines. Furthermore, the correctness of this conclusion has been verified in~\cite{liu2022coded,zhang2022performance}, where extensive simulations with state-of-art channel coding schemes are demonstrated.
The error performance of OTFS systems with carefully designed TF domain windows has been studied in~\cite{wei2020transmitter,wei2021performanceICC}, where the authors have shown that the TF domain windowing could effectively mitigate the Doppler effect from the channel and improve the channel estimation performance.
The achievable rate for CP/OFDM-based OTFS systems was studied in~\cite{rezazadehreyhani2018analysis}. In particular, it is shown that both CP/OFDM-based OTFS and conventional OFDM systems share the same achievable rate, due to the unitary transformation between TF and DD domains.
The pragmatic capacities of both OTFS and OFDM were compared in~\cite{Gaudio2021fair_comparison}, where the achievable rates of both OTFS and OFDM are computed under practical channel estimation and signal detection schemes. In particular, it is shown in~\cite{Gaudio2021fair_comparison} that the pragmatic capacity of OTFS outperforms the OFDM counterpart, especially in the presence of severe Doppler effects.
In~\cite{Surabhi2019peak}, the peak-to-average power ratio (PAPR) of OTFS modulation was investigated, which has shown that OTFS modulation has a better PAPR performance than that of OFDM modulation and GFDM modulation. In particular, the authors have shown that the PAPR of OTFS waveform is proportionate to the number of Doppler bins/time slots instead of the number of delay bins/sub-carriers. Therefore, it generally enjoys a relatively low PAPR in practical systems.
\subsection{Applications}
The success of OTFS has attracted many applications. In this section, we will mainly review some recent works on those applications, such as multiple access transmissions and ISAC.
\subsubsection{Multiple Access Transmissions}
In light of the advantages of OTFS modulation over the conventional OFDM modulation~\cite{Zhiqiang_magzine}, it is natural to investigate the multiple access schemes based on OTFS modulation to realize robust multi-user transmissions in high-mobility environments. For instance, a path-division multiple access scheme was proposed in~\cite{Muye2021new_path}, where the inter-user interference for downlink transmission is significantly suppressed by a low complexity beamforming design. Also, a study on DD domain multiple access schemes was reported in~\cite{surabhi2019multiple}, where the authors have shown the bit error rate (BER) advantages of the DD domain multiple access schemes over the conventional OFDM counterpart via simulations. The achievable rate upper-bounds for DD domain multiple access schemes have been studied in~\cite{Ruoxi2022achievable}, where the authors derived upper-bounds for both delay division multiple access scheme and Doppler division multiple access scheme. Furthermore, it is shown in~\cite{Ruoxi2022achievable,Ruoxi2022ICC} that the OTFS modulation has an achievable rate upper-bound that is independent from the delay and Doppler distributions, thanks to the ``channel hardening'' effect. However, the achievable rate upper-bound for the OFDM counterpart depends highly on the delay and Doppler distributions, due to the inevitable TF domain superposition among resolvable paths. Consequently, it is shown that DD domain multiple access schemes enjoy an improved achievable rate performance compared to the OFDM counterpart{\footnote{It should be noted that this conclusion does not contradict to the ones from~\cite{rezazadehreyhani2018analysis}, because the unitary transformation between different domains may not hold in the multiple access transmission, because each user only occupies limited resources~\cite{Ruoxi2022achievable}.}}.

\subsubsection{Integrated Sensing and Communications}
ISAC transmission has been widely recognized as an efficient approach to deal with the foreseeable coexistence between communication and radar~\cite{liu2020joint,liu2018mu}.
A key motivation for ISAC transmission designs is that both radar sensing and communication naturally have similar channel characteristics which can be exploited.
For example, let us consider a common downlink scenario in a mobile network, where the antennas for radar sensing and communication are co-located at the BS.
It is not hard to notice that the physical channel between the BS and user equipments (UEs) is the same for both radar sensing and communication, despite the fact that radar sensing is operated based on the received echoes at the BS after the \emph{round-trip} signal propagation, while signal detection for communication is based on the \emph{one-way} transmission from the BS to UEs.
It is worthwhile to notice that radar sensing carries out parameter estimations based on
the delay, Doppler, and angular features associated to resolvable paths, whose core idea aligns perfectly with the OTFS modulation for communication purposes.
The DD domain symbol multiplexing enables the direct interactions between the information symbols and the DD domain channel, whose channel response can be potentially inferred from the radar estimates in practice~\cite{yuan2021integrated}.
The synergistic ecosystem established by the needs for communication and radar sensing has motivated us to consider the ISAC transmission design based on OTFS modulation.
Various lines of research works have been conducted in the literature. For example, the effectiveness of OTFS modulation for ISAC transmission has been evaluated in~\cite{Gaudio2020on}, where the authors have shown that the estimation error lower bounds for radar sensing can be achieved by using OTFS signals while maintaining a satisfactory communication performance. This work has then been extended to the case of MIMO transmissions~\cite{gaudio2020hybrid}, where a hybrid digital-analog beamformer is devised for both radar sensing and communication. In addition, the authors in~\cite{gaudio2020hybrid} have also developed an efficient maximum-likelihood (ML) algorithm to facilitate target detection and parameter estimation.
In addition, a novel OTFS-based matched-filter algorithm for target range and velocity estimation for radar sensing has been proposed in~\cite{raviteja2019orthogonal}. Specifically, the proposed matched-filter algorithm takes advantage of the structures of the DD domain effective channel matrix and has shown better estimation performance compared to the OFDM counterpart.
Furthermore, an ISAC-assisted OTFS system has been proposed in~\cite{yuan2021integrated,WeijieWCNC2021}, where both uplink and downlink communications are considered. In particular, the authors proposed a novel DD domain channel estimation algorithm and introduced a message-passing-based detection algorithm for uplink transmission. On the other hand, the downlink communication transmission is designed based on the channel state information (CSI) obtained from radar sensing, such that it can bypass the need of channel estimation and equalization at the receiver side.

The application of OTFS modulation in ISAC transmissions will be discussed in detail in Chapter~7 of this thesis.

\section{Organizations of the Thesis}

This thesis is organized as follows. A review of wireless channels is provided in Chapter 2.
Furthermore, the detailed derivations for OTFS modulation appear in Chapter 3. Two OTFS signal detection approaches are introduced in Chapter 4 and Chapter 5, namely, the hybrid MAP and PIC detection and the cross domain iterative detection. In Chapter 6, we discuss the error performance of coded OTFS systems. The application of OTFS modulation in ISAC transmissions is studied in Chapter 7. Finally, the conclusion of this thesis is presented in Chapter 8, and some future research directions are also highlighted.

    \chapter{Wireless Channel Revisit}\label{c:channel}
Wireless channels play a fundamental role in communications over a wireless scenario.
Typical wireless communication systems may involve one or several transmitter(s) and receiver(s), which are equipped with at least one antenna capable of radiating or/and receiving electromagnetic waves.
Information is modulated onto the electromagnetic wave with a specific carrier frequency that is transmitted over the wireless channel between the transmitter(s) and receiver(s),
thereby enabling communication between transmitter(s) and receiver(s).
From a physical point of view, the propagation of electromagnetic wave is determined by Maxwell's equations according to the underlying wireless environment. Unfortunately, solving Maxwell's equations is usually infeasible due to the interactions between the electromagnetic wave and the scatterers (such as reflection, transmission, scattering,
and diffraction~\cite{hlawatsch2011wireless}), except for the ideal free-space propagation. However, from a communication point of view,
it is usually sufficient to describe the wireless channel in a stochastic manner with some reasonable simplifications.

In this chapter, we will introduce some fundamentals of wireless channels from the communication point of view. The ideas and discussions of this chapter follow the classic
textbook for wireless channels, e.g., Ref.~\cite{hlawatsch2011wireless,tse2005fundamentals,molisch2012wireless}.
In particular, we are mainly interested in the radio channels, despite of the fact that some of the related discussions are also relevant to acoustic channels.
\section{Wireless Channel Backgrounds}
Signals transmitted over wireless channels are generally affected by power attenuation, distortions, and the additive white Gaussian noise (AWGN).
In particular, the signal distortions may come from various channel impairments~\cite{hlawatsch2011wireless,tse2005fundamentals}, such as the multi-path effect and the Doppler effect, which are introduced as follows.

\subsection{Path Loss and Fading}
The received signal power is an important parameter determining the overall performance of wireless communications. In particular, the channel impairments may cause severe fluctuations of the receiver power, which can be characterized by three phenomena, namely, the pass loss, the large-scale fading, and the small-scale fading, respectively~\cite{hlawatsch2011wireless,molisch2012wireless}.

The path loss is caused by the power decay of the transmission, which is usually distance-dependent. In particular, the path loss is usually modeled by an exponential distribution. Let $\alpha$ be the path loss exponent{\footnote{The typical value of $\alpha$ is between $2$ to $4$. For example, the 3rd Generation Partnership Project (3GPP) suggests $\alpha=3.76$ for evolved universal terrestrial radio access (E-UTRA)~\cite{3GPP_Pathloss}.}. Then, the pass loss in decibels is given by ${\rm{PL}} \buildrel \Delta \over = {\rm{10}}\alpha {\rm{lo}}{{\rm{g}}_{10}}\left( d \right)$, where $d$ denotes the distance the electromagnetic wave has propagated. The path loss can be mitigated by transmit power control that is relevant to the link budget.
The transmit power control is usually enabled by a feedback loop from the receiver to the transmitter.

Large-scale fading usually comes from the block or attenuation of the propagation paths.
In specific, these two channel impairments are usually referred to as the \emph{shadowing} and \emph{absorption loss}, respectively~\cite{hlawatsch2011wireless}. Many experimental tests have shown that the large-scale fading can be accurately modeled by a random variable with the log-normal distribution that is related to the geographic characteristics~\cite{molisch2012wireless}. Similar to the path loss, the large-scale fading can be mitigated by the transmit power control, as the geographic characteristics are relatively constant.

Small-scale fading is the consequence of the constructive and destructive interference of multi-path transmission with respect to the corresponding delay and Doppler shifts. Unlike path loss and large-scale fading, small-scale fading varies fast and it is therefore usually modeled stochastically by Gaussian distributed channel coefficients.
Depending on whether there is an LoS path, the magnitude of the channel coefficient (channel gain) can be modeled by either Rayleigh distribution or Rice distribution.
Transmit power control cannot combat the small-scale fading as the received power fluctuates rapidly. The most effective method to mitigate the small-scale fading is the application of diversity techniques~\cite{tse2005fundamentals}.
In the following subsections, we will present common classifications for different types of small-scale fading.

\subsection{Flat Fading Channels}
We refer to a wireless channel as a flat fading channel, if the channel gives a (roughly) constant response for all time and frequency components during the transmission. A flat fading channel is neither time dispersive nor frequency dispersive, and the received signal is merely the multiplication of the transmitted signal with a specific fading coefficient in the noiseless regime.
A flat fading channel usually appears in indoor environments, where there are strong line-of-sight (LOS) components and few weak multi-path components.
For flat fading channels, the excess delays associated to different paths are usually small and the transmitter and receiver are relatively static or moving slowly.
\subsection{Time Dispersive Channels}
We refer to a channel as a time dispersive channel if it only has the multi-path effect.
The multi-path effect arises when the transmitted signal is picked up by the receive antenna after propagating through several different paths, which usually appears in the presence of multiple scatterers with relatively strong reflectivities~\cite{hlawatsch2011wireless,tse2005fundamentals}.
Consequently, the received signal is a superposition of several delayed, power attenuated replicas of the original transmitted signal. In other words, the original transmitted signal is smeared-out (dispersive) in the time domain after being transmitted over the wireless channels with multiple resolvable paths.
Time dispersive channels are frequency selective, which indicates that the strengths of channel responses are different among different frequency components. The frequency domain selectivity can be understood from the Fourier analysis,
where the time domain delays of different paths introduce different phase rotations to the frequency domain responses associated to different paths and therefore, frequency selectivity appears as the consequence of the non-coherent superposition of those responses.
A time dispersive channel could appear in large indoor halls or outdoor environments, where there are strong multi-path components. For time dispersive channels, the excess delay associated to different paths is usually large and the transmitter and receiver are relatively static or moving slowly.


\subsection{Frequency Dispersive Channels}
We refer to a channel as a frequency dispersive channel if it only has the Doppler effect.
The Doppler effects appear if there is a relative movement between the transmitter and receiver, in which case the emitted electromagnetic wave will experience frequency shifts.
The Doppler effect may cause different signal distortions subjected to the signal bandwidth and relative velocity. In particular, in the case where the relative velocity $v$ is much lower than the speed of light $c_0=3 * 10^8$ and the signal bandwidth is much smaller than the carrier frequency $f_{\rm c}$,
the Doppler effect can be approximated by a frequency shift $f_{\rm d}$, i.e.,~\cite{hlawatsch2011wireless,tse2005fundamentals}
\begin{align}
{f_{\rm{d}}} \buildrel \Delta \over = \frac{{v\cos \left( \theta  \right)}}{{{c_0}}}{f_{\rm{c}}},
\end{align}
where $ \theta $ is angle-of-arrival (AoA) of the electromagnetic wave relative to the direction of motion of the receiver.
Due to the Doppler effect, the received signal is a superposition of several frequency-shifted, power attenuated replicas of the original transmitted signal, resulting in frequency dispersion. Frequency dispersive channels are time selective, which again can be understood from the Fourier analysis.
A frequency dispersive channel could appear in outdoor environments, where there are strong LOS components and few weak multi-path components.
For frequency dispersive channels, the excess delay associated to different paths is usually small and the transmitter and receiver are relatively moving at a high speed.

\subsection{Doubly Selective Channels}
We refer to a channel as a frequency dispersive channel if it has both the multi-path effect and the Doppler effect.
Doubly selective channels are also called doubly dispersive channels or linear time-varying (LTV) channels, over which the received signal suffers from both time and frequency dispersions and selectivities relative to the transmitted signal.
The doubly selectivity is a direct consequence of signal transmission in the presence of the multi-path effect with non-negligible delay and Doppler shifts.
A doubly dispersive channel could appear in outdoor environments, where there are strong multi-path components. For doubly dispersive channels, the excess delay associated to different paths is usually large and the transmitter and receiver are relatively moving at a high speed.


Although doubly selective channels may impose great challenges for channel estimation and equalization from a communication perspective, recent advances in wireless communication have shown that both the time and frequency selectivities can be exploited to improve the communication performance~\cite{hlawatsch2011wireless,tse2005fundamentals}. In particular, doubly selectivity provides new degrees-of-freedom (DoFs), which are commonly known as the delay/frequency diversity and Doppler/time diversity. However, to fully exploit the diversity gain, the transmitted signal needs to be carefully designed and OTFS waveforms are a type of signals that can almost surely exploit the full time-frequency (TF) diversity.

We provide a table summarizing the aforementioned types of wireless channels. More details of doubly selective channels will be discussed in the following subsections.
\begin{table*}[htbp]
\caption{A Summary of Possible Conditions for Different Types of Wireless Channels}
\centering
\begin{tabular}{|c|c|c|c|c|c|}
\hline
Channel Type~&~Scenario~&~Multi-path~&~LOS~&~Delay~&~Doppler\\
\hline
Flat fading~&~Indoor~&~Weak~&~Yes~&~Small~&~Small \\
\hline
Time dispersive~&~Indoor/Outdoor~&~Strong~&~Yes~&~Large~&~Small \\
\hline
Frequency dispersive~&~Outdoor~&~Weak~&~Yes~&~Small~&~Large\\
\hline
Doubly dispersive~&~Outdoor~&~Strong~&~No~&~Large~&~Large\\
\hline
\end{tabular}
\label{C3_channel_summary}
\end{table*}

\section{Mathematical Representations of Wireless Channels: Deterministic Description}
In this section, we are interested in a deterministic scenario, where the physical attributes of the channel remain roughly unchanged during the signal transmission, including the number of scatterers, the relative velocity of the scatterers, the distances between the transmitter and the receiver relative to each scatterer, and the reflectivities associated to each scatterer, respectively. In particular, the deterministic scenario can be viewed as a ``snapshot" of the real channel, and how long can this snapshot accurately represent the channel genuinely depends on the real scenario~\cite{hlawatsch2011wireless,molisch2012wireless}. Some relevant discussions on this point will be given in the section. Without loss of generality, we are interested in the mathematical representations of wireless channels in the equivalent complex baseband domain.
In what follows, we focus on general LTV channels in a deterministic scenario with $P$ resolvable paths,
where $h_p$, $\tau_p$, $\nu_p$ are used to describe the complex fading coefficient, the time delay, and the Doppler frequency associated to the $p$-th path with $1 \le p \le P$, respectively.
\subsection{Time-Delay Representation of LTV Channels}
Assume that a signal $s(t)$ is transmitted over an LTV channel. Then, with the noiseless assumption, the received signal $r(t)$ can be modelled by{\footnote{Here, we assume that the channel takes the
Doppler shift first and the delay shift second. It should be noted that an equivalent interpretation could be derived by taking the delay shift first and the Doppler shift second. Some further discussions
on this ordering issue could be found in~\cite{Hadani2017orthogonal} and in Section 2.2.3 and Section 3.1.3.}}~\cite{hlawatsch2011wireless}
\begin{align}
r\left( t \right) = \sum\limits_{p = 1}^P {{h_p}} s\left( {t - {\tau _p}} \right){e^{j2\pi {\nu _p}t}}. \label{S4_io_relationship_TD1}
\end{align}
Particularly,~\eqref{S4_io_relationship_TD1} can be interpreted as the signal transmission over a wireless channel with $P$ distinct scatterers with respect to their own physical attributes, including the reflectivity (fading coefficient), the time delay, and the Doppler frequency.
According to~\eqref{S4_io_relationship_TD1}, it can be shown that the TD domain channel response of an LTV channel can be represented by~\cite{hlawatsch2011wireless}
\begin{align}
{h_{{\rm{TD}}}}\left( {t,\tau } \right) = \sum\limits_{p = 1}^P {{h_p}} {e^{j2\pi {\nu _p}t}}\delta \left( {\tau  - {\tau _p}} \right),\label{S4_TD_channel}
\end{align}
and the input-output relationship between $s(t)$ and $r(t)$ is given by
\begin{align}
r\left( t \right) = \int_{ - \infty }^\infty  {s\left( {t - \tau } \right){h_{{\rm{TD}}}}\left( {t,\tau } \right)} {\rm{d}}\tau .\label{S4_io_relationship_TD2}
\end{align}
Based on~\eqref{S4_TD_channel}, we notice that the TD domain channel only has at most $P$ separable responses along the delay domain, while it has responses for all the time domain components with respect to the Doppler frequency. For further clarification, let us consider the following example.

\textbf{Example 2-1} (\emph{Time-Delay Representation of LTV Channels}):
Let us consider the following example of the LTV channel, where the related parameters are given in Table~\ref{C3_Channel_Example_Parameters}. The corresponding TD representation of the channel is given in Fig.~\ref{Fig_TD_Channel1} and Fig.~\ref{Fig_TD_Channel2}.

\begin{table}[htbp]
\caption{Channel Parameters for Example 2-1, 2-2, and 2-3}
\centering
\begin{tabular}{|c|c|}
\hline
Number of paths $P$~&~$5$ \\
\hline
Time delays (ms)~&~$[0, 0.1, 0.3, 0.4, 0.5]$ \\
\hline
Doppler frequency (kHz)~&~$[-4, -2, 0, 2, 4]$ \\
\hline
Fading coefficients~&~Rayleigh fading\\
\hline
\end{tabular}
\label{C3_Channel_Example_Parameters}
\end{table}

\begin{figure}
\centering
\includegraphics[width=0.8\textwidth]{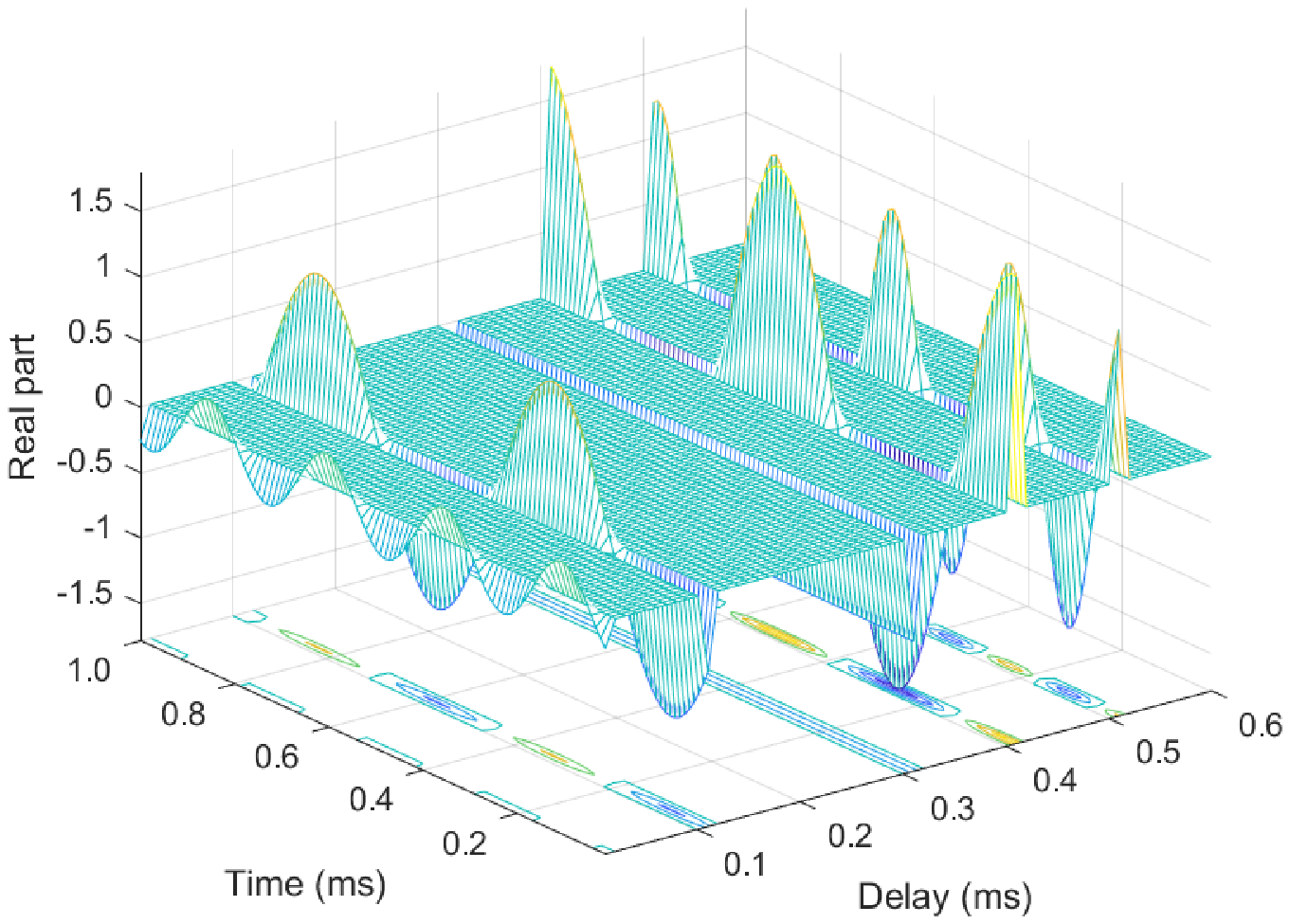}
\caption{TD domain representation of the LTV channel, where the channel parameters are given in Table~\ref{C3_Channel_Example_Parameters}. For a better clarification, only the real part of the complex channel response is shown.}
\label{Fig_TD_Channel1}
\centering
\end{figure}
\begin{figure}
\centering
\includegraphics[width=0.8\textwidth]{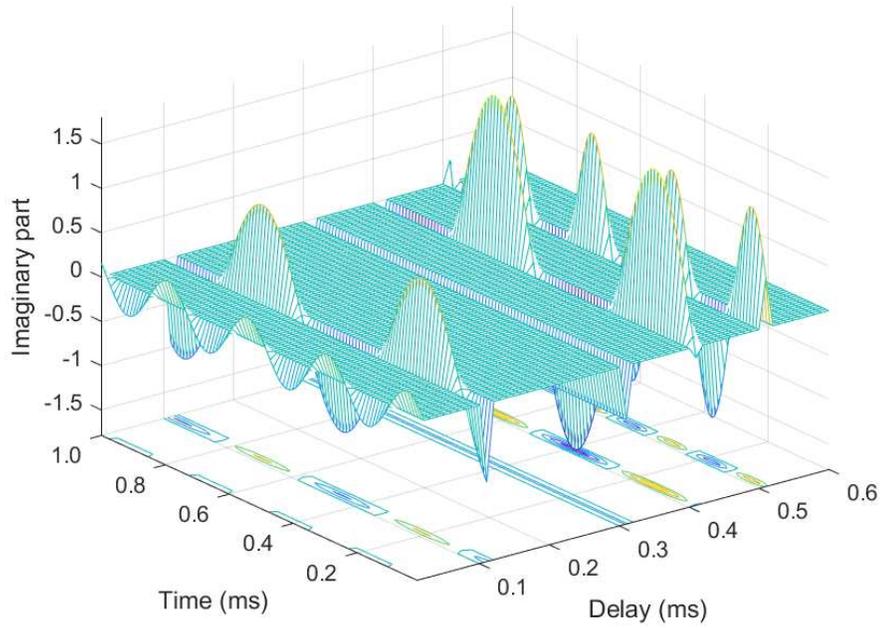}
\caption{TD domain representation of the LTV channel, where the channel parameters are given in Table~\ref{C3_Channel_Example_Parameters}. For a better clarification, only the imaginary part of the complex channel response is shown.}
\label{Fig_TD_Channel2}
\centering
\end{figure}
As indicated by both Fig.~\ref{Fig_TD_Channel1} and Fig.~\ref{Fig_TD_Channel2}, it can be shown that the TD representation of the channel is sparse in the delay domain, but dense in the time domain. In particular, the channel only has responses at the given time delays, while the channel response changes periodically with respect to the corresponding Doppler frequency along the time domain.
\subsection{Time-Frequency Representation of LTV Channels}
The TF domain channel can be obtained by performing the Fourier transform along the delay domain to~\eqref{S4_TD_channel}, i.e.,
\begin{equation}
  {h_{{\rm{TF}}}}\left( {t,f} \right) = \int_{ - \infty }^\infty  {{h_{{\rm{TD}}}}\left( {t,\tau } \right){e^{ - j2\pi f\tau }}} {\rm{d}}\tau  = \sum\limits_{p = 1}^P {{h_p}{e^{j2\pi {\nu _p}t}}} {e^{-j2\pi {\tau _p}f}} . \label{S4_TF_channel}
\end{equation}
The TF domain channel representation is the most known channel description in the field of wireless communications thanks to the success of orthogonal frequency-division multiplexing (OFDM). Some fundamentals of OFDM and its connections with OTFS will be discussed in the next chapter.
Similar to the TD domain channel representation, we use the following example to further clarify~\eqref{S4_TF_channel}.

\textbf{Example 2-2} (\emph{Time-Frequency Representation of LTV Channels}):
Let us consider the following example of the LTV channel, where the related parameters are given in Table~\ref{C3_Channel_Example_Parameters}. The corresponding TF domain representation of the channel is given in Fig.~\ref{Fig_TF_Channel}.

\begin{figure}
\centering
\includegraphics[width=0.8\textwidth]{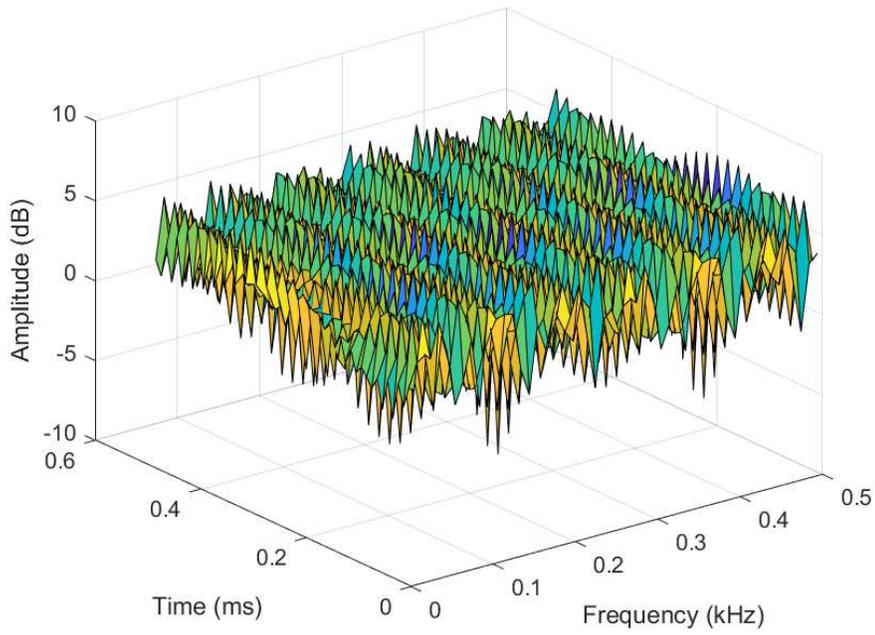}
\caption{TF domain representation of the LTV channel, where the channel parameters are given in Table~\ref{C3_Channel_Example_Parameters}. }
\label{Fig_TF_Channel}
\centering
\end{figure}

As can be observed from Fig.~\ref{Fig_TF_Channel}, the TF domain representation is dense over the whole TF plane.
This observation aligns with~\eqref{S4_TF_channel}, where we can notice that both delay and Doppler change
the phases of the channel responses. Although the TF domain channel response seems to be complex,
there generally exist regions within the TF plane, where the complex channel response remains roughly constant~\cite{hlawatsch2011wireless,molisch2012wireless}.
Such regions are usually referred to as the \emph{coherence region}~\cite{tse2005fundamentals}.
Some more discussions on the coherence region will appear in Section 2.3.3.

\subsection{Delay-Doppler Representation of LTV Channels}
The delay-Doppler (DD) domain channel response can be obtained by performing the Fourier transform along the time domain to~\eqref{S4_TD_channel}, i.e.,
\begin{equation}
    {h_{{\rm{DD}}}}\left( {\tau ,\nu } \right) = \int_{ - \infty }^\infty  {{h_{{\rm{TD}}}}\left( {t,\tau } \right){e^{ - j2\pi t \nu }}} {\rm{d}}t = \sum\limits_{p = 1}^P {{h_p}\delta \left( {\tau  - {\tau _p}} \right)} \delta \left( {\nu  - {\nu _p}} \right).\label{S4_DD_channel1}
\end{equation}
Equivalently, the DD domain channel response can also be obtained via the so-called \emph{symplectic finite Fourier transform} (SFFT) to the TF domain channel~\eqref{S4_TF_channel}~\cite{Hadani2017orthogonal}, i.e.,
\begin{equation}
    {h_{{\rm{DD}}}}\left( {\tau ,\nu } \right) = \int_{ - \infty }^\infty \int_{ - \infty }^\infty  {{h_{{\rm{TF}}}}\left( {t,f} \right){e^{ - j2\pi t \nu }}} {e^{j2\pi f\tau }}{\rm{d}}t{\rm{d}}f = \sum\limits_{p = 1}^P {{h_p}\delta \left( {\tau  - {\tau _p}} \right)} \delta \left( {\nu  - {\nu _p}} \right).\label{S4_DD_channel2}
\end{equation}
In particular, the SFFT can be viewed as the combination of a Fourier transform to the time domain and an inverse Fourier transform to the frequency domain~\cite{Hadani2017orthogonal}.
As indicated by~\eqref{S4_DD_channel2}, the dense TF domain channel responses become \emph{sparse} channel responses in the DD domain. Furthermore,~\eqref{S4_DD_channel2} implies that the responses associated to different paths are \emph{separable}, i.e., there is no interference between different paths' responses{\footnote{It should be noted that this only asymptotically holds given sufficiently large number of time and frequency resources~\cite{hlawatsch2011wireless}.}}.
On top of that, the DD domain channel response is \emph{compact} given the maximum delay and Doppler, which indicates that it only has responses within the region constrained by the maximum time delay and Doppler frequency. Similar to the previous subsections, let us consider the following example.

\textbf{Example 2-3} (\emph{Delay-Doppler Representation of LTV Channels}):
Let us consider the following example of the LTV channel, where the related parameters are given in Table~\ref{C3_Channel_Example_Parameters}. The corresponding DD domain representation of the channel is given in Fig.~\ref{Fig_DD_Channel}.

\begin{figure}
\centering
\includegraphics[width=0.8\textwidth]{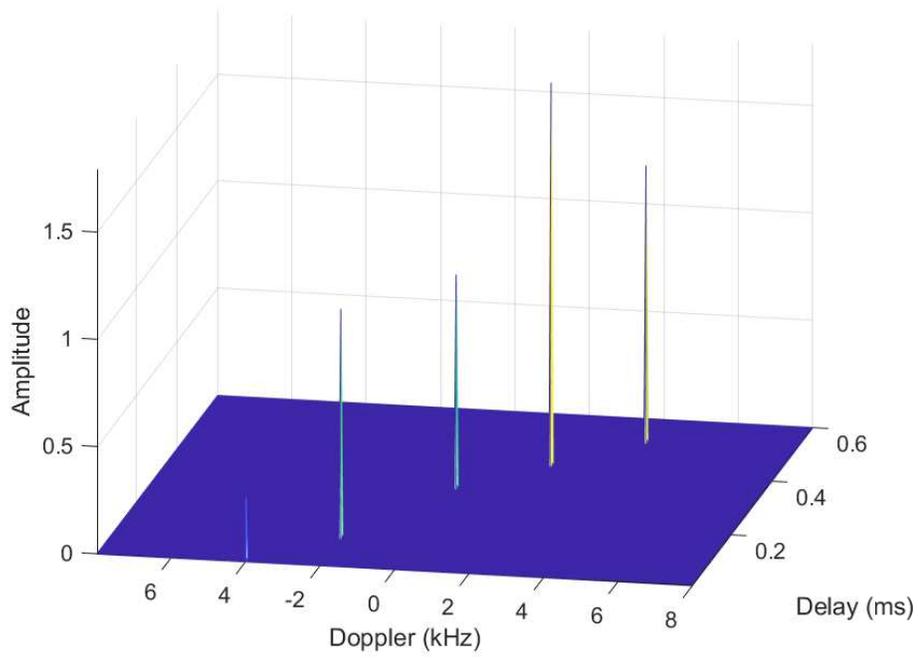}
\caption{DD domain representation of the LTV channel, where the channel parameters are given in Table~\ref{C3_Channel_Example_Parameters}.}
\label{Fig_DD_Channel}
\centering
\end{figure}

As can be observed from Fig.~\ref{Fig_DD_Channel}, the DD domain channel response only has five peaks over the whole DD plane, whose coordinations align with the parameters in Table~\ref{C3_Channel_Example_Parameters}. Furthermore, the channel responses from different paths are sufficiently separated as shown in Fig.~\ref{Fig_DD_Channel}. Those observations are consistent with our discussions above.

It is also of practical importance to discuss the DD domain input-output relationships for the purposes of communications. In specific, given the input signal $s\left( {t} \right)$, the corresponding
output signal $r\left( {t} \right)$ after transmitting over the DD domain channel ${h_{{\rm{DD}}}}\left( {\tau ,\nu } \right)$ can be written by~\cite{Hadani2017orthogonal,Raviteja2018interference}
\begin{equation}
r\left( t \right) = \int_{ - \infty }^\infty  {\int_{ - \infty }^\infty  {{h_{{\rm{DD}}}}\left( {\tau ,\nu } \right)} } {e^{j2\pi \nu \left( {t - \tau } \right)}}s\left( {t - \tau } \right){\rm{d}}\tau {\rm{d}}\nu ,
\label{S4_DD_channel_delay_first_Doppler_second}
\end{equation}
or
\begin{equation}
r\left( t \right) = \int_{ - \infty }^\infty  {\int_{ - \infty }^\infty  {{h_{{\rm{DD}}}}\left( {\tau ,\nu } \right)} } {e^{j2\pi \nu {t  } }}s\left( {t - \tau } \right){\rm{d}}\tau {\rm{d}}\nu.
\label{S4_DD_channel_Doppler_first_delay_second}
\end{equation}
The above two input-output relationships are the consequences of the ordering issue that is either we taking the delay shift first and Doppler shift second, or \emph{vice versa}.
Consequently, different orderings result in a phase difference ${e^{j2\pi \nu \tau}}$ between~\eqref{S4_DD_channel_delay_first_Doppler_second} and~\eqref{S4_DD_channel_Doppler_first_delay_second}.
Despite this phase difference, these two interpretations will
lead to equivalent results~\cite{Hadani2018OTFS_long}, as long as the notations are consistent.

\section{Mathematical Representations of Wireless Channels: Stochastic Description}
While the deterministic description of the wireless channel focuses on the ``snapshot" of the real channel, the stochastic description focuses on the statistical characteristics of the wireless channel, e.g., the second-order statistics. For any given linear time-varying channel, its second-order statistics can in general be represented by four different variables that are time, frequency, delay, and Doppler, respectively~\cite{hlawatsch2011wireless,bello1963characterization}.
In 1963, Bello has introduced the wide-sense stationary uncorrelated scattering (WSSUS) assumption, which greatly simplifies the above complex dependence from four variables to only two variables~\cite{bello1963characterization}. In what follows, we will introduce the concept of WSSUS channels first and then briefly extend our discussions to non-WSSUS channels.

\subsection{WSSUS Channels: Descriptions and Properties}
Without loss of generality, let us start our discussion from the TD domain channel in~\eqref{S4_TD_channel}. According to~\cite{hlawatsch2011wireless,bello1963characterization}, a \emph{wide-sense stationary} (WSS) channel should satisfy
\begin{align}
{\mathbb E}\left\{ {{h_{{\rm{TD}}}}\left( {t,\tau } \right)h_{{\rm{TD}}}^*\left( {t',\tau '} \right)} \right\} = {C_{{\rm{TD}}}}\left( {t - t';\tau ,\tau '} \right),
\label{S4_WSSUS_1_WSS}
\end{align}
i.e., the channel taps are jointly WSS with respect to the time variable, where ${C_{{\rm{TD}}}}$ denotes the correlation function of the TD domain channel responses.
On the other hand, if the LTV channel is said to feature \emph{uncorrelated scattering} (US), the following equation must hold~\cite{hlawatsch2011wireless,bello1963characterization}
\begin{align}
{\mathbb E}\left\{ {{h_{{\rm{TD}}}}\left( {t,\tau } \right)h_{{\rm{TD}}}^*\left( {t',\tau '} \right)} \right\} = {C_{{\rm{TD}}}}\left( {t, t';\tau} \right) \delta \left( {\tau  - \tau '} \right).
\label{S4_WSSUS_1_US}
\end{align}
By combining both~\eqref{S4_WSSUS_1_WSS} and~\eqref{S4_WSSUS_1_US}, we have
\begin{align}
{\mathbb E}\left\{ {{h_{{\rm{TD}}}}\left( {t,\tau } \right)h_{{\rm{TD}}}^*\left( {t',\tau '} \right)} \right\} = {C_{{\rm{TD}}}}\left( {t - t';\tau } \right)\delta \left( {\tau  - \tau '} \right),
\label{S4_WSSUS_1_WSSUS}
\end{align}
which holds for general WSSUS channels. The rationale of the WSSUS assumption is that in practical wireless channels, the channel impairments are introduced by physical scatterers with different reflectivities and any two distinct
scatterers generally have uncorrelated reflectivities.
In specific, the US assumption implies that the different delay shifts associated with different resolvable paths are uncorrelated. On the other hand, the WSS assumption indicates that the different Doppler shifts associated with different resolvable paths are uncorrelated such that the time domain channel response is jointly WSS with respect to the time variable.

Now let us turn our attention from the TD domain to the DD and TF domains. Based on the above discussions, let us define the ${C_{{\rm{DD}}}}\left( {\tau ;\nu } \right)$ as the \emph{channel scattering function}~\cite{hlawatsch2011wireless,bello1963characterization,molisch2012wireless}, which closely relates to the DD domain channel~\eqref{S4_DD_channel1} and~\eqref{S4_DD_channel2} by
\begin{align}
{\mathbb E}\left\{ {{h_{{\rm{DD}}}}\left( {\tau ,\nu } \right)h_{{\rm{DD}}}^*\left( {\tau ',\nu '} \right)} \right\} = {C_{{\rm{DD}}}}\left( {\tau ;\nu } \right)\delta \left( {\tau  - \tau '} \right)\delta \left( {\nu  - \nu '} \right).
\label{S4_WSSUS_DD_scattering}
\end{align}
The physical meaning of ${C_{{\rm{DD}}}}\left( {\tau ;\nu } \right)$ can be interpreted as the average strength of scatterers with time delay $\tau$ and Doppler frequency $\nu$ for the given LTV channel. As discussed in~\cite{hlawatsch2011wireless}, the channel scattering function is a white process but it is non-stationary.

The \emph{TF correlation function} is defined by
\begin{align}
{\mathbb E}\left\{ {{h_{{\rm{TF}}}}\left( {t,f} \right)h_{{\rm{TF}}}^*\left( {t',f'} \right)} \right\} = {C_{{\rm{TF}}}}\left( {t - t';f - f'} \right) = {C_{{\rm{TF}}}}\left( {\Delta t;\Delta f} \right),
\label{S4_WSSUS_TF_correlation}
\end{align}
which is a 2D stationary process with respect to the time variable and frequency variable. This interpretation of this 2D stationary process is straightforward. In specific, the WSS assumption implies that the channel is stationary in time by its definition, while the US assumption suggests that the channel is stationary in frequency as the Dopplers are uncorrelated~\cite{hlawatsch2011wireless}.
By observing the relationship between the TF and DD domain channel responses in~\eqref{S4_DD_channel2}, it can be shown that the TF correlation function and the channel scattering function can be calculated from each other via the SFFT and ISFFT~\cite{hlawatsch2011wireless}, e.g.,
\begin{align}
{C_{{\rm{DD}}}}\left( {\tau ;\nu } \right) = \int_{ - \infty }^\infty  {\int_{ - \infty }^\infty  {{C_{{\rm{TF}}}}\left( {\Delta t;\Delta f} \right){e^{ - j2\pi \left( { \Delta t \nu  - \Delta f \tau} \right)}}} } {\rm{d}}\Delta t{\rm{d}}\Delta f.
\label{S4_WSSUS_DD_TF}
\end{align}
What's more interesting is that~\eqref{S4_WSSUS_TF_correlation} and~\eqref{S4_WSSUS_DD_TF} together imply that the channel scattering function ${C_{{\rm{DD}}}}\left( {\tau ;\nu } \right)$ is essentially the 2D power spectral density (PSD) of the 2D stationary process ${h_{{\rm{TF}}}}$~\cite{hlawatsch2011wireless}. This relationship actually suggests that the channel estimation of either TF or DD domain can be carried out in its dual domain if the dual domain has more appealing properties, e.g., block-fading or a limited number of delay and Doppler responses.

Based on the discussions in this subsection, let us summarize the WSSUS channel characteristics in different domains in Table~\ref{C3_channel_summary_WSSUS}.
\begin{table*}[htbp]
\caption{WSSUS Channel Characteristics in Different Domains}
\centering
\begin{tabular}{|c|c|c|c|c|c|}
\hline
Domains~&~Statistics~&~Time~&~Frequency~&~Delay~&~Doppler\\
\hline
TD~&~${C_{{\rm{TD}}}}\left( {t - t';\tau } \right)$~&~Stationary~&~N/A~&~Uncorrelated~&~N/A \\
\hline
TF~&~${C_{{\rm{TF}}}}\left( {\Delta t ;\Delta f } \right)$~&~Stationary~&~Stationary~&~N/A~&~N/A \\
\hline
DD~&~${C_{{\rm{DD}}}}\left( {\tau ;\nu } \right)$~&~N/A~&~N/A~&~Uncorrelated~&~Uncorrelated\\
\hline
\end{tabular}
\label{C3_channel_summary_WSSUS}
\end{table*}
\subsection{WSSUS Channels: Delay and Doppler Profiles and Important Parameters}
As both delay and Doppler are two attributes of the scatterer, it is sometimes useful to focus on the ``marginal function" instead of the second order statistics. For example, the power-delay profile is defined by
\begin{align}
{C_{{\rm{Delay}}}}\left( \tau  \right) \buildrel \Delta \over = \int_{ - \infty }^\infty  {{C_{{\rm{DD}}}}\left( {\tau ;\nu } \right){\rm{d}}\nu } .
\label{S4_WSSUS_delay_profile}
\end{align}
Meanwhile, the power-Doppler profile is defined by
\begin{align}
{C_{{\rm{Doppler}}}}\left( \nu  \right) \buildrel \Delta \over = \int_{ - \infty }^\infty  {{C_{{\rm{DD}}}}\left( {\tau ;\nu } \right){\rm{d}}\tau } .
\label{S4_WSSUS_Doppler_profile}
\end{align}
Given the WSSUS assumption, ${C_{{\rm{Delay}}}}\left( \tau  \right)$ can be viewed as the mean power of the channel tap with delay $\tau$.
Similarly, ${C_{{\rm{Doppler}}}}\left( \tau  \right)$ can be viewed as the mean power of the channel tap with Doppler $\nu$. Exponential power-delay profile~\cite{molisch2012wireless} and the Jakes power-delay profile~\cite{jakes1994microwave} are two well-known models for ${C_{{\rm{Delay}}}}\left( \tau  \right)$ and ${C_{{\rm{Doppler}}}}\left( \nu  \right)$. In particular, the exponential power-delay profile is developed based on the distance-dependent decay of the signal power in the free space transmission, while the Jakes power-delay profile is derived based on the assumption of uniform distributed AoAs~\cite{jakes1994microwave}.

Based on the definition of power-delay profile and power-Doppler profile, some important parameters are ready to be discussed. For example, the pass loss is an important parameter implies the average power attenuation of the underlying channel, which directly links to the link budget. The pass loss ${\rho _{{\rm{PL}}}^2}$ can be derived by~\cite{hlawatsch2011wireless}
\begin{align}
{\rho _{{\rm{PL}}}^2} &= \int_{ - \infty }^\infty  {\int_{ - \infty }^\infty  {{C_{{\rm{DD}}}}\left( {\tau ;\nu } \right){\rm{d}}\tau {\rm{d}}\nu } }  = \int_{ - \infty }^\infty  {{C_{{\rm{Delay}}}}\left( \tau  \right){\rm{d}}\tau }  = \int_{ - \infty }^\infty  {{C_{{\rm{Doppler}}}}\left( \nu  \right){\rm{d}}\nu }\notag\\
&={\mathbb E}\left\{ {{{\left| {{h_{{\rm{TF}}}}\left( {t,f} \right)} \right|}^2}} \right\}.
\label{S4_WSSUS_PL}
\end{align}
Other than the pass loss, it is also useful for the practical system to have some knowledge of the channel attenuation with respect to the delay and Doppler. According to~\cite{hlawatsch2011wireless}, the mean delay and mean Doppler shift can be defined by
\begin{align}
\bar \tau  \buildrel \Delta \over = \frac{1}{{{\rho _{{\rm{PL}}}^2}}}\int_{ - \infty }^\infty  {\int_{ - \infty }^\infty  {\tau {C_{{\rm{DD}}}}\left( {\tau ;\nu } \right){\rm{d}}\tau {\rm{d}}\nu } }  = \frac{1}{{{\rho _{{\rm{PL}}}^2}}}\int_{ - \infty }^\infty  {\tau {C_{{\rm{Delay}}}}\left( \tau  \right){\rm{d}}\tau }
\label{S4_WSSUS_mean_delay},
\end{align}
and
\begin{align}
\bar \nu  \buildrel \Delta \over = \frac{1}{{{\rho _{{\rm{PL}}}^2}}}\int_{ - \infty }^\infty  {\int_{ - \infty }^\infty  {\nu {C_{{\rm{DD}}}}\left( {\tau ;\nu } \right){\rm{d}}\tau {\rm{d}}\nu } }  = \frac{1}{{{\rho _{{\rm{PL}}}^2}}}\int_{ - \infty }^\infty  {\nu {C_{{\rm{Doppler}}}}\left( \nu  \right){\rm{d}}\nu }
\label{S4_WSSUS_mean_Doppler},
\end{align}
respectively. The mean delay and mean Doppler shifts indicate the mean  absolute vales of the delay variable and Doppler variable that have a non-zero channel response{\footnote{Due to the causality of the LTV channel response, ${C_{{\rm{Delay}}}}\left( \tau  \right)=0$ for $\tau <0$.}}.
Furthermore, the \emph{delay spread} and \emph{Doppler spread} are also two terminologies widely used in the system design, which are typically defined by the root-mean-square (RMS) widths of power-delay profile and power-Doppler profile~\cite{hlawatsch2011wireless}, such as
\begin{align}
{\sigma _\tau } \buildrel \Delta \over = \frac{1}{{\sqrt {\rho _{{\rm{PL}}}^2} }}\sqrt {\int_{ - \infty }^\infty  {\int_{ - \infty }^\infty  {{{\left( {\tau  - \bar \tau } \right)}^2}{C_{{\rm{DD}}}}\left( {\tau ;\nu } \right){\rm{d}}\tau {\rm{d}}\nu } } }  = \frac{1}{{\sqrt {\rho _{{\rm{PL}}}^2} }}\sqrt {\int_{ - \infty }^\infty  {{{\left( {\tau  - \bar \tau } \right)}^2}{C_{{\rm{Delay}}}}\left( \tau  \right){\rm{d}}\tau } }
\label{S4_WSSUS_delay_spread},
\end{align}
and
\begin{align}
{\sigma _\nu } \buildrel \Delta \over = \frac{1}{{\sqrt {\rho _{{\rm{PL}}}^2} }}\sqrt {\int_{ - \infty }^\infty  {\int_{ - \infty }^\infty  {{{\left( {\nu  - \bar \nu } \right)}^2}{C_{{\rm{DD}}}}\left( {\tau ;\nu } \right){\rm{d}}\tau {\rm{d}}\nu } } }  = \frac{1}{{\sqrt {\rho _{{\rm{PL}}}^2} }}\sqrt {\int_{ - \infty }^\infty  {{{\left( {\nu  - \bar \nu } \right)}^2}{C_{{\rm{Doppler}}}}\left( \nu  \right){\rm{d}}\nu } }
\label{S4_WSSUS_Doppler_spread},
\end{align}
respectively. The delay and Doppler spread can be viewed as the ``effective" range of the delay and Doppler shifts that contains the most of the energy of the channel response. For a better illustration of the discussions in this subsection, let us consider the following example.

\textbf{Example 2-4} (\emph{Exponential Power-Delay Profile and Jakes Power-Doppler Profile~\cite{hlawatsch2011wireless,molisch2012wireless}}):
Let us discuss the WSSUS channel model considered in~\cite{hlawatsch2011wireless,molisch2012wireless}}, where the channel scattering function can be separated by
\begin{align}
{C_{{\rm{DD}}}}\left( {\tau ;\nu } \right) = \frac{1}{{\rho _{{\rm{PL}}}^2}}{C_{{\rm{Delay}}}}\left( \tau  \right){C_{{\rm{Doppler}}}}\left( \nu  \right).
\end{align}
In particular, we consider the exponential power-delay profile and Jakes power-Doppler profile, given by
\begin{equation}
{C_{{\rm{Delay}}}}\left( \tau  \right) = \left\{ \begin{array}{l}
\frac{{\rho _{\rm PL}^2}}{{{\tau _0}}}{e^{ - \frac{\tau }{{{\tau _0}}}}},\tau  \ge 0\\
0,\quad \quad  \quad \tau  < 0
\end{array} \right.
\label{S4_exponential_power_delay_profile}
\end{equation}
and
\begin{equation}
{C_{{\rm{Doppler}}}}\left( \nu  \right) = \left\{ \begin{array}{l}
\frac{{\rho _{\rm PL}^2}}{{\pi \sqrt {\nu _{\max }^2 - {\nu ^2}} }},\left| \nu  \right| < {\nu _{\max }}\\
0,\quad \quad \quad \quad\left| \nu  \right| > {\nu _{\max }}
\end{array} \right.
\label{S4_exponential_power_Doppler_profile}
\end{equation}
respectively.
In particular, the term $\tau_0$ is a delay parameter characterizing the decay exponent, while the term ${\nu _{\max }}$ is the maximum Doppler value.
For example, if we set ${\tau _0} = 5\; \mu s$, ${\nu _{\max }}=50 \;{\rm Hz}$, and ${\rho _{\rm PL}^2}=1$, the corresponding channel scattering function can be shown in Fig.~\ref{Fig_EXP_JAKES}.
\begin{figure}
\centering
\includegraphics[width=0.8\textwidth]{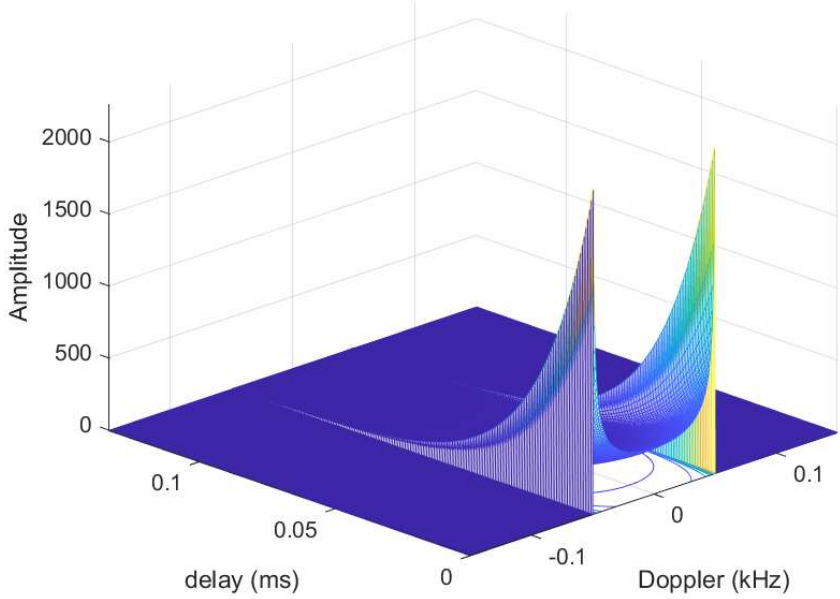}
\caption{Channel scattering function with exponential power-delay profile and Jakes power-Doppler profile, where the channel parameters are given in Example 2-4.}
\label{Fig_EXP_JAKES}
\centering
\end{figure}
As can be observed in Fig.~\ref{Fig_EXP_JAKES}, the channel scattering function behaves like a ``bowel-shaped" spectrum in the Doppler domain, while having an exponential decay in the delay domain.
In particular, it can be observed that the amplitudes of the channel scattering function tend to have large values for small delays and large Doppler shifts, while tend to have small values for large delays and small Doppler shifts.
Furthermore, it can be shown that $\bar \tau  = {\tau _0} = 20\; \mu s$, ${\bar \mu =\nu _{\max }}=50\; {\rm Hz}$, ${\sigma _\tau }= {\tau _0} = 20\; \mu s$, and $\bar \nu  = \frac{{{\nu _{\max }}}}{{\sqrt 2 }} = 35.36\; {\rm{Hz}}$, respectively.

\subsection{Underspread WSSUS Channels}
We focus on underspread WSSUS channels in the following. The underspread property is of particular interest for most of the wireless system designs, including both OFDM and OTFS systems~\cite{hlawatsch2011wireless}.
For an underspread channel, there exists a TF region that is no smaller than one, where the channel response is roughly constant. Specifically, the corresponding time domain and frequency domain intervals are commonly referred to as the \emph{coherence time} $T_{\rm c}$ and \emph{coherence bandwidth} $F_{\rm c}$, respectively.
In particular, the value of $T_{\rm c}$ is inversely proportional to the Doppler spread, while the value of $F_{\rm c}$ is inversely proportional to the delay spread~\cite{hlawatsch2011wireless}. Typically, we have
\begin{align}
{T_{\rm{c}}} \buildrel \Delta \over = \frac{1}{{{\sigma _\nu }}}
\label{S4_WSSUS_coherence_time},
\end{align}
and
\begin{align}
{F_{\rm{c}}} \buildrel \Delta \over = \frac{1}{{{\sigma _\tau }}}
\label{S4_WSSUS_coherence_bandwidth},
\end{align}
respectively. The approximation accuracy corresponding to the coherence time and bandwidth can be examined via the calculation of the mean squared difference
between ${h_{{\rm{TF}}}}\left( {t ,f } \right)$ and ${h_{{\rm{TF}}}}\left( {t + \Delta t,f + \Delta f} \right)$,
for $\left| {\Delta t} \right| \le T_{\rm c}$ and $\left| {\Delta f} \right| \le F_{\rm c}$. In specific, it can be shown that~\cite{hlawatsch2011wireless}.
\begin{align}
\frac{1}{{\rho _{{\rm{PL}}}^2}}{\mathbb E}\left\{ {{{\left| {{h_{{\rm{TF}}}}\left( {t + \Delta t,f + \Delta f} \right) - {h_{{\rm{TF}}}}\left( {t,f} \right)} \right|}^2}} \right\}
\le 2\pi \left[ {{{\left( {\frac{{\Delta t}}{{{T_{\rm{c}}}}}} \right)}^2} + {{\left( {\frac{{\Delta f}}{{{F_{\rm{c}}}}}} \right)}^2}} \right]
\label{S4_WSSUS_coherence_region_exam}.
\end{align}
Meanwhile, the Taylor expansion of ${C_{{\rm{TF}}}}\left( {\Delta t;\Delta f} \right)$ and can also be used to show the approximation accuracy and some relevant details can be found in~\cite{hlawatsch2011wireless}.

According to the definitions of coherence time and coherence bandwidth, a channel is said to be underspread, if ${T_{\rm{c}}}{F_{\rm{c}}} \ge  1$, ${T_{\rm{c}}}\ge  {\sigma _\tau }$, and ${F_{\rm{c}}}\ge  {\sigma _\nu }$~\cite{hlawatsch2011wireless}.
Alternative descriptions can also be given based on the delay and Doppler spreads. For example, for general underspread channel, ${\sigma _\tau }{\sigma _\nu } \le 1$ holds. A more relaxed but insightful description of the underspread channel relying on the maximum delay $\tau_{\rm max}$ and Doppler $\nu_{\rm max}$ values.
In specific, the maximum Doppler value is given by ${\nu_{\max }} = \frac{v}{c_0}{f_c}$, where $v$ is the relative speed, $c_0$ is the speed of light, and $f_c$ is the carrier frequency, respectively. The maximum delay shift is given by ${\tau_{\max }} = \frac{d_{\rm{max}}}{c_0}$, where $d_{\rm{max}}$ is the maximum distance difference among the $P$ channel paths.
Then, general underspread channels always have ${\rm{4}}{\tau _{{\rm{max}}}}{\nu _{{\rm{max}}}} \le {\rm{1}}$~\cite{hlawatsch2011wireless}.

The underspread condition implies that the channel cannot be both strong time dispersion and strong frequency dispersion~\cite{hlawatsch2011wireless}. Equivalently, this indicates that the channel cannot be both strongly time-selective and strongly frequency-selective~\cite{hlawatsch2011wireless}.
It should be noted that most of the real-world wireless (radio) channels are virtually always underspread. However, these underspread channels may not hold for some specific communication scenarios, such as the underwater acoustic channels~\cite{hlawatsch2011wireless}. For a better understanding of the underspread channel, let us consider the following example.

\textbf{Example 2-5} (\emph{Underspread Channel~\cite{hlawatsch2011wireless}}):
Following the channel parameters and profiles given in \textbf{Example 2-4},
it can be shown from~\eqref{S4_exponential_power_delay_profile} and~\eqref{S4_exponential_power_Doppler_profile} that ${T_{\rm{c}}} = \frac{{\sqrt 2 }}{{{\nu _{\max }}}} = 28.28 \; {\rm ms}$ and ${F_{\rm{c}}} = \frac{{1 }}{{{\tau _0}}} = 50.00 \; {\rm kHz}$.
Let us further consider a signal of duration $1 \; {\rm ms}$ and bandwidth $5 \; {\rm kHz}$. Then, based on~\eqref{S4_WSSUS_coherence_region_exam}, we have
\begin{align}
&\frac{1}{{\rho _{{\rm{PL}}}^2}}{\mathbb E}\left\{ {{{\left| {{h_{{\rm{TF}}}}\left( {t + \Delta t,f + \Delta f} \right) - {h_{{\rm{TF}}}}\left( {t,f} \right)} \right|}^2}} \right\} \notag\\
\le& 2\pi \left[ {{{\left( {\frac{{\Delta t}}{{{T_{\rm{c}}}}}} \right)}^2} + {{\left( {\frac{{\Delta f}}{{{F_{\rm{c}}}}}} \right)}^2}} \right]\notag\\
\le& 2\pi \left[ {{{\left( {\frac{{1 \times {{10}^{ - 3}}}}{{28.28 \times {{10}^{ - 3}}}}} \right)}^2} + {{\left( {\frac{{5 \times {{10}^3}}}{{50 \times {{10}^3}}}} \right)}^2}} \right]\notag\\
\approx &7.07\% ,
\end{align}
which implies that the mean squared difference of the TF domain channel responses for any two TF grids is at most $7.07\%$. In fact, the underlying can be viewed as a flat fading channel and it is strongly underspared,
whose TF domain channel response is roughly constant for the transmission of the whole block, i.e., it is a block-fading channel~\cite{hlawatsch2011wireless}.
\subsection{Extensions to Non-WSSUS Channels}
We have discussed the important characteristics of WSSUS channels in the previous subsections. Now let us turn our attention to non-WSSUS channels~\cite{matz2005non}.
In contrast to the WSSUS channels, non-WSSUS channels have not been widely
studied in the literature, despite its practical interests.
In specific, the non-WSSUS channels require the time, frequency, delay, and Doppler together to describe the channel characteristics. According to~\cite{hlawatsch2011wireless},
the \emph{channel correlation function} can be used to describe the correlations between the DD domain parameters and between the TF domain parameters, which is defined by
\begin{align}
{\cal R}\left( {\Delta t,\Delta f,\Delta \tau ,\Delta \nu } \right) \buildrel \Delta \over = &\int_{ - \infty }^\infty  {\int_{ - \infty }^\infty  {E\left\{ {{h_{{\rm{TF}}}}\left( {t,f + \Delta f} \right)h_{{\rm{TF}}}^*\left( {t - \Delta t,f} \right)} \right\}} } {e^{ - j2\pi \left( {t\Delta \nu  - f\Delta \tau } \right)}}{\rm{d}}t{\rm{d}}f\notag\\
=&\int_{ - \infty }^\infty  {\int_{ - \infty }^\infty  {E\left\{ {{h_{{\rm{DD}}}}\left( {\tau ,\nu  + \Delta \nu } \right)h_{{\rm{DD}}}^*\left( {\tau  - \Delta \tau ,\nu } \right)} \right\}} } {e^{j2\pi \left( {\nu \Delta t - \tau \Delta f} \right)}}{\rm{d}}\tau {\rm{d}}\nu .
\label{Non_WSSUS_CCF}
\end{align}
In particular, the channel correlation function can be viewed as a description for the correlations of the multi-path components separated by $\Delta t$ in time, $\Delta f$ in frequency,
$\Delta \tau$ in delay, and $\Delta \mu$ in Doppler. Further discussions regarding the channel correlation function can be found in~\cite{hlawatsch2011wireless,matz2005non}.

In general WSSUS channels, the stationarity does not hold for the whole TF plane. However, it can be shown that there exist regions in the TF plane, described by the time and frequency intervals, where the stationarity of the channel response roughly holds.
Following the descriptions in~\cite{hlawatsch2011wireless}, the \emph{stationarity time} and \emph{stationarity bandwidth} are defined by
\begin{align}
{T_{\rm{s}}} \buildrel \Delta \over = \frac{1}{{\overline {\Delta \nu } }},
\label{S4_stationarity time}
\end{align}
and
\begin{align}
{F_{\rm{s}}} \buildrel \Delta \over = \frac{1}{{\overline {\Delta \tau } }},
\label{S4_stationarity bandwidth}
\end{align}
respectively, where ${\overline {\Delta \nu } }$ and ${\overline {\Delta \tau } }$ are the delay and Doppler lags within which there are significant correlations~\cite{hlawatsch2011wireless}.
We are not going to introduce the details of stationarity time and stationarity bandwidth here. But it can be shown that for general LTV channels, the \emph{stationarity region} defined by the
both the \emph{stationarity time} and \emph{stationarity bandwidth} are much larger than the \emph{coherence region}~\cite{matz2005non}.
The intuition of this fact is not hard to understand. The stationarity region quantifies the TF region where the channel can be roughly approximated by a WSSUS channel, while the coherence
region describes the TF region where the channel response is roughly constant for a WSSUS channel.
Therefore, the coherence region must be no larger than the region where the channel
is roughly WSSUS by definition and consequently, there could be several coherence regions with potentially different channel responses within one stationarity region.

\textbf{Remarks 2-1}:
OTFS modulation essentially takes advantage of the deterministic DD domain channel response given in~\eqref{S4_DD_channel1}, where it is generally assumed that the channel response holds
for at least one block transmission~\cite{Hadani2017orthogonal}. However, the channel response may change after several blocks' transmission in practical scenarios.
But how often will the DD domain channel change is still not fully understood in the literature. An intuitive understanding of this question may
be developed based on the stationarity region of the non-WSSUS channels.
The multi-path transmission comes from the signal reflections of certain scatterers. In practical systems, the reflectivity for each scatter could hold for several blocks of
transmissions due to the fact that the change of physical geometry of the channel is so slow compared to the speed of
light. Therefore, the channel responses for adjacent blocks may be quite similar
because they are very likely to be generated from the same scatterers. This observation can be viewed as a sufficient condition for having the stationarity region in the non-WSSUS channel.
If the physical attributes of the scatterers remain unchanged, the delay and Doppler responses associated to different paths are roughly the same and therefore independent, then the channel can be approximated
as a WSSUS channel. The duration of this WSSUS approximation may be dependent on the stationarity region,
which is much larger than the coherence region.

\section{Sampling and Reconstruction}
In practical systems, sampling is of great importance, because the baseband equivalent system model essentially looks at the discrete channel response instead of the continuous one.
Therefore, it is necessary to discuss how to sample the channel response such that the actual response can
be reconstructed with acceptable errors.
\subsection{Sampling in the TF Domain}
Conventional sampling usually takes place in the TF domain. An example could be the well-known OFDM modulation. Without loss of generality, let us consider the sampling using the
\emph{sinc} function. Thus, the corresponding interpolation can be described by~\cite{hlawatsch2011wireless}
\begin{align}
{\hat h_{{\rm{TF}}}}\left( {t,f} \right) = \sum\limits_{k =  - \infty }^\infty  {\sum\limits_{l =  - \infty }^\infty  {{h_{{\rm{TF}}}}\left( {kT,lF} \right)} } {\rm{sinc}}\left( {\frac{\pi }{T}\left( {t - kT} \right)} \right){\rm{sinc}}\left( {\frac{\pi }{F}\left( {f - lF} \right)} \right),
\label{TF_sampling}
\end{align}
where $T$ and $F$ are the sampling time and frequency, respectively.
In order to achieve the perfect reconstruction in the mean squared sense, the channel scattering function ${C_{{\rm{DD}}}}\left( {\tau ;\nu } \right)$ must have a compact support within the range of
$\left[ { - {\tau _{\max }},{\tau _{\max }}} \right] \times \left[ { - {\nu _{\max }},{\nu _{\max }}} \right]$ and the sampling time and frequency need to satisfy
$F \le \frac{1}{{2{\tau _{\max }}}}$ and $T \le \frac{1}{{2{\nu _{\max }}}}$~\cite{hlawatsch2011wireless}.
On the other hand, if the above condition is not met, ${\hat h_{{\rm{TF}}}}\left( {t,f} \right) $ and ${h_{{\rm{TF}}}}\left( {t,f} \right) $ will inevitably be different due to the
aliasing. However, it is be shown that the normalized mean squared reconstruction error can be upper-bounded by~\cite{hlawatsch2011wireless}
\begin{align}
\frac{1}{{\rho _{{\rm{PL}}}^2}}{\mathbb E}\left\{ {{{\left| {{{\hat h}_{{\rm{TF}}}}\left( {t,f} \right) - {h_{{\rm{TF}}}}\left( {t,f} \right)} \right|}^2}} \right\} \le 2\left( {\sigma _\tau ^2{F^2} + \sigma _\nu ^2{T^2}} \right).
\label{TF_sampling_error_bound}
\end{align}
Some interesting insights can be obtained based on this upper-bound. Firstly, we notice that the sampling error relates to both the RMSs of the power-delay and power-Doppler profiles.
This observation suggests that the sampling error is generally small for underspread channels, where the RMSs are small.
Secondly, we also notice that smaller sampling time and frequency will also be helpful
for an accurate channel reconstruction.
Furthermore, it can be shown from~\eqref{TF_sampling_error_bound} that the upper-bound is minimized by choosing
\begin{align}
\frac{T}{F} = \frac{{{\sigma _\tau }}}{{{\sigma _\nu }}},
\end{align}
and we have
\begin{align}
\frac{1}{{\rho _{{\rm{PL}}}^2}}{\mathbb E}\left\{ {{{\left| {{{\hat h}_{{\rm{TF}}}}\left( {t,f} \right) - {h_{{\rm{TF}}}}\left( {t,f} \right)} \right|}^2}} \right\}
\le 2TF{\sigma _\tau }{\sigma _\nu }.
\label{TF_sampling_error_bound2}
\end{align}

Based on the above discussions, we summarize that the TF sampling could be more accurate if the sampling time and frequency are sufficiently small and the channel is more underspread.

\subsection{Sampling in the DD Domain}
Similar to the sampling in the TF domain, the channel response can be sampled in the DD domain by certain pulses. However, direct sampling in the DD domain has not been widely
discussed in the literature. To the best knowledge of the author, the DD domain sampling has only received much attention recently since the invention of OTFS modulation.
Conventionally, the information is carried in the TF domain for most of the communication schemes. In this case, the transmitted signal directly faces the TF domain channel, where the
equivalent discrete system model relies on the TF domain sampling of the channel response.
On the other hand, for OTFS systems, the information is carried in the DD domain instead of the TF domain. Thus, the OTFS signal faces the DD domain channel and the corresponding
equivalent discrete system model needs to be developed based on the DD domain sampling.
As the DD domain sampling has a close relationship to the pulse shaping for OTFS transmissions, we leave the related discussions in the following chapter.

\section{Summary of the Chapter}
In this chapter, we have briefly summarized the characteristics and properties of general wireless channels. Furthermore, we have discussed the mathematical descriptions of
wireless channels including both the deterministic channel description and the stochastic description, where the difference and connections between these two types of descriptions
are discussed. We have also introduced some key properties of the wireless channels, including the WSSUS channels, non-WSSUS channels, and underspread channels. The discussions
for the related channels are also provided with the help of some examples.
Finally, we have briefly discussed the sampling of the channels in different domains.

    \chapter{OTFS Concepts and Principles}\label{c:literature}
In this chapter, we present the concepts and principles of OTFS modulation~\cite{Hadani2017orthogonal,Hadani2018OTFS_long,Zhiqiang_magzine}.
Although OTFS modulation was only formally introduced in 2017, it has stimulated a wide range of research from different viewpoints. In specific, this chapter will discuss some of the key components of OTFS modulation in a systematic way. It should be noted that OTFS modulation inherits the merits of ZT/DZT by multiplexing the information symbols in the DD domain (Zak domain)~\cite{Hadani2017orthogonal,Hadani2018OTFS_long,Zhiqiang_magzine}. Therefore, before formally introducing the definition of OTFS modulation, we will first review the definitions and properties of ZT and DZT in the coming section.
After discussing the related contexts of ZT and DZT, we will introduce the concepts of OTFS modulation.
It should be noted that there are several implementations of OTFS modulation in the literature~\cite{Hadani2017orthogonal,mohammed2021derivation,Raviteja2019practical,Raviteja2018interference,lampel2021orthogonal,rezazadehreyhani2018analysis}. Without loss of generality, we will only focus on one of the popular implementations, namely DZT-based OTFS~\cite{Raviteja2018interference}, and discuss the corresponding system models{\footnote{There are also other popular choices of OTFS implementation, such as the OFDM-based OTFS~\cite{rezazadehreyhani2018analysis}. However, the details of those implementations are out of the scope of this thesis. }}.

\section{A Summary on the Zak Transform and Discrete Zak Transform}
The ZT~\cite{zak1967finite} (also known as Weil-Brezin mapping and Gelfand mapping) was originally introduced in the field of solid state physics and then studied by Janssen from a signal processing point of view~\cite{janssen1988zak}. Later, some advanced results of ZT and its corresponding discrete form were developed by B\"olcskei \emph{et. al.} in~\cite{Bolcskei1994Gabor,bolcskei1997discrete}. The ZT has a close relationship to the related concepts of Gabor expansion, which is introduced in the coming subsection.

\subsection{Gabor Expansion and Gabor Coefficients}
\begin{figure}
\centering
\includegraphics[width=0.8\textwidth]{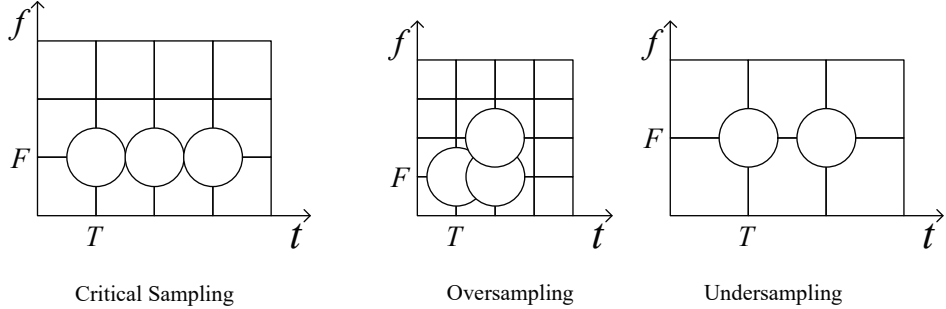}
\caption{An illustration for Gabor expansions with critical sampling, oversampling, and undersampling.}
\label{C4_Fig_Gabor}
\centering
\end{figure}

In the pioneering paper of Gabor in 1946~\cite{gabor1946theory}, he suggested expanding signals into TF shifted versions of an elementary Gaussian function (pulse). By doing so, the TF plane is partitioned into rectangles of equal sizes, which is then referred to as the \emph{Gabor expansion}. The local content on each TF grid point is called the \emph{Gabor coefficient} or the \emph{expansion coefficient}, which relates to the way of sampling for the underlying signal. Let $T$ and $F$ be the time-shift and frequency-shift parameters. With respect to the choices of $T$ and $F$, there are generally three types of sampling are of interest, namely, the critical sampling for $TF=1$, oversampling for $TF<1$, and undersampling for $TF>1$, as indicated in Fig.~\ref{C4_Fig_Gabor}~\cite{Bolcskei1994Gabor}. Critical sampling is the most commonly used sampling technique. However, it may have poor numerical stability~\cite{Bolcskei1994Gabor}, while the oversampling could have a better numerical stability at the cost of redundant and non-unique Gabor coefficients{\footnote{An example of oversampling signals could be the type of faster-than-Nyquist signals~\cite{General-anderson2013faster,li2020code,li2017reduced,li2020time,li2018superposition,Shuangyang2020codebase}. However, it should be noted that the DoF improvement of faster-than-Nyquist signaling comes from the excess bandwidth of the signaling pulse and a higher transmit symbol rate, instead of the oversampling itself at the receiver side~\cite{General-anderson2013faster,rusek2009constrained,li2021faster}.}}. On the other hand, the Gabor expansion generally does not exist for arbitrary signals in the case of undersampling~\cite{Bolcskei1994Gabor}.

A frame of Gabor type is commonly referred to as the \emph{Weyl-Heisenberg} frame and the corresponding Gabor expansion is a common tool for filterbank analysis~\cite{Bolcskei1994Gabor}. For example, the impulse responses of individual filters in filterbank analysis are simply the modulated versions of the same signaling pulse. Consequently, the resultant spectral decomposition is a ``constant-bandwidth'' decomposition~\cite{Bolcskei1994Gabor}. Some more details regarding these could be found in~\cite{Bolcskei1994Gabor,bolcskei1997discrete}.  In the context of Gabor expansion, the ZT is a commonly used signal transform to obtain the Gabor expansion, which will be introduced in the coming subsection.
\subsection{Definitions of Zak Transform and Discrete Zak Transform}
The ZT is a certain operation that takes a function of one variable as input and produces a function of two variables as output.
Several definitions of ZT and DZT have appeared in the literature. In this thesis, we focus on the ones in~\cite{janssen1988zak,mohammed2021derivation,Bolcskei1994Gabor,bolcskei1997discrete,Mohammed2021DDTVT} as stated below.

\textbf{Definition 3-1} (\emph{Zak Transform~\cite{Bolcskei1994Gabor,janssen1988zak,mohammed2021derivation,bolcskei1997discrete}}):
Let $x\left( t \right)$ be a complex time-continuous function and $T$ be a positive constant. Then, the ZT is defined by
\begin{align}
{{\cal Z}_x}\left( {\tau ,\nu } \right) \buildrel \Delta \over = \sqrt T \sum\limits_{k =  - \infty }^\infty  {x\left( {\tau  + kT} \right){e^{ - j2\pi k\nu T}}} ,
\label{C4_Zak}
\end{align}
for $ -\infty   < \tau  < \infty $ and $ -\infty  < \nu  < \infty$.

\textbf{Definition 3-2} (\emph{Discrete Zak Transform~\cite{bolcskei1997discrete,Bolcskei1994Gabor,janssen1988zak,mohammed2021derivation}}):
Let $x\left[ n \right]$ be a function of integer variable $n$ and $M$ and $N$ be two positive constants. Then, the DZT is defined by
\begin{align}
{\cal D}{{\cal Z}_x}\left[ {l,k} \right] \buildrel \Delta \over = \frac{1}{{\sqrt N }}\sum\limits_{n = 0}^{N - 1} {x\left[ {l + nM} \right]{e^{ - j2\pi \frac{n}{N}k}}} ,
\label{C4_Discrete_Zak}
\end{align}
for $ 0 \le l \le M - 1 $ and $ 0 \le k \le N - 1$.

\begin{figure}
\centering
\includegraphics[width=0.8\textwidth]{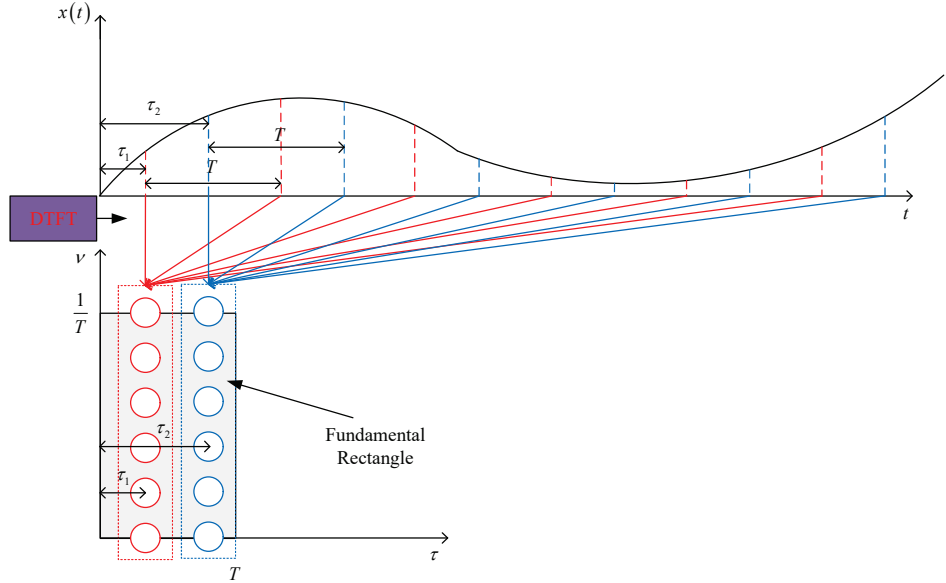}
\caption{A graphical interpretation of the ZT.}
\label{C4_Fig_Zak}
\centering
\end{figure}

As implied by Fig.~\ref{C4_Fig_Zak}, the ZT may be interpreted as the discrete-time Fourier transform (DTFT) of a periodically sampled signal with sample period $T$ and time shift $\tau$.
In the context of ZT, the so-called ``\emph{fundamental rectangle}" is a range of size $\tau  \in \left[ {0,T} \right)$ and $\nu  \in \left[ {0,\frac{1}{T}} \right)$, which is used to sufficiently describe the characteristics of ZT of the signal $x\left( t \right)$~\cite{Bolcskei1994Gabor}. The rationale of this comes from the quasi-periodicity property of the ZT, which will be explained in the coming subsection.

On the other hand, the DZT is the corresponding discrete form (discrete in both delay and Doppler) of ZT with $T=1$, which is defined in the fundamental rectangle. For completeness, we may assume that $x\left[ n \right]$ is a periodic sequence with a period of $MN$ and thus the definition of DZT is also applicable outside the fundamental rectangle{\footnote{It should be noted that such a periodic assumption is usually enabled by adding CP in communication systems. Some of the related discussions can be found in Section~3.2.1.}}~\cite{Bolcskei1994Gabor}. In specific, in some references, the DZT with an $MN$-periodic sequence $x\left[ n \right]$ is called ``\emph{extended DZT}''~\cite{bolcskei1997discrete}.
Other than the ZT and DZT, the discrete-time ZT (discrete in delay and continues in Doppler) is also of importance for some applications~\cite{bolcskei1997discrete}. However, the details of discrete-time ZT are out of the scope of this thesis and we refer the interested readers to~\cite{bolcskei1997discrete} for more details.
\subsection{Properties and Relations of the Zak Transform}
In this subsection, we will summarize some useful properties of ZT based on the results from~\cite{Bolcskei1994Gabor,Hadani2017orthogonal}.

\textbf{Proposition 3-1} (\emph{Quasi-Periodicity~\cite{Bolcskei1994Gabor}}):
The ZT is quasi-periodic along the delay axis with period $T$ and periodic along the Doppler axis with period ${\frac{1}{T}}$, i.e.,
\begin{align}
{{\cal Z}_x}\left( {\tau  + T,\nu } \right) = {e^{j2\pi T\nu }}{{\cal Z}_x}\left( {\tau ,\nu } \right)\label{C4_Zak_quasi_periodicity_delay},
\end{align}
and
\begin{align}
{{\cal Z}_x}\left( {\tau ,\nu  + \frac{1}{T}} \right) = {{\cal Z}_x}\left( {\tau ,\nu } \right)  \label{C4_Zak_periodicity_Doppler}.
\end{align}

\textbf{Proof}: The related proofs can be directly obtained by based on the definition of ZT in Definition 3-1. \hfill $\blacksquare$

As implied by Proposition 3-1, the fundamental rectangle of size $\tau  \in \left[ {0,T} \right)$ and $\nu  \in \left[ {0,\frac{1}{T}} \right)$ is sufficient to describe the characteristics of signal $x\left( t \right)$~\cite{Bolcskei1994Gabor}, because the DD representations of the signal outside of the fundamental rectangle are determined by periodically extending the representations within the fundamental rectangle.

\textbf{Proposition 3-2} (\emph{Shift Properties~\cite{Bolcskei1994Gabor}}):
Given the original signal $x\left( t \right)$, and its time-delayed version ${x_1}\left( t \right) = x\left( {t - {\tau _0}} \right)$ and frequency-shifted version ${x_2}\left( t \right) = x\left( t \right){e^{j2\pi {\nu _0}t}}$, the following holds
\begin{align}
{{\cal Z}_{{x_1}}}\left( {\tau ,\nu } \right) = {{\cal Z}_x}\left( {\tau-\tau_0 ,\nu } \right)\label{C4_Zak_shift_time},
\end{align}
and
\begin{align}
{{\cal Z}_{{x_2}}}\left( {\tau ,\nu } \right) = {e^{j2\pi {\nu _0}\tau }}{{\cal Z}_x}\left( {\tau ,\nu  - {\nu _0}} \right)  \label{C4_Zak_shift_frequency}.
\end{align}

\textbf{Proof}: The related proofs can be directly obtained by based on the definition of ZT in Definition 3-1.
\hfill $\blacksquare$

\textbf{Proposition 3-3} (\emph{Inverse Zak Transform~\cite{Bolcskei1994Gabor}}):
Given a signal $x\left( t \right)$, whose ZT is given by ${{\cal Z}_x}\left( {\tau ,\nu } \right)$, the inverse ZT (IZT) is represented by
\begin{align}
x\left( t \right) = {\cal IZT}\left( {{{\cal Z}_x}\left( {t,\nu } \right)} \right)\buildrel \Delta \over = {\sqrt{T}}\int_0^{\frac{1}{T}} {{{\cal Z}_x}\left( {t ,\nu } \right)} {\rm{d}}\nu  \label{C4_Inverse_ZT}.
\end{align}
\textbf{Proof}:
\begin{align}
{\sqrt{T}}\int_0^{\frac{1}{T}} {{{\cal Z}_x}\left( {t ,\nu } \right)} {\rm{d}}\nu  =& T\int_0^{\frac{1}{T}} {\sum\limits_{k =  - \infty }^\infty  {x\left( {t + kT} \right){e^{ - j2\pi k\nu T}}} } {\rm{d}}\nu  \notag\\=& T\sum\limits_{k =  - \infty }^\infty  {x\left( {t + kT} \right)\int_0^{\frac{1}{T}} {{e^{ - j2\pi k\nu T}}} } {\rm{d}}\nu .
\end{align}
By noticing $\int_0^{\frac{1}{T}} {{e^{ - j2\pi k\nu T}}} {\rm{d}}\nu  = \frac{1}{T}\delta \left[ k \right]$,~\eqref{C4_Inverse_ZT} holds.
\hfill $\blacksquare$

\textbf{Proposition 3-4} (\emph{Connections to Fourier transform~\cite{Bolcskei1994Gabor}}):
Given a signal $x\left( t \right)$, whose ZT and Fourier transform are given by ${{\cal Z}_x}\left( {\tau ,\nu } \right)$ and $X\left( f \right)$, respectively, the following holds
\begin{align}
X\left( f \right) = \frac{1}{{\sqrt T }}\int_0^T {{{\cal Z}_x}\left( {\tau ,f } \right){e^{ - j2\pi f \tau }}} {\rm{d}}\tau  \label{C4_ZT_and_FT}.
\end{align}
\textbf{Proof}:
\begin{align}
&\frac{1}{{\sqrt T }}\int_0^T {{{\cal Z}_x}\left( {\tau ,f} \right){e^{ - j2\pi f\tau }}} {\rm{d}}\tau\notag\\
=&\int_0^T {\sum\limits_{k =  - \infty }^\infty  {x\left( {\tau  + kT} \right){e^{ - j2\pi \left( {\tau  + kT} \right)f}}} } {\rm{d}}\tau
= \int_{ - \infty }^\infty  {x\left( t \right){e^{ - j2\pi ft}}} {\rm{d}}t=X\left( f \right).
\end{align}
\hfill $\blacksquare$

\textbf{Proposition 3-5} (\emph{Multiplication Property~\cite{lampel2021orthogonal}}):
Let $z\left( t \right)=x\left( t \right)y\left( t \right)$ be the modulated signal of the signal $x\left( t \right)$ with respect to $y\left( t \right)$. Then, the ZTs of $z\left( t \right)$, $x\left( t \right)$, and $y\left( t \right)$ satisfy
\begin{align}
{{\cal Z}_z}\left( {\tau ,\nu } \right) = {\sqrt T}\int_0^{\frac{1}{T}} {{{\cal Z}_x}\left( {\tau ,\nu '} \right){{\cal Z}_y}\left( {\tau ,\nu  - \nu '} \right)} d\nu '. \label{C4_ZT_modulation}
\end{align}
\textbf{Proof}:
\begin{align}
{{\cal Z}_z}\left( {\tau ,\nu } \right) &= \sqrt T \sum\limits_{n =  - \infty }^\infty  {x\left( {\tau  + nT} \right)y\left( {\tau  + nT} \right)} {e^{ - j2\pi n\nu T}}\notag\\
&= \sqrt T \sum\limits_{n =  - \infty }^\infty  {\sqrt T\int_0^{\frac{1}{T}} {{{\cal Z}_x}\left( {\tau  + nT,\nu '} \right)} d\nu 'y\left( {\tau  + nT} \right)} {e^{ - j2\pi n\nu T}}\notag\\
&= \sqrt T \sum\limits_{n =  - \infty }^\infty  {\sqrt T\int_0^{\frac{1}{T}} {{{\cal Z}_x}\left( {\tau ,\nu '} \right)} d\nu 'y\left( {\tau  + nT} \right)} {e^{ - j2\pi n\left( {\nu  - \nu '} \right)T}}\notag\\
&= \sqrt T\int_0^{\frac{1}{T}} {{{\cal Z}_x}\left( {\tau ,\nu '} \right)} d\nu '\sqrt T \sum\limits_{n =  - \infty }^\infty  {y\left( {\tau  + nT} \right)} {e^{ - j2\pi n\left( {\nu  - \nu '} \right)T}}\notag\\
&= \sqrt T\int_0^{\frac{1}{T}} {{{\cal Z}_x}\left( {\tau ,\nu '} \right){{\cal Z}_y}\left( {\tau ,\nu  - \nu '} \right)} d\nu '
\end{align}
\hfill $\blacksquare$

\textbf{Proposition 3-6} (\emph{DD Domain Twisted Convolution~\cite{Hadani2017orthogonal}}):
Let ${{\cal Z}_r}\left( {\tau ,\nu } \right)$ be the ZT of the received signal $r\left( t \right)$ and ${{\cal Z}_s}\left( {\tau ,\nu } \right)$ be the ZT of the transmitted signal $s\left( t \right)$. Then, with the DD domain channel response ${h_{{\rm{DD}}}}\left( {\tau ,\nu } \right)$, the DD domain twisted convolution can be derived
according to~\eqref{S4_DD_channel_delay_first_Doppler_second}
and~\eqref{S4_DD_channel_Doppler_first_delay_second} by,
\begin{align}
{{\cal Z}_r}\left( {\tau ,\nu } \right) = \int_{ - \infty }^\infty  {\int_{ - \infty }^\infty  {{{\cal Z}_s}\left( {\tau  - \tau ',\nu  - \nu '} \right)} } {h_{{\rm{DD}}}}\left( {\tau ',\nu '} \right){e^{j2\pi \nu '\left(\tau-\tau '\right) }}{\rm{d}}\tau '{\rm{d}}\nu '.
\label{C4_DD_twisted_conv_delay_first_Doppler_second}
\end{align}
or
\begin{align}
{{\cal Z}_r}\left( {\tau ,\nu } \right) = \int_{ - \infty }^\infty  {\int_{ - \infty }^\infty  {{{\cal Z}_s}\left( {\tau  - \tau ',\nu  - \nu '} \right)} } {h_{{\rm{DD}}}}\left( {\tau ',\nu '} \right){e^{j2\pi \nu '\tau }}{\rm{d}}\tau '{\rm{d}}\nu '.
\label{C4_DD_twisted_conv_Doppler_first_delay_second}
\end{align}

\textbf{Proof}: According to~\eqref{C4_Zak} and~\eqref{S4_DD_channel_delay_first_Doppler_second}, we have
\begin{align}
{{\cal Z}_r}\left( {\tau ,\nu } \right) &= \sqrt T \sum\limits_{k =  - \infty }^\infty  {r\left( {\tau  + kT} \right){e^{ - j2\pi k\nu T}}} \notag\\
&= \sqrt T \sum\limits_{k =  - \infty }^\infty  {\int_{ - \infty }^\infty  {\int_{ - \infty }^\infty  {{h_{{\rm{DD}}}}\left( {\tau ',\nu '} \right){e^{j2\pi \nu '\left( {\tau  + kT - \tau '} \right)}}s\left( {\tau  + kT - \tau '} \right){e^{ - j2\pi k\nu T}}} } {\rm{d}}\tau '{\rm{d}}\nu '}  \notag\\
& = \sqrt T \int_{ - \infty }^\infty  {\int_{ - \infty }^\infty  {\sum\limits_{k =  - \infty }^\infty  {s\left( {\tau  - \tau ' + kT} \right)} } } {e^{ - j2\pi k\left( {\nu  - \nu '} \right)T}}{h_{{\rm{DD}}}}\left( {\tau ',\nu '} \right){e^{j2\pi \nu '\left( {\tau  - \tau '} \right)}}{\rm{d}}\tau '{\rm{d}}\nu ' \notag\\
& = \int_{ - \infty }^\infty  {\int_{ - \infty }^\infty  {{{\cal Z}_s}\left( {\tau  - \tau ',\nu  - \nu '} \right)} } {h_{{\rm{DD}}}}\left( {\tau ',\nu '} \right){e^{j2\pi \nu '\left( {\tau  - \tau '} \right)}}{\rm{d}}\tau '{\rm{d}}\nu '.
\end{align}
On the other hand, based on~\eqref{C4_Zak} and~\eqref{S4_DD_channel_delay_first_Doppler_second}, we have
\begin{align}
{{\cal Z}_r}\left( {\tau ,\nu } \right) &= \sqrt T \sum\limits_{k =  - \infty }^\infty  {r\left( {\tau  + kT} \right){e^{ - j2\pi k\nu T}}} \notag\\
&= \sqrt T \sum\limits_{k =  - \infty }^\infty  {\int_{ - \infty }^\infty  {\int_{ - \infty }^\infty  {{h_{{\rm{DD}}}}\left( {\tau ',\nu '} \right){e^{j2\pi \nu '\left( {\tau  + kT } \right)}}s\left( {\tau  + kT - \tau '} \right){e^{ - j2\pi k\nu T}}} } {\rm{d}}\tau '{\rm{d}}\nu '}  \notag\\
& = \sqrt T \int_{ - \infty }^\infty  {\int_{ - \infty }^\infty  {\sum\limits_{k =  - \infty }^\infty  {s\left( {\tau  - \tau ' + kT} \right)} } } {e^{ - j2\pi k\left( {\nu  - \nu '} \right)T}}{h_{{\rm{DD}}}}\left( {\tau ',\nu '} \right){e^{j2\pi \nu '{\tau }}}{\rm{d}}\tau '{\rm{d}}\nu ' \notag\\
& = \int_{ - \infty }^\infty  {\int_{ - \infty }^\infty  {{{\cal Z}_s}\left( {\tau  - \tau ',\nu  - \nu '} \right)} }
{h_{{\rm{DD}}}}\left( {\tau ',\nu '} \right){e^{j2\pi \nu '{\tau}}}{\rm{d}}\tau '{\rm{d}}\nu '.
\end{align}
\hfill $\blacksquare$

Note that the two interpretations of the twisted convolution properties of the ZT align with the two interpretations of the DD domain channel input-output relationships in~\eqref{S4_DD_channel_delay_first_Doppler_second}
and~\eqref{S4_DD_channel_Doppler_first_delay_second}, where the channel performs the delay operation and the Doppler operation in different orders.

%

\subsection{Properties and Relations of the Discrete Zak Transform}
In this subsection, we will summarize some useful properties of DZT based on the results from~\cite{bolcskei1997discrete,Bolcskei1994Gabor}, where we assume that $x\left[ n \right]$ is an $MN$-periodic sequence for the completeness of definition. As the following propositions are the discrete representations of the corresponding propositions of ZT, we omit some of the related proofs here and refer the interested readers to~\cite{bolcskei1997discrete,Bolcskei1994Gabor} for more details.

\textbf{Proposition 3-7} (\emph{Quasi-Periodicity~\cite{bolcskei1997discrete}}):
The DZT is quasi-periodic along the delay axis with period $M$ and periodic along the Doppler axis with period $N$, i.e.,
\begin{align}
{\cal D}{\cal Z}_x\left[ {l + M,k} \right] = {e^{j2\pi \frac{k}{N}}}{\cal D}{\cal Z}_x\left[ {l,k} \right]\label{C4_discrete_Zak_quasi_periodicity_delay},
\end{align}
and
\begin{align}
{\cal D}{\cal Z}_x\left[ {l,k+N} \right] = {\cal D}{\cal Z}_x\left[ {l,k} \right] \label{C4_discrete_Zak_periodicity_Doppler}.
\end{align}

\textbf{Proposition 3-8} (\emph{Shift Properties}):
Given the $MN$-periodic $x\left[ n \right]$, and its time-delayed version ${x_1}\left[ n \right] = x\left[ {n - {l_0}} \right]$ and frequency-shifted version ${x_2}\left[ n \right] = x\left[ n \right]{e^{j2\pi \frac{{{k_0}}}{{MN}}n}}$, the following holds
\begin{align}
{{\cal DZ}_{{x_1}}}\left[ {l ,k } \right] = {{\cal DZ}_x}\left[ {l-l_0 ,k } \right]\label{C4_discrete_Zak_shift_time},
\end{align}
and
\begin{align}
{{\cal DZ}_{{x_2}}}\left[ {l,k} \right] = {e^{j2\pi \frac{{{k_0}l}}{{MN}}}}{{\cal DZ}_x}\left[ {l,k - {k_0}} \right]  \label{C4_discrete_Zak_shift_frequency}.
\end{align}

\textbf{Proposition 3-9} (\emph{Inverse Discrete Zak Transform~\cite{bolcskei1997discrete}}):
Given the $MN$-periodic sequence $x\left[ n \right]$, whose DZT is given by ${{\cal DZ}_x}\left[ {l ,k} \right]$, the inverse DZT (IDZT) is represented by
\begin{align}
x\left[ n \right] = {\cal IDZT}\left( {{{\cal DZ}_x}\left[ {n,k } \right]} \right) \buildrel \Delta \over = \frac{1}{{\sqrt N }}\sum\limits_{k = 0}^{N - 1} {{{\cal DZ}_x}\left[ {n,k} \right]}   \label{C4_Inverse_discrete_ZT}.
\end{align}
\textbf{Proof}:
\begin{align}
\frac{1}{{\sqrt N }}\sum\limits_{k = 0}^{N - 1} {{{\cal DZ}_x}\left[ {n,k} \right]}
& = \frac{1}{N}\sum\limits_{k = 0}^{N - 1} {\sum\limits_{l = 0}^{N - 1} {x\left[ {n + lM} \right]{e^{ - j2\pi \frac{l}{N}k}}} }\notag\\
&= \frac{1}{N}\sum\limits_{l = 0}^{N - 1} {x\left[ {n + lM} \right]\sum\limits_{k = 0}^{N - 1} {{e^{ - j2\pi \frac{l}{N}k}}} } \notag\\
&= \frac{1}{N}\sum\limits_{l = 0}^{N - 1} {x\left[ {n + lM} \right]N\sum\limits_{k =  - \infty }^\infty  {\delta \left[ {l - kN} \right]} } \notag\\
&=\sum\limits_{l = 0}^{N - 1} {\sum\limits_{k =  - \infty }^\infty  {x\left[ {n + kMN} \right]} }  = x\left[ n \right].
\end{align}
\hfill $\blacksquare$

\textbf{Proposition 3-10} (\emph{Connections to Discrete Fourier transform~\cite{bolcskei1997discrete}}):
Given a $MN$-periodic $x\left[ n \right]$, whose DZT and discrete Fourier transform (DFT) are given by ${{\cal DZ}_x}\left[ {l ,k} \right]$ and $X\left[ k \right]$, respectively, the following holds
\begin{align}
X\left[ k \right] = \sqrt N \sum\limits_{l = 0}^{M - 1} {{{\cal DZ}_x}\left[ {l,k} \right]} {e^{ - j2\pi \frac{{kl}}{N}}} \label{C4_DZT_and_FT},
\end{align}
for $0 \le k \le K-1$.

\textbf{Proof}:
The proof relates to the decimation-in-time fast Fourier transform (FFT) algorithm and the Cooley-Tukey algorithm~\cite{bolcskei1997discrete}. Some details could be found in~\cite{bolcskei1997discrete} and~\cite{Bolcskei1994Gabor}. \hfill $\blacksquare$

%

\textbf{Proposition 3-11} (\emph{Multiplication Property~\cite{lampel2021orthogonal}}):
Let $z\left[ n \right]=x\left[ n \right]y\left[ n \right]$ be the modulated signal of the $MN$-periodic sequence $x\left[ n \right]$ with respect to $y\left[ n \right]$. Then, the DZTs of $z\left[ n \right]$, $x\left[ n \right]$, and $y\left[ n \right]$ satisfy
\begin{align}
{\cal D}{{\cal Z}_z}\left[ {l,k} \right] = \frac{1}{{\sqrt N }}\sum\limits_{k' = 0}^{N - 1} {{\cal D}{{\cal Z}_x}\left[ {l,k'} \right]{\cal D}{{\cal Z}_y}\left[ {l,k - k'} \right]} . \label{C4_DZT_modulation_property}
\end{align}

\textbf{Proof}:
\begin{align}
{\cal D}{{\cal Z}_z}\left[ {l,k} \right] =& \frac{1}{{\sqrt N }}\sum\limits_{n = 0}^{N - 1} {x\left[ {l + nM} \right]} y\left[ {l + nM} \right]{e^{ - j2\pi \frac{n}{N}k}}\notag\\
=& \frac{1}{N}\sum\limits_{n = 0}^{N - 1} {\sum\limits_{k' = 0}^{N - 1} {{\cal D}{{\cal Z}_x}\left[ {l + nM,k'} \right]} } y\left[ {l + nM} \right]{e^{ - j2\pi \frac{n}{N}k}}\notag\\
=& \frac{1}{N}\sum\limits_{n = 0}^{N - 1} {\sum\limits_{k' = 0}^{N - 1} {{\cal D}{{\cal Z}_x}\left[ {l,k'} \right]} } y\left[ {l + nM} \right]{e^{ - j2\pi \frac{n}{N}\left( {k - k'} \right)}}\notag\\
=& \frac{1}{{\sqrt N }}\sum\limits_{k' = 0}^{N - 1} {{\cal D}{{\cal Z}_x}\left[ {l,k'} \right]{\cal D}{{\cal Z}_y}\left[ {l,k - k'} \right]}.
\end{align}
\hfill $\blacksquare$

\textbf{Proposition 3-12} (\emph{Alternative Definition}):
Given the $MN$-periodic sequence $x\left[ n \right]$, the DZT ${{\cal DZ}_x}\left[ {l ,k} \right]$ given in~\eqref{C4_Discrete_Zak} can be alternatively defined by
\begin{align}
{\cal D}{{\cal Z}_x}\left[ {l,k} \right] \buildrel \Delta \over = \frac{1}{{\sqrt N }}\sum\limits_{n = 0}^{N - 1} {x\left[ {l - nM} \right]{e^{  j2\pi \frac{n}{N}k}}} ,
\label{C4_Discrete_Zak_alternative}
\end{align}
for $ 0 \le l \le M - 1 $ and $ 0 \le k \le N - 1$.

\textbf{Proof}:
\begin{align}
\frac{1}{{\sqrt N }}\sum\limits_{n = 0}^{N - 1} {x\left[ {l - nM} \right]{e^{j2\pi \frac{n}{N}k}}} & = \frac{1}{{\sqrt N }}\sum\limits_{n = 0}^{N - 1} {x\left[ {l - nM + NM} \right]{e^{j2\pi \frac{n}{N}k}}} \notag\\
& = \frac{1}{{\sqrt N }}\sum\limits_{n = 0}^{N - 1} {x\left[ {l + \left( {N - n} \right)M} \right]{e^{ - j2\pi \frac{{N - n}}{N}k}}} \notag\\
&= \frac{1}{{\sqrt N }}\sum\limits_{n' = 0}^{N - 1} {x\left[ {l + n'M} \right]{e^{ - j2\pi \frac{{n'}}{N}k}}} ={\cal D}{{\cal Z}_x}\left[ {l,k} \right].
\end{align}
\hfill $\blacksquare$
\section{Concepts of OTFS Modulation: DZT-based OTFS}
In this section, we will discuss the related contexts of DZT-based OTFS.
In comparison to OFDM-based OTFS, the key difference of DZT-based OTFS lies in the way of cyclic prefix (CP) insertion.
Although in the pioneering papers from Hadani \emph{et. al.}, the details of how to append CPs in the frame have been omitted, the CP insertion play a fundamental role for distinguishing different types of OTFS. Specifically, different ways of CP insertion may result in different \emph{effective channel matrix} structures for OTFS transmissions. Therefore, before introducing the details of DZT-based OTFS, we will first review the reduced-CP frame structure that was, to the best of knowledge of the author, introduced by Raviteja \emph{et. al.} in~\cite{Raviteja2019practical}.

For notational clarity, we henceforth use $M$ and $N$ to denote the number of delay bins/sub-carriers and the number of Doppler bins/time slots for the related discussions.
Furthermore, following the well-known definitions in OFDM transmissions, we denote by $\Delta f$ the subcarrier spacing and $T$ the time slot duration. It should be noted that as both sub-carrier spacing and time slot are TF domain definitions, we will also use the corresponding definitions in the DD domain interchangeably, i.e., Doppler period and delay period, for the related descriptions in the DD domain.
We restrict ourselves to consider the case of \emph{critical sampling}, i.e., $T\Delta f=1$. Some discussions for OTFS transmissions that are not critical sampled could be found in~\cite{Pfadler2020pulse}.
\subsection{Reduced-CP Frame Structure}
In practical multi-carrier transmissions, only limited resources can be used to convey information. In this case, CPs are of importance for the signal transmission, because the asymptotical equivalence between Toeplitz matrix and circulant matrix does not hold with finite frame lengths~\cite{zhu2017asymptotic}. Therefore, the insertion of CP is necessary to make sure that the corresponding Fourier-based kernels could diagonalize/sparsify the underlying channel matrix for related receiver designs.

\begin{figure}
\centering
\includegraphics[width=0.8\textwidth]{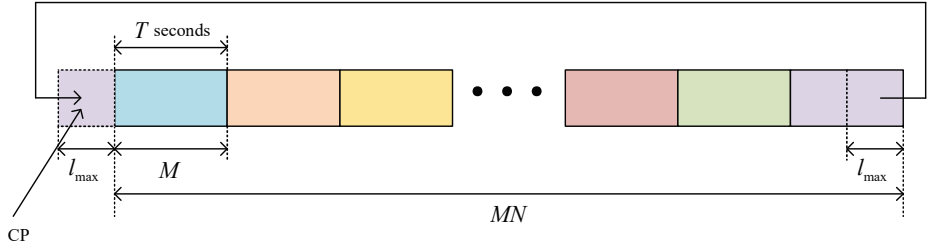}
\caption{A diagram of the CP insertion for DZT-based OTFS transmissions.}
\label{C4_CP_DZT_OTFS}
\centering
\end{figure}

To preserve the properties of DZT, it is required that the transmitted sequence is $MN$-periodic in TD domain. Given ${\tau _{\max }}$ the maximum delay of the underlying channel, it can be shown that the
TD domain channel response is of length $l_{\max}+1$ according to the symbol rate, where $l_{\max}$ is the \emph{maximum delay index} of the channel given by ${l_{\max }} \buildrel \Delta \over = \left\lceil {\frac{{M{\tau _{\max }}}}{T}} \right\rceil $. With such a channel and a transmitted sequence containing $MN$ samples, the total number of samples in the receive sequence is $MN+l_{\max}$. This implies that the
length of CP should be no smaller than the maximum delay index ${l_{\max }}$.
A brief diagram of the considered CP insertion is illustrated in Fig.~\ref{C4_CP_DZT_OTFS}. As implied by the figure, only $l_{\max}$ samples copied from the end of the frame are  inserted at the beginning of the frame, yielding a total frame length of $l_{\max}+MN$.
Such a CP arrangement ensures that the sequence is $MN$-periodic from the range $-l_{\max} \le n \le MN-1$ after CP removal at the receiver, which is sufficient for maintaining the properties of DZT over a channel with a maximum delay ${\tau _{\max }}$. Correspondingly, this reduced-CP frame structure maintains the elegant structures of the corresponding representations in the DD domain (Zak domain) even after possible delay and Doppler shifts due to the channel response, where the DD domain (Zak domain) symbols in the fundamental rectangle could fully characterize the DD domain received signal. In order words, the reduced-CP frame maintains the \emph{symmetric} property of the ZT for an arbitrary channel response, because the symbols in the fundamental rectangle can always fully represent the signals under any possible delay and Doppler shifts of interests, e.g., $0 \le {\tau _p} \le T$ and $0 \le {\nu _p} \le \Delta f$, for $1 \le p\le P$.
Furthermore, it is shown in~\cite{Raviteja2019practical} that adding CP in such a way allows the effective TD domain channel matrix to have a cyclic structure, yielding some appealing properties for the effective TD domain channel matrix. Some more details regarding this issue could be found in Section~3.3.

An interesting discussion at this point could be the variations of this reduced-CP frame structure. For example, it is reported in~\cite{Raviteja2018interference,pandey2021low} that padding zeros instead of adding CP could result in a power loss for the effective DD domain channel. As a final remark to the reduced-CP frame structure, it is necessary to point out that it allows a much smaller signaling overhead compared to CP insertion for conventional OFDM modulation, where one CP is appended at the beginning of each OFDM symbol in order to eliminate the influence of ISI.

\subsection{Descriptions of DZT-based OTFS: Transmitter Side}
There are multiple ways to implement the DZT-based OTFS modulation and perhaps the most well-known one is from the first paper of OTFS by Hadani \emph{et. al.}~\cite{Hadani2017orthogonal}.
In this subsection, we will build a general understanding of the transmitter design for DZT-based OTFS, where the original OTFS transmitter from~\cite{Hadani2017orthogonal} and more involved understandings from~\cite{Raviteja2018interference} will be discussed.

\begin{figure}
\centering
\includegraphics[width=0.8\textwidth]{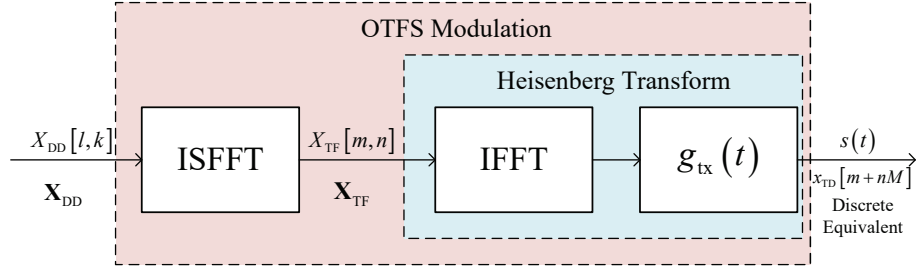}
\caption{A transmitter diagram of the DZT-based OTFS modulation according to~\cite{Hadani2017orthogonal,Raviteja2018interference}.}
\label{C4_Transmitter_diagram_Original}
\centering
\end{figure}

Without loss of generality, let us consider the transmitter diagram of the DZT-based OTFS modulation~\cite{Hadani2017orthogonal,Raviteja2018interference} as shown in Fig.~\ref{C4_Transmitter_diagram_Original}. Let ${\bf{x}}_{\rm DD} \in {{\mathbb{A}}^{M N}}$ be the DD domain information symbol vector of length $MN$ that is to be transmitted, where ${\mathbb{A}}$ denotes an energy normalized constellation set.
In particular, the information symbol vector ${\bf{x}}_{\rm DD}$ can be arranged as a 2D information symbol matrix ${\bf{X}}_{\rm DD} \in {{\mathbb{A}}^{M \times N}}$, i.e., ${\bf{x}}_{\rm DD} \buildrel \Delta \over = \textrm{vec}\left( {\bf{X}}_{\rm DD} \right)$, and the $(l,k)$-th element of ${\bf{X}}_{\rm DD}$, ${X_{\rm DD}\left[ {l,k} \right]}$, is the modulated pulse at the $k$-th Doppler and $l$-th delay grid point~\cite{Hadani2017orthogonal}, for $0 \le k \le N-1,0 \le l \le M-1$.
As indicated by Fig.~\ref{C4_Transmitter_diagram_Original}, the TF domain transmitted symbol $X_{\rm TF}\left[ {m,n} \right], 0 \le m \le M-1, 0 \le n \le N-1$ is obtained according to ${\bf{X}}_{\rm DD}$ via the ISFFT~\cite{Hadani2017orthogonal}, i.e.,
\begin{equation}
X_{\rm TF}\left[ {m,n} \right] = \frac{1}{{\sqrt {MN} }}\sum\limits_{k = 0}^{N - 1} {\sum\limits_{l = 0}^{M - 1} {X_{\rm DD}\left[ {l,k} \right]} } {e^{j2\pi \left( {\frac{{nk}}{N} - \frac{{ml}}{M}} \right)}}  . \label{C4_ISFFT}
\end{equation}
\begin{figure}
\centering
\includegraphics[width=0.7\textwidth]{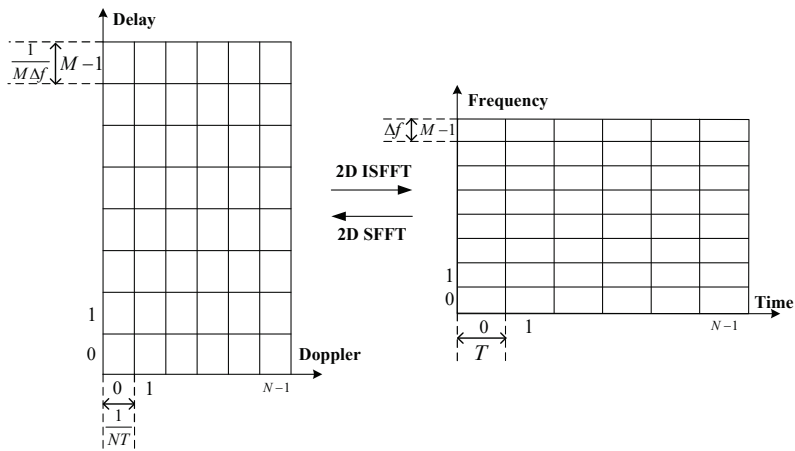}
\caption{The diagram of the transformation between the DD domain and the TF domain.}
\label{C4_Grid}
\centering
\end{figure}

A brief diagram regarding the DD and TF domain transformation is shown in Fig.~\ref{C4_Grid}, where $\Delta f$ is the frequency spacing between adjacent sub-carriers and $T = {1 \mathord{\left/
 {\vphantom {1 {\Delta f}}} \right.
 \kern-\nulldelimiterspace} {\Delta f}}$ is the corresponding TF domain time slot duration.
On the other hand, the sampling time $1/(M \Delta  f)$ and sampling frequency $1/(N T)$ are referred to as the \emph{delay resolution} and the \emph{Doppler resolution} of the DD grid, respectively \cite{Raviteja2018interference}, which indicate how precise the acquisition of the channel delay and Doppler can be for the underlying OTFS system. Some details on the delay and Doppler resolutions will be discussed in Section~3.2.4 .

The transmitted OTFS signal $s\left( t \right)$ can be obtained by performing the Heisenberg transform~\cite{Hadani2017orthogonal} to ${\bf{X}}_{\rm TF}$ with the transmitter shaping pulse $g_{{\rm{tx}}}(t)$.
In particular, the Heisenberg transform could be interpreted as a multi-carrier modulator, and a popular choice for implementing the Heisenberg transform is to apply the conventional OFDM modulator~\cite{Hadani2017orthogonal,Raviteja2018interference,Zhiqiang_magzine}.
In specific, according to the conventional OFDM modulation, the Heisenberg transform could be implemented by an inverse fast Fourier transform (IFFT) module and transmit pulse shaping, in which case the resultant transmitted OTFS signal $s\left( t \right)$ is written by
\begin{equation}
s\left( t \right) = \sum\limits_{n = 0}^{N - 1} {\sum\limits_{m = 0}^{M - 1} {X_{\rm TF}\left[ {m,n} \right]{g_{{\rm{tx}}}}\left( {t - nT} \right){e^{j2\pi m\Delta f\left( {t - nT} \right)}}} }.\label{C4_OTFS_signal}
\end{equation}
Based on~\eqref{C4_OTFS_signal}, it is also convenient to define TD domain OTFS signal samples ${\bf{s}}$ of length $MN$ after pulse shaping, which is in fact the Nyquist rate sampled version of~\eqref{C4_OTFS_signal} with a sample period $T/M$, i.e.,
\begin{align}
{s}\left[ {m + nM} \right] \buildrel \Delta \over = &s\left( {nT + \frac{m}{M}T} \right) \notag\\
=& \sum\limits_{n' = 0}^{N - 1} {\sum\limits_{m' = 0}^{M - 1} {{X_{{\rm{TF}}}}\left[ {n',m'} \right]{g_{{\rm{tx}}}}\left( {\left( {n - n'} \right)T + \frac{m}{M}T} \right){e^{j2\pi \left( {n - n'} \right)m'}}{e^{j2\pi \frac{{mm'}}{M}}}} }. \label{C4_OTFS_signal_sampled}
\end{align}
In particular, according to~\eqref{C4_OTFS_signal_sampled}, we define the \emph{periodically extended} sampled pulse shaping $g\left[ k \right]$, for $k \in {\mathbb Z}$, by
\begin{align}
g\left[ k \right] \buildrel \Delta \over = {g_{{\rm{tx}}}}\left( {\frac{{{{\left[ k \right]}_{MN}}}}{M}T} \right).\label{C4_sampled_shaping_pulse}
\end{align}
Then, by substituting~\eqref{C4_ISFFT} and~\eqref{C4_sampled_shaping_pulse} into~\eqref{C4_OTFS_signal_sampled}, and noticing that ${e^{j2\pi \left( {n - n'} \right)m'}}=1$, we have
\begin{align}
{s}\left[ {m + nM} \right] &= \frac{1}{{\sqrt {MN} }}\sum\limits_{k = 0}^{N - 1} {\sum\limits_{l = 0}^{M - 1} {{X_{{\rm{DD}}}}\left[ {l,k} \right]} } \sum\limits_{n' = 0}^{N - 1} {\sum\limits_{m' = 0}^{M - 1} {{e^{j2\pi \left( {\frac{{n'k}}{N} - \frac{{m'l}}{M}} \right)}}} } g\left[ {\left( {n - n'} \right)M + m} \right]{e^{j2\pi \frac{{mm'}}{M}}}\notag\\
&= \frac{1}{{\sqrt {MN} }}\sum\limits_{k = 0}^{N - 1} {\sum\limits_{l = 0}^{M - 1} {{X_{{\rm{DD}}}}\left[ {l,k} \right]} } \sum\limits_{n' = 0}^{N - 1} {g\left[ {\left( {n - n'} \right)M + m} \right]} {e^{j2\pi \frac{{n'k}}{N}}}\sum\limits_{m' = 0}^{M - 1} {{e^{j2\pi \frac{m'}{M}\left( {m - l} \right)}}} .\label{C4_s_TD_der1}
\end{align}
Noticing $\sum\limits_{m' = 0}^{M - 1} {{e^{j2\pi \frac{m'}{M}\left( {m - l} \right)}}}  = M\sum\limits_{m' =  - \infty }^\infty  {\delta \left[ {m - l - m'M} \right]} $, we can further expand~\eqref{C4_s_TD_der1} by
\begin{align}
{s}\left[ {m + nM} \right] = \sqrt {\frac{M}{N}} \sum\limits_{k = 0}^{N - 1} {{X_{{\rm{DD}}}}\left[ {m,k} \right]} \sum\limits_{n' = 0}^{N - 1} {g\left[ {\left( {n - n'} \right)M + m} \right]} {e^{j2\pi \frac{{n'k}}{N}}}. \label{C4_s_TD_der2_old}
\end{align}
In fact,~\eqref{C4_s_TD_der2_old} has a direct connection to the IDZT defined in~\eqref{C4_Inverse_discrete_ZT}. In particular, we can arrange~\eqref{C4_s_TD_der2_old} as
\begin{align}
{s}\left[ {m + nM} \right] = \sqrt {M} \sum\limits_{k = 0}^{N - 1} {{X_{{\rm{DD}}}}\left[ {m,k} \right]}  {\frac{1}{\sqrt N}}\sum\limits_{n' = 0}^{N - 1} {g\left[ {nM + m - n'M} \right]} {e^{j2\pi \frac{{n'k}}{N}}}. \label{C4_s_TD_der2}
\end{align}
According to the alternative definition of DZT in~\eqref{C4_Discrete_Zak_alternative}, we have
\begin{align}
{s}\left[ {m + nM} \right] &= \sqrt M \sum\limits_{k = 0}^{N - 1} {{X_{{\rm{DD}}}}\left[ {m,k} \right]} {{\cal DZ}_g}\left[ {nM + m,k} \right] \label{C4_s_TD_der3} \\
& = \sqrt M \sum\limits_{k = 0}^{N - 1} {{X_{{\rm{DD}}}}\left[ {m,k} \right]} {{\cal DZ}_g}\left[ {m,k} \right]{e^{j2\pi \frac{n}{N}k}},
\label{C4_s_TD_der4}
\end{align}
where the derivation from~\eqref{C4_s_TD_der3} to~\eqref{C4_s_TD_der4} is due to the quasi-periodicity shown in~\eqref{C4_discrete_Zak_quasi_periodicity_delay}.
It is interesting to see from~\eqref{C4_s_TD_der4} that the TD domain OTFS signal has a direct connection to the DD domain information symbol matrix ${\bf{X}}_{\rm DD}$ and the corresponding DD domain expression of the shaping pulse ${{\cal DZ}_g}$.
In particular, let us apply the DZT to~\eqref{C4_s_TD_der4}, yielding
\begin{align}
{\cal D}{{\cal Z}_{{s}}}\left[ {l,k} \right] &= \frac{1}{{\sqrt N }}\sum\limits_{n = 0}^{N - 1} {{s}\left[ {l + nM} \right]{e^{ - j2\pi \frac{n}{N}k}}} \notag\\
&= \sqrt {\frac{M}{N}} \sum\limits_{n = 0}^{N - 1} {\sum\limits_{k' = 0}^{N - 1} {{X_{{\rm{DD}}}}\left[ {l,k'} \right]{\cal D}{{\cal Z}_g}\left[ {l,k'} \right]{e^{j2\pi \frac{n}{N}k'}}{e^{ - j2\pi \frac{n}{N}k}}} } \notag\\
&= \sqrt {\frac{M}{N}} \sum\limits_{k' = 0}^{N - 1} {{X_{{\rm{DD}}}}\left[ {l,k'} \right]{\cal D}{{\cal Z}_g}\left[ {l,k'} \right]\sum\limits_{n = 0}^{N - 1} {{e^{j2\pi \frac{n}{N}\left( {k' - k} \right)}}} } \notag\\
&= \sqrt {\frac{M}{N}} \sum\limits_{k' = 0}^{N - 1} {{X_{{\rm{DD}}}}\left[ {l,k'} \right]{\cal D}{{\cal Z}_g}\left[ {l,k'} \right]N\sum\limits_{m =  - \infty }^\infty  {\delta \left( {k' - k + mN} \right)} } \notag\\
&= \sqrt {MN} {X_{{\rm{DD}}}}\left[ {l,k} \right]{\cal D}{{\cal Z}_g}\left[ {l,k} \right].\label{C4_s_TD_DZT}
\end{align}
It is not hard to notice from~\eqref{C4_s_TD_DZT} that pulse shaping has certain effects on the equivalent DD domain expression of the OTFS signal. More importantly, the shaping pulse has a multiplication relationship with the DD domain information symbols, which is quite different from the conventional understanding of pulse shaping based on applying convolution.

A popular shaping pulse is the rectangular pulse, which has been widely applied in the related OTFS literature, because it can enable a simple and straightforward input-output relationship~\cite{Raviteja2019practical}. Based on~\eqref{C4_sampled_shaping_pulse}, the samples of the rectangular pulse can be defined by
\begin{align}
g\left[ k \right] = \left\{ \begin{array}{l}
\frac{1}{{\sqrt M }},\quad\quad\quad\! 0 \le {\left[ k \right]_{MN}} \le M - 1\\
0,\quad\quad\quad\quad{\rm{else}}.
\end{array} \right. \label{C4_Rect_pulse}
\end{align}
It is useful to specifically define the ${\bf{s}}$ with rectangular pulse by ${\bf{x}}_{\rm TD}$, which is commonly referred to as the TD domain transmitted symbol vector in the related literature, because it can demonstrate the characteristics of the OTFS signal in an insightful way. By substituting~\eqref{C4_Rect_pulse} into~\eqref{C4_s_TD_der1}, we have
\begin{align}
{x_{{\rm{TD}}}}\left[ {m + nM} \right]= \frac{1}{{\sqrt N }}\sum\limits_{k = 0}^{N - 1} {{X_{{\rm{DD}}}}\left[ {m,k} \right]{e^{j2\pi \frac{{nk}}{N}}}}. \label{C4_OTFS_TD_symbol}
\end{align}
The correctness of~\eqref{C4_OTFS_TD_symbol} can also be verified based on~\eqref{C4_s_TD_der4}. With the rectangular shaping pulse in~\eqref{C4_Rect_pulse}, the ${{\cal DZ}_g}$ can be obtained by
\begin{align}
{{\cal DZ}_g}\left[ {l,k} \right] = \frac{1}{{\sqrt {MN} }},\label{C4_DD_rect}
\end{align}
for $0 \le l \le M$ and $0 \le k \le N$.
By substituting~\eqref{C4_DD_rect} into~\eqref{C4_s_TD_der4}, we can also obtain~\eqref{C4_OTFS_TD_symbol}.

\begin{figure}
\centering
\includegraphics[width=0.8\textwidth]{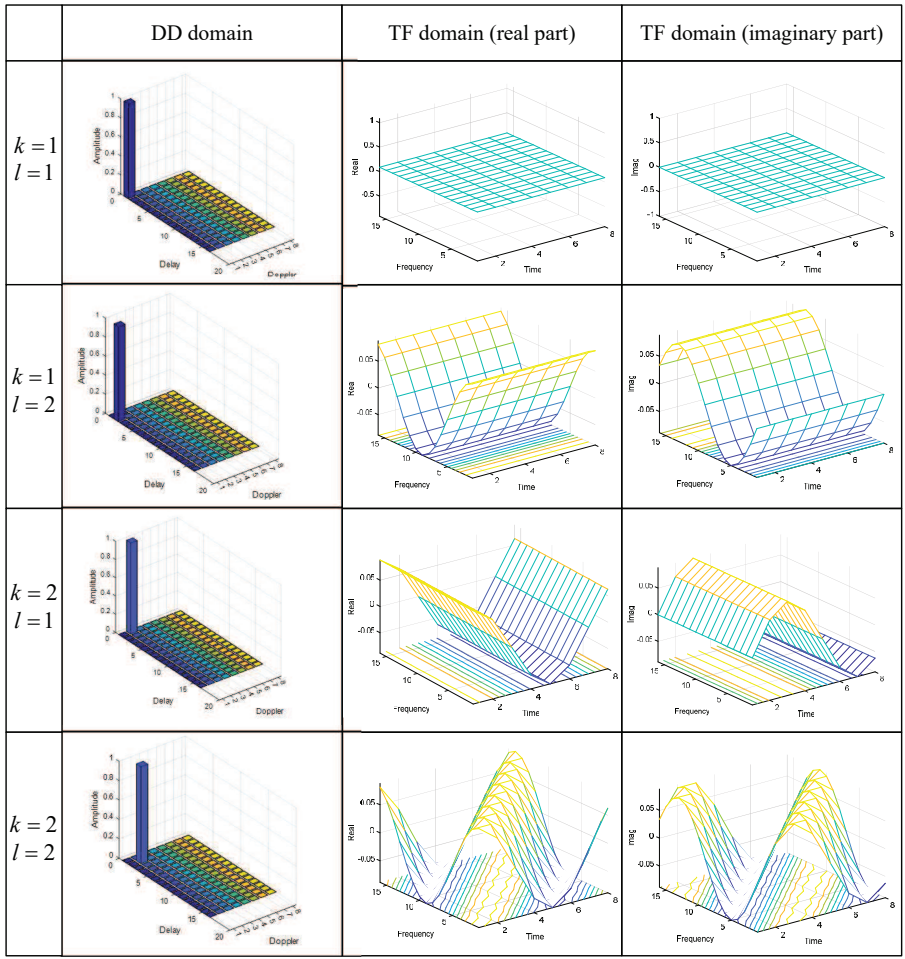}
\caption{An illustration of DD domain symbols and their corresponding TF domain OTFS symbols.}
\label{C4_OTFS_signal_TF}
\centering
\end{figure}

Next, we will demonstrate the characteristics of OTFS modulation with rectangular pulse in different domains based on~\eqref{C4_ISFFT},~\eqref{C4_OTFS_signal}, and~\eqref{C4_OTFS_TD_symbol}, where the connections among the DD domain symbols, TF domain symbols, and TD domain OTFS signals will be illustrated via simulations{\footnote{Note that the related figures are plotted by assuming $1\le k\le N$ and $1\le l\le M$, which is slightly different from the calculations in~\eqref{C4_ISFFT},~\eqref{C4_OTFS_signal}, and~\eqref{C4_OTFS_TD_symbol}. However, it should be noted that this setting will not change the overall characteristics of OTFS modulation. }}.
We demonstrate the TF domain OTFS symbols with corresponding DD domain information symbols in Fig.~\ref{C4_OTFS_signal_TF}. As shown in Fig.~\ref{C4_OTFS_signal_TF}, different DD domain symbol positions will result in different values of the TF domain symbols. In particular, we observe that each DD domain symbol will spread onto the whole TF plane before transmission. This observation implies that the DD domain symbol principally experiences the whole TF domain channel fluctuation, leading to a potential of achieving full channel diversity~\cite{Zhiqiang_magzine,Li2020on,li2021performance_analysis,Raviteja2019effective,Surabhi2019on}. Furthermore, we also observe that the values of TF domain symbols change in a periodic fashion with respect to the underlying values of $k$ and $l$, which indicates that the spreading from the DD domain to the TF domain could be interpreted as a code-division multiple access (CDMA)-type of processing with a Fourier transform-based kernel.

\begin{figure}
\centering
\includegraphics[width=0.8\textwidth]{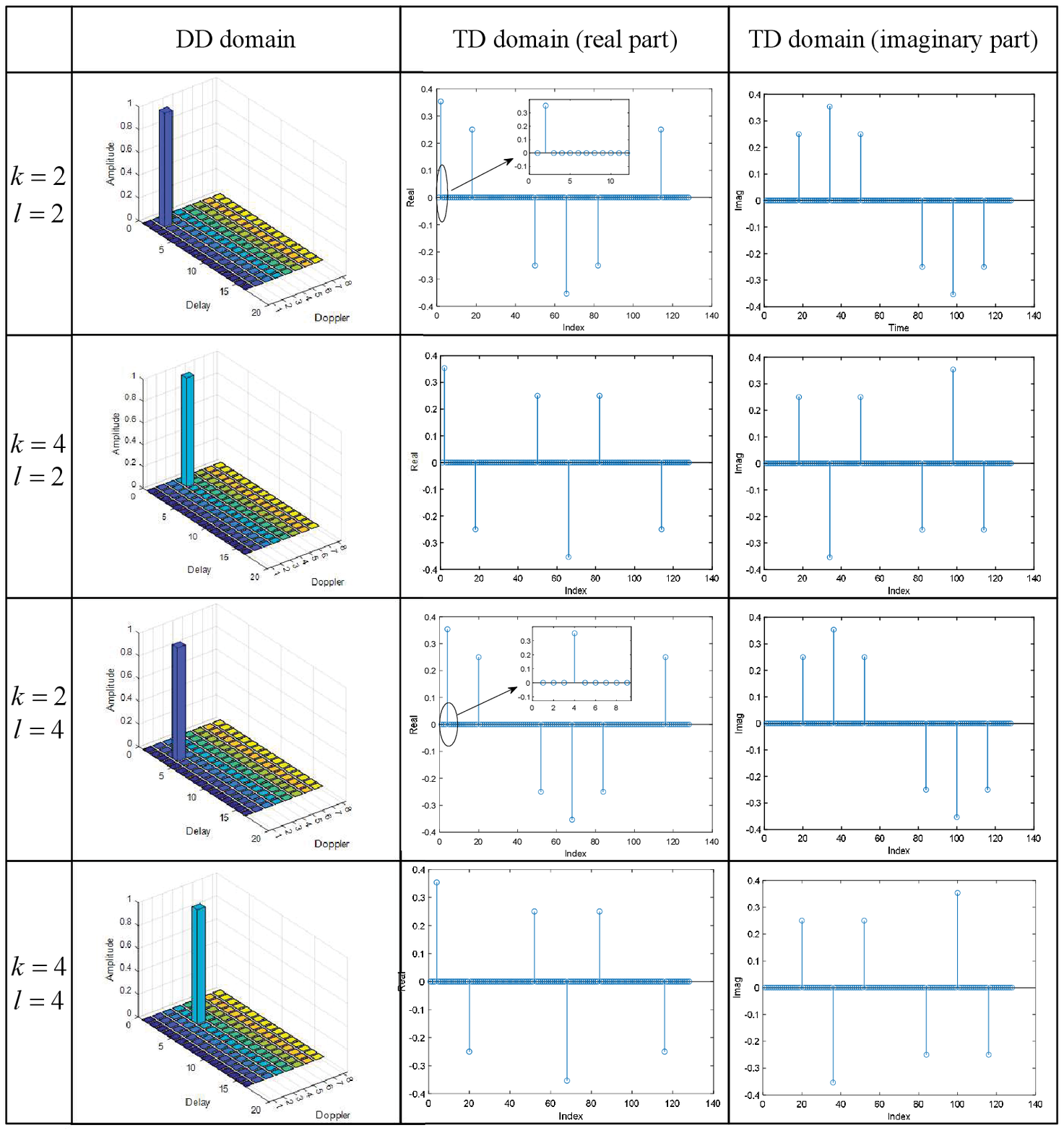}
\caption{An illustration of DD domain symbols and their corresponding TD domain OTFS signals.}
\label{C4_OTFS_signal_TD}
\centering
\end{figure}

The TD domain OTFS signals with corresponding DD domain information symbols are shown in Fig.~\ref{C4_OTFS_signal_TD}. As can be observed from Fig.~\ref{C4_OTFS_signal_TD}, the information symbol positions on the DD grid will change the signal characteristics of the TD domain OTFS signal.
In particular, we observe that the TD domain OTFS signal samples have values that are changed periodically with respect to the underlying Doppler index of the information symbol in the DD domain.
On the other hand,  we also notice that different delay indices of the information symbol will cause time shifts in the TD domain OTFS signal. In fact, these two interesting observations could be interpreted by the properties of ZT as implied by Fig.~\ref{C4_Fig_Zak}, where the DD domain symbols that have different delay indices are separated in the TD domain in a TDMA manner, while the DD domain symbols that have different Doppler indices are separated in the TD domain in an OFDM manner~\cite{Hadani2017orthogonal,Hadani2018OTFS_long,hadani2018whitebook}.

Based on the above discussions of OTFS signals, we can interpret the OTFS modulation as a combination of TDMA, OFDM, and CDMA transmissions. In particular, the TD domain OTFS signal behaves locally like TDMA (localized pulses in the TD domain), globally like OFDM (orthogonal tones in the TF domain), and spreading like CDMA (2D
spreading from the DD domain), thus inheriting their beneficial properties. In fact, both TDMA and OFDM could be seen as two special cases of the OTFS family with the delay period (time slot duration) or the Doppler period (sub-carrier spacing) tending to infinity~\cite{Hadani2018OTFS_long,hadani2018whitebook,Zhiqiang_magzine}.

\subsection{Descriptions of DZT-based OTFS: Receiver Side}
\begin{figure}
\centering
\includegraphics[width=0.8\textwidth]{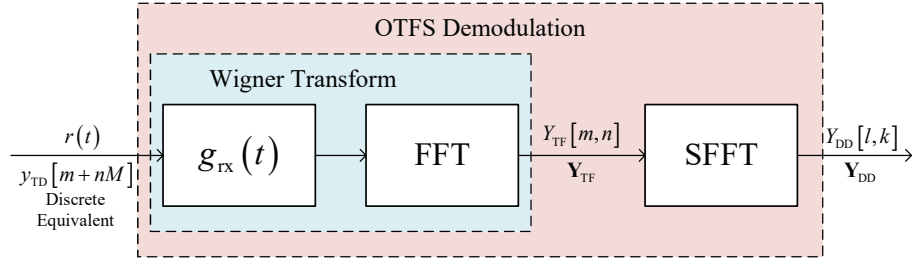}
\caption{A receiver diagram of the DZT-based OTFS modulation according to~\cite{Hadani2017orthogonal,Raviteja2018interference}.}
\label{C4_Receiver_diagram_Original}
\centering
\end{figure}

Now let us turn our attention to the receiver side.
Let $r\left( t \right)$ be the TD domain received signal corresponding to the transmitted signal $s\left( t \right)$. The relationship between $s\left( t \right)$ and $r\left( t \right)$ will be discussed in detail in the coming subsection.
With $r\left( t \right)$, the OTFS demodulator first performs the Wigner transform with the receive shaping pulse $g_{\rm rx}\left( t \right)$ (matched-filtering) and then apply the SFFT as shown in Fig.~\ref{C4_Receiver_diagram_Original}.
In particular, the Wigner transform could be interpreted as a multi-carrier demodulator, and a popular choice for implementing the Wigner transform is to apply the conventional OFDM demodulator~\cite{Hadani2017orthogonal,Raviteja2018interference,Zhiqiang_magzine}.
In specific, according to the conventional OFDM demodulation, the Wigner transform could be implemented by the receive pulse shaping and the FFT module, in which case the resultant TF domain received symbol ${Y_{{\rm{TF}}}}\left[ {m,n} \right]$ for $0 \le m \le M-1$ and $0 \le n \le N-1$ is written by{\footnote{Here, we assume that the appended CP has been removed.}}
\begin{equation}
{Y_{{\rm{TF}}}}\left[ {m,n} \right] = {\left. {{A_{{g_{{\rm{rx}}}},r}}\left( {t,f} \right)} \right|_{t = nT,f = m\Delta f}},\label{C4_Y_TF}
\end{equation}
where ${{A_{{g_{{\rm{rx}}}},r}}\left( {t,f} \right)}$ is the so-called ``cross-ambiguity function'' defined by
\begin{equation}
{A_{{g_{{\rm{rx}}}},r}}\left( {t,f} \right) \buildrel \Delta \over = \int_{ 0 }^{NT}  {g_{{\rm{rx}}}^*\left( {t' - t} \right)} r\left( {t'} \right){e^{ - j2\pi f\left( {t' - t} \right)}}{\rm{d}}t' .\label{C4_cross_ambiguity}
\end{equation}

Based on ${Y_{{\rm{TF}}}}\left[ {m,n} \right]$, the SFFT can be performed to obtain the DD domain received symbol ${Y_{{\rm{DD}}}}\left[ {l,k} \right]$ for $0 \le l \le M-1$ and $0 \le k \le N-1$, such as
\begin{align}
{Y_{{\rm{DD}}}}\left[ {l,k} \right] = \frac{1}{{\sqrt {MN} }}\sum\limits_{n = 0}^{N - 1} {\sum\limits_{m = 0}^{M - 1} {{Y_{{\rm{TF}}}}\left[ {m,n} \right]{e^{ - j2\pi \left( {\frac{{nk}}{N} - \frac{{ml}}{M}} \right)}}} } .\label{C4_Y_DD}
\end{align}

Similar to the previous subsection, we can characterize the above receiver operations using the DZT. By substituting~\eqref{C4_Y_TF} into~\eqref{C4_Y_DD}, we have
\begin{align}
{Y_{{\rm{DD}}}}\left[ {l,k} \right] &= \frac{1}{{\sqrt {MN} }}\sum\limits_{n = 0}^{N - 1} {\sum\limits_{m = 0}^{M - 1} {\int_0^{NT} {g_{{\rm{rx}}}^*\left( {t' - nT} \right)} r\left( {t'} \right){e^{ - j2\pi m\Delta f\left( {t' - nT} \right)}}{e^{ - j2\pi \left( {\frac{{nk}}{N} - \frac{{ml}}{M}} \right)}}} } {\rm{d}}t'\notag\\
& = \frac{1}{{\sqrt {MN} }}\sum\limits_{n = 0}^{N - 1} {\int_0^{NT} {g_{{\rm{rx}}}^*\left( {t' - nT} \right)} r\left( {t'} \right){e^{ - j2\pi \frac{{nk}}{N}}}\sum\limits_{m = 0}^{M - 1} {{e^{ - j2\pi m\left( {\Delta ft' - \frac{l}{M}} \right)}}} } {\rm{d}}t' .\label{C4_Y_DD_der1}
\end{align}
Notice that $\sum\limits_{m = 0}^{M - 1} {{e^{ - j2\pi m\left( {\Delta ft' - \frac{l}{M}} \right)}}}  = M\sum\limits_{n' =  - \infty }^\infty  {\delta \left( {\Delta ft' - \frac{l}{M} - n'} \right)} $,~\eqref{C4_Y_DD_der1} can be further simplified by
\begin{align}
{Y_{{\rm{DD}}}}\left[ {l,k} \right] &= \sqrt {\frac{M}{N}} \sum\limits_{n = 0}^{N - 1} {\int_0^{NT} {\sum\limits_{n' =  - \infty }^\infty  {g_{{\rm{rx}}}^*\left( {t' - nT} \right)} } r\left( {t'} \right){e^{ - j2\pi \frac{{nk}}{N}}}\delta \left( {\Delta ft' - \frac{l}{M} - n'} \right)} {\rm{d}}t'\notag\\
&= \sqrt {\frac{M}{N}} \sum\limits_{n = 0}^{N - 1} {\sum\limits_{n' = 0}^{N - 1} {g_{{\rm{rx}}}^*\left( {\frac{l}{M}T + n'T - nT} \right)} r\left( {\frac{l}{M}T + n'T} \right){e^{ - j2\pi \frac{{nk}}{N}}}}
.\label{C4_Y_DD_der2}
\end{align}
As implied by~\eqref{C4_Y_DD_der2}, it is sufficient to use discrete samples of ${g_{{\rm{rx}}}}\left( t \right)$ and $r\left( t \right)$ with a sampling period ${\frac{T}{M}}$ to obtain ${Y_{{\rm{DD}}}}\left[ {l,k} \right]$. Without loss of generality, we consider the same shaping pulse for both the transmitter side and the receiver side, i.e., ${g_{{\rm{tx}}}}\left( t \right)={g_{{\rm{rx}}}}\left( t \right)={g}\left( t \right)$ and denote by $r\left[ n \right]$ the $n$-th sample of TD domain received signal $r\left( t \right)$, for $0 \le n \le MN-1$. Note that $r\left[ n \right]$ in fact represents the TD domain received symbols $y_{\rm TD} \left[ n \right]$, corresponding to $x_{\rm TD} \left[ n \right]$.
With $g\left[ k \right]$, for $k \in {\mathbb Z}$, being the \emph{periodically extended} sampled pulse shaping ${g}\left( t \right)$,
we can rewrite~\eqref{C4_Y_DD_der2} by
\begin{align}
{Y_{{\rm{DD}}}}\left[ {l,k} \right] &= \sqrt {\frac{M}{N}} \sum\limits_{n = 0}^{N - 1} {\sum\limits_{n' = 0}^{N - 1} {{g^*}\left[ {l + \left( {n' - n} \right)M} \right]} r\left[ {l + n'M} \right]{e^{ - j2\pi \frac{{nk}}{N}}}} \notag\\
& = \sqrt M \sum\limits_{n' = 0}^{N - 1} {r\left[ {l + n'M} \right]} {\left( {\frac{1}{{\sqrt N }}\sum\limits_{n = 0}^{N - 1} {g\left[ {l + n'M - nM} \right]{e^{j2\pi \frac{{nk}}{N}}}} } \right)^*}\notag\\
&= \sqrt M \sum\limits_{n' = 0}^{N - 1} {r\left[ {l + n'M} \right]} {\cal D}{\cal Z}_g^*\left[ {l + n'M,k} \right]
.\label{C4_Y_DD_der3}
\end{align}
Furthermore, according to the quasi-periodicity shown in~\eqref{C4_discrete_Zak_quasi_periodicity_delay},~\eqref{C4_Y_DD_der3} can be simplified by
\begin{align}
{Y_{{\rm{DD}}}}\left[ {l,k} \right] &= \sqrt M \sum\limits_{n' = 0}^{N - 1} {r\left[ {l + n'M} \right]} {\left( {{\cal D}{{\cal Z}_g}\left[ {l,k} \right]{e^{j2\pi \frac{{n'k}}{N}}}} \right)^*}\notag\\
&= \sqrt M \sum\limits_{n' = 0}^{N - 1} {r\left[ {l + n'M} \right]} {e^{ - j2\pi \frac{{n'k}}{N}}}{\cal D}{\cal Z}_g^*\left[ {l,k} \right]\notag\\
&= \sqrt {MN} {\cal D}{\cal Z}_{r}\left[ {l,k} \right]{\cal D}{\cal Z}_g^*\left[ {l,k} \right],\label{C4_Y_DD_Zak}
\end{align}
where ${\cal D}{\cal Z}_{r}\left[ {l,k} \right]$ denotes the DZT of $r\left[ n \right]$, for $0 \le n \le MN-1$.
Similar to the transmitter side, in the case of rectangular pulse for the receiver processing, by substituting~\eqref{C4_DD_rect} into~\eqref{C4_Y_DD_Zak}, we have
\begin{align}
{Y_{{\rm{DD}}}}\left[ {l,k} \right] ={\cal D}{\cal Z}_{r}\left[ {l,k} \right].\label{C4_Y_DD_Zak_rect}
\end{align}
It is not hard to see from~\eqref{C4_Y_DD_Zak_rect} that, in the case of rectangular pulse for the receiver processing, the resultant DD domain received symbols are exactly the same as the DZT of the samples TD domain received signal $r\left[ n \right]$. This fact is consistent with our derivations about the processing at the transmitter side in the previous subsection.

\subsection{Descriptions of DZT-based OTFS: Equivalent Model, Pulse Shaping, and DD Domain Channel Sampling}
We have discussed the corresponding transmitter and receiver processing for OTFS transmissions. Based on those discussions, we found that the OTFS transmission has a close relationship to the IDZT/DZT. In particular, we noticed from~\eqref{C4_s_TD_der4} and \eqref{C4_Y_DD_Zak} that the transmit pulse shaping and the receive matched filtering can be expressed in a concise manner via the IDZT/DZT. This motivates us to consider the pulse shaping issue for OTFS systems in the DD domain from an IDZT/DZT point of view.

Based on~\eqref{C4_s_TD_DZT} and~\eqref{C4_Y_DD_Zak}, we notice that the OTFS transmission can be equivalently expressed by the DD domain representations of each module based on IDZT/DZT. 
In fact, this observation is not new, and has been discussed in the previous studies~\cite{mohammed2021derivation,Mohammed2021DDTVT,lampel2021orthogonal}.

\begin{figure}
\centering
\includegraphics[width=0.8\textwidth]{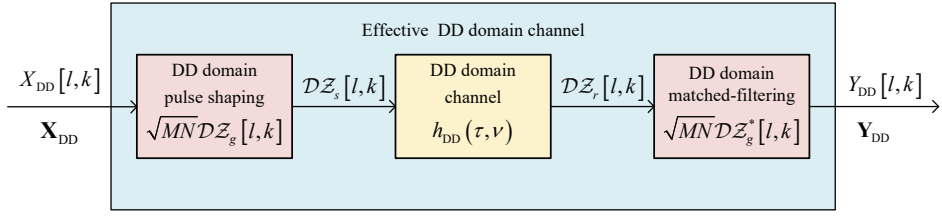}
\caption{A diagram of the equivalent model for DZT-based OTFS transmission.}
\label{C4_equivalent_model_Original}
\centering
\end{figure}

This observation aligns with the previous works on 
Let us consider the equivalent model depicted in Fig.~\ref{C4_equivalent_model_Original}. As shown in Fig.~\ref{C4_equivalent_model_Original}, both the pulse shaping and matched-filtering can be equivalently implemented in the DD domain in a multiplication manner, corresponding to ${\cal D}{\cal Z}_g\left[ {l,k} \right]$.

Now let us consider the impact of the DD domain channel ${h_{{\rm{DD}}}}\left( {\tau ,\nu } \right)$.
For the ease of presentation, we ignore the noise terms in the corresponding equations in the following derivations of this subsection.
Without loss of generality, we assume that the channel performs the delay operation first and the Doppler operation second, where the corresponding channel operations are characterized by~\eqref{S4_DD_channel_delay_first_Doppler_second}, and~\eqref{C4_DD_twisted_conv_delay_first_Doppler_second}~\cite{Hadani2017orthogonal,Raviteja2018interference}.
Therefore, according to the twisted convolution property in~\eqref{C4_DD_twisted_conv_delay_first_Doppler_second}, the following theorem holds.

\textbf{Theorem 3-1} (\emph{Input-Output Relationships based on ZT}):
Let ${u_p}\left( t \right) \buildrel \Delta \over = {e^{j2\pi {\nu _p}t}}$ and let ${s_p}\left( t \right) \buildrel \Delta \over = s\left( t \right){u_p}\left( t \right)$, for $1 \le p \le P$. Then, for $\tau  \in \left[ {0,T} \right)$ and $\nu  \in \left[ {0,\frac{1}{T}} \right)$, we have
\begin{align}
{{\cal Z}_r}\left( {\tau ,\nu } \right) = \sum\limits_{p = 1}^P {{h_p}} {{\cal Z}_{{s_p}}}\left( {\tau  - {\tau _p},\nu } \right) . \label{C4_IO_ZT_T4_1}
\end{align}
\textbf{Proof}: This theorem can be derived based on~\eqref{C4_DD_twisted_conv_delay_first_Doppler_second}. By substituting~\eqref{S4_DD_channel1} into~\eqref{C4_DD_twisted_conv_delay_first_Doppler_second}, we have
\begin{align}
{{\cal Z}_r}\left( {\tau ,\nu } \right) &= \sum\limits_{p = 1}^P {{h_p}} {{\cal Z}_s}\left( {\tau  - {\tau _p},\nu  - {\nu _p}} \right){e^{j2\pi {\nu _p}\left( {\tau  - {\tau _p}} \right)}}\notag\\
&= \sum\limits_{p = 1}^P {{h_p}} \sqrt T \sum\limits_{n =  - \infty }^\infty  {s\left( {\tau  - {\tau _p} + nT} \right){e^{ - j2\pi n\left( {\nu  - {\nu _p}} \right)T}}} {e^{j2\pi {\nu _p}\left( {\tau  - {\tau _p}} \right)}}\notag\\
&= \sum\limits_{p = 1}^P {{h_p}} \sqrt T \sum\limits_{n =  - \infty }^\infty  {\underbrace {s\left( {\tau  - {\tau _p} + nT} \right){e^{j2\pi {\nu _p}\left( {\tau  - {\tau _p} + nT} \right)}}}_{{s_p}\left( {\tau  - {\tau _p} + nT} \right)}} {e^{ - j2\pi n\nu T}}\notag\\
& = \sum\limits_{p = 1}^P {{h_p}} {{\cal Z}_{{s_p}}}\left( {\tau  - {\tau _p},\nu } \right).
\end{align}
This completes the proof of the theorem. \hfill $\blacksquare$

Theorem 3-1 characterizes the input-output relationship of OTFS from a ZT point of view, which is valid for time-continuous signals $s\left( t \right)$ and $r\left( t \right)$. As implied by Theorem 3-1, the DD domain received signal could be interpreted as the superposition of $P$ delayed DD domain ``modulated'' signals of the transmitted signal $s\left( t \right)$, with respect to the Doppler effect ${u_p}\left( t \right)$ and the fading coefficient ${h_p}$ for each resolvable path.
On the other hand, in practical systems, a discrete system model is always of interest from an implementation perspective. To derive the corresponding discrete system model based on~\eqref{C4_IO_ZT_T4_1}, we need to consider the DD domain sampling and interpolation.

To align our previous derivations, we consider the critical sampling for the derivation of the discrete system model.
With the given TF resources, it can be shown that the DD domain sample periods (resolution) along the delay and Doppler domains are $\frac{T}{M}$ and $\frac{1}{{NT}}$, respectively. Therefore, let us define the delay and Doppler indices associated to the $p$-th path by $l_p$ and $k_p$, respectively, such as
\begin{equation}
{\tau _p} = \frac{{{l_p} + {\iota _p}}}{{M\Delta f}},\quad
{\nu _p} = \frac{{{k_p} + {\kappa _p}}}{{NT}}.
\label{C4_delay_Doppler_resolution}
\end{equation}
Notice that the term $ - {1 \mathord{\left/
 {\vphantom {1 2}} \right.
 \kern-\nulldelimiterspace} 2} \le{\iota _p}\le {1 \mathord{\left/
 {\vphantom {1 2}} \right.
 \kern-\nulldelimiterspace} 2}$ and $- {1 \mathord{\left/
 {\vphantom {1 2}} \right.
 \kern-\nulldelimiterspace} 2} \le {\kappa _p} \le {1 \mathord{\left/
 {\vphantom {1 2}} \right.
 \kern-\nulldelimiterspace} 2}$ denote the fractional delay and Doppler shifts which correspond to the fractional shifts from the nearest delay and Doppler indices~\cite{Li2020on,li2021performance_analysis,Raviteja2018interference}.
It should be noted that the typical value of the sampling time ${1 \mathord{\left/
 {\vphantom {1 {M\Delta f}}} \right.
 \kern-\nulldelimiterspace} {M\Delta f}}$ in the delay domain is usually sufficiently small. Therefore, the impact of
fractional delays in typical wideband systems can be neglected{\footnote {It should be noted that the effect of fractional Doppler shifts can be mitigated by adding TF domain windows~\cite{wei2020transmitter}.}}~\cite{tse2005fundamentals}.
However, we will still consider the fractional delay case here in order to obtain some general conclusions on OTFS modulation.
As for the Doppler shift, we have $-{k_{\max }} \le k_p+ {\kappa _p} \le {k_{\max }}$, where $k_{\max }$ is the maximum Doppler index satisfying ${k_{\max }} = \left\lceil {NT{\nu _{\max }}} \right\rceil $~\cite{Raviteja2018interference}.

Given the DD domain resolutions, we can approximate the ZT by using the DZT. Without loss of generality, we focus on the DD domain interpolation using the rectangular kernel, such as
\begin{equation}
{{\cal Z}_x}\left( {\tau ,\nu } \right) \approx \sqrt{NT} {\cal D}{{\cal Z}_x}\left[ {l,k} \right], \quad{\rm for}\;\tau  \in \left[ {\frac{T}{M}\left( {l - 0.5} \right),\frac{T}{M}\left( {l + 0.5} \right)} \right),\nu  \in \left[ {\frac{ {k - 0.5} }{{NT}},\frac{{k + 0.5}}{{NT}}} \right), \label{C4_ZT_DZT_approximation_interpolation}
\end{equation}
where the constant $\sqrt{NT}$ comes from the definitions of ZT and DZT given in~\eqref{C4_Zak} and~\eqref{C4_Discrete_Zak}.
Then, we can approximate ${{\cal Z}_r}\left( {\frac{T}{M}l,\frac{k}{{NT}}} \right)$ by
\begin{align}
{{\cal Z}_r}\left( {\frac{T}{M}l,\frac{k}{{NT}}} \right)& = \sum\limits_{p = 1}^P {{h_p}} {{\cal Z}_{{s_p}}}\left( {\frac{T}{M}\left( {l - {l_p} - {\iota _p}} \right),\frac{k}{{NT}}} \right)\notag\\
&\approx \sum\limits_{p = 1}^P \sqrt{NT} {{h_p}} {\cal D}{{\cal Z}_{s_p}}\left[ {l - {l_p},k} \right] ,\label{C4_ZT_DZT_io_der1}
\end{align}
which implies that
\begin{align}
{\cal D}{{\cal Z}_r}\left[ {l,k} \right]\approx \sum\limits_{p = 1}^P {{h_p}} {\cal D}{{\cal Z}_{s_p}}\left[ {l - {l_p},k} \right],\label{C4_ZT_DZT_io_der2}
\end{align}
according to~\eqref{C4_ZT_DZT_approximation_interpolation}.
As implied by~\eqref{C4_ZT_DZT_io_der1}, the above approximation ignores the potential effect from the fractional delay, which could only be valid with a given sufficient large bandwidth. However, we will continue our derivations based on~\eqref{C4_ZT_DZT_io_der1}, as the bandwidth is usually sufficient in practice~\cite{Raviteja2018interference}.
Based on~\eqref{C4_DZT_modulation_property}, we can expand ${\cal D}{{\cal Z}_{{s_p}}}\left[ {l,k} \right]$
by
\begin{align}
{\cal D}{{\cal Z}_{{s_p}}}\left[ {l,k} \right] = \frac{1}{{\sqrt N }}\sum\limits_{k' = 0}^{N - 1} {{\cal D}{{\cal Z}_s}\left[ {l,k'} \right]} {\cal D}{{\cal Z}_{{u_p}}}\left[ {l,k - k'} \right],\label{C4_DZT_s_p}
\end{align}
where
\begin{align}
{\cal D}{{\cal Z}_{{u_p}}}\left[ {l,k} \right] &= \frac{1}{{\sqrt N }}\sum\limits_{n = 0}^{N - 1} {{e^{j2\pi {\nu _p}\left( {l + nM} \right)\frac{T}{M}}}{e^{ - j2\pi \frac{n}{N}k}}} \notag\\
&= \frac{1}{{\sqrt N }}{e^{j2\pi \frac{{{k_p} + {\kappa _p}}}{{MN}}l}}\sum\limits_{n = 0}^{N - 1} {{e^{j2\pi \frac{n}{N}\left( {{k_p} + {\kappa _p} - k} \right)}}} . \label{C4_DZT_z_p}
\end{align}
Here, it is crucial to note that the $\frac{T}{M}$-sampled ${u_p}\left( t \right)$, i.e., $\left[ {{u_p}\left( {\frac{T}{M}} \right),{u_p}\left( {\frac{{2T}}{M}} \right),...,{u_p}\left( {NT} \right)} \right]$ is not a $MN$-periodic sequence, when $\kappa_p \ne 0$. Thus, the DZT concept in~\eqref{C4_DZT_z_p} is not strictly valid in the case of fractional Doppler. However, for notational consistency, we slightly relax the definition of DZT, such that~\eqref{C4_DZT_z_p} holds.
Then, by substituting~\eqref{C4_DZT_s_p} into~\eqref{C4_ZT_DZT_io_der2}, we obtain
\begin{align}
&{\cal D}{{\cal Z}_r}\left[ {l,k} \right]\notag\\
=& \sum\limits_{p = 1}^P {\frac{{{h_p}}}{{\sqrt N }}} \sum\limits_{k' = 0}^{N - 1} {{\cal D}{{\cal Z}_s}\left[ {l - {l_p},k'} \right]} {\cal D}{{\cal Z}_{{u_p}}}\left[ {l - {l_p},k - k'} \right]\notag\\
=& \sum\limits_{p = 1}^P {\frac{{{h_p}}}{{\sqrt N }}} \sum\limits_{k' = 0}^{N - 1} {{\cal D}{{\cal Z}_s}\left[ {{{\left[ {l - {l_p}} \right]}_M},k'} \right]} {\alpha _{l - {l_p},k'}}{\cal D}{{\cal Z}_{{u_p}}}\left[ {{{\left[ {l - {l_p}} \right]}_M},k - k'} \right]{\alpha _{l - {l_p},k - k'}}, \label{C4_ZT_DZT_io_der3}
\end{align}
where ${\alpha _{l,k}}$ is a phase offset as the result of quasi-periodicity property in~\eqref{C4_discrete_Zak_quasi_periodicity_delay}, and it is given by
\begin{align}
{\alpha _{l,k}} = \left\{ \begin{array}{l}
1,\quad \quad \quad \quad \quad \quad l \ge 0,\\
{e^{ - j2\pi \frac{{k}}{N}}},\quad\quad \quad\!\! \!\!\quad l < 0.
\end{array} \right. \label{C4_Phase_offset_io_integer_PS}
\end{align}
It should be noted that since the term ${\cal D}{{\cal Z}_{u_p}}\left[ {l,k} \right]$ only has the same value of the actual DZT of ${u_p}\left( t \right)$ samples, because the samples of ${u_p}\left( t \right)$ may not be $MN$-periodic. Therefore, we apply the quasi-periodicity property to ${\cal D}{{\cal Z}_{{u_p}}}\left[ {{{\left[ {l - {l_p}} \right]}_M},k - k'} \right]$ along the delay domain in order to maintain the mathematical consistency. Substitute~\eqref{C4_DZT_z_p} into~\eqref{C4_ZT_DZT_io_der3}, yielding
\begin{align}
{\cal D}{{\cal Z}_r}\left[ {l,k} \right] &= \sum\limits_{p = 1}^P {\frac{{{h_p}}}{{\sqrt N }}} {\alpha _{l - {l_p},k}}\sum\limits_{k' = 0}^{N - 1} {{\cal D}{{\cal Z}_s}\left[ {{{\left[ {l - {l_p}} \right]}_M},k'} \right]} {\cal D}{{\cal Z}_{{u_p}}}\left[ {{{\left[ {l - {l_p}} \right]}_M},k - k'} \right]\notag\\
& = \sum\limits_{p = 1}^P {\frac{{{h_p}}}{N}} {\alpha _{l - {l_p},k}}\sum\limits_{k' = 0}^{N - 1} {{\cal D}{{\cal Z}_s}\left[ {{{\left[ {l - {l_p}} \right]}_M},k'} \right]} {e^{j2\pi \frac{{{k_p} + {\kappa _p}}}{{MN}}{{\left[ {l - {l_p}} \right]}_M}}}\sum\limits_{n = 0}^{N - 1} {{e^{j2\pi \frac{n}{N}\left( {{k_p} + {\kappa _p} - \left( {k - k'} \right)} \right)}}} \notag\\
&= \sum\limits_{p = 1}^P {\frac{{{h_p}}}{N}} {\alpha _{l - {l_p},k - \left( {{k_p} + {\kappa _p}} \right)}}\sum\limits_{k' = 0}^{N - 1} {{\cal D}{{\cal Z}_s}\left[ {{{\left[ {l - {l_p}} \right]}_M},k'} \right]} {e^{j2\pi \frac{{{k_p} + {\kappa _p}}}{{MN}}\left( {l - {l_p}} \right)}}\sum\limits_{n = 0}^{N - 1} {{e^{j2\pi \frac{n}{N}\left( {{k_p} + {\kappa _p} - \left( {k - k'} \right)} \right)}}} .\label{C4_ZT_DZT_io_der4}
\end{align}


For the ease of presentation, we in the following derive the DD domain input-output relationship separately for both the integer and fractional Doppler cases with respect to the underlying pulse shaping.
Notice that in the case of integer Doppler, we have $\sum\limits_{n = 0}^{N - 1} {{e^{j2\pi \frac{n}{N}\left( {{k_p} + {\kappa _p} - \left( {k - k'} \right)} \right)}}}  = N\sum\limits_{n =  - \infty }^\infty  {\delta \left[ {k' - \left( {k - {k_p}} \right) + nN} \right]} $.
Then, based on~\eqref{C4_s_TD_DZT},~\eqref{C4_ZT_DZT_io_der4}, and~\eqref{C4_Y_DD_Zak}, we can obtain
\begin{align}
{Y_{{\rm{DD}}}}\left[ {l,k} \right]=& \sqrt {MN} \sum\limits_{p = 1}^P {{h_p}} {e^{j2\pi \frac{{{k_p} }}{{MN}}\left( {l - {l_p}} \right)}} {{\cal D}{{\cal Z}_s}\left[ {l - {l_p},k - {k_p}} \right]} {\cal D}{\cal Z}_g^*\left[ {l,k} \right]\notag\\
=& MN\sum\limits_{p = 1}^P {{h_p}} {\alpha _{l - {l_p},k - {k_p}}}{e^{j2\pi \frac{{{k_p}}}{{MN}}\left( {l - {l_p}} \right)}}{X_{{\rm{DD}}}}\left[ {{{\left[ {l - {l_p}} \right]}_M},{{\left[ {k - {k_p}} \right]}_N}} \right]\notag\\
&\quad\quad{\cal D}{{\cal Z}_g}\left[ {{{\left[ {l - {l_p}} \right]}_M},{{\left[ {k - {k_p}} \right]}_N}} \right]{\cal D}{\cal Z}_g^*\left[ {l,k} \right]. \label{C4_DZT_io_integer_PS}
\end{align}
On the other hand, in the case of fractional Doppler, based on~\eqref{C4_s_TD_DZT},~\eqref{C4_ZT_DZT_io_der4}, and~\eqref{C4_Y_DD_Zak}, we can obtain
\begin{align}
{Y_{{\rm{DD}}}}\left[ {l,k} \right] =& \sqrt {\frac{M}{N}} \sum\limits_{p = 1}^P h_p {\alpha _{l - {l_p},k - \left( {{k_p} + {\kappa _p}} \right)}}{e^{j2\pi \frac{{{k_p} + {\kappa _p}}}{{MN}}\left( {l - {l_p}} \right)}}\sum\limits_{k' = 0}^{N - 1} {\sum\limits_{n = 0}^{N - 1} {{e^{j2\pi \frac{n}{N}\left( {{k_p} + {\kappa _p} - \left( {k - k'} \right)} \right)}}} } \notag\\
&\quad\quad{\cal D}{{\cal Z}_s}\left[ {{{\left[ {l - {l_p}} \right]}_M},k'} \right]{\cal D}{\cal Z}_g^*\left[ {l,k} \right]\notag\\
=& M \sum\limits_{p = 1}^P h_p {\alpha _{l - {l_p},k - \left( {{k_p} + {\kappa _p}} \right)}}{e^{j2\pi \frac{{{k_p} + {\kappa _p}}}{{MN}}\left( {l - {l_p}} \right)}}\sum\limits_{k' = 0}^{N - 1} {\sum\limits_{n = 0}^{N - 1} {{e^{j2\pi \frac{n}{N}\left( {{k_p} + {\kappa _p} - \left( {k - k'} \right)} \right)}}} }  \notag\\
&\quad\quad{X_{{\rm{DD}}}}\left[ {{{\left[ {l - {l_p}} \right]}_M},k'} \right]{\cal D}{{\cal Z}_g}\left[ {{{\left[ {l - {l_p}} \right]}_M},k'} \right]{\cal D}{\cal Z}_g^*\left[ {l,k} \right]\notag\\
=& M\sum\limits_{p = 1}^P h_p {\alpha _{l - {l_p},k - \left( {{k_p} + {\kappa _p}} \right)}}{e^{j2\pi \frac{{{k_p} + {\kappa _p}}}{{MN}}\left( {l - {l_p}} \right)}}\sum\limits_{k' = 0}^{N - 1} {{e^{j\pi \frac{{N - 1}}{N}\left( {{k_p} + {\kappa _p} - \left( {k - k'} \right)} \right)}}} \notag\\
&\quad\quad\frac{{\sin \pi \left( {{k_p} + {\kappa _p} - \left( {k - k'} \right)} \right)}}{{\sin \frac{\pi }{N}\left( {{k_p} + {\kappa _p} - \left( {k - k'} \right)} \right)}}{X_{{\rm{DD}}}}\left[ {{{\left[ {l - {l_p}} \right]}_M},k'} \right]{\cal D}{{\cal Z}_g}\left[ {{{\left[ {l - {l_p}} \right]}_M},k'} \right]{\cal D}{\cal Z}_g^*\left[ {l,k} \right].
\label{C4_DZT_io_fractional_PS}
\end{align}
Comparing with~\eqref{C4_DZT_io_integer_PS}, it can be seen from~\eqref{C4_DZT_io_fractional_PS} that the fractional Doppler could cause power leakage in the DD domain, which results in the ISI along the Doppler domain. This observation is consistent with the related literature on the fractional Doppler~\cite{Raviteja2018interference}.
Furthermore, it is interesting to see from both~\eqref{C4_DZT_io_integer_PS} and~\eqref{C4_DZT_io_fractional_PS} that for each DD domain received symbol $Y_{\rm DD}\left[ {l,k} \right]$, its value is related to the DD domain transmitted symbols, the DD domain pulse shaping, the DD domain matched-filtering, and the DD domain channel response ${h_p}$ of each path.
This observation gives us an intuition for the pulse design for OTFS transmission that is to optimize the pulse shape based on the values of ${\cal D}{{\cal Z}_g}\left[ {{{\left[ {l - l_p} \right]}_M},k'} \right] {\cal D}{\cal Z}_g^*\left[ {l,k} \right]$.
For example, the related optimization could be carried out based on the underlying channel characteristics, e.g., the sparsity of the DD domain channel response. Although the details of the pulse design are out of the scope of this thesis, we offer some examples with typical pulses as follows.

\textbf{Example 3-1} (\emph{Rectangular Shaping Pulse with Integer Delay and Doppler}):
In the case of rectangular shaping pulse with integer delay and Doppler, the DD domain input-output relationship can be derived by substituting~\eqref{C4_DD_rect} into~\eqref{C4_DZT_io_integer_PS}, such as
\begin{align}
{Y_{{\rm{DD}}}}\left[ {l,k} \right] =\sum\limits_{p = 1}^P {{h_p}} {\alpha _{l - {l_p},k - {k_p}}}{e^{j2\pi \frac{{{k_p} }}{{MN}}\left( {l - {l_p}} \right)}}{X_{{\rm{DD}}}}\left[ {{{\left[ {l - {l_p}} \right]}_M},{{\left[ {k - {k_p}} \right]}_N}} \right].
\label{C4_DD_io_relationship_integer_delay_Doppler_rect_pulse}
\end{align}

\textbf{Example 3-2} (\emph{Rectangular Shaping Pulse with Integer Delay and Fractional Doppler}):
In the case of rectangular shaping pulse with integer delay and fractional Doppler, the DD domain input-output relationship can be derived by substituting~\eqref{C4_DD_rect} into~\eqref{C4_DZT_io_fractional_PS}, such as
\begin{align}
{Y_{{\rm{DD}}}}\left[ {l,k} \right] =& {\frac{{{1}}}{N}}\sum\limits_{p = 1}^P h_p {\alpha _{l - {l_p},k - \left( {{k_p} + {\kappa _p}} \right)}}{e^{j2\pi \frac{{{k_p} + {\kappa _p}}}{{MN}}\left( {l - {l_p}} \right)}}\sum\limits_{k' = 0}^{N - 1} {{e^{j\pi \frac{{N - 1}}{N}\left( {{k_p} + {\kappa _p} - \left( {k - k'} \right)} \right)}}} \notag\\
&\quad\quad\frac{{\sin \pi \left( {{k_p} + {\kappa _p} - \left( {k - k'} \right)} \right)}}{{\sin \frac{\pi }{N}\left( {{k_p} + {\kappa _p} - \left( {k - k'} \right)} \right)}}{X_{{\rm{DD}}}}\left[ {{{\left[ {l - {l_p}} \right]}_M},k'} \right].
\label{C4_DD_io_relationship_integer_delay_fractional_Doppler_rect_pulse}
\end{align}

From the above two examples, we can notice that rectangular pulse shaping cannot mitigate the power leakage induced by fractional Doppler. To characterize the effect of pulse shaping on fractional Doppler, we could study the interference pattern associated to ${Y_{{\rm{DD}}}}\left[ {l,k} \right]$ and ${X_{{\rm{DD}}}}\left[ {{{\left[ {l - {l_p}} \right]}_M},k'} \right]$, for $0 \le k, k' \le N-1$.
Recall~\eqref{C4_DZT_io_fractional_PS}, it can be noticed that the corresponding interference energy depends on the value of
\begin{align}
V\left[ {k - k'} \right] = \left| {M\frac{{\sin \pi \left( {{k_p} + {\kappa _p} - \left( {k - k'} \right)} \right)}}{{\sin \frac{\pi }{N}\left( {{k_p} + {\kappa _p} - \left( {k - k'} \right)} \right)}}{\cal D}{{\cal Z}_g}\left[ {{{\left[ {l - {l_p}} \right]}_M},k'} \right]{\cal D}{\cal Z}_g^*\left[ {l,k} \right]} \right|.\label{C4_DD_interference_amplitude}
\end{align}
In particular, we denote by $V\left[ {k - k'} \right]$ the amplitude of the interference from ${X_{{\rm{DD}}}}\left[ {{{\left[ {l - {l_p}} \right]}_M},k'} \right]$ to ${Y_{{\rm{DD}}}}\left[ {l,k} \right]$, which could be used to optimize the shaping pulse in order to have a desired interference pattern.
However, the details of pulse shaping design is beyond the scope of this thesis, and we leave this interesting research problem as our future work.

\section{Matrix Form Representation of the DZT-based OTFS}
We have studied the DZT-based OTFS in the previous section. Although the mathematical derivations are complete, it is often convenient to use the equivalent matrix form representation to describe the corresponding processing. In this section, we will focus on the matrix form representation of the DZT-based OTFS transmissions. We will follow the same notations as the previous subsection for the ease of presentation.

According to~\cite{Raviteja2019practical}, two normalized DFT matrices ${\bf{F}}_M$ and ${\bf{F}}_N$ of size $M \times M$ and $N \times N$ can be used to characterize the SFFT in~\eqref{C4_ISFFT}. Thus, given the DD domain information symbol matrix ${\bf{X}}_{\rm DD} $, the TF domain information symbol matrix ${\bf{X}}_{\rm TF} $ can be written by
\begin{equation}
{{\bf{X}}_{{\rm{TF}}}} = {{\bf{F}}_M}{\bf{X}}_{\rm DD}{\bf{F}}_N^{\rm{H}}.
\label{C4_TF_symbol_matrix}
\end{equation}
Then, by considering the rectangular pulse for the transmitter shaping pulse, the time domain transmitted symbol matrix can be obtained by~\cite{Raviteja2019practical}
\begin{equation}
{\bf{X}}_{\rm TD} = {{\bf{I}}_M}{\bf{F}}_M^{\rm{H}}{{\bf{X}}_{{\rm{TF}}}} = {\bf{X}}_{\rm DD}{\bf{F}}_N^{\rm{H}}.
\label{C4_TD_symbol_matrix}
\end{equation}
It should be noted that~\eqref{C4_TD_symbol_matrix} is consistent with~\eqref{C4_s_TD_der4}, i.e., ${{\bf{x}}_{{\rm{TD}}}} = {\bf{s}}$, where ${{\bf{x}}_{{\rm{TD}}}} \buildrel \Delta \over = {\rm{vec}}\left( {{{\bf{X}}_{{\rm{TD}}}}} \right)$, by inserting the DZT of the rectangular pulse. In summary, the corresponding information symbol vectors for DD, TF, and TD domains are given as follows.
\begin{align}
{\bf{x}}_{\rm DD} &\buildrel \Delta \over = {\rm{vec}}\left( {\bf{X}}_{\rm DD} \right), \label{C4_DD_transmitted_symbol_vec}\\
{{\bf{x}}_{{\rm{TF}}}} &\buildrel \Delta \over = {\rm{vec}}\left( {{{\bf{X}}_{{\rm{TF}}}}} \right) = \left( {{\bf{F}}_N^{\rm{H}} \otimes {{\bf{F}}_M}} \right){\bf{x}}_{\rm DD}, \; {\rm and } \quad\label{C4_TF_transmitted_symbol_vec}\\
{\bf{x}}_{\rm TD} &\buildrel \Delta \over = {\rm{vec}}\left( {\bf{X}}_{\rm TD} \right) = \left( {{\bf{F}}_N^{\rm{H}} \otimes {{\bf{I}}_M}} \right){\bf{x}}_{\rm DD}. \label{C4_TD_transmitted_symbol_vec}
\end{align}
Corresponding to the TD domain transmitted symbol vector ${\bf{x}}_{\rm TD}$, the time domain effective channel ${\bf{H}}_{\rm{T}}^{{\rm{eff}}}$ with a reduced CP frame format can be written by{\footnote{It should be noted that the following equation is exact in the case of integer delay and Doppler~\cite{Raviteja2019practical} but only approximates the input-output relationship in the fractional Doppler case. In particular, the approximation comes from the fact that the fractional Doppler could destroy the $MN$-periodicity of the received signal after CP removal, which causes a slight phase difference associated to the first few symbols.}}~\cite{Raviteja2019practical}
\begin{align}
{\bf{H}}_{\rm{TD}}^{{\rm{eff}}} = \sum\limits_{p = 1}^P {{h_p}} {{\bm{\Pi }}^{{l_p}+{\iota _p}}}{{\bm{\Delta}} ^{{k_p}+{\kappa _p}}}, \label{C4_TD_domain_channel}
\end{align}
where ${\bm{\Pi }}$ is the permutation matrix (forward cyclic shift), i.e.,
\begin{equation}
{\bm{\Pi }} = {\left[ {\begin{array}{*{20}{c}}
0& \cdots &0&1\\
1& \ddots &0&0\\
 \vdots & \ddots & \ddots & \vdots \\
0& \cdots &1&0
\end{array}} \right]_{MN \times MN}},   \label{C4_TD_forward_cyclic_shift_matrix}
\end{equation}
and ${\bm{\Delta}}=\textrm{diag}\{{\gamma}^0,{\gamma}^1,...,{\gamma}^{MN-1}\} $ is a diagonal matrix with $\gamma \buildrel \Delta \over = {e^{\frac{{j2\pi }}{{MN}}}}$~\cite{Raviteja2019practical}. Here. we note again that the fractional delay ${\iota _p}$ can be safely neglected in practical systems~\cite{Raviteja2019practical}.
Based on~\eqref{C4_TD_domain_channel}, the TD domain received symbol vector $\bf r$ is given by
\begin{equation}
{\bf{r}} = {\bf{H}}_{\rm{TD}}^{{\rm{eff}}}{\bf{x}}_{\rm TD} + {\bf{w}} \label{C4_TD_io_matrix_form},
\end{equation}
where ${\bf{w}}$ is the corresponding AWGN sample vectors in the TD domain with one-sided PSD $N_0$.
Specifically, $\bf r$ can be rearranged into the 2D time domain received symbol matrix $\bf R$.
By applying the rectangular pulse as the receiver filtering pulse, we can obtain the corresponding TF domain and DD domain received symbol matrices as
\begin{align}
{{\bf{Y}}_{{\rm{TF}}}} &= {{\bf{F}}_M}{{\bf{I}}_M}{\bf{R}} = {{\bf{F}}_M}{\bf{R}}, \notag\\
{\bf{Y}}_{{\rm{DD}}} &= {\bf{F}}_M^{\rm{H}}{{\bf{Y}}_{{\rm{TF}}}}{{\bf{F}}_N} = {\bf{R}}{{\bf{F}}_N}.
\end{align}
Therefore, we can derive the corresponding vector form of the received symbols in the TF domain and DD domain by
 \begin{align}
{{\bf{y}}_{{\rm{TF}}}} &\buildrel \Delta \over = {\rm{vec}}\left( {{{\bf{Y}}_{{\rm{TF}}}}} \right) = \left( {{{\bf{I}}_N} \otimes {{\bf{F}}_M}} \right){\bf{r}},\label{C4_TF_received_symbol_vec}\\
{\bf{y}}_{\rm DD} &\buildrel \Delta \over = {\rm{vec}}\left( {\bf{Y}}_{\rm DD} \right) = \left( {{{\bf{F}}_N} \otimes {{\bf{I}}_M}} \right){\bf{r}}. \label{C4_DD_received_symbol_vec}
\end{align}
Based on the previous analysis, we are ready to demonstrate the input-output relationship of OTFS modulation with respect to different domains in matrix forms.
Let us denote the effective channel matrices in TF domain and DD domain by ${\bf{H}}_{\rm{TF}}^{{\rm{eff}}} $ and ${\bf{H}}_{\rm{DD}}^{{\rm{eff}}}$, respectively.
In specific, based on~\eqref{C4_TF_transmitted_symbol_vec},~\eqref{C4_TD_domain_channel}, and~\eqref{C4_TF_transmitted_symbol_vec}, we have
\begin{align}
{\bf{H}}_{{\rm{TF}}}^{{\rm{eff}}} = \sum\limits_{i = 1}^P {{h_i}} \left( {{{\bf{I}}_N} \otimes {{\bf{F}}_M}} \right){{\bf{\Pi }}^{{l_i+{\iota_i}}}}{{\bf{\Delta }}^{{k_i} + {\kappa _i}}}\left( {{{\bf{I}}_N} \otimes {\bf{F}}_M^{\rm{H}}} \right).
\label{C4_TF_channel_fractional}
\end{align}
Similarly, based on~\eqref{C4_DD_transmitted_symbol_vec},~\eqref{C4_TD_domain_channel}, and~\eqref{C4_DD_received_symbol_vec}, we have
\begin{align}
{\bf{H}}_{\rm{DD}}^{{\rm{eff}}} = \sum\limits_{i = 1}^P {{h_i}\left( {{{\bf{F}}_N} \otimes {{\bf{I}}_M}} \right)} {{\bf{\Pi }}^{{l_i+{\iota_i}}}}{{\bf{\Delta }}^{{k_i} + {\kappa _i}}}\left( {{\bf{F}}_N^{\rm{H}} \otimes {{\bf{I}}_M}} \right).
\label{C4_DD_channel_fractional}
\end{align}

Next, we use some examples to demonstrate the differences between the effective channels in different domains, where the corresponding effective channel matrices are generated based on~\eqref{C4_TD_domain_channel},~\eqref{C4_TF_channel_fractional}, and~\eqref{C4_DD_channel_fractional}, respectively. Without loss of generality, let us consider the channel parameters given in Table~\ref{C4_channel_parameters}.
\begin{table}[htbp]
\caption{Channel Parameters for Effective Channels}
\centering
\begin{tabular}{|c|c|}
\hline
Parameter~&~Values\\
\hline
Number of delay bins (sub-carriers)~&~$M=32$\\
\hline
Number of Doppler bins(time slots)~&~$N=16$\\
\hline
Number of paths~&~$P=4$\\
\hline
Fading coefficients~&~$\left[ {0.1 + 0.1i,0.2 + 0.2i,0.3 + 0.3i,0.4 + 0.4i} \right]$\\
\hline
Delay indices (integer)~&~$\left[ 0, 1, 2, 3\right]$\\
\hline
Delay indices (fractional)~&~$\left[ 0.5, 1.5, 2.5, 3.5\right]$\\
\hline
Doppler indices (integer)~&~$\left[ 0, 1, 2, 3\right]$\\
\hline
Doppler indices (fractional)~&~$\left[ 0.5, 1.5, 2.5, 3.5\right]$\\
\hline
\end{tabular}
\label{C4_channel_parameters}
\end{table}

\begin{figure}
\centering
\includegraphics[width=0.7\textwidth]{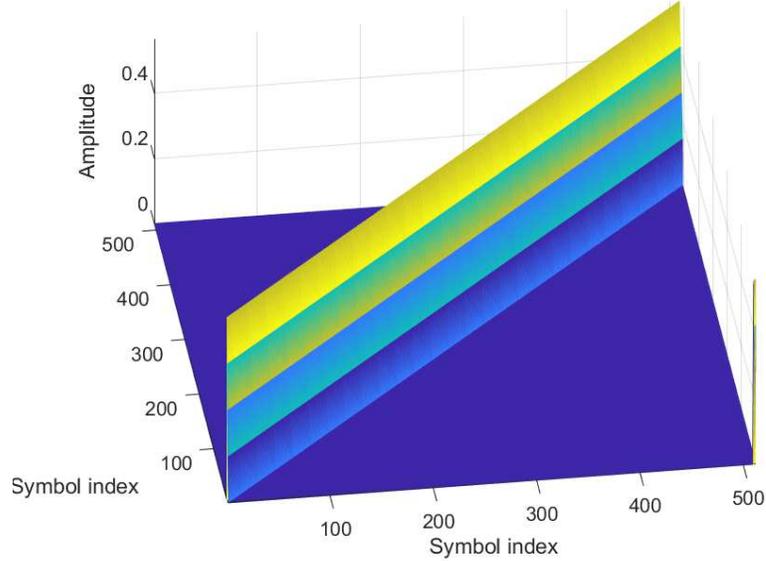}
\caption{TD domain effective channel matrix for OTFS transmissions with integer delay and Doppler indices, where the corresponding channel parameters are given in Table~\ref{C4_channel_parameters}.}
\label{C4_TD_channel_integer}
\centering
\end{figure}

\begin{figure}
\centering
\includegraphics[width=0.7\textwidth]{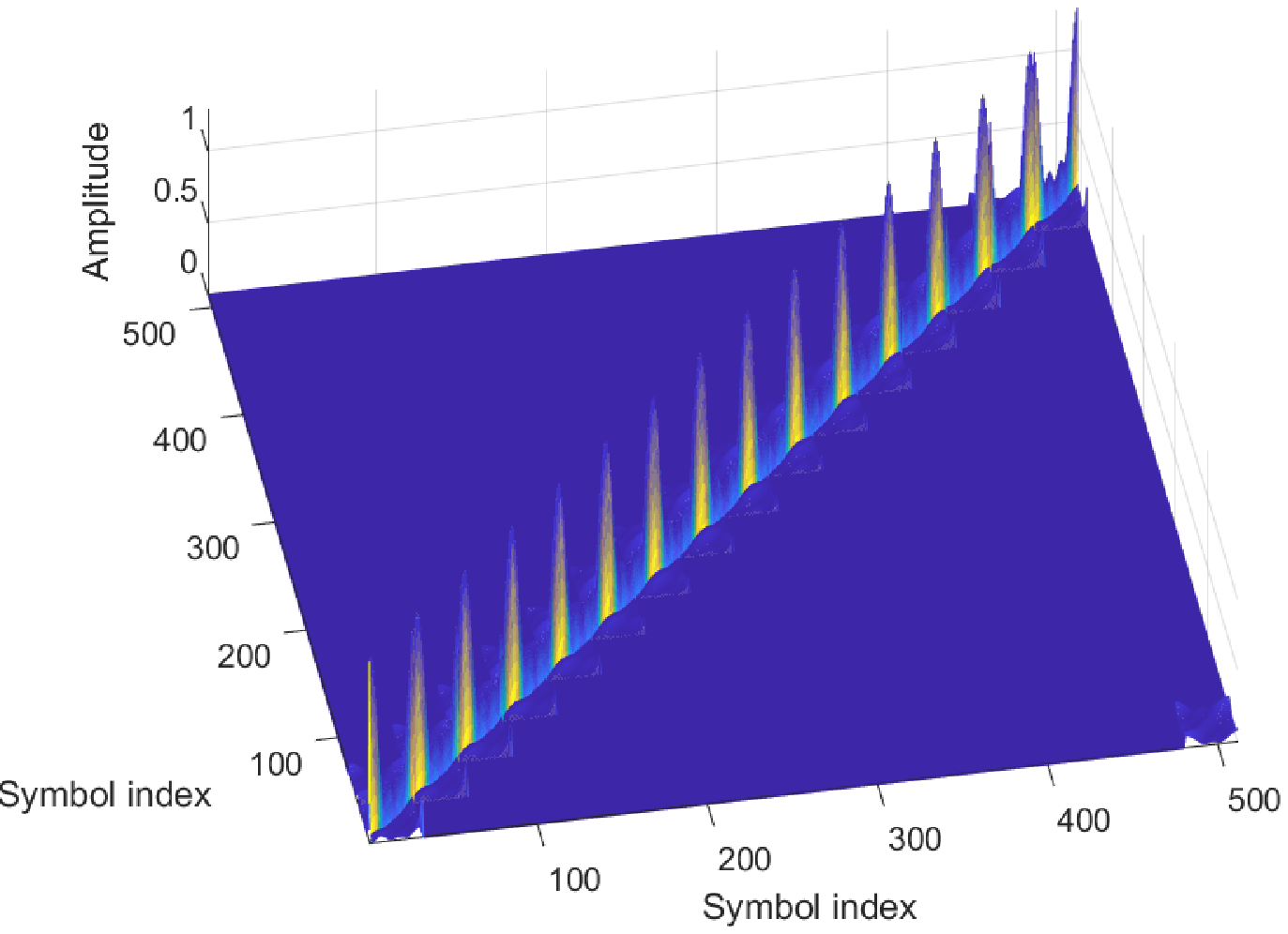}
\caption{TF domain effective channel matrix for OTFS transmissions with integer delay and Doppler indices, where the corresponding channel parameters are given in Table~\ref{C4_channel_parameters}.}
\label{C4_TF_channel_integer}
\centering
\end{figure}

\begin{figure}
\centering
\includegraphics[width=0.7\textwidth]{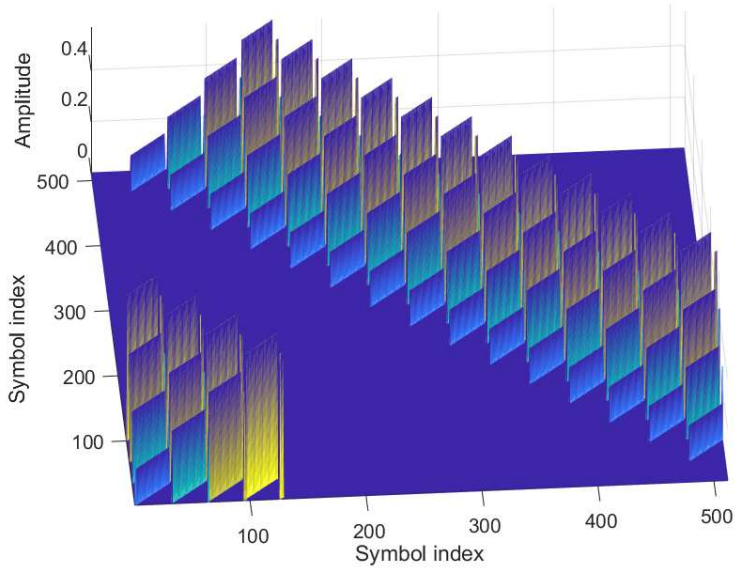}
\caption{DD domain effective channel matrix for OTFS transmissions with integer delay and Doppler indices, where the corresponding channel parameters are given in Table~\ref{C4_channel_parameters}.}
\label{C4_DD_channel_integer}
\centering
\end{figure}

Let us first consider the integer delay and Doppler case, where the corresponding channel matrices for TD, TF, and DD domains are given in Fig.~\ref{C4_TD_channel_integer},~\ref{C4_TF_channel_integer}, and~\ref{C4_DD_channel_integer}, respectively. As implied by Figs.~\ref{C4_TD_channel_integer}, we notice that the non-zero responses in the TD domain channel matrix are banded around the matrix diagonal, where we can see that channel responses associated to four paths are separated. However, it should be noted that the TD domain channel sparsity only holds when different paths have different delay indices. In the case of the same delay indices, the associated paths will inevitably introduce overlapped responses. On the other hand, we notice from Fig.~\ref{C4_TF_channel_integer} that the TF domain channel matrix is a block-diagonal matrix, where the non-zero responses are concentrated around the diagonal blocks and the blocks next to them. This observation can be explained by the fact that there are both ISI and inter-carrier interference (ICI) in the TF domain, where the diagonal blocks contain the ICI, while their adjacent block contains the ISI. We also observe from~\ref{C4_DD_channel_integer} that the DD domain channel matrix is a block matrix. Furthermore, it can be shown that there are at most $P$ non-zero elements in each column/row of the DD domain channel matrix, which indicates that the matrix is sparse and that the paths are also separable. In comparison to the TD domain channel separability, different paths have separate responses as long as either their delay or Doppler indices are different. All the above discussions here are consistent with our understanding in Chapter~3.

\begin{figure}
\centering
\includegraphics[width=0.7\textwidth]{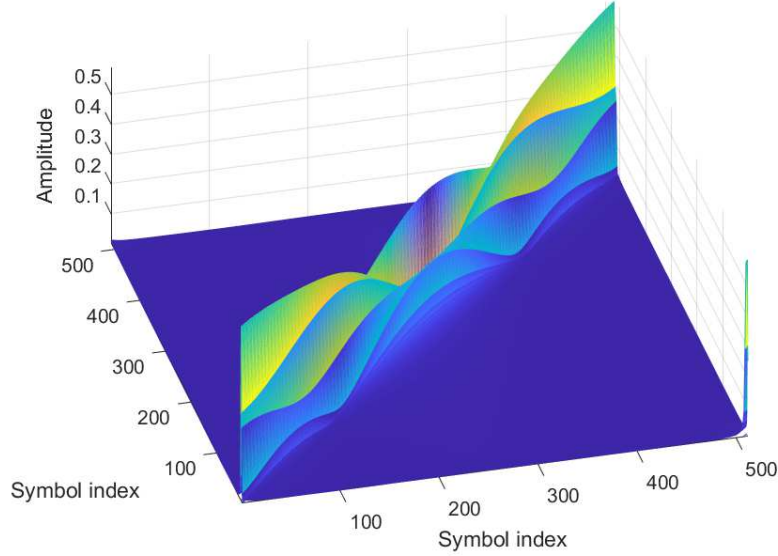}
\caption{TD domain effective channel matrix for OTFS transmissions with fractional delay and integer Doppler indices, where the corresponding channel parameters are given in Table~\ref{C4_channel_parameters}.}
\label{C4_TD_channel_fractional_delay_integer_Doppler}
\centering
\end{figure}

\begin{figure}
\centering
\includegraphics[width=0.7\textwidth]{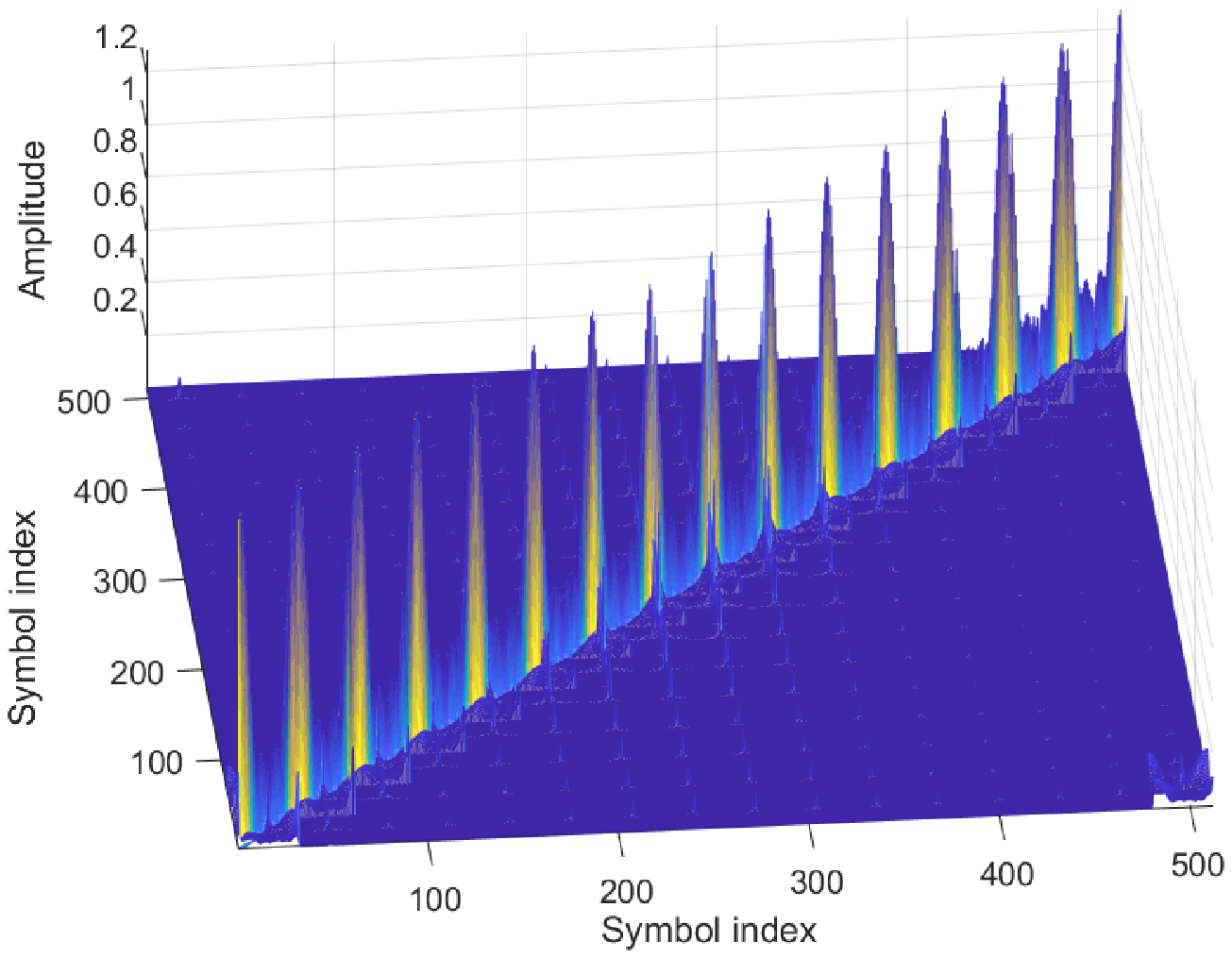}
\caption{TF domain effective channel matrix for OTFS transmissions with fractional delay and integer Doppler indices, where the corresponding channel parameters are given in Table~\ref{C4_channel_parameters}.}
\label{C4_TF_channel_fractional_delay_integer_Doppler}
\centering
\end{figure}

\begin{figure}
\centering
\includegraphics[width=0.7\textwidth]{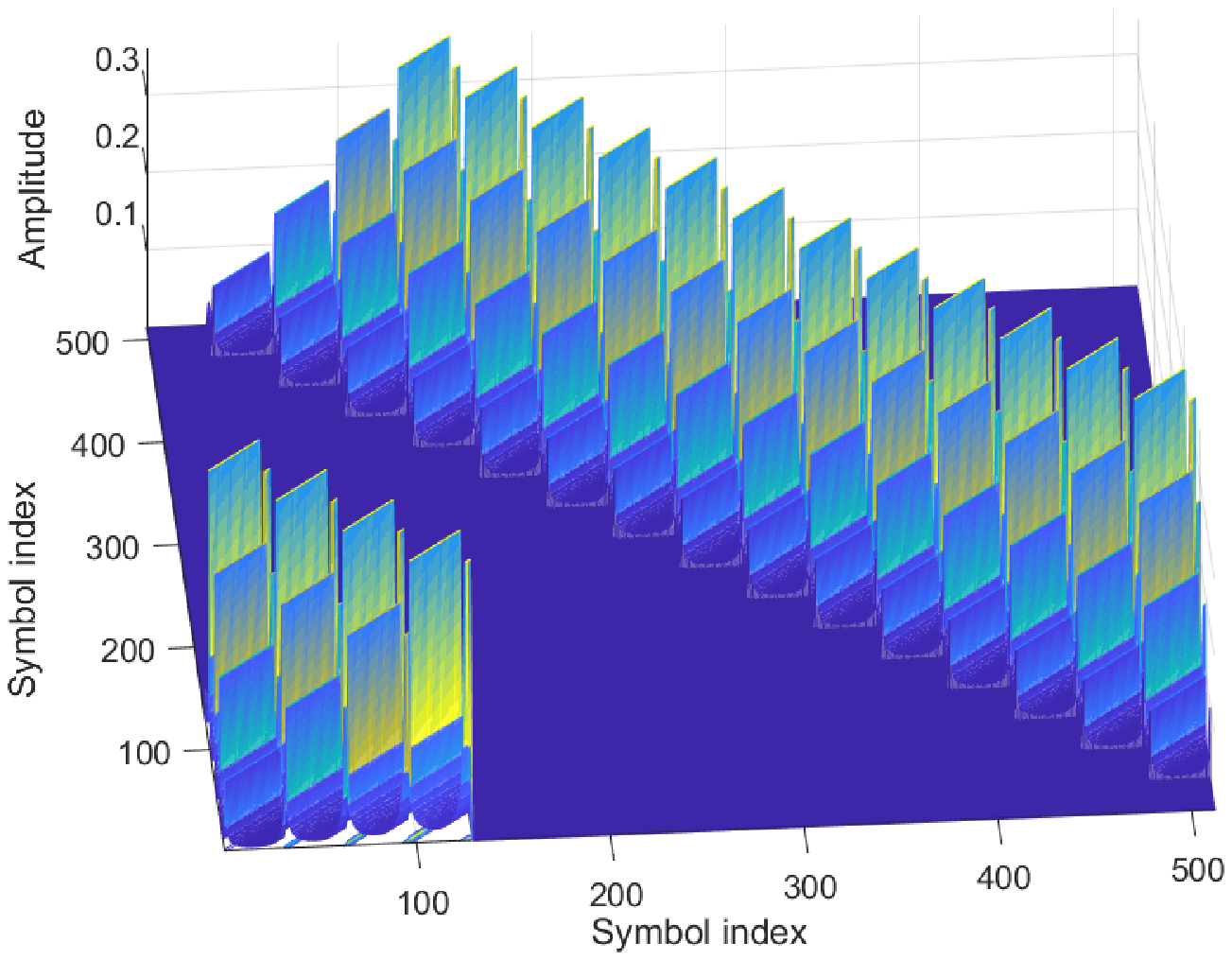}
\caption{DD domain effective channel matrix for OTFS transmissions with fractional delay and integer Doppler indices, where the corresponding channel parameters are given in Table~\ref{C4_channel_parameters}.}
\label{C4_DD_channel_fractional_delay_integer_Doppler}
\centering
\end{figure}

Now let us consider the non-integer delay or Doppler cases. The channel matrices for TD, TF, and DD domains with fractional delay and integer Doppler are given in Figs.~\ref{C4_TD_channel_fractional_delay_integer_Doppler},~\ref{C4_TF_channel_fractional_delay_integer_Doppler}, and~\ref{C4_DD_channel_fractional_delay_integer_Doppler}, respectively. We observe that the overall responses of the effective channel matrices are roughly the same as the corresponding integer cases. However, there are also some differences. In specific, from Fig.~\ref{C4_TD_channel_fractional_delay_integer_Doppler}, we notice that the presence of fractional delay causes power fluctuations in the TD domain channel matrix. Furthermore, we see from Fig.~\ref{C4_TF_channel_fractional_delay_integer_Doppler} that the TF domain channel suffers from power leakage, e.g. there are responses away from the diagonal blocks. Moreover, we also notice that there is power leakage in the DD domain channel matrix from Fig.~\ref{C4_DD_channel_fractional_delay_integer_Doppler}. However, the power leakage is not very obvious compared to the TF domain channel matrix.

\begin{figure}
\centering
\includegraphics[width=0.7\textwidth]{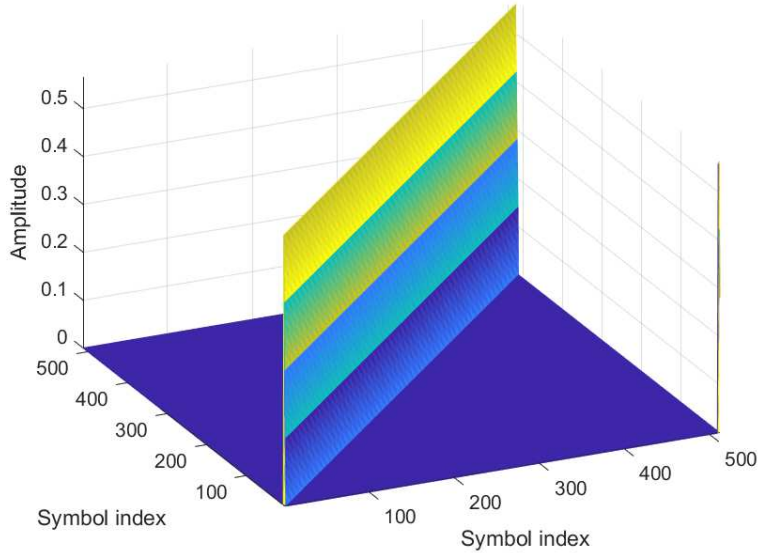}
\caption{TD domain effective channel matrix for OTFS transmissions with integer delay and fractional Doppler indices, where the corresponding channel parameters are given in Table~\ref{C4_channel_parameters}.}
\label{C4_TD_channel_integer_delay_fractional_Doppler}
\centering
\end{figure}

\begin{figure}
\centering
\includegraphics[width=0.7\textwidth]{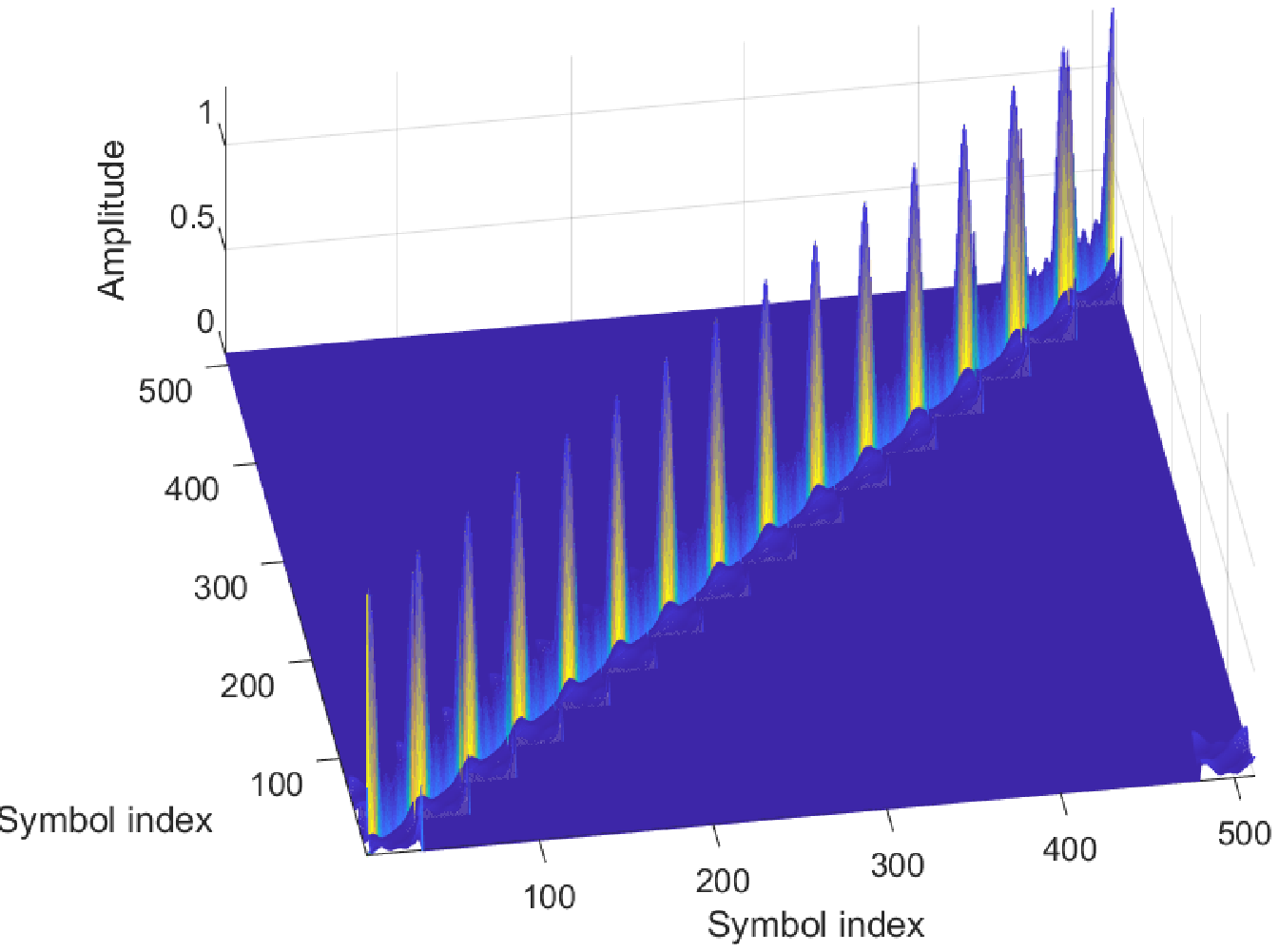}
\caption{TF domain effective channel matrix for OTFS transmissions with integer delay and fractional Doppler indices, where the corresponding channel parameters are given in Table~\ref{C4_channel_parameters}.}
\label{C4_TF_channel_integer_delay_fractional_Doppler}
\centering
\end{figure}

\begin{figure}
\centering
\includegraphics[width=0.7\textwidth]{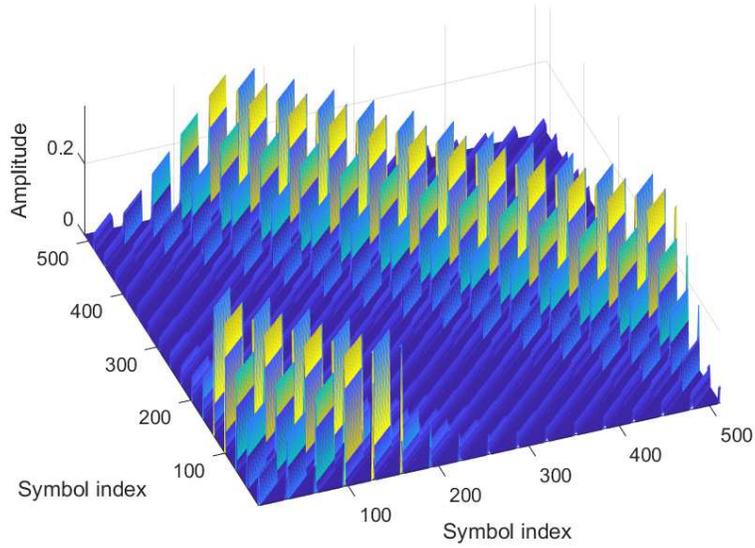}
\caption{DD domain effective channel matrix for OTFS transmissions with integer delay and fractional Doppler indices, where the corresponding channel parameters are given in Table~\ref{C4_channel_parameters}.}
\label{C4_DD_channel_integer_delay_fractional_Doppler}
\centering
\end{figure}

We present the channel matrices for TD, TF, and DD domains with integer delay and fractional Doppler in Figs.~\ref{C4_TD_channel_integer_delay_fractional_Doppler},~\ref{C4_TF_channel_integer_delay_fractional_Doppler}, and~\ref{C4_DD_channel_integer_delay_fractional_Doppler}, respectively.
As can be observed from Fig.~\ref{C4_TD_channel_integer_delay_fractional_Doppler}, the amplitude of the TD domain channel matrix remains the same as the integer case with the considered channel parameters. This is because the fractional Doppler results in phase differences in the TD domain and therefore, the amplitude does not change. On the other hand, we notice that the TF domain channel matrix with fractional Doppler also suffers power leakage as shown in Fig.~\ref{C4_TF_channel_integer_delay_fractional_Doppler}. However, this power leakage is not significant. As implied by Fig.~\ref{C4_DD_channel_integer_delay_fractional_Doppler}, the DD domain channel matrix can be very dense due to the power leakage in the presence of fractional Doppler, although the block matrix structure may still hold.

As the fractional Doppler is commonly presented in practical systems, it is necessary to further discuss the related subjects in this case. It should be noted that the effect of fractional Doppler can be effectively mitigated by adding TF domain windows~\cite{wei2020transmitter}. However, adding TF domain windows can potentially reduce the received SNR~\cite{wei2020transmitter}. As an alternative, it is possible to carry out the signal detection in a cross domain manner~\cite{li2021cross} by exploiting the unitary transformation between different domains. Some of the related details for cross domain iterative OTFS detection will be provided in Chapter~6.

\section{Parameter Selection}
We have discussed the concept of OTFS in the previous sections. In this section, we aim to provide a brief discussion on the modulation parameter selection. It is understood that the parameters for OTFS modulation need to align with the definition of DZT and they should also be chosen according to the underlying channel conditions.
In particular, it can be noticed from the definitions of both ZT and DZT that critical sampling is preferred for the mathematical completeness, which indicates that $T \Delta f=1$.
Furthermore, from the channel point of view, certain delay and Doppler resolutions need to be achieved in order to meet the performance requirements. As such, we summarize the related constraints on the parameter selection in Table~\ref{C4_Parameter_Selection}.

\bgroup
\def\arraystretch{1.5}
\begin{table}[htbp]
\caption{Parameter Selection for OTFS Modulation}
\centering
\begin{tabular}{|c|c|c|}
\hline
Parameter~&~Notation~&~Constraints\\
\hline
Time slot duration~&~$T$~&~$T = \frac{1}{{\Delta f}} = \frac{{{T_{{\rm{frame}}}}}}{N}$\\
\hline
Sub-carrier spacing~&~$\Delta f$~&~$\Delta f = \frac{1}{{ T}} = \frac{{{\rm BW}}}{M}$\\
\hline
Number of delay bins (sub-carriers)~&~$M$~&~$M = \frac{{{\rm{BW}}}}{{\Delta f}} = \frac{T}{{\Delta \tau }}$\\
\hline
Number of Doppler bins(time slots)~&~$N$~&~$N = \frac{{{T_{{\rm{frame}}}}}}{T} = \frac{1}{{T\Delta \nu }}$\\
\hline
Bandwidth~&~${{\rm{BW}}}$~&~${\rm{BW}} = M\Delta f = \frac{1}{{\Delta \tau }}$\\
\hline
Frame duration~&~${T_{{\rm{frame}}}}$~&~${T_{{\rm{frame}}}} = NT = \frac{1}{{\Delta \nu }}$\\
\hline
Delay resolution~&~$\Delta \tau $~&~$\Delta \tau  = \frac{T}{M} = \frac{1}{{{\rm{BW}}}}$\\
\hline
Doppler resolution~&~$\Delta \nu $~&~$\Delta \nu  = \frac{{\Delta f}}{N} = \frac{1}{{{T_{{\rm{frame}}}}}}$\\
\hline
\end{tabular}
\label{C4_Parameter_Selection}
\end{table}
\egroup

To further explain the parameter selection, let us consider the following example of the frame format from Cohere's website~\cite{Cohere_website}.

\textbf{Example 3-3} (\emph{Parameter Selection}):
According to the contents on the Cohere's website, an OTFS frame could consist of $M=500$ sub-carriers and $N=20$ time slots. In the case, where the delay resolution and Doppler resolution of the wireless channel are given by
$\Delta \tau =100 {\rm ns}$ and $\Delta \nu =1 {\rm KHz}$, it can be shown that the total bandwidth ${\rm BW}=10 {\rm MHZ}$ and the frame duration ${T_{{\rm{frame}}}}=1 {\rm ms}$. Based on these parameters, the time slot duration $T$ and the sub-carrier spacing $\Delta f$ can be calculated as $T=50 {\rm{\mu s}}$ and $\Delta f=20 {\rm{KHz}}$.

The above example shows a specific frame format that is suitable for applications operating in wireless channels with $\Delta \tau =100 {\rm ns}$ and $\Delta \nu =1 {\rm KHz}$. On the other hand, it should be noted that with different channel conditions or available TF resources, the above frame format may need to be changed in order to fit the specific transmission scenario. However, this example could be valid for a wide range of applications~\cite{Cohere_website}.

\section{Summary of the Chapter}

In this chapter, we have studied the concept of OTFS modulation from a DZT point of view. In specific, we summarized some important properties of both ZT and DZT. Based on these properties, we derived the OTFS concept, including its transmitter processing, receiver processing, and the input-output relationship. Furthermore, we have also provided the matrix form representation corresponding to our derivations and discussed the properties of effective channel matrices in different domains with various channel conditions. In the end, we have also provided some brief discussions on parameter selection.

    \chapter{Hybrid MAP and PIC Detection for OTFS Modulation}\label{c:literature}
Channel detection is an important part to unleash the potential of OTFS modulation.
In this Chapter, we propose a novel hybrid MAP and
PIC detection (Hybrid-MAP-PIC) algorithm for OTFS modulation~\cite{li2021hybrid}. The proposed Hybrid-MAP-PIC algorithm is motivated by the fact that path gains associated to different paths can be highly diverse and therefore, we can only select the paths with large gains for MAP detection, while performing PIC detection for the remaining paths, in order to reduce the detection complexity without significantly degrading the error performance. In particular, the proposed Hybrid-MAP-PIC algorithm can be viewed as a combination of the
\emph{symbol-wise} MAP algorithm (enumerates all possible interference combinations) and the messaging passing algorithm in~\cite{Raviteja2018interference} (treats all interferences as Gaussian variables) with a reduced complexity.
Specifically, we first derive the near-optimal detection algorithm in the \emph{symbol-wise} MAP sense, according to the framework of SPA~\cite{Kschischang2001factor}.
Then, we propose a partitioning rule to separate the interference into two subsets based on the path gains of the channel.
Based on the partitioning rule, the Hybrid-MAP-PIC algorithm is naturally introduced to exploit the power discrepancy of each subset by performing both MAP and PIC detections.
Simulation results show that the proposed Hybrid-MAP-PIC algorithm outperforms the existing OTFS detection algorithm and only has a marginal performance loss (less than 1 dB) to the near-optimal \emph{symbol-wise} MAP algorithm.
The main contributions of this chapter are summarized as follows:
\begin{itemize}
\item Based on the OTFS system model, we derive the near-optimal \emph{symbol-wise} MAP algorithm according to the framework of the SPA~\cite{Kschischang2001factor}.
\item We propose a Hybrid-MAP-PIC algorithm based on the proposed partitioning rule. Furthermore, we also investigate the error performance of the Hybrid-MAP-PIC algorithm and show that the Hybrid-MAP-PIC algorithm can achieve the same exponential error performance as the near-optimal \emph{symbol-wise} MAP algorithm, but enjoys a reduced-complexity at the cost of an SNR degradation. Meanwhile, we also discuss the selection of the algorithm parameter.
\end{itemize}
\section{Near-Optimal Symbol-Wise MAP Detection}
In this section, we aim to derive the near-optimal \emph{symbol-wise} MAP detection for OTFS systems. Without loss of generality, we consider the OTFS transmissions with integer delay and Doppler and ideal pulse~\cite{Raviteja2018interference}, and the corresponding DD domain input-output relationship can be written in a symbol-wise manner, such as
\begin{equation}
Y_{\rm DD}\left[ {l,k} \right] = \sum\limits_{i = 1}^P {{h_i}{e^{ - j2\pi {\nu _i}{\tau _i}}}X_{\rm DD}\left[ {{{\left[ {l - l_i} \right]}_M},{{\left[ {k - k_i} \right]}_N}} \right]}  + w \left[ {l,k} \right],
\label{C5_DD_model}
\end{equation}
where the related parameters are defined in Chapter~3.
Based on~\eqref{C5_DD_model}, the OTFS detection can be carried out according to the \emph{symbol-wise} MAP rule, i.e.,
\begin{equation}
{\hat X}_{\rm DD}\left[ {l,k} \right] = \arg \mathop {\max }\limits_{{X}_{\rm DD}\left[ {l,k} \right]  \in {\mathbb A}} \Pr \left\{ {{ X}_{\rm DD}\left[ {l,k} \right] |{\bf{Y}_{\rm DD}}} \right\},
\label{C5_MAP_rule}
\end{equation}
where ${\hat X}_{\rm DD}\left[ {l,k} \right] $ is the element at the $l$-th row and $k$-th column in the 2D estimated symbol matrix ${\bf{\hat X}}_{\rm DD}$.
For notational brevity, let us define the following sets.
\begin{align}
\mathbb{H}^{\left( i \right)}&\buildrel \Delta \over = \left\{ {{h_j}\left| {1 \le j \le P,j \ne i} \right.} \right\},\notag\\
\mathbb{Y}_{l,k} &\buildrel \Delta \over =  \left\{ {Y_{\rm DD}\left[{{\left[ {l + l_i} \right]}_M},{{\left[ {k + k_i} \right]}_N}\right]\big| {1 \le i \le P} } \right\},\notag\\
\mathbb{X}_{l,k}^{\left( i \right)}&\!\buildrel \Delta \over = \!\!
 \left\{ {{X_{{\rm{DD}}}}\left[ {\left[ {l + {l_i} - {l_j}} \right]_M,\left[ {k + {k_i} - {k_j}} \right]_N} \right]\left| {1 \le j \le P,j \ne i} \right.} \right\}.\notag
\end{align}
According to~\eqref{C5_DD_model}, it can be shown that the set $\mathbb{Y}_{l,k}$ contains the $P$ received symbols that are associated to the DD domain transmitted symbol $X_{\rm DD}\left[ {l,k} \right]$,
while the set $\mathbb{X}_{l,k}^{\left( i \right)}$ contains $P-1$ DD domain transmitted symbols that are related to the received symbol $\mathbb{Y}_{l,k}\left[ i \right]$.
In particular, the probability $\Pr \left\{ X_{\rm DD}\left[ {l,k} \right]|{{\bf Y}_{\rm DD}} \right\}$ can be factorized with respect to $\mathbb{Y}_{l,k}$ and $\mathbb{X}_{l,k}^{\left( i \right)}$, for which we have the following Theorem.

\textbf{Theorem 4-1} \emph{(Probability Factorization)}:
Assuming that the transmitted symbols in ${\bf{X}}_{\rm DD}$ are independently taking values in the constellation set ${\mathbb{A }}$ with equal probabilities, the \emph{a posteriori} probability of $\Pr \left\{ X_{\rm DD}\left[ {l,k} \right]|{{\bf Y}_{\rm DD}} \right\}$ can be approximated as
\begin{align}
\Pr \left\{ {{X_{{\rm{DD}}}}\left[ {l,k} \right]|{{\bf{Y}}_{{\rm{DD}}}}} \right\} \approx& \prod\limits_{i = 1}^P {\sum\limits_{{\mathbb X}_{l,k}^{\left( i \right)}} {\Pr } } \left\{ {{{\mathbb Y}_{l,k}}\left[ i \right]\left| {{\mathbb X}_{l,k}^{\left( i \right)},{X_{{\rm{DD}}}}\left[ {l,k} \right]} \right.} \right\}\Pr \left\{ {{\mathbb X}_{l,k}^{\left( i \right)}\left| {{\bf{Y}}_{{\rm{DD}}}^{{\notin {{\mathbb Y}_{l,k}}\left[ i \right]}}} \right.} \right\}\notag\\
&\quad\quad\quad\Pr \left\{ {{X_{{\rm{DD}}}}\left[ {l,k} \right]} \right\},
\label{C5_sum_product}
\end{align}
where ${{\bf{Y}}_{{\rm{DD}}}^{{\notin {{\mathbb Y}_{l,k}}\left[ i \right]}}}$ denotes the set of ${\bf{Y}}_{\rm DD}$ excluding the element ${\mathbb{Y}_{l,k}}\left[ i \right]$.

\emph{Proof}: The proof follows the standard SPA. According to the Bayes's rule,~\eqref{C5_MAP_rule} can be expanded as
\begin{equation}
\Pr \left\{ {{X_{{\rm{DD}}}}\left[ {l,k} \right]\left| {{{\bf{Y}}_{{\rm{DD}}}}} \right.} \right\} \propto \Pr \left\{ {{{\bf{Y}}_{{\rm{DD}}}}\left| {{X_{{\rm{DD}}}}\left[ {l,k} \right]} \right.} \right\}\Pr \left\{ {{X_{{\rm{DD}}}}\left[ {l,k} \right]} \right\} .
\label{C5_derivation1}
\end{equation}
Let $\left. {{\mathbb{Y}_{l,k}}} \right|_{i + 1}^P$ denote the vector of the $(i+1)$-th element ${\mathbb{Y}_{l,k}\left[ {i+1} \right]}$ to the $P$-th element ${\mathbb{Y}_{l,k}\left[ {P} \right]}$ of ${\mathbb{Y}_{l,k}}$.
By observing~\eqref{C5_DD_model},~\eqref{C5_derivation1} can be further derived according to the chain rule, which yields
\begin{align}
&\Pr \left\{ {{{\bf{Y}}_{{\rm{DD}}}}\left| {{X_{{\rm{DD}}}}\left[ {l,k} \right]} \right.} \right\}\Pr \left\{ {{X_{{\rm{DD}}}}\left[ {l,k} \right]} \right\}\notag\\
 =& \prod\limits_{i = 1}^P {\Pr \left\{ {{{\mathbb Y}_{l,k}}\left[ i \right]\left| {\left. {{{\mathbb Y}_{l,k}}} \right|_{i + 1}^P,{{\bf{Y}}_{{\rm{DD}}}}\backslash {{\mathbb Y}_{l,k}},{X_{{\rm{DD}}}}\left[ {l,k} \right]} \right.} \right\}} \Pr \left\{ {{X_{{\rm{DD}}}}\left[ {l,k} \right]} \right\}\notag\\
 =&\prod\limits_{i = 1}^P {\sum\limits_{{\mathbb X}_{l,k}^{\left( i \right)}} {\Pr \left\{ {{{\mathbb Y}_{l,k}}\left[ i \right],{\mathbb X}_{l,k}^{\left( i \right)}\left| {\left. {{{\mathbb Y}_{l,k}}} \right|_{i + 1}^P,{{\bf{Y}}_{{\rm{DD}}}}\backslash {{\mathbb Y}_{l,k}},{X_{{\rm{DD}}}}\left[ {l,k} \right]} \right.} \right\}} } \Pr \left\{ {{X_{{\rm{DD}}}}\left[ {l,k} \right]} \right\},
\label{C5_derivation2}
\end{align}
where ${\bf{Y}}\backslash{\mathbb{Y}_{l,k}}$ denotes the complementary set of ${\mathbb{Y}_{l,k}}$ with respect to ${\bf{Y}}_{\rm DD}$.
By considering the chain rule again, we obtain
\begin{align}
&\Pr \left\{ {{{\bf{Y}}_{{\rm{DD}}}}\left| {{X_{{\rm{DD}}}}\left[ {l,k} \right]} \right.} \right\}\Pr \left\{ {{X_{{\rm{DD}}}}\left[ {l,k} \right]} \right\}\notag\\
=&\prod\limits_{i = 1}^P {\sum\limits_{{\mathbb{X}_{l,k}^{\left( i \right)}}} {\Pr \left\{ {{\mathbb{Y}_{l,k}}\left[ i \right]\left| {{\mathbb{X}_{l,k}^{\left( i \right)}},{X_{\rm DD}\left[ {k,l} \right]}} \right.} \right\}} } \Pr \left\{ {\left. {\mathbb{X}_{l,k}^{\left( i \right)}} \right|\left. {{\mathbb{Y}_{l,k}}} \right|_{i + 1}^P,{\bf{Y}}\backslash{\mathbb{Y}_{l,k}}} \right\}\Pr \left\{ {X_{\rm DD}\left[ {k,l} \right]} \right\}.
\label{C5_derivation4}
\end{align}
Finally, by assuming that the elements from ${\mathbb{X}_{l,k}^{\left( i \right)}}$ are independent to the elements from $\left.{{\mathbb{Y}_{l,k}}} \right|_{1}^{i-1}$, we arrive at the conclusion given in Theorem 4-1. Note that the approximation becomes exact when the above assumption is valid, i.e., the corresponding graphical model does not contain any cycles.\hfill $\blacksquare$

\begin{figure}
\centering
\includegraphics[width=0.7\textwidth]{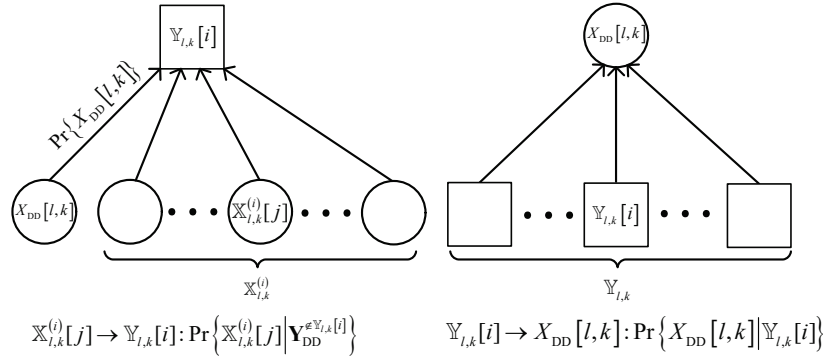}
\caption{The probabilistic graphical model of the near-optimal \emph{symbol-wise} MAP algorithm.}
\label{C5_Factor_graph1}
\centering
\end{figure}

It can be observed that the probability factorization given in~\eqref{C5_sum_product} can be fully characterized by a probabilistic graphical model as shown in Fig.~\ref{C5_Factor_graph1},
according to the SPA{\footnote{According to the frame work of SPA, the approximation in~\eqref{C5_sum_product} becomes exact when the corresponding model does not contain any cycles.}}~\cite{Kschischang2001factor}.
In specific, the function node ${\mathbb{Y}_{l,k}}\left[ i \right]$ receives the messages passed from the variable nodes $X_{\rm DD}\left[ {l,k} \right]$ and ${\mathbb{X}_{l,k}^{\left( i \right)}}[j]$, for ${1 \le j \le P}, j \ne i$, as implied in Fig.~\ref{C5_Factor_graph1}, where the \emph{a prior} probability $\Pr\left\{{X_{\rm DD}\left[ {l,k} \right]}\right\}$ is passed from $X_{\rm DD}\left[ {l,k} \right]$ and the probability $\Pr \left\{ {{\mathbb{X}_{l,k}^{\left( i \right)}[j]}\left| {{\bf{Y}}_{{\rm{DD}}}^{{\notin {{\mathbb Y}_{l,k}}\left[ i \right]}}} \right.} \right\}$ is passed from ${\mathbb{X}_{l,k}^{\left( i \right)}}[j]$.
On the other hand, the message of probability $\Pr \left\{ {\left. {X_{\rm DD}\left[ {l,k} \right]} \right|{\mathbb{Y}_{l,k}}\left[ i \right]} \right\}$ is passed from the
${\mathbb{Y}_{l,k}}\left[ i \right], {1 \le i \le P}$, to the variable node $X_{\rm DD}\left[ {l,k} \right]$.
In particular, we have
\begin{align}
\Pr \left\{ {\big. {{\mathbb{Y}_{l,k}}\left[ i \right]} \big|\mathbb{X}_{l,k}^{\left( i \right)},X_{\rm DD}\left[ {l,k} \right]} \right\} = \frac{1}{{\sqrt {\pi {N_0}} }}
\exp \left( {\! -\! {{\left| {{\mathbb{Y}_{l,k}}\left[ i \right] \!-\! \sum\limits_{\scriptstyle j = 1\hfill\atop
}^{P-1} {{\mathbb{H}^{\left( i \right)}[j]}\mathbb{X}_{l,k}^{\left( i \right)}\left[ j \right] \!- \!{h_i}X_{\rm DD}\left[ {l,k} \right]} } \right|}^2}\!\Bigg /\!\!{N_0}} \right),
\label{C5_probability1}
\end{align}
and
\begin{equation}
\Pr \left\{ {X_{\rm DD}\left[ {l,k} \right]\left| {{\bf{Y}}_{{\rm{DD}}}^{{\notin {{\mathbb Y}_{l,k}}\left[ i \right]}}} \right.} \right\} \propto \prod\limits_{\scriptstyle j = 1\hfill\atop
\scriptstyle j \ne i\hfill}^P {\Pr \left\{ {\left. {X_{\rm DD}\left[ {l,k} \right]} \right|{\mathbb{Y}_{l,k}}\left[ j \right]} \right\}} .
\label{C5_probability2}
\end{equation}
The detailed procedures for the near-optimal \emph{symbol-wise} MAP algorithm are summarized in Algorithm~4-1.

\begin{algorithm}[htb]
\caption*{\textbf{Algorithm} \textbf{4-1} $\quad$Symbol-by-symbol Detection for DD Domain Symbols}
\hspace*{0.02in} {\bf Input:}
${\bf{Y}}_{\rm DD}$, ${\mathbb{A}}$, $M$, $N$, $P$, the maximum number of iteration $I_{\max}$, \emph{a prior} probability $\Pr\left\{{X_{\rm DD}\left[ {l,k} \right]}\right\}$, and the channel state information $h_i$, $k_i$, $l_i$, for $1\le i \le P$.\\
\hspace*{0.02in} {\bf Output:}
${\bf{\hat X}}_{\rm DD}$ and $\Pr \left\{ {\left. {X_{\rm DD}\left[ {l,k} \right]} \right|{\bf{Y}}_{\rm DD}} \right\}$.\\
\hspace*{0.02in} {\bf Steps:}
\begin{algorithmic}[1]
\State \textbf{for} $I=1;I \le I_{\max }$ \textbf{do}
\State $\quad$\textbf{for} $i=1;i \le P$ \textbf{do}
\State $\quad$$\quad$\textbf{for} $k=0;k \le N-1$ \textbf{do}
\State $\quad$$\quad$$\quad$\textbf{for} $l=0;l \le M-1$ \textbf{do}

\State $\quad$$\quad$$\quad$$\quad$ Enumerate all combinations of ${\mathbb{X}_{l,k}^{\left( i \right)}}$.
\State $\quad$$\quad$$\quad$$\quad$ For each possible combination of ${\mathbb{X}_{l,k}^{\left( i \right)}}$, compute~\eqref{C5_probability1} and $\Pr \left\{ {{\mathbb{X}_{l,k}^{\left( i \right)}[j]}\left| {{\bf{Y}}_{{\rm{DD}}}^{{\notin {{\mathbb Y}_{l,k}}\left[ i \right]}}} \right.} \right\}$ based on~\eqref{C5_probability2}.
\State $\quad$$\quad$$\quad$$\quad$ Compute $\Pr \left\{ {X_{\rm DD}\left[ {l,k} \right]|{\bf{Y}}_{\rm DD}} \right\}$ by using~\eqref{C5_sum_product}.
\State $\quad$$\quad$$\quad$$\quad$ Make hard decision of $X_{\rm DD}\left[ l,k \right]$ based on~\eqref{C5_MAP_rule}.
\State $\quad$$\quad$$\quad$\textbf{end for}
\State $\quad$$\quad$\textbf{end for}
\State $\quad$\textbf{end for}
\State \textbf{end for}
\State \Return ${\bf{\hat X}}_{\rm DD}$ and $\Pr \left\{ {\left. {X_{\rm DD}\left[ {l,k} \right]} \right|{\bf{Y}}_{\rm DD}} \right\}$.
\end{algorithmic}
\end{algorithm}
\textbf{Remarks 1:}
Since the algorithm is derived from the \emph{symbol-wise} MAP sense, in principle, it is able to achieve the optimal error performance of OTFS systems in terms of the BER, if the corresponding graphical model does not contain any cycles.
Meanwhile, it can be observed from lines $5$ and $6$ of Algorithm~4-1 that the detection complexity of the near-optimal \emph{symbol-wise} MAP algorithm is exponential to the number of paths $P$. However, such a complexity becomes prohibitive when the number of paths is significantly large. Therefore, we propose a reduced-complexity detection method based on the near-optimal \emph{symbol-wise} MAP algorithm in the following section in order to strike a balance between the detection complexity and performance.

\section{Hybrid MAP and PIC Detection}
It can be observed from~\eqref{C5_sum_product} that the detection complexity mainly arises from the enumeration of all possible combinations of ${\mathbb{X}_{l,k}^{\left( i \right)}}$.
To reduce the detection complexity, we intend to separate the set ${\mathbb{X}_{l,k}^{\left( i \right)}}$ into two subsets and only enumerate the combinations of one subset.
Let $L$ be the size of the subset, whose total combinations are to be enumerated. Then, we have the following Proposition.

\begin{figure}
\centering
\includegraphics[width=0.5\textwidth]{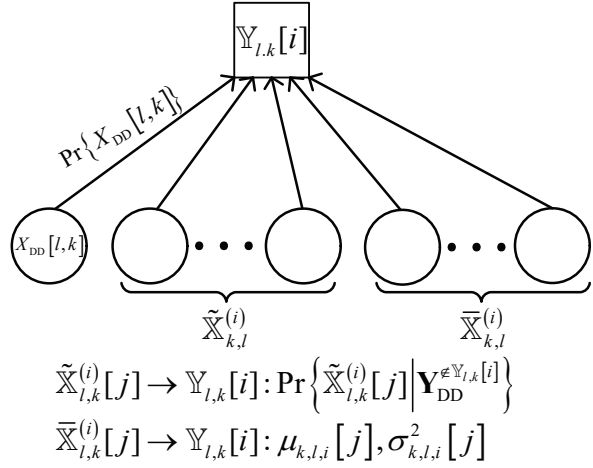}
\caption{The probabilistic graphical model of the Hybrid-MAP-PIC algorithm.}
\label{C5_Factor_graph2}
\centering
\end{figure}

\textbf{Proposition 1} \emph{(Partitioning Rule)}:
Assuming that the path gains in $\mathbb{H}^{\left( i \right)}$ are sorted in descending order according to its power, i.e., $|h_k|^2>|h_j|^2$, if $k<j$, the two subsets of ${\mathbb{X}_{l,k}^{\left( i \right)}}$ are defined as
\begin{align}
\tilde{\mathbb{X}}_{l,k}^{\left( i \right)}&\buildrel \Delta \over = \left\{ {{\mathbb{X}}_{l,k}^{\left( i \right)}\left[ j \right]\left| {1 \le j \le L} \right.} \right\}
\end{align}
and
\begin{align}
\bar{\mathbb{X}}_{l,k}^{\left( i \right)}&\buildrel \Delta \over = \left\{ {{\mathbb{X}}_{l,k}^{\left( i \right)}\left[ j \right]\left| {L+1 \le j \le P-1} \right.} \right\},
\end{align}
respectively.
Naturally, we propose to perform MAP detection for the subset $\tilde{\mathbb{X}}_{l,k}^{\left( i \right)}$, while perform PIC for the subset $\bar{\mathbb{X}}_{l,k}^{\left( i \right)}$, since the subset $\bar{\mathbb{X}}_{l,k}^{\left( i \right)}$ may have less impact on the overall error performance compared with that of the subset $\tilde{\mathbb{X}}_{l,k}^{\left( i \right)}$. More specifically, we assume that the elements in  $\bar{\mathbb{X}}_{l,k}^{\left( i \right)}$ are Gaussian variables \cite{Raviteja2018interference}, i.e., $\bar{\mathbb{X}}_{l,k}^{\left( i \right)}\left[ j \right]$ has a mean ${\mu _{k,l,i}}\left[ j \right]$ and variance $\sigma _{k,l,i}^2\left[ j \right]$.
In particular, the values of ${\mu _{k,l,i}}\left[ j \right]$ and $\sigma _{k,l,i}^2\left[ j \right]$ can be derived from the \emph{a posteriori} probabilities in the previous iteration.
Thus, we can modify~\eqref{C5_probability1} as
\begin{align}
&\Pr \left\{ {{\mathbb{Y}_{l,k}}\left[ i \right]\left| {\tilde{\mathbb{X}}_{l,k}^{\left( i \right)},\bar{\mathbb{X}}_{l,k}^{\left( i \right)},X_{\rm DD}\left[ {l,k} \right]} \right.} \right\} \notag\\
= &\frac{1}{{\sqrt {\pi \left( {{N_0} + {\sigma ^2}} \right)} }}\exp \left( { - \left| {{\mathbb{Y}_{l,k}}\left[ i \right] - {h_i}X_{\rm DD}\left[ {l,k} \right] - \sum\nolimits_{j = 1}^L {{{\mathbb H}^{\left( i \right)}}\left[ j \right]\tilde{\mathbb{X}}_{l,k}^{\left( i \right)}[j]} } \right.} \right.\notag\\
&\left. {{{{{\left. { - \!\!\sum\nolimits_{j = 1}^{P - L - 1} \!\!{{{\mathbb H}^{\left( i \right)}}\!\left[ {j + L} \right]\mathbb{E}\left\{ \bar{\mathbb{X}}_{l,k}^{\left( i \right)}[j] \right\}} } \right|}^2\!}} \mathord{\left/
 {\vphantom {{{{\left. { - \!\!\sum\nolimits_{j = 1}^{P - L - 1}\!\! {{{\mathbb H}^{\left( i \right)}}\!\left[ {j + L} \right]\mathbb{E}\left\{ \bar{\mathbb{X}}_{l,k}^{\left( i \right)}[j] \right\}} } \right|}^2\!}} {\left( {{N_0} + {\sigma ^2}}\!\! \right)}}} \right.
 \!\kern-\nulldelimiterspace}\!\! {\left( {{N_0} + {\sigma ^2}} \right)}}} \right),
\label{C5_probability3}
\end{align}
where ${\sigma ^2} = \sum\nolimits_{j = 1}^{P - L - 1} {{{\left| {{{\mathbb{H}}^{\left( i \right)}}\left[ {j + L} \right]} \right|}^2}\sigma _{k,l,i}^2\left[ j \right]} $.
In particular, according to~\eqref{C5_probability3}, the corresponding graphical model{\footnote{Since changing~\eqref{C5_probability1} to~\eqref{C5_probability3} does not alter the message passing from function nodes to variable nodes, we only show the message passing from variable nodes to function nodes in Fig.~\ref{C5_Factor_graph2}.}} is given in Fig.~\ref{C5_Factor_graph2}.
Specifically, for each variable node ${\tilde{\mathbb{X}}_{l,k}^{\left( i \right)}}[j]$ that contributes major interference to the detected symbol $X_{\rm DD}[l,k]$, the probability $\Pr \left\{ {{\tilde{\mathbb{X}}_{l,k}^{\left( i \right)}[j]}\left| {{\bf{Y}}_{{\rm{DD}}}^{{\notin {{\mathbb Y}_{l,k}}\left[ i \right]}}} \right.} \right\}$ is passed to the function node ${\mathbb{Y}_{l,k}}\left[ i \right]$. On the other hand, for each variable node ${\bar{\mathbb{X}}_{l,k}^{\left( i \right)}}[j]$ that contributes less interference to the detected symbol $X_{\rm DD}[l,k]$, the corresponding mean ${\mu _{k,l,i}}\left[ j \right]$ and variance $\sigma _{k,l,i}^2\left[ j \right]$ are passed to the function node ${\mathbb{Y}_{l,k}}\left[ i \right]$.
The details of the Hybrid-MAP-PIC algorithm are summarized in Algorithm~2.
\begin{algorithm}[htb]
\caption*{\textbf{Algorithm} \textbf{4-2} $\quad$Hybrid MAP and PIC Detection Algorithm}
\hspace*{0.02in} {\bf Input:}
$\bf{Y}$, ${\mathbb{A}}$, $M$, $N$, $P$, the maximum number of iteration $I_{\max}$, \emph{a prior} probability $\Pr\left\{{x\left[ {k,l} \right]}\right\}$, and the channel state information $h_i$, ${l_\nu ^{\left( i \right)}}$, ${l_\tau ^{\left( i \right)}}$, for $1\le i \le P$.\\
\hspace*{0.02in} {\bf Output:}
${\bf{\hat X}}$ and $\Pr \left\{ {\left. {x\left[ {k,l} \right]} \right|\bf{Y}} \right\}$.
\hspace*{0.02in} {\bf Steps:}
\begin{algorithmic}[1]
\State \textbf{for} $I=1;I \le I_{\max }$ \textbf{do}
\State $\quad$\textbf{for} $i=1;i \le P$ \textbf{do}
\State $\quad$$\quad$\textbf{for} $k=0;k \le N-1$ \textbf{do}
\State $\quad$$\quad$$\quad$\textbf{for} $l=0;l \le M-1$ \textbf{do}

\State $\quad$$\quad$$\quad$$\quad$ Calculate the mean and variance of each elements in $\bar{\mathbb{X}}_{l,k}^{\left( i \right)}$.
\State $\quad$$\quad$$\quad$$\quad$ Enumerate all combinations of $\tilde{\mathbb{X}}_{l,k}^{\left( i \right)}$.
\State $\quad$$\quad$$\quad$$\quad$ For each possible combination of $\tilde{\mathbb{X}}_{l,k}^{\left( i \right)}$, compute~\eqref{C5_probability3} and $\Pr \left\{ {{\tilde{\mathbb{X}}_{l,k}^{\left( i \right)}[j]}\left| {{\bf{Y}}_{{\rm{DD}}}^{{\notin {{\mathbb Y}_{l,k}}\left[ i \right]}}} \right.} \right\}$ based on~\eqref{C5_probability2}.
\State $\quad$$\quad$$\quad$$\quad$ Compute $\Pr \left\{ {X_{\rm DD}\left[ {l,k} \right]|{\bf{Y}}} \right\}$ by using~\eqref{C5_sum_product}.
\State $\quad$$\quad$$\quad$$\quad$ Make hard decision of $X_{\rm DD}\!\left[ {k,l} \right]$ based on~\eqref{C5_MAP_rule}.
\State $\quad$$\quad$$\quad$\textbf{end for}
\State $\quad$$\quad$\textbf{end for}
\State $\quad$\textbf{end for}
\State \textbf{end for}
\State \Return ${\bf{\hat X}}_{\rm DD}$ and $\Pr \left\{ {\left. {X_{\rm DD}\left[ {k,l} \right]} \right|{\bf{Y}}_{\rm DD}} \right\}$.
\end{algorithmic}
\end{algorithm}

\subsection{Error Performance Analysis}
Let us first focus on the error performance of uncoded systems.
As the proposed Hybrid-MAP-PIC algorithm performs MAP detection for $\tilde{\mathbb{X}}_{l,k}^{\left( i \right)}$ and performs PIC for
$\bar{\mathbb{X}}_{l,k}^{\left( i \right)}$, it is obvious that a genie-aided error performance lower bound can be derived by assuming that the PIC perfectly cancels out the interference caused by the transmitted symbols corresponding to $\bar{\mathbb{X}}_{l,k}^{\left( i \right)}$, i.e., ${\mu _{k,l,i}}\left[ j \right]={X_{{\rm{DD}}}}\left[ {{{\left[ {l + {l_i} - {l_j}} \right]}_M},{{\left[ {k + {k_i} - {k_j}} \right]}_N}} \right]$ and $\sigma _{k,l,i}^2\left[ j \right]=0$.
With this assumption, the corresponding error performance should be the same as that for the near-optimal
\emph{symbol-wise} MAP algorithm as indicated by~\eqref{C5_probability3}.

On the other hand, an error performance upper-bound can be derived by considering the worst case scenario, where PIC cannot cancel any interference at all, i.e., ${\mu _{k,l,i}}\left[ j \right]=0$ and $\sigma _{k,l,i}^2\left[ j \right]=E_s$, with $E_s$ being the average symbol energy.
With this assumption,~\eqref{C5_probability3} becomes
\begin{align}
&\Pr \left\{ {{\mathbb{Y}_{l,k}}\left[ i \right]\left| {\tilde{\mathbb{X}}_{l,k}^{\left( i \right)},\bar{\mathbb{X}}_{l,k}^{\left( i \right)},X_{\rm DD}\left[ {l,k} \right]} \right.} \right\} = \frac{1}{{\sqrt {\pi \left( {{N_0} + {\sigma ^2}} \right)} }}\notag\\
&\exp \left( { - \frac{{{{\left| {{{\mathbb Y}_{l,k}}\left[ i \right] - {h_i}X_{\rm DD}\left[ {l,k} \right] - \sum\nolimits_{j = 1}^L {{{\mathbb H}^{\left( i \right)}}\left[ j \right]\tilde {\mathbb X}_{l,k}^{\left( i \right)}\left[ j \right]} } \right|}^2}}}{{{N_0} + {\sigma ^2}}}} \right),
\label{C5_probability4}
\end{align}
where ${\sigma ^2} = \sum\nolimits_{j = 1}^{P - L - 1} {{{\left| {{{\mathbb{H}}^{\left( i \right)}}\left[ {j + L} \right]} \right|}^2}E_s} $.
Comparing~\eqref{C5_probability1} and~\eqref{C5_probability4}, we notice that in the worst case,
Hybrid-MAP-PIC algorithm suffers from some SNR degradations inevitably due to the inaccuracy of the mean and variance of the symbol estimates.
To quantify the SNR degradation, we define the effective signal-to-interference-plus-noise ratio (SINR) associated to ${\mathbb{Y}_{l,k}}\left[ i \right]$ by
\begin{align}
{\rm{SINR}}_{\rm eff}^{\left( i \right)} = \frac{{ {{{\left| {{h_i}} \right|}^2}{E_s}} }}{{{N_0} + \sum\nolimits_{j = 1}^{P - L - 1} {{{\left| {{{\mathbb H}^{\left( i \right)}}\left[ {j + L} \right]} \right|}^2}{E_s}} }}, \label{C5_SINR}
\end{align}
where we can see that the effective SINR can be different for different ${\mathbb{Y}_{l,k}}\left[ i \right]$ depending on ${{{\mathbb H}^{\left( i \right)}}}$.
Therefore, selecting an appropriate $L$ can ensure a good performance-complexity trade-off, which will be discussed in the following subsection.
However, it should be noted that Hybrid-MAP-PIC algorithm only changes the probability calculation in~\eqref{C5_probability1} but does not change the \emph{a posteriori} probability factorization in~\eqref{C5_sum_product}. Therefore, the proposed Hybrid-MAP-PIC algorithm can achieve the same diversity gain as the near-optimal \emph{symbol-wise} MAP algorithm despite the value of $L$, because the Hybrid-MAP-PIC algorithm still detects each symbol from $P$ received symbols.

We note that the message passing algorithm{\footnote{Here, we consider the message passing algorithm in~\cite{Raviteja2018interference} without damping.}} proposed in~\cite{Raviteja2018interference} is a special case of the proposed Hybrid-MAP-PIC algorithm with $L=0$. Therefore, with $L>0$,
Hybrid-MAP-PIC algorithm can outperform the message passing algorithm in~\cite{Raviteja2018interference}.
Furthermore, the Hybrid-MAP-PIC algorithm does not need \emph{damping}, which is different from the algorithm in~\cite{Raviteja2018interference}.
In specific, the algorithm in~\cite{Raviteja2018interference} requires damping because the Gaussian approximation is applied to all $P-1$ related symbols for detecting each symbol, which can cause severe error propagation. In contrast, the proposed Hybrid-MAP-PIC algorithm collects the most of the interference energy from the $L$ related symbols. Therefore, the error prorogation is significantly reduced and thus damping is not necessary.

As for coded OTFS systems, the performance loss of the Hybrid-MAP-PIC algorithm compared with the near-optimal
\emph{symbol-wise} MAP algorithm is expected to be marginal, since the applied channel code usually can provide reliable estimates of the transmitted symbols. In general, iterations between the detector and channel-decoder are required in order to feed back useful information from the decoder to the detector.

\subsection{Complexity Analysis and Parameter Selection}
According to~\eqref{C5_probability3}, it can be shown that the detection complexity of the proposed Hybrid-MAP-PIC algorithm only increases exponential to $L$ instead of $P$ for the near-optimal \emph{symbol-wise} MAP algorithm. A detailed detection complexity analysis for both Hybrid-MAP-PIC algorithm and near-optimal \emph{symbol-wise} MAP algorithm is given in Table~\ref{C5_Complexity}, where ${\cal X}$ denotes the cardinality of the constellation set.
\begin{table}[htbp]
\caption{Detection Complexity Analysis}
\centering
\begin{tabular}{|c|c|}
\hline
Detection method~&~Asymptotical complexity\\
\hline
Hybrid-MAP-PIC~&~$MN{\cal O}\left( {{{\cal X}^L} + \left( {P - L} \right){\cal X}} \right)$\\
\hline
\emph{symbol-wise} MAP~&~$MN{\cal O}\left( {{{\cal X}^P}} \right)$\\
\hline
\end{tabular}
\label{C5_Complexity}
\end{table}
As impled by Table~\ref{C5_Complexity}, the proposed Hybrid-MAP-PIC algorithm generally enjoys a lower detection complexity compared to the near-optimal \emph{symbol-wise} MAP algorithm. Furthermore, in the case where $L=P$, both two algorithms share the same detection complexity.

In order to achieve a good performance-complexity trade-off, we can select different $L$ for the algorithm depending on the channel conditions.
In particular,~\eqref{C5_SINR} shows that the error performance degradation is mainly due to the SNR reduction, where a small $L$ and a large $P$ can decrease the effective SINR.
Therefore, for a good error performance, $L$ should increase with increased value of $P$, in order to collect more interference energy. Our simulation results show that $L=P/2$ usually performs quite well in terms of the error performance. On top of that, in the low SNR regime, we can select a large                                                                                                                    $L$, because the error performance is dominated by the received SNR. On the other hand, in the high SNR regime, a small $L$ is usually sufficient,
as the overall error performance is dominated by the corresponding diversity gain that is irrelevant with $L$.


\section{Numerical Results}
Without loss of generality, we set $N=100$ and $M=150$ for OTFS modulation, where the DD domain transmitted symbols are binary phase shift keying (BPSK) modulated.
In our simulations, all the detection algorithms have a maximum iteration number as $I_{\rm{max}}=10$, and we assume the perfect CSI at the receiver.
Specifically, we set the maximum delay index as $l_{\max }=10$ and the maximum Doppler index as $k_{\max }=6$,
which is corresponding to a relative user equipment speed around $250$ km/h with $4$ GHz carrier frequency and $15$ kHz sub-carrier spacing, and the total bandwidth is $2.25$ MHz. For each channel realization, we randomly select the
delay and Doppler indices such that $ - {k_{\max }} \le k_i \le {k_{\max }}$ and $0 \le l_i \le {l_{\max }}$. The fading coefficients are generated according to an exponential power-delay profile with an exponent $0.1$.

\begin{figure}
\centering
\includegraphics[width=0.7\textwidth]{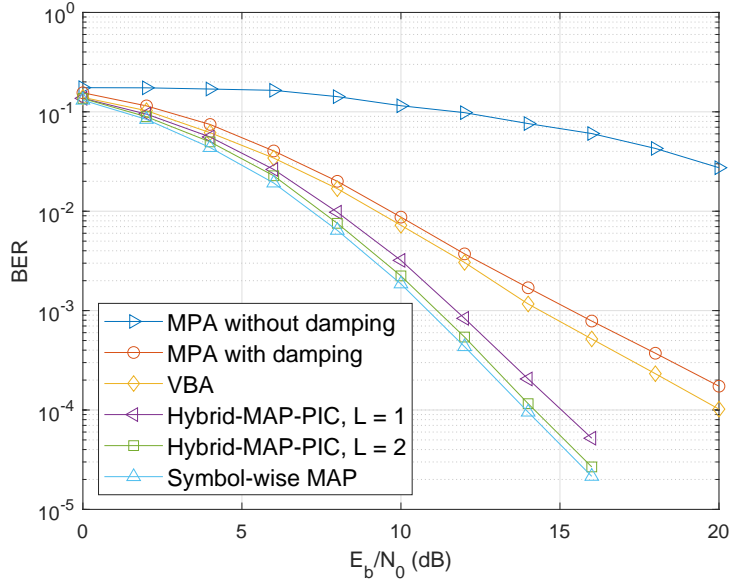}
\caption{BER performance of the proposed Hybrid-MAP-PIC algorithm for uncoded OTFS systems with $P=4$, compared with the message passing algorithm in~\cite{Raviteja2018interference}, the variational Bayes algorithm in~\cite{Yuan2019simple}, and the near-optimal \emph{symbol-wise} MAP algorithm.}
\label{C5_uncoded}
\centering
\end{figure}


Fig.~\ref{C5_uncoded} demonstrates the BER performance of the proposed Hybrid-MAP-PIC algorithm for uncoded OTFS systems with $P=4$, compared with that of the message passing algorithm~\cite{Raviteja2018interference} (denoted by ``'MPA'' in Fig.~\ref{C5_uncoded}), the variational Bayes algorithm~\cite{Yuan2019simple} (denoted by ``VBA'' in Fig.~\ref{C5_uncoded}), and the near-optimal \emph{symbol-wise} MAP algorithm, respectively.
As can be observed from the figure, the message passing algorithm without damping cannot achieve a good error performance even with high SNRs.
Yet, with a damping factor $0.7$, the message passing algorithm can obtain a BER around $1 \times {10^{ - 4}}$ at the SNR at $20$ dB, which is similar to that of the variational Bayes algorithm. Furthermore, we observe that the proposed Hybrid-MAP-PIC algorithm outperforms the message passing algorithm and the variational Bayes algorithm by roughly $6$ dB and $4$ dB
with $L=1$, respectively, at a BER around $1 \times {10^{ - 4}}$.
Furthermore, we notice that the BER performance of the proposed Hybrid-MAP-PIC algorithm improves with the increase of $L$ and approaches that of the near-optimal \emph{symbol-wise} MAP algorithm.
More importantly, we observe that the Hybrid-MAP-PIC algorithm and the near-optimal \emph{symbol-wise} MAP algorithm share roughly the same BER slope. This fact indicates that the Hybrid-MAP-PIC algorithm
obtains the same diversity gain as the near-optimal \emph{symbol-wise} MAP algorithm, which is consistent with our analysis.


\begin{figure}
\centering
\includegraphics[width=0.7\textwidth]{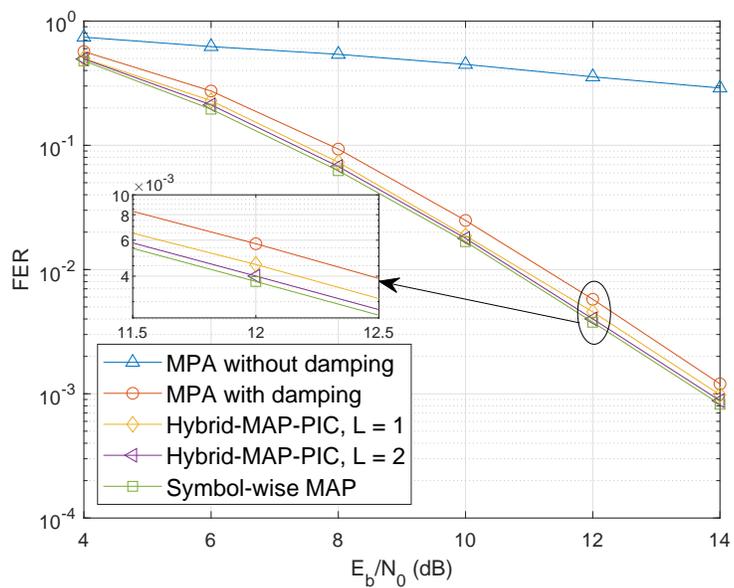}
\caption{FER performance of the proposed Hybrid-MAP-PIC algorithm for convolutionally coded OTFS systems with $P=4$, comparing with the message passing algorithm in~\cite{Raviteja2018interference} and the near-optimal \emph{symbol-wise} MAP algorithm.}
\label{C5_conv_coded}
\centering
\end{figure}

Fig.~\ref{C5_conv_coded} demonstrates the frame error rate (FER) performance of the proposed Hybrid-MAP-PIC algorithm for convolutionally coded OTFS systems with $P=4$, where the
 rate-$1/2$~recursive~(7, 5) convolutional code is applied and Turbo equalization with $2$ iterations is performed. For comparison, we also include the FER performance of the message passing algorithm in~\cite{Raviteja2018interference} and the near-optimal \emph{symbol-wise} MAP algorithm. Similar to the previous figure, the message passing algorithm cannot obtain a good error performance without damping even with Turbo equalization.
However, with a damping factor $0.7$, the message passing algorithm exhibits a good error performance approaching that of Hybrid-MAP-PIC algorithm with $L=1$. Furthermore, the
proposed Hybrid-MAP-PIC algorithm with $L=2$ achieves almost the same error performance as the near-optimal \emph{symbol-wise} MAP algorithm. These observations are consistent with our discussions about the error performance of the Hybrid-MAP-PIC algorithm for coded OTFS systems.
In particular, we notice that the Hybrid-MAP-PIC algorithm has only a marginal performance loss compared with the near-optimal \emph{symbol-wise} MAP algorithm even with $L=1$.
This indicates that the proposed Hybrid-MAP-PIC algorithm can provide a good trade-off between the error performance and complexity.

\section{Summary of the Chapter}
In this section, we proposed a Hybrid-MAP-PIC algorithm for OTFS modulation. We first derived the near-optimal \emph{symbol-wise} MAP algorithm. Then, we proposed a partitioning rule which divides the related symbols into two sets according to their associated channel path gains.
The Hybrid-MAP-PIC algorithm was proposed to exploit the power discrepancy between the two subsets.
Simulation results verified the our analysis of the error performance of the Hybrid-MAP-PIC algorithm and showed that the Hybrid-MAP-PIC algorithm can provide a good trade-off between the error performance and detection complexity.

    \chapter{Cross Domain Iterative Detection for OTFS Modulation}\label{c:literature}

In this chapter, we propose a novel cross domain iterative detection algorithm for OTFS modulation, where the extrinsic information is passed between the TD domain and DD domain via the corresponding unitary transformations~\cite{li2021cross}.
This special detection structure is motivated by the connection between the orthogonality and message passing based on the view from the orthogonal AMP (OAMP) algorithm~\cite{Ma2017orthogonal} and AMP with a unitary transformation (UAMP)~\cite{guo2015approximate,yuan2021amp,luo2021unitary}.
In particular, the rationale behind our work is that the unitary transformation between the TD domain and DD domain allows the detection/estimation errors in one domain to be principally \emph{orthogonal} to the detection/estimation errors in the other domain, which can suppress the correlation between the input errors and the output errors for each domain detection/estimation~\cite{Ma2015Turbo,Ma2017orthogonal,Haifeng2022ICC}.
Therefore, the detection/estimation instability due to the positive feedback effect, which usually caused by the correlation between the input and output errors for the iterative receivers, can be greatly reduced during iterations, and this stability in return improves the error performance~\cite{Ma2015Turbo,Ma2017orthogonal}.
In specific, we apply conventional L-MMSE estimator for equalization in the TD domain, while adopt a simple \emph{symbol-by-symbol} detection algorithm in the DD domain. Interestingly, by combining such two basic methods, the proposed algorithm shows promising error performance even in very severe and complex fractional Doppler cases.
Note that conventional iterative receiver improves the error performance by iteratively exchanging the extrinsic information between two disjoint components, such as the detector and decoder, via interleaving/de-interleaving. In contrast, our proposed iterative algorithm exchanges the extrinsic information within the same component (detector) but between two orthogonal domains via unitary transformation. In other words, we separate the OTFS detection problem into two parts, corresponding to the TD domain (performing de-correlation) and the DD domain (performing de-noising) and iteratively exchange the extrinsic information via unitary transformation.
In particular, we provide a detailed proof that explains the advantage of the proposed algorithm and briefly discuss its detection complexity.
The main contributions of this chapter are summarized as follows.
\begin{itemize}
\item Based on the TD domain channel sparsity, we propose a cross domain iterative detection algorithm that operates in both the TD domain and DD domain. Furthermore, according to the unitary property of the domain transformation, the details of cross domain message passing are discussed. We also prove that the symbol-by-symbol DD domain detection cannot provide any extrinsic information itself and therefore the extrinsic information need to be calculated in the TD domain.
\item We provide theoretical analysis for the MSE performance of the proposed algorithm based on the state evolution~\cite{Ma2015Turbo}. In particular, we derive the average MSE per iteration and prove that the proposed algorithm can converge after a few iterations. Furthermore, we show that the average MSE for both the TD domain and the DD domain share the same value in the convergence and the error performance improvement brought by the cross domain message passing is originated from the non-Gaussian distribution of practical DD domain signal constellations.
\item We investigate the effective DD domain signal-to-noise ratio (SNR) in order to study the error performance in the convergence. We prove that the corresponding effective SNR can approach the maximum receiver SNR for a given fading channel, i.e., almost all the energy from different paths are collected and coherently combined, which indicates that the proposed algorithm can theoretically approach the error performance of the optimal maximum-likelihood sequence estimation (MLSE) detection.
    On the other hand, we show that the computational complexity of the proposed algorithm is much lower compared to that of the MLSE detection thanks to the efficient implementation based on the DFT kernel for cross domain message passing between the TD domain and DD domain.
    We also show that the overall detection complexity of the proposed algorithm does not increase in the presence of fractional Doppler.
\item The error performance of the proposed algorithm is evaluated by numerical simulations. Simulation results agree with our theoretical analysis and demonstrate a significant performance improvement compared to conventional OTFS detection methods.
\end{itemize}
It should be noted that although the proposed algorithm builds upon the core idea of OAMP and UAMP, and has a similar structure as that in~\cite{Ma2015Turbo}, it has important contributions compared to~\cite{Ma2015Turbo} making it suitable for solving the unique challenges of OTFS detection.
\begin{itemize}
\item As OAMP, UAMP, and the work~\cite{Ma2015Turbo} are specifically designed for compressed sensing issues, they cannot be directly extended to the considered OTFS detection.
Compared to~\cite{Ma2015Turbo}, the proposed algorithm applies ML-based DD domain detection with respect to the DD domain constellation and is able to produce both hard-decision outputs and high quality soft information, which is more suitable for practical communications, especially for coded systems.
\item The work~\cite{Ma2015Turbo} requires the transmitted symbols are Bernoulli-Gaussian distributed with a sufficient number of zeros (sparsity constraint), which is often unrealistic for practical communication systems. Compared to~\cite{Ma2015Turbo}, the proposed algorithm does not have this requirement. As a matter of fact, we prove that the performance improvement of the cross domain message passing actually comes from the practical non-Gaussian DD domain constellation constraint.
\item The algorithm in~\cite{Ma2015Turbo} only works with a partial DFT matrix. Instead of intentionally introducing a partial DFT matrix, the proposed algorithm takes advantages of the unitary transformation associated with the OTFS modulation for message passing. By doing so, the proposed algorithm exploits the channel characteristics from both
    the TD domain and DD domain and is thus robust for various channel conditions.
\end{itemize}

\section{Cross Domain Iterative Detection for OTFS Modulation}
Without loss of generality, we focus on the rectangular pulse shaped point-to-point OTFS transmissions with integer delay and fractional Doppler.
For notational brevity, let us slightly abuse the notations in the following presentation. Let us denote the DD domain and TD domain information symbol vectors by $\bf x$ and $\bf z$, respectively.
Meanwhile, the TD domain and DD domain received symbol vectors are denoted by $\bf r$ and $\bf y$, respectively.
Notice that information symbols are multiplexed on the DD domain while the channel sparsity holds in the TD domain. Therefore, we propose an iterative detector which performs de-correlation in the TD domain to eliminate the effects of fading, multi-path, and Doppler, while performing de-noising in the DD domain to reduce the inaccuracy due to the noise.
In particular, we assume that the entries in $\bf{x}$ independently take values from a normalized constellation set ${\mathbb{A}}$ with equal probabilities, and thus we have ${\mathbb E}\left[ {{\bf{x}}{{\bf{x}}^{\rm{H}}}} \right] = {{\bf{I}}_{MN}}$.
Then, it can be shown that entries in the TD domain OTFS signal vector $\bf{z}$ are also independent from each other, i.e.,
\begin{align}
{\mathbb E}\left[ {{\bf{z}}{{\bf{z}}^{\rm{H}}}} \right] = \left( {{\bf{F}}_N^{\rm{H}} \otimes {{\bf{I}}_M}} \right){\mathbb E}\left[ {{\bf{x}}{{\bf{x}}^{\rm{H}}}} \right]\left( {{{\bf{F}}_N} \otimes {{\bf{I}}_M}} \right)={{\bf{I}}_{MN}}.\label{C6_E_z_z}
\end{align}
\begin{figure*}
\centering
\includegraphics[width=0.8\textwidth]{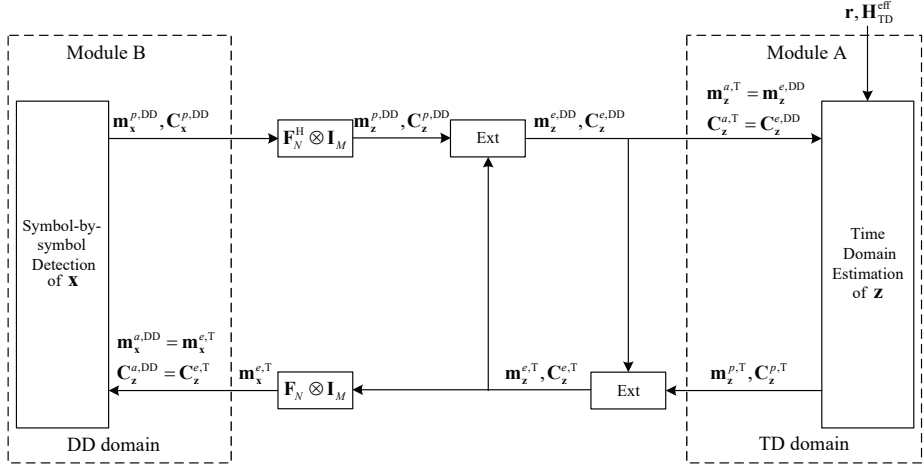}
\caption{The diagram of the proposed OTFS detector.}
\label{C6_Detection_diagram}
\centering
\end{figure*}
Furthermore, recalling~\eqref{C4_TD_domain_channel}, it is not hard to notice that each TD domain OTFS signal is a superposition of $N$ DD domain OTFS symbols with specific phase rotations, due to the spreading effect of IFFT. Therefore, in practical OTFS systems, where $N$ is sufficiently large, the TD domain OTFS signal behaves like a Gaussian variable according to the central limit Theorem.
We assume that the entries in $\bf{z}$ are independent and identically distributed~(i.i.d.) Gaussian variables with a unit variance according to~\eqref{C6_E_z_z}.
We consider a detection structure that consists of two individual modules corresponding to the TD domain and the DD domain, namely, module A and module B, as shown in Fig.~\ref{C6_Detection_diagram}.
In specific, module A aims to estimate the TD domain OTFS signals $\bf{z}$ and passes the estimates to module B for the detection of DD domain OTFS symbols $\bf{x}$.
On the other hand, module B carries out a simple symbol-by-symbol detection for the DD domain transmitted symbol vector $\bf{x}$ according to the estimates of TD domain OTFS signal $\bf{z}$.
In Fig.~\ref{C6_Detection_diagram}, ``Ext'' denotes the calculation of extrinsic information, while ``${{\bf{F}}_N \otimes {{\bf{I}}_M}}$'' and ``${{\bf{F}}_N^{\rm{H}} \otimes {{\bf{I}}_M}}$'' denote the corresponding unitary transformation from the TD domain to the DD domain and from the DD domain to the TD domain, respectively.
For notational brevity, we apply superscripts denoting the \emph{a priori}, \emph{a posteriori}, and extrinsic information of the means (denoted by $\bf m$) and covariance matrices (denoted by $\bf C$) for the TD domain OTFS signal vector $\bf{z}$ and DD domain symbol vector $\bf{x}$, respectively, as shown in Table~\ref{C6_Detection_parameters}.
\begin{table*}[htbp]
\centering
\small
\caption{Notations of Parameters for Cross Domain Iterative Detection}
\begin{tabular}{|l|l|l|l|l|l|l|}
\hline
  & \multicolumn{3}{c|}{TD Domain}           & \multicolumn{3}{c|}{DD Domain}             \\ \cline{2-7}
  & \emph{a priori} & \emph{a posteriori} & extrinsic (from) & \emph{a priori} & \emph{a posteriori} & extrinsic (from) \\ \hline
$\bf z$ & ${\bf{m}}_{\bf{z}}^{a,{\rm{T}}},{\bf{C}}_{\bf{z}}^{a,{\rm{T}}}$& ${\bf{m}}_{\bf{z}}^{p,{\rm{T}}},{\bf{C}}_{\bf{z}}^{p,{\rm{T}}}$&${\bf{m}}_{\bf{z}}^{e,{\rm{T}}},{\bf{C}}_{\bf{z}}^{e,{\rm{T}}}$&${\bf{m}}_{\bf{z}}^{a,{\rm{DD}}},{\bf{C}}_{\bf{z}}^{a,{\rm{DD}}}$& ${\bf{m}}_{\bf{z}}^{p,{\rm{DD}}},{\bf{C}}_{\bf{z}}^{p,{\rm{DD}}}$&${\bf{m}}_{\bf{z}}^{e,{\rm{DD}}},{\bf{C}}_{\bf{z}}^{e,{\rm{DD}}}$\\ \hline
$\bf x$ & ${\bf{m}}_{\bf{x}}^{a,{\rm{T}}},{\bf{C}}_{\bf{x}}^{a,{\rm{T}}}$& ${\bf{m}}_{\bf{x}}^{p,{\rm{T}}},{\bf{C}}_{\bf{x}}^{p,{\rm{T}}}$&${\bf{m}}_{\bf{x}}^{e,{\rm{T}}},{\bf{C}}_{\bf{x}}^{e,{\rm{T}}}$&${\bf{m}}_{\bf{x}}^{a,{\rm{DD}}},{\bf{C}}_{\bf{x}}^{a,{\rm{DD}}}$& ${\bf{m}}_{\bf{x}}^{p,{\rm{DD}}},{\bf{C}}_{\bf{x}}^{p,{\rm{DD}}}$&${\bf{m}}_{\bf{x}}^{e,{\rm{DD}}},{\bf{C}}_{\bf{x}}^{e,{\rm{DD}}}$\\ \hline
\end{tabular}
\label{C6_Detection_parameters}
\end{table*}

Before we introduce the details, we note that the proposed algorithm is based on the idea of iteratively updating the extrinsic information based on a unitary transformation,
where the multiplication of a diagonal covariance matrix of the variables of interest and the underlying unitary matrix is usually performed for each iteration. However, an observation for this type of algorithms is that the diagonal covariance matrix usually becomes a non-diagonal matrix after the multiplication, even if the corresponding variables are independently distributed. This is because the diagonal entries of the covariance matrix may not be of the same value due to the channel conditions, such as distortion and noise. A common solution for this issue is to simply discard the non-diagonal entries (treated as zeros)~\cite{Ma2015Turbo,guo2011concise}, according to the fact that unitary transformation does not affect independency between variables. In the proposed algorithm, we will adopt this approach and will also use a different approach (as will be discussed in Section~5.1.3 and Section~5.5.1) for the related unitary transformations.

\subsection{Module A: L-MMSE Estimator for TD Domain OTFS Signal}
To estimate the TD domain OTFS signal $\bf{z}$, we apply the conventional L-MMSE estimator in module A.
In specific, the estimation is based on the TD domain received symbol vector $\bf r$ and the TD domain
effective channel ${{\bf{H}}_{\rm{TD}}^{{\rm{eff}}}}$ with the aid of the \emph{a priori} information, ${\bf{m}}_{\bf{z}}^{a,{\rm{T}}}$ and ${\bf{C}}_{\bf{z}}^{a,{\rm{T}}}$, that is fed back from module B.
It should be noted that ${\bf{C}}_{\bf{z}}^{a,{\rm{T}}}$ is a diagonal matrix due to the i.i.d. assumption and it is initialized as ${\bf I}_{MN}$ for the first iteration.
Therefore, it is straightforward to obtain the L-MMSE estimation matrix ${{\bf{W}}_{{\rm{MMSE}}}}$ as~\cite{kay1993fundamentals}
\begin{align}
{{\bf{W}}_{{\rm{MMSE}}}} = {\bf{C}}_{\bf{z}}^{a,{\rm{T}}}{\left( {{\bf{H}}_{\rm{TD}}^{{\rm{eff}}}} \right)^{\rm{H}}}{\left( {{\bf{H}}_{\rm{TD}}^{{\rm{eff}}}{\bf{C}}_{\bf{z}}^{a,{\rm{T}}}{{\left( {{\bf{H}}_{\rm{TD}}^{{\rm{eff}}}} \right)}^{\rm{H}}} + {N_0}{{\bf{I}}_{MN}}} \right)^{ - 1}}.
\label{C6_MMSE_W}
\end{align}
Furthermore, it can be shown that the \emph{a posteriori} estimation output ${\bf{m}}_{\bf{z}}^{p, {\rm T}}$ of $\bf z$ is given by
\begin{align}
{\bf{m}}_{\bf{z}}^{p,{\rm{T}}}
= &{\bf{m}}_{\bf{z}}^{a,{\rm{T}}} + {{\bf{W}}_{{\rm{MMSE}}}}\left( {{\bf{r}} - {\bf{H}}_{\rm{TD}}^{{\rm{eff}}}{\bf{m}}_{\bf{z}}^{a,{\rm{T}}}} \right) \notag\\
= & {\bf{m}}_{\bf{z}}^{a,{\rm{T}}}\! + \!{\bf{C}}_{\bf{z}}^{a,{\rm{T}}}{\left( {{\bf{H}}_{\rm{TD}}^{{\rm{eff}}}} \right)^{\rm{H}}}{\left( {{\bf{H}}_{\rm{TD}}^{{\rm{eff}}}{\bf{C}}_{\bf{z}}^{a,{\rm{T}}}{{\left( {{\bf{H}}_{\rm{TD}}^{{\rm{eff}}}} \right)}^{\rm{H}}} + {N_0}{{\bf{I}}_{MN}}} \right)^{ - 1}}
\left( {{\bf{r}} - {\bf{H}}_{\rm{TD}}^{{\rm{eff}}}{\bf{m}}_{\bf{z}}^{a,{\rm{T}}}} \right),
\label{C6_MMSE_output_mean}
\end{align}
and the \emph{a posteriori} covariance matrix ${\bf{C}}_{\bf{z}}^{p, {\rm T}}$ associated with $\bf z$ is given by
\begin{align}
{\bf{C}}_{\bf{z}}^{p, {\rm T}}= {\bf{C}}_{\bf{z}}^{a,{\rm{T}}}-{\bf{C}}_{\bf{z}}^{a,{\rm{T}}}{\left( {{\bf{H}}_{\rm{TD}}^{{\rm{eff}}}} \right)^{\rm{H}}}{\left( {{\bf{H}}_{\rm{TD}}^{{\rm{eff}}}{\bf{C}}_{\bf{z}}^{a,{\rm{T}}}{{\left( {{\bf{H}}_{\rm{TD}}^{{\rm{eff}}}} \right)}^{\rm{H}}} + {N_0}{{\bf{I}}_{MN}}} \right)^{ - 1}}{\bf{H}}_{\rm{TD}}^{{\rm{eff}}}{\bf{C}}_{\bf{z}}^{a,{\rm{T}}}.
\label{C6_MMSE_output_variance}
\end{align}
It should be noted that the diagonal entries of ${\bf{C}}_{\bf{z}}^{p, {\rm T}}$ are the \emph{a posteriori} MSEs of the estimates of $\bf z$ after the L-MMSE estimation,
while the non-diagonal entries can be discarded (treated as zeros), because only the diagonal entries are of interest according to the i.i.d. assumption~\cite{Ma2015Turbo}.
The details of the L-MMSE estimation are summarized in Algorithm~5-1.

\begin{algorithm}[htb]
\caption*{\textbf{Algorithm}~5-1 L-MMSE Estimation for TD Domain OTFS Signal $\bf z$}
\hspace*{0.02in} {\bf Input:}
$\bf r$, ${{\bf{H}}_{\rm{TD}}^{{\rm{eff}}}}$, ${\bf{m}}_{\bf{z}}^{a,{\rm{T}}}$ and ${\bf{C}}_{\bf{z}}^{a,{\rm{T}}}$\\
\hspace*{0.02in} {\bf Steps:}
\begin{algorithmic}[1]
\State Compute the L-MMSE estimator matrix ${{\bf{W}}_{{\rm{MMSE}}}}$ by~\eqref{C6_MMSE_W}.
\State Calculate the estimation output ${\bf{m}}_{\bf{z}}^{p,{\rm{T}}}$ by~\eqref{C6_MMSE_output_mean}.
\State Calculate the MSE matrix ${\bf{C}}_{\bf{z}}^{p,{\rm{T}}}$ by~\eqref{C6_MMSE_output_variance}.
\State \Return ${\bf{m}}_{\bf{z}}^{p,{\rm{T}}}$ and ${\bf{C}}_{\bf{z}}^{p,{\rm{T}}}$.
\end{algorithmic}
\label{C6_L_MMSE_algorithm}
\end{algorithm}

\subsection{Cross Domain Message Passing: from TD Domain to DD Domain}
Based on the TD domain L-MMSE estimation, we can obtain the \emph{a posteriori} information of TD domain OTFS signal vector $\bf z$. However, in order to perform the iterative detection, it is important to pass the extrinsic information rather than the \emph{a posteriori} information between two the modules.
Let us define ${\bf{m}}_{\bf{z}}^{e,{\rm{T}}}$ and ${{\bf{C}}_{\bf{z}}^{e,{\rm{T}}}}$ as the extrinsic mean and covariance matrix from the L-MMSE estimation. Then, according to the concise formulas in~\cite{guo2011concise}, we have
\begin{align}
{\bf{C}}_{\bf{z}}^{e,{\rm{T}}}&= {\left( {{{\left( {{\bf{C}}_{\bf{z}}^{p,{\rm{T}}}} \right)}^{ - 1}} - {{\left( {{\bf{C}}_{\bf{z}}^{a,{\rm{T}}}} \right)}^{{\rm{ - 1}}}}} \right)^{ - 1}} ,\label{C6_c_z_ext_A}\\
{\bf{m}}_{\bf{z}}^{e,{\rm{T}}} &= {\bf{C}}_{\bf{z}}^{e,{\rm{T}}}\left( {{{\left( {{\bf{C}}_{\bf{z}}^{p,{\rm{T}}}} \right)}^{ - 1}}{\bf{m}}_{\bf{z}}^{p,{\rm{T}}} - {{\left( {{\bf{C}}_{\bf{z}}^{a,{\rm{T}}}} \right)}^{ - 1}}{\bf{m}}_{\bf{z}}^{a,{\rm{T}}}} \right) . \label{C6_m_z_ext_A}
\end{align}
According to our DD domain detection formulation as will be discussed in the coming subsection, we need to obtain the DD domain \emph{a priori} mean of $\bf x$, i.e., ${\bf{m}}_{\bf{x}}^{a,{\rm{DD}}}$,
and DD domain \emph{a priori} covariance matrix of $\bf z$, i.e., ${\bf{C}}_{\bf{z}}^{a,{\rm{DD}}}$.
In specific, based on the TD domain extrinsic mean of $\bf z$, we can obtain the TD domain extrinsic mean of $\bf x$, which will then be forwarded to module B as the \emph{a priori} information, i.e.,
\begin{align}
{\bf{m}}_{\bf{x}}^{a,{\rm{DD}}} ={\bf{m}}_{\bf{x}}^{e,{\rm{T}}}= \left( {{{\bf{F}}_N} \otimes {{\bf{I}}_M}} \right){\bf{m}}_{\bf{z}}^{e,{\rm{T}}}. \label{C6_m_x_pri}
\end{align}
On the other hand, the extrinsic covariance matrix ${{\bf{C}}_{\bf{z}}^{e,{\rm{T}}}}$ will be directly passed to module B as \emph{a priori} information as well.
We have ${\bf{C}}_{\bf{z}}^{a,{\rm{DD}}}={\bf{C}}_{\bf{z}}^{e,{\rm{T}}}$.
With ${\bf{m}}_{\bf{x}}^{a,{\rm{DD}}}$ and  ${\bf{C}}_{\bf{z}}^{a,{\rm{DD}}}$ in hand, the DD domain detection is now ready to perform.

\subsection{Module B: Symbol-by-symbol Detection for DD Domain Symbols}
By considering the relationship between the TD domain OTFS signals $\bf z$ and DD domain OTFS symbols $\bf x$ in~\eqref{C4_TD_transmitted_symbol_vec}, we can
formulate the DD domain detection problem by
\begin{align}
{\bf{m}}_{\bf{z}}^{e,{\rm{T}}} = \left( {{\bf{F}}_N^{\rm{H}} \otimes {{\bf{I}}_M}} \right){\bf{x}} + {\bf{\hat w}}, \label{C6_DD_detection_from_TD}
\end{align}
where $\bf{\hat w}$ are white Gaussian noise samples with zero mean and the variance of the $k$-th entry in $\bf{\hat w}$ is ${{C}}_{\bf{z}}^{a,{\rm{DD}}}[k,k]$.
The justification of~\eqref{C6_DD_detection_from_TD} is necessary and it is discussed in the Section~5.5.1~of this chapter.
According to the ML detection rule, the detection output $\bf {\hat x}$ should satisfy
\begin{align}
{\bf{\hat x}} = \arg \mathop {\max }\limits_{\bf{x}} \Pr \left( {{\bf{m}}_{\bf{z}}^{e,{\rm{T}}}\left| {\bf{x}} \right.} \right),
\label{C6_ML_rule}
\end{align}
where
\begin{align}
\Pr \left( {{\bf{m}}_{\bf{z}}^{e,{\rm{T}}}\left| {\bf{x}} \right.} \right) \! \!
\propto &\exp \left( { - {{\left( {{\bf{m}}_{\bf{z}}^{e,{\rm{T}}} - \left( {{\bf{F}}_N^{\rm{H}} \otimes {{\bf{I}}_M}} \right){\bf{x}}} \right)}^{\rm{H}}}} \right.\left. {{{\left( {{\bf{C}}_{\bf{z}}^{a,{\rm{DD}}}} \right)}^{ - 1}} \! \!\left( {{\bf{m}}_{\bf{z}}^{e,{\rm{T}}} - \left( {{\bf{F}}_N^{\rm{H}} \otimes {{\bf{I}}_M}} \right){\bf{x}}} \right)} \right) . \label{C6_MAP_rule_Forney}
\end{align}
In particular, the probability factorization based on the form of~\eqref{C6_MAP_rule_Forney} is referred to as the Forney observation model, which was firstly introduced by Forney in~\cite{forney1972maximum}. The MLSE detection complexity of the Forney observation model based on~\eqref{C6_MAP_rule_Forney} is exponential to the number of non-zero entries in each row of the matrix ${{\bf{F}}_N^{\rm{H}} \otimes {{\bf{I}}_M}}$. However, as the DFT matrix ${\bf{F}}_N$ is a dense matrix, the detection complexity of the Forney observation model can be very high.
Therefore, we consider a different probability factorization that is equivalent to~\eqref{C6_MAP_rule_Forney} but only requires a linear detection complexity by taking advantage of the unitary property of the matrix ${{\bf{F}}_N^{\rm{H}} \otimes {{\bf{I}}_M}}$.

Let us separate the ISFFT kernel ${{\bf{F}}_{{\rm{ISFFT}}}} \buildrel \Delta \over ={{\bf{F}}_N^{\rm{H}} \otimes {{\bf{I}}_M}}$ into row vectors, i.e.,
\begin{align}
{{\bf{F}}_{{\rm{ISFFT}}}}= {\left[ {{{\left( {{{\bf{F}}_{{\rm{ISFFT}}}}\left[ 0 \right]} \right)}^{\rm{T}}},{{\left( {{{\bf{F}}_{{\rm{ISFFT}}}}\left[ 1 \right]} \right)}^{\rm{T}}},...,{{\left( {{{\bf{F}}_{{\rm{ISFFT}}}}\left[ MN-1 \right]} \right)}^{\rm{T}}}} \right]^{\rm{T}}}.
\end{align}
Noticing that ${{\bf{C}}_{\bf{z}}^{a,{\rm{DD}}}}$ is a diagonal matrix according to its definition and ${\bf{m}}_{\bf{x}}^{a,{\rm{DD}}} = \left( {{{\bf{F}}_N} \otimes {{\bf{I}}_M}} \right){\bf{m}}_{\bf{z}}^{e,{\rm{T}}}$, \eqref{C6_MAP_rule_Forney} can be equivalently expressed by~\eqref{C6_MAP_rule_Ungerboeck},
\begin{align}
&\Pr \left( {{\bf{m}}_{\bf{z}}^{e,{\rm{T}}}\left| {\bf{x}} \right.} \right)\notag\\
 \propto& \exp \left( { - \sum\limits_{k = 0}^{MN - 1} {\frac{1}{{C_{\bf{z}}^{a,{\rm{DD}}}\left[ {k,k} \right]}}{{\left| {{{\bf{F}}_{{\rm{ISFFT}}}}\left[ k \right]{\bf{m}}_{\bf{x}}^{a,{\rm{DD}}} - {{\bf{F}}_{{\rm{ISFFT}}}}\left[ k \right]{\bf{x}}} \right|}^2}} } \right)\notag\\
\propto &\exp \left( { \sum\limits_{k = 0}^{MN - 1} {\frac{1}{{C_{\bf{z}}^{a,{\rm{DD}}}\left[ {k,k} \right]}}\left( {{-{\left| {{{\bf{F}}_{{\rm{ISFFT}}}}\left[ k \right]{\bf{m}}_{\bf{x}}^{a,{\rm{DD}}}} \right|}^2} + 2{\mathop{\rm Re}\nolimits} \left\{ {{x^*}\left[ k \right]m_{\bf{x}}^{a,{\rm{DD}}}\left[ k \right]} \right\} - {{\left| {x\left[ k \right]} \right|}^2}} \right)} } \right)\notag\\
\propto&\prod\limits_{k = 0}^{MN - 1} {\Pr \left( {{\bf{m}}_{\bf{z}}^{e,{\rm{T}}}\left| {x\left[ k \right]} \right.} \right)},\label{C6_MAP_rule_Ungerboeck}
\end{align}
where
\begin{align}
\Pr \left( {{\bf{m}}_{\bf{z}}^{e,{\rm{T}}}\left| {x\left[ k \right]} \right.} \right)
= &\Pr \left( {m_x^{a,{\rm{DD}}}\left[ k \right]\left| {x\left[ k \right]} \right.} \right) \notag\\
\propto&  {\exp \Big( {\frac{{\rm{1}}}{{C_{\bf{z}}^{a,{\rm{DD}}}\left[ k,k \right]}}\big( {2{\mathop{\rm Re}\nolimits} \left\{ {{x^*}\left[ k \right]m_{\bf{x}}^{a,{\rm{DD}}}\left[ k \right]} \right\} - {{\left| {x\left[ k \right]} \right|}^2}} \big)} \Big)} . \label{C6_DD_domain_detection_metric}
\end{align}
In particular, the probability factorization in the form of~\eqref{C6_MAP_rule_Ungerboeck} and~\eqref{C6_DD_domain_detection_metric} is referred to as the Ungerboeck observation model, which was firstly introduced by Ungerboeck in~\cite{ungerboeck1974adaptive}.
Both Forney and Ungerboeck observation models have been widely used for data detection~\cite{li2017reduced,li2020code,li2020time}, and the OTFS detection based on Forney and Ungerboeck observation models has also been considered in the previous works~\cite{Raviteja2018interference,Gaudio2020on}.
It can be observed from~\eqref{C6_DD_domain_detection_metric} that, the optimal MLSE detection for DD domain symbols can be carried out in a simple symbol-by-symbol form, and the
corresponding inputs for detection are the DD domain symbol estimates ${\bf{m}}_{\bf{x}}^{a,{\rm{DD}}}$ and the covariance matrix ${\bf{C}}_{\bf{z}}^{a,{\rm{DD}}}$ of the TD domain OTFS signal $\bf z$.

With the i.i.d. assumption of $x[k]$, the \emph{a posteriori} probability $\Pr \left( {x\left[ k \right]\left| {{\bf{m}}_{\bf{z}}^{e,{\rm{T}}}} \right.} \right)$ is essentially the same as that of ${\Pr \left( {{\bf{m}}_{\bf{z}}^{e,{\rm{T}}}\left| {x\left[ k \right]} \right.} \right)}$, i.e., $\Pr \left( {x\left[ k \right]\left| {{\bf{m}}_{\bf{z}}^{e,{\rm{T}}}} \right.} \right) \propto \Pr \left( {{\bf{m}}_{\bf{z}}^{{e},{\rm{T}}}\left| {x\left[ k \right]} \right.} \right)$.
Therefore, we can obtain the \emph{a posteriori} mean ${{m}}_{\bf{x}}^{{p,\rm{DD}}}\left[ k \right]$ of $x\left[ k \right]$ by
\begin{align}
{m}_{\bf{x}}^{p,{\rm{DD}}}\left[ k \right]  = {\mathbb E}\left[ {x\left[ k \right]|{\bf{m}}_{\bf{z}}^{e,{\rm{T}}}} \right]=\sum\limits_{i = 1}^{{\cal X}\left( {\mathbb{A}} \right)} {\Pr \left( {x\left[ k \right] = {\mathbb A}\left[ i \right]\left| {{\bf{m}}_{\bf{z}}^{e,{\rm{T}}}} \right.} \right)}  \times {\mathbb{A}}\left[ i \right],\label{C6_m_x_post_B_element}
\end{align}
where ${\mathbb{A}}\left[ i \right]$ is the $i$-th DD domain constellation point.
Meanwhile, the \emph{a posteriori} covariance matrix ${{\bf C}}_{\bf{x}}^{p, {\rm{DD}}}$ of $\bf x$ is a diagonal matrix
due to the i.i.d. assumption, whose $k$-th diagonal entry is the \emph{a posteriori} variance of $x[k]$ and is given by
\begin{align}
{C}_{\bf{x}}^{p,{\rm{DD}}}\left[ {k,k} \right] &={\mathbb E}\left[ {{{\left| {x\left[ k \right] - {\mathbb E}\left[ {x\left[ k \right]|{\bf{m}}_{\bf{z}}^{e,{\rm{T}}}} \right]} \right|}^2}} \right] \notag\\
&= \sum\limits_{i = 1}^{{\cal X}\left( {\mathbb{A}} \right)} {\Pr \left( {x\left[ k \right] = {\mathbb A}\left[ i \right]\left| {{\bf{m}}_{\bf{z}}^{e,{\rm{T}}}} \right.} \right)} \notag\\
&\quad \times {\left| {{\mathbb{A}}\left[ i \right]} \right|^2} - {\left| {m}_{\bf{x}}^{p,{\rm{DD}}}\left[ k \right] \right|^2}. \label{C6_C_x_post_B_element}
\end{align}
Based on the \emph{a priori} and \emph{a posteriori} information, the extrinsic information of the DD domain detection is ready to be computed. However, we notice that the
DD domain detection with~\eqref{C6_DD_domain_detection_metric} is a component-wise operation, and therefore in principle, the DD domain detection cannot provide any extrinsic information. The following Proposition clarifies this problem.

\textbf{Proposition~5-1} \emph{(Extrinsic Information from the DD Domain Detection)}:
The DD domain detection with~\eqref{C6_DD_domain_detection_metric} is a component-wise operation and thus it cannot provide any extrinsic information.

\emph{Proof}: The proof is given in Appendix of this chapter.

According to Proposition 5-1, we know that directly computing the extrinsic information based on the DD domain detection is not a good choice. Therefore, we firstly convert the \emph{a posteriori} probability of the DD domain symbols $\bf x$ to the \emph{a posteriori} probability of TD domain signals $\bf z$ and then compute the extrinsic information.
In specific, the \emph{a posteriori} mean ${\bf{m}}_{\bf{x}}^{p,{\rm{DD}}}$ and covariance matrix ${{\bf C}}_{\bf{x}}^{p,{\rm{DD}}}$ will be served as the
outputs of module B. The details of how to compute the extrinsic information based on the outputs of module B will be discussed in the coming subsection.
With the above discussion, we summarize the symbol-by-symbol detection for DD domain symbols in Algorithm~5-2.
\begin{algorithm}[htb]
\caption*{\textbf{Algorithm} 5-2 Symbol-by-symbol Detection for DD Domain Symbols}
\hspace*{0.02in} {\bf Input:}
${\bf{m}}_{\bf{x}}^{a,{\rm{DD}}}$, ${\bf{C}}_{\bf{z}}^{a,{\rm{T}}}$, and $\mathbb A$\\
\hspace*{0.02in} {\bf Steps:}
\begin{algorithmic}[1]
\State \textbf{for} $k$ from 0 to $MN-1$ \textbf{do}
\State $\quad$ Calculate the \emph{a posteriori} probability of $x_k$ by~\eqref{C6_DD_domain_detection_metric}.
\State $\quad$ Make hard decision on $x_k$, which is denoted as ${\hat x}[k]$.
\State $\quad$ Compute the \emph{a posteriori} mean ${m}_{\bf{x}}^{p,{\rm{DD}}}\left[ k \right]$ of $x_k$ by~\eqref{C6_m_x_post_B_element}.
\State $\quad$ Compute the \emph{a posteriori} variance of $x_k$ by~\eqref{C6_C_x_post_B_element}.
\State \textbf{end for}
\State Compute  ${\bf m}_{\bf{x}}^{p,{\rm{DD}}}$ of $\bf x$, based on ${m}_{\bf{x}}^{p,{\rm{DD}}}\left[ k \right]$, $0 \le k \le MN-1$.
\State Compute  ${\bf C}_{\bf{x}}^{p,{\rm{DD}}}$ of $\bf x$, based on ${C}_{\bf{x}}^{p,{\rm{DD}}}\left[ k,k \right]$, $0 \le k \le MN-1$.
\State \Return ${\bf m}_{\bf{x}}^{p,{\rm{DD}}}$, ${\bf C}_{\bf{x}}^{p,{\rm{DD}}}$, and $\bf {\hat x}$.
\end{algorithmic}
\label{C6_DD_detection_algorithm}
\end{algorithm}

\subsection{Cross Domain Message Passing: from DD Domain to TD Domain}
Based on the description in the previous subsection, we will discuss the message passing from DD domain to TD domain.
According to~\eqref{C4_TD_transmitted_symbol_vec}, ${{\bf m}}_{\bf{x}}^{p,{\rm{DD}}}$ and ${{\bf C}}_{\bf{x}}^{p,{\rm{DD}}}$ are converted to the \emph{a posteriori} mean ${\bf{m}}_{\bf{z}}^{p,{\rm{DD}}}$ and covariance matrix ${\bf{C}}_{\bf{z}}^{p,{\rm{DD}}}$ of the TD domain OTFS signal $\bf{z}$ by
\begin{align}
{\bf{m}}_{\bf{z}}^{p,{\rm{DD}}} &= \left( {{\bf{F}}_N^{\rm{H}} \otimes {{\bf{I}}_M}} \right){{\bf m}}_{\bf{x}}^{p,{\rm{DD}}} ,\label{C6_m_z_post_B}\\
{\bf{C}}_{\bf{z}}^{p,{\rm{DD}}} &= \left( {{\bf{F}}_N^{\rm{H}} \otimes {{\bf{I}}_M}} \right){{\bf C}}_{\bf{x}}^{p,{\rm{DD}}}\left( {{{\bf{F}}_N} \otimes {{\bf{I}}_M}} \right). \label{C6_C_z_post_B}
\end{align}
Again, we note that ${\bf{C}}_{\bf{z}}^{p,{\rm{DD}}}$ can be a dense matrix if the diagonal entries of ${{\bf C}}_{\bf{x}}^{p,{\rm{DD}}}$ are not of the same value.
In this case, we can discard the non-diagonal entries (treated as zeros), according to the i.i.d. assumption~\cite{Ma2015Turbo}.
Similar to~\eqref{C6_c_z_ext_A} and~\eqref{C6_m_z_ext_A} , we can compute the extrinsic information of $\bf{z}$ in terms of the mean and covariance matrix, which are given by
\begin{align}
{\bf{C}}_{\bf{z}}^{e,{\rm{DD}}}&= {\left( {{{\left( {{\bf{C}}_{\bf{z}}^{p,{\rm{DD}}}} \right)}^{ - 1}} - {{\left( {\bf{C}}_{\bf{z}}^{a,{\rm{DD}}} \right)}^{{\rm{ - 1}}}}} \right)^{ - 1}} ,\label{C6_c_z_pri_A}\\
{\bf{m}}_{\bf{z}}^{e,{\rm{DD}}} &= {\bf{C}}_{\bf{z}}^{e,{\rm{DD}}}\left( {{{\left( {{\bf{C}}_{\bf{z}}^{p,{\rm{DD}}}} \right)}^{ - 1}}{\bf{m}}_{\bf{z}}^{p,{\rm{DD}}} - {{\left( {{\bf{C}}_{\bf{z}}^{a,{\rm{DD}}}} \right)}^{ - 1}}{\bf{m}}_{\bf{z}}^{e,{\rm{T}}}} \right) . \label{C6_m_z_pri_A}
\end{align}
Finally, the extrinsic information of $\bf{z}$ is fed back to module A as the \emph{a priori} mean and covariance matrix for the next iteration, i.e., ${\bf{m}}_{\bf{z}}^{a,{\rm{T}}} = {\bf{m}}_{\bf{z}}^{e,{\rm{DD}}}$ and ${\bf{C}}_{\bf{z}}^{a,{\rm{T}}} = {\bf{C}}_{\bf{z}}^{e,{\rm{DD}}}$.

\begin{algorithm}[htb]
\caption*{\textbf{Algorithm} 5-3 Cross Domain Iterative Detection for OTFS Modulation}
\hspace*{0.02in} {\bf Input:}
$\bf r$, ${{\bf{H}}_{\rm{TD}}^{{\rm{eff}}}}$, $L_{\rm max}$, and $\mathbb A$\\
\hspace*{0.02in} {\bf Initialization:}
Set ${m}_{\bf{z}}^{a,{\rm{T}}}[k]=0$, for $0 \le k \le MN-1$ and ${\bf{C}}_{\bf{z}}^{{a},{\rm{T}}} = {{\bf{I}}_{MN}}$.\\
\hspace*{0.02in} {\bf Steps:}
\begin{algorithmic}[1]
\State \textbf{for} $l$ from 1 to $L_{\rm max}$ \textbf{do}
\State $\quad$ Perform the L-MMSE estimation for TD domain OTFS signal according to Algorithm~5-1.
\State $\quad$ Compute ${\bf{C}}_{\bf{z}}^{a,{\rm{DD}}}$ based on~\eqref{C6_c_z_ext_A} and ${\bf{m}}_{\bf{x}}^{a,{\rm{DD}}}$ based on~\eqref{C6_m_z_ext_A}.
\State $\quad$ Perform symbol-by-symbol detection for DD domain symbols according to Algorithm~5-2 based on~\eqref{C6_m_z_post_B} and~\eqref{C6_C_z_post_B}.
\State $\quad$ Compute ${\bf{C}}_{\bf{z}}^{a,{\rm{T}}}$ and ${\bf{m}}_{\bf{z}}^{a,{\rm{T}}}$ based on~\eqref{C6_c_z_pri_A} to~\eqref{C6_m_z_pri_A}.
\State \textbf{end for}
\State \Return $\bf {\hat x}$.
\end{algorithmic}
\label{C6_OTFS_detection_algorithm}
\end{algorithm}

Based on the above discussion, we summarize the proposed OTFS detection in Algorithm~5-3, where the term $L_{\rm max}$ is referred to as the maximum number of iterations. For notational brevity, we drop the iteration index $l$ of the corresponding matrices in Algorithm~5-3. So far, we have introduced the proposed cross domain iterative detection. In the following section, we will investigate its error performance and the computation complexity.
\section{Performance Analysis}
In this section, we will investigate the asymptotical error performance of the proposed detection algorithm with $MN \to \infty $ and its detection complexity.
Since the proposed algorithm involves several iterations between two modules, we characterize the error performance by the recursion of two states corresponding to each module.
Meanwhile, we will also discuss the detection complexity corresponding to each module.

\subsection{MSE Performance Analysis via State Evolution}
Without loss of generality, we first investigate the MSE performance with a given TD domain effective channel ${\bf{H}}_{\rm{TD}}^{{\rm{eff}}}$ and the convergence behaviour of the proposed algorithm.
To this end, we can investigate the average \emph{a priori} variance of the inputs to module A and module B during each iteration.
In particular, by noticing the i.i.d. assumption of both the DD domain OTFS symbols and the TD domain OTFS signal,
we define the two states for module A and module B at the $l$-th iteration by
\begin{align}
v_z^{{a},{\rm{T}}}\left( l \right)\buildrel \Delta \over = &{\mathbb E}\left[ {C_{\bf{z}}^{e,{\rm{DD}}}\left[ {k,k} \right]} \right]=\mathop {\lim }\limits_{MN \to \infty }  \frac{1}{{MN}}{\rm{Tr}}\left( {{\bf{C}}_{\bf{z}}^{{e},{\rm{DD}}}} \right),\\
v_z^{{a},{\rm{DD}}}\left( l \right) \buildrel \Delta \over = &{\mathbb E}\left[ {C_{\bf{z}}^{{e},{\rm{T}}}\left[ {k,k} \right]} \right] = \mathop {\lim }\limits_{MN \to \infty } \frac{1}{{MN}}{\rm{Tr}}\left( {{\bf{C}}_{\bf{z}}^{{e},{\rm{T}}}} \right),
\end{align}
where the expectation is with respect to the index $k$. In specific, these two states can be viewed as the average MSEs of the inputs to module A and module B at the $l$-th iteration, respectively.
For notational brevity, we further define the ratios between the
OTFS signal energy and the average \emph{a priori} variance of the inputs to each module, i.e., the effective SNRs for the TD domain and DD domain, by
\begin{align}
{\eta _{\rm{T}}}\left( l \right) \buildrel \Delta \over = &\frac{1}{v_{z}^{a,{\rm{T}}}\left( l \right)},\\
{\eta _{\rm{DD}}}\left( l \right) \buildrel \Delta \over = &\frac{1}{v_{z}^{a,{\rm{DD}}}\left( l \right)}. \label{C6_eta_DD_def}
\end{align}
Now, we focus on the evolution between the two states $v_z^{{a},{\rm{T}}}\left( l \right)$ and $v_z^{{a},{\rm{DD}}}\left( l \right)$. We will investigate the corresponding average variance of the inputs and outputs for each module.
In particular, we consider the following assumption.

\textbf{Assumption 5-1}:
For the $l$-th iteration, the main diagonal entries of \emph{a priori} covariance matrices ${\bf{C}}_{\bf{z}}^{a,{\rm{T}}}$ and ${\bf{C}}_{\bf{z}}^{a,{\rm{DD}}}$ are of the same value as ${v_{z}^{a,{\rm{T}}}\left( l \right)}$ and ${v_{z}^{a,{\rm{DD}}}\left( l \right)}$, respectively.

It should be noted that the above assumption is reasonable with a sufficiently large number of $MN$, due to the law of large numbers.
With this assumption, we will discuss the connection between the two states based on Algorithm~5-3.
In specific, we can first derive the \emph{a posteriori} covariance matrix ${\bf{C}}_{\bf{z}}^{p,{\rm{T}}}$ according to~\eqref{C6_MMSE_output_variance} and further derive the extrinsic covariance matrix ${\bf{C}}_{\bf{z}}^{e,{\rm{T}}}$ according to~\eqref{C6_c_z_ext_A}, i.e.,
\begin{align}
v_z^{{a},{\rm{DD}}}\left( l \right) = \frac{1}{{\frac{1}{{v_z^{{p},{\rm{T}}}\left( l \right)}} - \frac{1}{{v_z^{{a},{\rm{T}}}\left( l \right)}}}}, \label{C6_state_evolution_A_B}
\end{align}
where
\begin{align}
&v_z^{{p},{\rm{T}}}\left( l \right)\notag\\
=&v_z^{{a},{\rm{T}}}\left( l \right) - \frac{{{{\left( {v_z^{{a},{\rm{T}}}\left( l \right)} \right)}^2}}}{{MN}}{\rm{Tr}}\left( {{{\left( {{\bf{H}}_{\rm{TD}}^{{\rm{eff}}}} \right)}^{\rm{H}}}{{\left( {v_z^{{a},{\rm{T}}}\left( l \right){\bf{H}}_{\rm{TD}}^{{\rm{eff}}}{{\left( {{\bf{H}}_{\rm{TD}}^{{\rm{eff}}}} \right)}^{\rm{H}}} + {N_0}{{\bf{I}}_{MN}}} \right)}^{ - 1}}{\bf{H}}_{\rm{TD}}^{{\rm{eff}}}} \right).     \label{C6_a_posteriori_MMSE}
\end{align}
The above equations demonstrate the connection between the state $v_z^{{a},{\rm{T}}}\left( l \right)$ and $v_z^{{a},{\rm{DD}}}\left( l \right)$ at the $l$-th iteration. In the following, we will consider the update of the state $v_z^{a,{\rm{T}}}\left( l +1\right)$ based on the state $v_z^{{a},{\rm{DD}}}\left( l \right)$.
Let us define the MSE of the DD domain symbol detection, given an AWGN observation with an SNR $\eta$ by~\cite{Ma2015Turbo}
\begin{align}
MSE\left( \eta  \right) = {\mathbb E}\left[ {{{\left| {x - {\mathbb E}\left[ {x|x + \xi } \right]} \right|}^2}} \right], \label{C6_Evolution_MMSE}
\end{align}
where $x$ is an arbitrary DD domain OTFS symbol and $\xi$ is an AWGN sample with zero mean and variance $1/{\eta}$.

It should be noted that the DD domain detection for~\eqref{C6_DD_detection_from_TD} is derived from the MLSE rule based on the Ungerboeck observation model.
When all the noise terms in~\eqref{C6_DD_detection_from_TD} share the same variance, the MLSE detections based on both the Ungerboeck observation model and the Forney observation model will share the same error performance~\cite{ungerboeck1974adaptive}. Therefore, according to Assumption~1, the MSE performance for the applied DD domain detection is the same as the MLSE detection based on the Forney observation model, which is given by~\cite{Ma2015Turbo}
\begin{align}
v_x^{{p},{\rm{DD}}}\left( l \right) \buildrel \Delta \over =& {\mathbb E}\left[ {C_{\bf{z}}^{{p},{\rm{DD}}}\left[ {k,k} \right]} \right] = \mathop {\lim }\limits_{MN \to \infty } \frac{1}{{MN}}{\rm{Tr}}\left( {{\bf{C}}_{\bf{x}}^{{p},{\rm{DD}}}} \right) \notag\\
=& MSE\left( {{\eta _{\rm{DD}}}\left( l \right)} \right).\label{C6_Average_trace}
\end{align}
Since the extrinsic information is calculated in the TD domain, we need to convert the DD domain \emph{a posteriori} covariance matrix ${{\bf C}_{\bf{x}}^{p,{\rm{DD}}}}$ to the TD domain \emph{a posteriori} covariance matrix ${{\bf C}_{\bf{z}}^{p,{\rm{DD}}}}$, according to~\eqref{C6_C_z_post_B}.
Denote by ${v_z^{{p},{\rm{DD}}}\left( l \right)}$ the average of the main diagonal entries of ${{\bf C}_{\bf{z}}^{p,{\rm{DD}}}}$. According to the law of large numbers, we can show that ${v_z^{{p},{\rm{DD}}}\left( l \right)}=v_x^{{p},{\rm{DD}}}\left( l \right)$, due to the unitary transformation between the DD domain and the TD domain.
Thus, according to~\eqref{C6_c_z_pri_A}, we have
\begin{align}
v_z^{{a},{\rm{T}}}\left( {l + 1} \right) = \frac{1}{{\frac{1}{{v_z^{{p},{\rm{DD}}}}\left( l \right)}} - \frac{1}{{v_z^{a,{\rm{DD}}}\left( l \right)}}}. \label{C6_v_z_pri_A}
\end{align}
Based on~\eqref{C6_v_z_pri_A}, the state evolution from the state $v_z^{{a},{\rm{DD}}}\left( l \right)$ to the state $v_z^{{a},{\rm{T}}}\left( l +1\right)$ is now explicit.

We notice the above state evolution requires the calculation of MSE~\eqref{C6_Evolution_MMSE}. In order to calculate the MSE, we need to compute the \emph{a posteriori} mean of $x$. However, the calculation of the \emph{a posteriori} mean is in general a nonlinear function of the observation $x + \xi$ with respect to the specific constellation shape, unless $x$ is Gaussian distributed~\cite{lozano2005mercury}.
Therefore, in order to obtain some general conclusions regarding the MSE characteristics, we consider a Monte Carlo approach to calculate the MSE.
In particular, by considering a sufficiently large value of $MN$, we produce ${m}_{\bf{x}}^{a,{\rm{DD}}}$ by using the Monte Carlo approach with a given constellation set $\mathbb A$ according to~\eqref{C6_DD_detection_from_TD},
where the variance of $\bf{\hat w}$ is set to be $1/{\eta _{\rm{DD}}}\left( l \right)$ due to the law of large numbers.
Therefore, based on the generated ${m}_{\bf{x}}^{a,{\rm{DD}}}$, we can obtain the MSE value based on~\eqref{C6_Evolution_MMSE}.

According to the above analysis, we notice that there exists a fixed point in the state evolution given a sufficiently high SNR, where the overall MSE performance of the proposed algorithm does not change anymore with the increase of number of iterations $l$, i.e., the algorithm is converged. In particular,
we consider the converged MSE performance and
denote the corresponding average \emph{a posteriori} variance with respect to the TD domain estimates and to the DD domain detection outputs by $v_z^{p,{\rm{T}}}$ and $v_x^{p,{\rm{DD}}}$. Then, we have the following Theorem.

\textbf{Theorem 5-1} \emph{(Fixed Point of State Evolution)}:
There exists a fixed point in the state evolution given a sufficiently high SNR, where the algorithm is converged. Furthermore, in the convergence, the average \emph{a posteriori} variance with respect to the TD domain estimates and the DD domain detection outputs share the same value, i.e.,
\begin{align}
v_z^{p,{\rm{T}}} = v_x^{p,{\rm{DD}}}.\label{C6_Fixed_point}
\end{align}

\emph{Proof}: Note that the values of the states $v_z^{a,{\rm{T}}}\left( l \right)$ and $v_z^{{a},{\rm{DD}}}\left( l \right)$ will not change with the increase of the iteration number when the algorithm is converged. Therefore, we combine~\eqref{C6_state_evolution_A_B} and~\eqref{C6_v_z_pri_A} and drop the iteration index $l$, yielding
\begin{align}
\frac{{\rm{1}}}{{\frac{1}{{v_z^{p,{\rm{T}}}}} - \frac{1}{{v_z^{a,{\rm{T}}}}}}} = \frac{{\rm{1}}}{{\frac{1}{{v_z^{p,{\rm{DD}}}}} - \frac{1}{{v_z^{a,{\rm{DD}}}}}}}.
\end{align}
After some straightforward manipulations, we can obtain~\eqref{C6_Fixed_point}. Let us denote by $f\left( {v_z^{a,{\rm{T}}}} \right)$ the function of ${v_z^{a,{\rm{T}}}}$ in the form of~\eqref{C6_a_posteriori_MMSE}. Then, by combining~\eqref{C6_a_posteriori_MMSE},~\eqref{C6_Average_trace}, and~\eqref{C6_Fixed_point}, it can be shown that the following equation must hold, when the algorithm is converged,
\begin{align}
f\left( {\frac{1}{{\frac{1}{{MSE\left( {{\eta _{{\rm{DD}}}}} \right)}} - {\eta _{{\rm{DD}}}}}}} \right) = MSE\left( {{\eta _{{\rm{DD}}}}} \right).\label{C6_Fixed_point_der1}
\end{align}
According to the definition of ${\eta _{{\rm{DD}}}}$, it is not hard to notice that $0 < {\eta _{{\rm{DD}}}} < \frac{1}{{MSE\left( {{\eta _{{\rm{DD}}}}} \right)}}$ holds. After some mathematical derivations regarding the second order derivatives with respect to~\eqref{C6_Fixed_point_der1}, it can be shown that with a sufficiently high SNR, i.e., $N_0$ is sufficiently small, there exists only one solution of ${\eta _{{\rm{DD}}}} $ satisfying~\eqref{C6_Fixed_point_der1} with $0 < {\eta _{{\rm{DD}}}} < \frac{1}{{MSE\left( {{\eta _{{\rm{DD}}}}} \right)}}$. Thus, the convergence is guaranteed. Unfortunately, due to the space limitation, we omit the detailed mathematical derivations for the above discussions.
This completes the proof of Theorem~5-1.$\hfill\blacksquare$

Theorem 1 indicates that the proposed algorithm can converge and in the convergence, both TD domain estimation and DD domain detection can provide the same accuracy regarding the data recovery. Other than the convergence behavior of the proposed algorithm, it is also important to derive the corresponding error performance, when the proposed algorithm is converged.
According to Theorem 1, we can evaluate this in either TD domain or DD domain, when the proposed algorithm is converged. This issue will be discussed in detail in the coming subsection.
In the following, we investigate a special case where the DD domain symbols are assumed to be Gaussian distributed. We note that such a case is not practically important but it can provide some interesting insights for the analysis of the proposed algorithm. In particular, we have the following Proposition.

\textbf{Proposition 5-2} \emph{(Detection Performance with Gaussian Constellation in DD Domain)}:
For the case where the DD domain symbols are Gaussian distributed, the DD domain detection cannot provide any error performance improvement, i.e., $v_z^{{p},{\rm{DD}}}\left( l \right) = v_z^{{a},{\rm{DD}}}\left( l \right)$.

\emph{Proof}: With the Gaussian assumption, we have
\begin{align}
v_z^{p,{\rm{DD}}}\left( l \right)&=MSE\left( \eta  \right) = {\mathbb E}\left[ {{{\left| {x - {\mathbb E}\left[ {x|x + \xi } \right]} \right|}^2}} \right] \notag\\
&= {\mathbb E}\left[ {{{\left| \xi  \right|}^2}} \right] = \frac{1}{\eta }=v_z^{{a},{\rm{DD}}}\left( l \right).
\end{align}
This completes the proof of Proposition 5-2.$\hfill\blacksquare$

Proposition 5-2 suggests that, if the DD domain constellation is Gaussian distributed, iteratively updating the extrinsic information between the TD domain and DD domain will not introduce any error performance gain. In other words, the error performance improvement is due to the non-Gaussian constellation constraint in the DD domain.
Intuitively speaking, the DD domain detection can be viewed as a de-noising operation. If the constellation is Gaussian distributed, the ML detection will give the detection output as $x + \xi$ because $ {\mathbb E}\left[ {x|x + \xi } \right]=x + \xi$ always holds, thus it cannot correct any error induced by the noise.
On the other hand, when the DD domain constellation is not Gaussian distributed, applying iterations crossing time and DD domain can potentially improve the error performance. This is because
the TD domain estimation in module A assumes that the $\bf z$ is a Gaussian vector due to the spreading effect of ISFFT, which does not take advantage of the DD domain constellation constraint. Therefore, by performing DD domain symbol detection, the DD domain constellation constraint is exploited by the proposed algorithm, which can lead to a potential error performance improvement.

\begin{figure}
\centering
\includegraphics[width=0.7\textwidth]{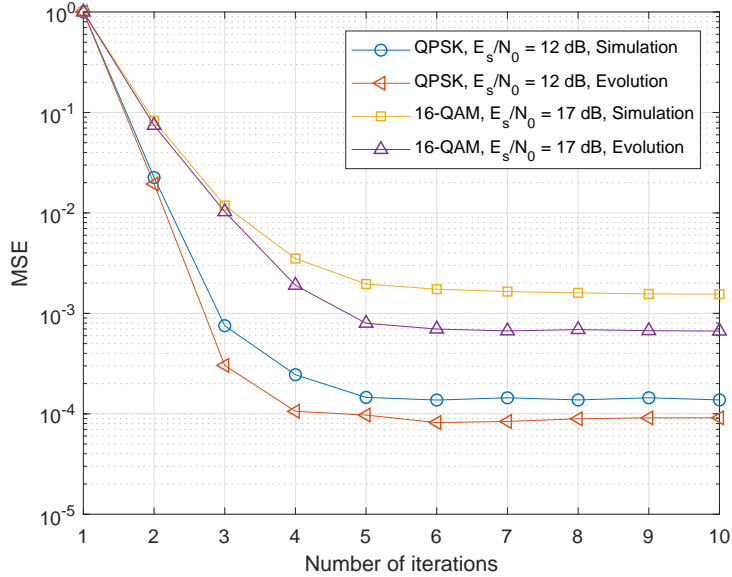}
\caption{TD domain MSE performance for OTFS modulation with $P=4$, where the frame contains $2048$ DD domain QPSK or 16-QAM symbols. In specific, the SNR for the QPSK case is
${E_s}/{N_0}=12$ dB, while that for the 16-QAM case is ${E_s}/{N_0}=17$ dB.}
\label{C6_MSE_QPSK_16QAM}
\centering
\end{figure}

Fig.~\ref{C6_MSE_QPSK_16QAM} shows the TD domain MSE performance of the proposed algorithm, where $N=32$ and $M=64$, respectively. Without loss of generality, the TD domain effective channel ${\bf{H}}_{\rm{TD}}^{{\rm{eff}}}$ is generated according to~\eqref{C4_TD_domain_channel}, where $P=4$ and the channel coefficients are $[-0.27+0.35i, 0.17+0.01i, 0.56-0.33i,0.31-0.56i]$, the delay indices are $[0, 9, 4, 7]$, and the Doppler indices are $[4.70, -2.26, 1.23, -3.46]$, respectively.
In specific, we consider two constellation mappings at different SNRs, including the quadrature phase shift keying (QPSK) at ${E_s}/{N_0}=12$ dB and 16-quadrature amplitude modulation (16-QAM) at ${E_s}/{N_0}=17$ dB and show state $v_z^{a,{\rm{T}}}\left( {l} \right)$ and the corresponding MSE values in Fig.~\ref{C6_MSE_QPSK_16QAM}.
As observed from the figure, with an increased number of iterations, the
MSE performance of the proposed algorithm first decreases and then saturates at MSEs around $1.3 \times 10^{-4}$ for the QPSK case and around $1.6 \times 10^{-3}$ for the 16-QAM case.
Meanwhile, the derived state evolution shows a close match to the actual MSE performance. This observation indicates that the derived state evolution is consistent with the simulation results.

\subsection{Analysis of Effective DD Domain SNR}
In this subsection, we will investigate the converged error performance.
In particular, we are interested in the effective DD domain SNR $\eta_{\rm DD}(l)$, because the DD domain detection is a simple component-wise operation, where $\eta_{\rm DD}(l)$ determines the BER performance for a given constellation.
Let us define the second-order TD domain OTFS channel matrix ${\bf{G}}_{\rm{TD}}^{{\rm{eff}}} \buildrel \Delta \over = {\bf{H}}_{\rm{TD}}^{{\rm{eff}}}{\left( {{\bf{H}}_{\rm{TD}}^{{\rm{eff}}}} \right)^{\rm{H}}}$.
Noticing that ${{\bf{G}}_{\rm{TD}}^{{\rm{eff}}}}$ is a Hermitian matrix by its definition, there exists a unitary matrix $\bf U$ such that ${\bf{G}}_{\rm{TD}}^{{\rm{eff}}} = {\bf{U\Lambda }}{{\bf{U}}^{\rm{H}}}$, where ${\bf{\Lambda }}$ is a diagonal matrix whose $(k,k)$-th entry is the $k$-th eigenvalue ${\lambda _k}$ of ${\bf{G}}_{\rm{TD}}^{{\rm{eff}}}$.
Thus, we can rewrite the \emph{a posteriori} variance $v_z^{p,{\rm{T}}}\left( l \right)$ in~\eqref{C6_a_posteriori_MMSE} according to the eigenvalues ${\lambda _k}$ as shown in the following Lemma.

\textbf{Lemma 5-1} \emph{(TD domain a posteriori variance)}:
The TD domain \emph{a posteriori} variance $v_z^{p,{\rm{T}}}\left( l \right)$ in~\eqref{C6_a_posteriori_MMSE} can be simplified by
\begin{align}
v_z^{p,{\rm{T}}}\left( l \right) = v_z^{{a},{\rm{T}}}\left( l \right) - \frac{{v_z^{{a},{\rm{T}}}\left( l \right)}}{{MN}}\sum\limits_{k = 1}^{MN} {\frac{{v_z^{{a},{\rm{T}}}\left( l \right){\lambda _k}}}{{v_z^{{a},{\rm{T}}}\left( l \right){\lambda _k} + {N_0}}}} . \label{C6_Lemma1}
\end{align}

\emph{Proof}: The proof is given in the Appendix of this chapter.

In order to obtain some important insights of the effective DD domain SNR $\eta_{\rm DD}(l)$, we need to investigate the property of the eigenvalues ${\lambda _k}$. For the ease of further derivation, let us consider the following assumption.

\textbf{Assumption 5-2}:
The delay index associated to each resolvable path is different to each other, i.e., ${l_i} \ne {l_j},\forall i \ne j, 1 \le i,j \le P$.

It should be noted that the value of the delay index depends on the specific reflectors corresponding to each resolvable path. Furthermore, in the case where the maximum delay index $l_{\rm max}$ is much larger than $P$, it is unlikely to have a channel realization where different paths share the same delay index. However, later we will show that even without the above assumption, our following derivation can still provide an accurate estimate for the effective DD domain SNR $\eta_{\rm DD}(l)$.
With Assumption 5-2 in hand, we can derive the following Lemma for ${\bf G}_{\rm TD}^{\rm eff}$.

\textbf{Lemma 5-2} \emph{(Main diagonal entries of ${\bf G}_{\rm TD}^{\rm eff}$)}:
Under Assumption 5-2, the main diagonal entries of ${\bf G}_{\rm TD}^{\rm eff}$ are of the same value ${\left\| {\bf{h}} \right\|^2}$, where ${\bf{h}} = {\left[ {{h_1},{h_2},...,{h_P}} \right]^{\rm{T}}}$ is the path gain vector.

\emph{Proof}: The Lemma can be derived by noticing that ${{\bm \Pi} ^{{l_i}}}{{\bm\Delta} ^{{k_i} + {\kappa _i}}}{{\bm\Delta} ^{ - {k_i} - {\kappa _i}}}{{\bm\Pi} ^{ - {l_i}}} = {{\bf{I}}_{MN}}$.$\hfill\blacksquare$

According to Lemma 5-1 and Lemma 5-2, we can now derive the lower-bound of the DD domain \emph{a priori} variance $v_z^{a,{\rm{DD}}}\left( l \right)$. The corresponding results are summarized in the following Theorem.

\textbf{Theorem 5-2} \emph{(Lower-bound of $v_z^{a,{\rm{DD}}}\left( l \right)$)}:
Under Assumption 2, the DD domain \emph{a priori} variance $v_z^{a,{\rm{DD}}}\left( l \right)$ is lower-bounded by $\frac{{N_0}}{{{{{\left\| {\bf{h}} \right\|}^2}}}}$, where the lower bound becomes tighter if the TD domain \emph{a priori} variance $v_z^{a,{\rm{T}}}\left( l \right)$ tends to zero and the lower bound is achieved when $v_z^{a,{\rm{T}}}\left( l \right)=0$.

\emph{Proof}: The proof is given in the Appendix of this chapter.

Immediately, we can derive the upper-bound of $\eta_{\rm DD}(l)$ based on Theorem 5-2.

\textbf{Corollary 5-1} \emph{(Upper-bound of $\eta_{\rm DD}(l)$)}:
Under Assumption 5-2, the DD domain effective SNR ${\eta _{\rm{DD}}}\left( {l} \right)$ is upper-bounded by $\frac{{{{\left\| {\bf{h}} \right\|}^2}}}{{{N_0}}}$, and the upper-bound becomes tighter with more iterations.

\emph{Proof}: The proof can be derived from~\eqref{C6_eta_DD_def} and Theorem~5-2.$\hfill\blacksquare$

It is interesting to see from Theorem 5-2 and Corollary 5-1 that the effective DD domain SNR of the proposed algorithm can theoretically approach the maximum receiver SNR for a given fading channel~\cite{tse2005fundamentals}, with a sufficient number of iterations. Equivalently, this observation indicates that the proposed algorithm can approach the error performance of MLSE theoretically given a sufficient number of iterations.
This is because the applied DD domain detection in~\eqref{C6_MAP_rule_Ungerboeck} is derived from the ML detection rule, which is optimal in the sense of minimizing the symbol error rate given the effective DD domain SNR. Therefore, as the effective DD domain SNR approaches its maximum with an increased number of iterations, the proposed algorithm gradually approaches the MLSE performance given the channel conditions.
It should be noted that the MLSE can provide the optimal ML error performance, but usually requires a prohibitively high complexity~\cite{tse2005fundamentals}.
Therefore, the proposed algorithm can be viewed as a type of reduced-complexity detection algorithm that can potentially approach the optimal error performance.
On the other hand, we note that the above analysis is based on Assumption 5-2. For the case where different resolvable paths share the same delay index, we will demonstrate that the
effective DD domain SNR also follows Corollary 5-1 by numerical simulations.

\begin{figure}
\centering
\includegraphics[width=0.7\textwidth]{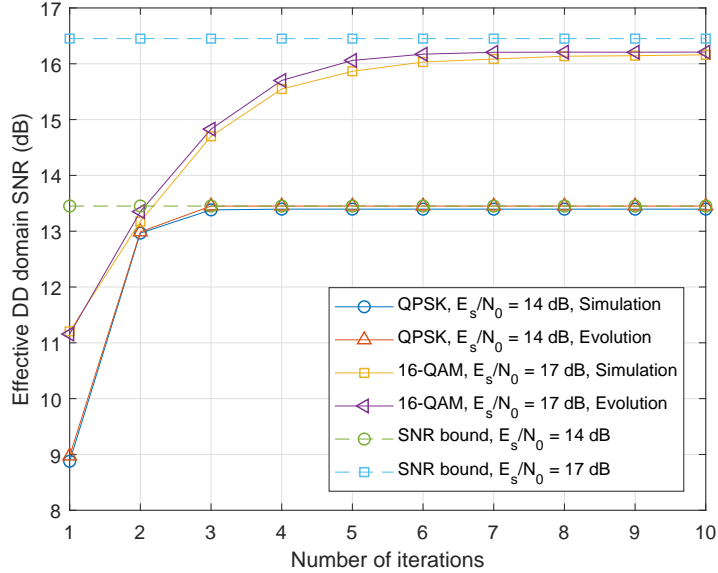}
\caption{Effective DD domain SNRs for OTFS modulation with different number of iterations, where the frame contains $2048$ DD domain QPSK or 16-QAM symbols. In specific, the SNR for the QPSK case is
${E_s}/{N_0}=14$ dB, while that for the 16-QAM case is ${E_s}/{N_0}=17$ dB. The considered wireless channel contains $P=4$ paths with different delay indices for each path. }
\label{C6_SNR_Analysis_different_delay}
\centering
\end{figure}

Fig.~\ref{C6_SNR_Analysis_different_delay} shows the effective DD domain SNRs with respect to number of iterations.
Without loss of generality, the TD domain effective channel ${\bf{H}}_{\rm{TD}}^{{\rm{eff}}}$ is generated according to~\eqref{C4_TD_domain_channel}, where $P=4$. Specifically, the channel coefficients are given by $[-0.04-0.31i, 0.40-0.11i, -0.43+0.18i, 0.59+0.21i]$, the delay indices are given by $[0, 5, 2, 8]$, and the Doppler indices are given by $[-3.08, 3.45, -3.94, -0.72]$, respectively.
Similarly, we consider both QPSK and 16-QAM constellations with ${E_s}/{N_0}=14$ dB and ${E_s}/{N_0}=17$ dB, respectively. Meanwhile, we also plot the SNR derived based on the state evolution and the
corresponding SNR upper bound, i.e., $\frac{{{{\left\| {\bf{h}} \right\|}^2}}}{{{N_0}}}$.
As observed from the figure, with an increased number of iterations, the effective DD domain SNR increases.
In specific, the derived SNR based on the state evolution shows a close match to the actual SNR performance based on the simulation.
More importantly, the derived SNR upper bound agrees with the simulation results and state evolution, and the bound becomes tighter as the number of iteration increases, which is consistent with the above analysis.

\begin{figure}
\centering
\includegraphics[width=0.7\textwidth]{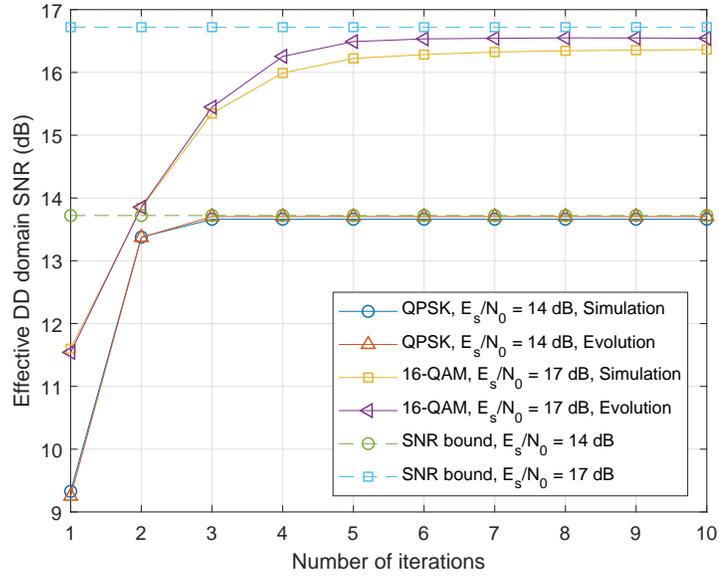}
\caption{Effective DD domain SNRs for OTFS modulation with different number of iterations, where the frame contains $2048$ DD domain QPSK or 16-QAM symbols. In specific, the SNR for the QPSK case is
${E_s}/{N_0}=14$ dB, while that for the 16-QAM case is ${E_s}/{N_0}=17$ dB. The considered wireless channel contains $P=4$ paths and the first two paths share the same delay index.}
\label{C6_SNR_Analysis_same_delay}
\centering
\end{figure}

To have a fair comparison, we show the effective DD domain SNRs with respect to the number of iterations with a specific channel realization in Fig.~\ref{C6_SNR_Analysis_same_delay}, where different resolvable paths share the same delay index.
In specific, the channel contains $P=4$ paths and the channel coefficients are $[-0.21+0.27i, 0.17+0.01i, 0.22-0.66i, 0.31-0.46i]$, the delay indices are $[0, 0, 4, 7]$, and the Doppler indices are $[-3.28, 4.45, -1.94, -0.23]$, respectively.
We observe that both QPSK case and 16-QAM case have similar SNR performance to the previous figure. This observation indicates that the above analysis is also valid when different resolvable paths share the same delay index.

\subsection{Analysis of Detection Complexity}
We first consider the computational complexity of module A. It is obvious that the most computation relates to the matrix inverse in~\eqref{C6_MMSE_W}. Generally, the computation complexity order of the matrix inversion is ${\cal O}\left( {{{\left( {MN} \right)}^3}} \right)$.
However, this complexity can be further reduced by considering the sparse and banded structure of the TD domain effective channel ${\bf{H}}_{\rm{TD}}^{{\rm{eff}}}$.
For example, the banded structure enables an LU decomposition-based method to calculate the matrix inverse with a reduced complexity~\cite{kilicc2013inverse}. Furthermore,
with some appropriate approximations, some algorithms with an even lower complexity can also be used for the calculation, such as the algorithm proposed in~\cite{bickel2012approximating}. Due to the space limitation, we leave the details of reduced-complexity TD domain L-MMSE detection designs for our future work.
As for module B, it is obvious that the computational complexity is of order ${\cal O}\left( {{MN}} \right)$ because of the component-wise operation. On the other hand,
the computational complexity for the domain transformation can be low by considering the special structure of the corresponding kernels. In particular, the Kronecker products ${\bf{F}}_N \otimes {{\bf{I}}_M}$ and ${\bf{F}}_N^{\rm{H}} \otimes {{\bf{I}}_M}$ in~\eqref{C6_m_x_pri} and~\eqref{C6_m_z_post_B} can be efficiently calculated based on FFT and IFFT and the corresponding computation complexity will be ${\cal O}\left( {MN\log N} \right)$.
Therefore, we can calculate the total detection complexity of the proposed algorithm per iteration, which is given by ${\cal O}\left( {{{\left( {MN} \right)}^3} + 2MN\log N + MN} \right)$.
It should be noted that the detection complexity does not increase in the presence of the fractional Doppler indices. In comparison of the proposed algorithm, the detection complexity of the optimal MLSE detection is exponential to the number of non-zero entries per row/column of the corresponding channel matrix, which can be as high as ${\cal O}\left( {{\cal X}{{\left( {\mathbb A} \right)}^{MN}}} \right)$ when the fractional Doppler shift exists.
Therefore, the proposed algorithm
can significantly reduce the detection complexity compared to the optimal MLSE detection, especially in the presence of fractional Doppler shifts.

\section{Numerical Results}
In this section, we will evaluate the BER performance of the proposed algorithm by assuming perfect CSI. We consider the average BER performance with a sufficient number of realizations of the channel. Specifically, the corresponding channel matrix is generated according to~\eqref{C4_TD_domain_channel}. The channel coefficients are randomly generated based on a uniform power delay profile, and the delay and Doppler indices are randomly generated within the range of $[0, l_{\rm max}]$ and $[-k_{\rm max}, k_{\rm max}]$, where $l_{\rm max}=10$ and $k_{\rm max}=5$, unless otherwise specified. We note that, as mentioned before, the delay index can only be an integer number, while the Doppler index can be a fractional number~\cite{Raviteja2018interference}.
Without loss of generality, we consider the QPSK modulated OTFS system with different numbers of paths, where $M=64$ and $N=32$, respectively, unless otherwise specified.
Specifically, we set the subcarrier spacing as $15$ kHz, and the total bandwidth of the transmission is $960$ kHz, unless otherwise specified.
We also provide other detection methods for comparison that include the MMSE detection based on the DD domain effective channel ${\bf H}_{\rm DD}^{\rm eff}$,
DD domain detection based on the SPA~\cite{li2021hybrid}, and the DD domain message passing algorithm in~\cite{Raviteja2018interference}.
The considered SPA detection is derived based on the graphical model corresponding to the DD domain effective channel, whose computational complexity
can be ${\cal O}\left( {{\cal X}{{\left( {\mathbb A} \right)}^{MN}}} \right)$ in the case of complex fractional Doppler shifts~\cite{li2021hybrid}. In particular, the
considered SPA detection can theoretically approach the error performance of the optimal MLSE detection and achieve the same performance when the graphical model does not contain any cycle~\cite{li2021hybrid}.
However, since the DD domain SPA detection requires a very high detection complexity in the fractional Doppler case, we only consider the integer Doppler case for simplicity.

\begin{figure}
\centering
\includegraphics[width=0.7\textwidth]{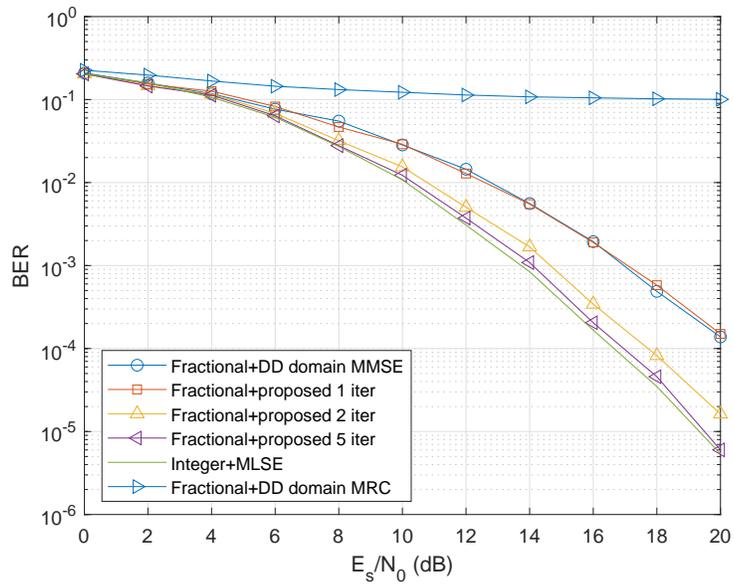}
\caption{BER performance for OTFS modulation with $P=4$, where the error performance of the proposed algorithm is compared with that of DD domain MRC and MMSE detections, as well as the SPA detection with integer Doppler shifts.}
\label{C6_P4_BER}
\centering
\end{figure}
Fig.~\ref{C6_P4_BER} shows the BER performance with $P=4$.
As observed from the figure, the proposed algorithm with only one iteration outperforms the maximum ratio combining (MRC) scheme, where the MRC scheme is implemented by multiplexing the Hermitian conjugate of the effective DD domain channel matrix with the DD domain received symbol vector.
Meanwhile, the error performance of the proposed algorithm with one iteration is almost the same as that with the MMSE detection. This observation indicates that applying the same detection method for OTFS modulation over different domains can result in the similar error performance.
Furthermore,
with the increased number of iterations, the proposed algorithm outperforms the MMSE detection and almost achieves the performance of SPA detection with only integer Doppler indices.
In specific, with BER below $1 \times 10^{-4}$, the proposed algorithm with $5$ iterations has only a marginal performance gap (around $0.2$ dB) compared to the SPA detection.
This observation clearly substantiates our theoretical analysis in the previous section.

\begin{figure}
\centering
\includegraphics[width=0.7\textwidth]{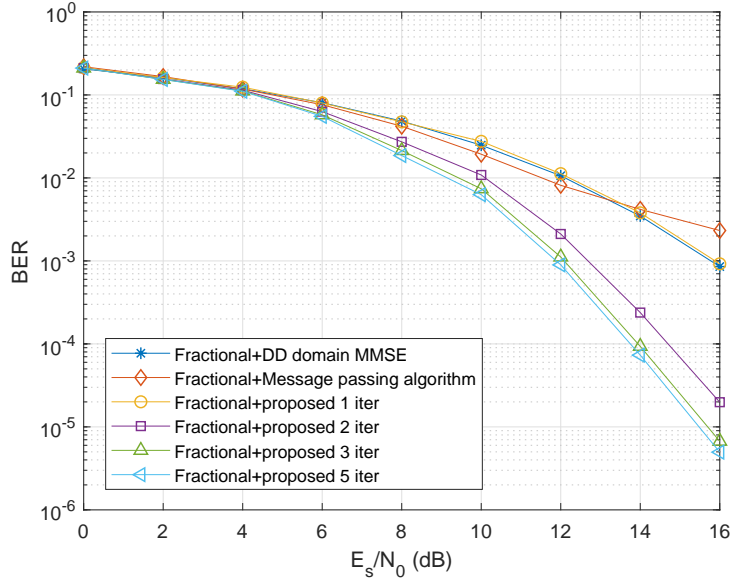}
\caption{BER performance for OTFS modulation with fractional Doppler shifts, where $P=10$.}
\label{C6_P10_BER}
\centering
\end{figure}
Fig.~\ref{C6_P10_BER} shows the BER performance with $P=10$. Under such a complex channel condition, the DD domain effective channel ${\bf H}_{\rm DD}^{\rm eff}$ can be very dense and conventional detection methods based on ${\bf H}_{\rm DD}^{\rm eff}$ may not have a good error performance.
As shown in the figure, the proposed algorithm with one iteration performs almost the same with the DD domain MMSE detection, which is consistent with the observation in the previous figure. Meanwhile, the DD domain message passing algorithm~\cite{Raviteja2018interference} also shows a similar error performance to both the DD domain MMSE detection and the proposed algorithm with one iteration.
However, with the increased number of iterations, the proposed algorithm significantly outperforms the MMSE detection and message passing algorithm. Specifically, at BER $ \approx 1 \times {10^{ - 3}}$, with only $2$ iterations, the error performance of the proposed algorithm shows an around $3.2$ dB gain compared to that of the MMSE detection and the gain to the message passing algorithm is even more.
Furthermore, with $5$ iterations, the SNR gain to the MMSE detection performance is increased to around $4.1$ dB. This observation shows the advantage of the proposed algorithm over conventional detection algorithms, which also agrees with our previous theoretical analysis.

\begin{figure}
\centering
\includegraphics[width=0.7\textwidth]{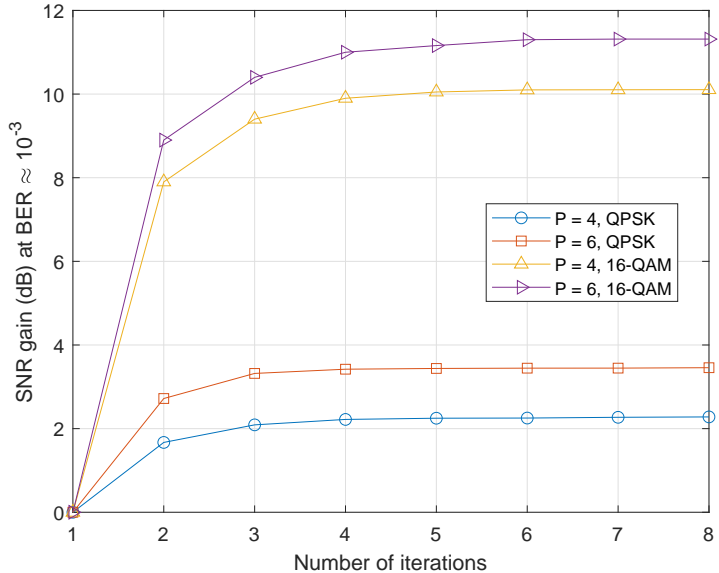}
\caption{SNR gains vs. number of iterations, where both QPSK and 16-QAM signals are considered with $P =4$ and $P =6$.}
\label{C6_SNR_Gain}
\centering
\end{figure}
In Fig.~\ref{C6_SNR_Gain}, we provide the simulation results of the SNR gains of the proposed algorithm relative to the MMSE detection in the presence of fractional Doppler shifts, at BER $ \approx {10^{ - 3}}$. We set $M=32$, $N=16$, $l_{\rm max}=3$, and $k_{\rm max}=6$, where both QPSK and 16-QAM signals are considered. As shown in the figure, the SNR gains increase with iterations until saturating at their maximum values. We observe that $5$ iterations are usually sufficient for acquiring the maximum SNR gains. We also notice that the SNR gains increase with more paths for both QPSK and 16-QAM. Furthermore, we observe that the proposed algorithm enjoys a larger SNR gain with a larger constellation size, which is consistent with our discussions in Section~5.2.2. This is because the error performance gap between the MMSE detection and ML detection increases with an increased size of constellations, and the proposed algorithm can substantially reduce this performance loss with more iterations.

\section{Summary of the Chapter}
In this chapter, we proposed a novel cross domain iterative detection for OTFS modulation. We derived the state evolution for the proposed algorithm and investigated its detection performance including both the MMSE performance and the DD domain effective SNR. In particular, we show that the proposed algorithm can approach the error performance of
MLSE detection even in the presence of complex fractional Doppler shifts, but only requires a much lower detection complexity. Our analytical results are explicitly verified by simulation results, where a significant performance improvement can be observed compared to the conventional detection algorithms for channels with fractional Doppler shifts.
The cross domain signal processing may be a new research direction for OTFS modulation and general multi-carrier modulation schemes.
Our future work will focus on the extension of the proposed algorithm to general wireless channels. An interesting starting point for the extension could be the application of the eigenvalue decomposition for the underlying channel matrix, where two unitary matrices are comprised by left and right eigenvectors. The details of this extension will be studied in the future.

\section{Appendices of the Chapter}
In this section, we provide the related proofs from this chapter.
\subsection{Justification of Problem Formulation~\eqref{C6_DD_detection_from_TD}}
The problem formulation~\eqref{C6_DD_detection_from_TD} is based on the following relationship between the extrinsic mean of $\bf z$ and the exact value of $\bf z$, i.e.,
\begin{align}
{\bf{m}}_{\bf{z}}^{e,{\rm{T}}} ={\bf{z}}+ {{\bf{\hat w}}}, \label{C6_Time_domain_estimation}
\end{align}
where ${{\bf{\hat w}}}$ is a white Gaussian noise sample vector with zero mean and a diagonal covariance matrix ${\bf{C}}_{\bf{z}}^{a,{\rm{DD}}}={\bf{C}}_{\bf{z}}^{e,{\rm{T}}}$ standing for the inaccuracy of the MMSE estimation in the TD domain.
The rational behind~\eqref{C6_Time_domain_estimation} is that unitary transformation ${{\bf{F}}_N^{\rm{H}} \otimes {{\bf{I}}_M}} $ can suppress the residual correlations due to the spreading effect of ${\bf{F}}_N^{\rm{H}}$~\cite{Ma2015Turbo}.

We note that the considered problem formulation is not the only way for DD domain detection. A more straightforward way is
to first convert the extrinsic information ${\bf{m}}_{\bf{z}}^{e,{\rm{T}}}$ of the TD domain signal $\bf z$ to DD domain and then compare it with the DD domain symbol constellation.
In specific, we can directly apply the transformation ${{{\bf{F}}_N} \otimes {{\bf{I}}_M}}$ onto~\eqref{C6_Time_domain_estimation} and obtain the corresponding DD domain symbol estimates, i.e.,
\begin{align}
{\bf{m}}_{\bf{x}}^{a,{\rm{DD}}}= \left( {{{\bf{F}}_N} \otimes {{\bf{I}}_M}} \right){\bf{m}}_{\bf{z}}^{e,{\rm{T}}}={\bf{x}} + \left( {{{\bf{F}}_N} \otimes {{\bf{I}}_M}} \right){{\bf{\hat w}}},
\end{align}
where the term $\left( {{{\bf{F}}_N} \otimes {{\bf{I}}_M}} \right){{\bf{\hat w}}}$ is the equivalent noise in the DD domain and its covariance matrix is denoted by ${\bf{C}}_{\bf{x}}^{a,{\rm{DD}}}$.
Recalling~\eqref{C4_TD_transmitted_symbol_vec}, we have
\begin{align}
{\bf{C}}_{\bf{x}}^{a,{\rm{DD}}} &= \left( {{{\bf{F}}_N} \otimes {{\bf{I}}_M}} \right){\bf{C}}_{\bf{z}}^{a,{\rm{DD}}}\left( {{\bf{F}}_N^{\rm{H}} \otimes {{\bf{I}}_M}} \right) .\label{C6_c_x_pri}
\end{align}
It should be noted that in the asymptotical regime, i.e., $MN$ tends to infinity, the diagonal entries of the diagonal matrix ${\bf{C}}_{\bf{z}}^{a,{\rm{DD}}}$ tend to be of the same value owing to the law of large numbers, i.e., ${\bf{C}}_{\bf{z}}^{a,{\rm{DD}}} \propto {{\bf{I}}_{MN}}$. In this case, we have ${\bf{C}}_{\bf{x}}^{a,{\rm{DD}}}={\bf{C}}_{\bf{z}}^{a,{\rm{DD}}}$ and the DD domain symbol detection can be done in a straightforward symbol-by-symbol fashion. However, in practice, the diagonal entries of ${\bf{C}}_{\bf{z}}^{a,{\rm{DD}}}$ may not be of the same value due to the specific noise values.
In this case, ${\bf{C}}_{\bf{x}}^{a,{\rm{DD}}}$ can be non-diagonal and dense. As mentioned before, a common approach for this issue is to discard the non-diagonal entries, which is simple but may introduce performance loss.
In contrast to the above solution, we can consider the DD domain detection problem formulation in~\eqref{C6_DD_detection_from_TD}. As can be noticed from Section~5.1.3, the proposed DD domain detection algorithm based on~\eqref{C6_DD_detection_from_TD} bypasses the problem that the diagonal entries of ${\bf{C}}_{\bf{z}}^{a,{\rm{DD}}}$ are of different values and can still be efficiently carried out in a symbol-by-symbol fashion.
\subsection{Proof of Proposition 5-1}
According to the definition of extrinsic information, we can derive the extrinsic mean $m_{\bf{x}}^{e,{\rm{DD}}}\left[ k \right]$ by excluding the contribution of $x[k]$ in~\eqref{C6_m_x_post_B_element}.
By noticing that $m_{\bf{x}}^{p,{\rm{DD}}}\left[ k \right] = {\mathbb E}\left[ {x\left[ k \right]|{\bf{m}}_{\bf{z}}^{e,{\rm{T}}}} \right] = {\mathbb E}\left[ {x\left[ k \right]|{\bf{m}}_{\bf{x}}^{a,{\rm{DD}}}} \right]$, we have
\begin{align}
m_{\bf{x}}^{e,{\rm{DD}}}\left[ k \right] = {\mathbb E}\left[ {x\left[ k \right]|{\bf{m}}_{\bf{x}}^{a,{\rm{DD}}}/m_{\bf{x}}^{a,{\rm{DD}}}\left[ k \right]} \right],
\end{align}
where ${{\bf{m}}_{\bf{x}}^{a,{\rm{DD}}}/m_{\bf{x}}^{a,{\rm{DD}}}\left[ k \right]}$ denotes ${\bf m}_{\bf{x}}^{e,{\rm{DD}}}$ excluding the entry of ${m_{\bf{x}}^{a,{\rm{DD}}}\left[ k \right]}$.
Furthermore, it can be noticed that the \emph{a posteriori} probability of $x[k]$ only relates to ${m_{\bf{x}}^{a,{\rm{DD}}}\left[ k \right]}$ instead of other entries in ${\bf m}_{\bf{x}}^{e,{\rm{DD}}}$.
Therefore, it can be shown that $m_{\bf{x}}^{e,{\rm{DD}}}\left[ k \right] =0$~\cite{Ma2015Turbo}. Similarly, it can be shown that DD domain detection cannot provide any extrinsic information in terms of the covariance matrix either, due to the component-wise operation~\cite{Ma2015Turbo}.
This completes the proof of Proposition 1.$\hfill\blacksquare$

\subsection{Proof of Lemma 5-2}
According to~\eqref{C6_a_posteriori_MMSE}, we have
\begin{align}
v_z^{{p},{\rm{T}}}\left( l \right)=&v_z^{{a},{\rm{T}}}\left( l \right) - \frac{{{{\left( {v_z^{{a},{\rm{T}}}\left( l \right)} \right)}^2}}}{{MN}}{\rm{Tr}}\left( {{{\left( {{\bf{H}}_{\rm{TD}}^{{\rm{eff}}}} \right)}^{\rm{H}}}{{\left( {v_z^{{a},{\rm{T}}}\left( l \right){\bf{H}}_{\rm{TD}}^{{\rm{eff}}}{{\left( {{\bf{H}}_{\rm{TD}}^{{\rm{eff}}}} \right)}^{\rm{H}}} + {N_0}{{\bf{I}}_{MN}}} \right)}^{ - 1}}{\bf{H}}_{\rm{TD}}^{{\rm{eff}}}} \right) \notag\\
=&v_z^{{a},{\rm{T}}}\left( l \right) - \frac{{{{\left( {v_z^{{a},{\rm{T}}}\left( l \right)} \right)}^2}}}{{MN}}{\rm{Tr}}\left( {{{\left( {v_z^{{a},{\rm{T}}}\left( l \right){\bf{G}}_{\rm{TD}}^{{\rm{eff}}} + {N_0}{{\bf{I}}_{MN}}} \right)}^{ - 1}}{\bf{G}}_{\rm{TD}}^{{\rm{eff}}}} \right). \label{C6_Lemma2_der1}
\end{align}
Furthermore, by considering ${\bf{G}}_{\rm{TD}}^{{\rm{eff}}} = {\bf{U\Lambda }}{{\bf{U}}^{\rm{H}}}$, ~\eqref{C6_Lemma2_der1} can be further simplified by~\eqref{C6_Lemma2_der2}, as shown at the top of next page.
\begin{align}
v_z^{{p},{\rm{T}}}\left( l \right)
=&v_z^{{a},{\rm{T}}}\left( l \right) - \frac{{{{\left( {v_z^{{a},{\rm{T}}}\left( l \right)} \right)}^2}}}{{MN}}{\rm{Tr}}\left( {{{\left( {v_z^{{a},{\rm{T}}}\left( l \right){\bf{U\Lambda }}{{\bf{U}}^{\rm{H}}} + {N_0}{{\bf{I}}_{MN}}} \right)}^{ - 1}}{\bf{U\Lambda }}{{\bf{U}}^{\rm{H}}}} \right)\notag\\
=&v_z^{{a},{\rm{T}}}\left( l \right) - \frac{{{{\left( {v_z^{{a},{\rm{T}}}\left( l \right)} \right)}^2}}}{{MN}}{\rm{Tr}}\left( {{\bf{U}}{{\left( {v_z^{{a},{\rm{T}}}\left( l \right){\bf{\Lambda }} + {N_0}{{\bf{I}}_{MN}}} \right)}^{ - 1}}{\bf{\Lambda }}{{\bf{U}}^{\rm{H}}}} \right)\notag\\
=&v_z^{{a},{\rm{T}}}\left( l \right) - \frac{{v_z^{{a},{\rm{T}}}\left( l \right)}}{{MN}}\sum\limits_{k = 1}^{MN} {\frac{{v_z^{{a},{\rm{T}}}\left( l \right){\lambda _k}}}{{v_z^{{a},{\rm{T}}}\left( l \right){\lambda _k} + {N_0}}}} .\label{C6_Lemma2_der2}
\end{align}
This completes the proof of Lemma 5-2.$\hfill\blacksquare$

\subsection{Proof of Theorem 5-2}
The proof is based on the application of Jensen's inequality. Let us consider the following function $f\left( \lambda  \right) = {\frac{{v\lambda }}{{v\lambda  + {N_0}}}}$,
whose second-order derivative with respect to $\lambda$ is of the form ${f^{''}}\left( \lambda  \right) =  - \frac{{2{v^2}{N_0}}}{{{{\left( {v\lambda  + {N_0}} \right)}^3}}}$.
It can be shown that with $v$ and $N_0$ strictly above zero, the above function is a concave function.
Furthermore, with $v$ close to zero, ${f^{''}}\left( \lambda  \right)$ tends to be a zero, which indicates that $f\left( \lambda  \right)$ tends to be a linear function.
Therefore, according to Jensen's inequality, it is obvious that
\begin{align}
\frac{1}{{MN}}\sum\limits_{k = 1}^{MN} {f\left( {{\lambda _k}} \right)}  \le f\left( {\frac{1}{{MN}}\sum\limits_{k = 1}^{MN} {{\lambda _k}} } \right) = \frac{{\frac{v}{{MN}}\sum\limits_{k = 1}^{MN} {{\lambda _k}} }}{{\frac{v}{{MN}}\sum\limits_{k = 1}^{MN} {{\lambda _k}}  + {N_0}}},
\end{align}
where the bound becomes tighter with decreasing $v$ and the equality is achieved when $v=0$.
Therefore, we have
\begin{align}
v_z^{p,{\rm{T}}}\left( l \right) &= v_z^{{a},{\rm{T}}}\left( l \right) -  \frac{{v_z^{{a},{\rm{T}}}\left( l \right)}}{{MN}}\sum\limits_{k = 1}^{MN} {\frac{{v_z^{{a},{\rm{T}}}\left( l \right){\lambda _k}}}{{v_z^{{a},{\rm{T}}}\left( l \right){\lambda _k} + {N_0}}}} \notag\\
&\ge v_z^{{a},{\rm{T}}}\left( l \right) - \frac{{{{\left( {v_z^{{a},{\rm{T}}}\left( l \right)} \right)}^2} \frac{1}{{MN}}\sum\limits_{k = 1}^{MN} {{\lambda _k}} }}{{v_z^{{a},{\rm{T}}}\left( l \right) \frac{1}{{MN}}\sum\limits_{k = 1}^{MN} {{\lambda _k}}  + {N_0}}}, \label{C6_Theorem2_der1}
\end{align}
where the lower bound becomes tighter with the decrease of $v_z^{{a},{\rm{T}}}\left( l \right)$ and the equality is achieved when $v_z^{{a},{\rm{T}}}\left( l \right)$ becomes zero.
Notice that $\sum\limits_{k = 1}^{MN} {{\lambda _k}}  = {\rm{Tr}}\left( {{\bf{G}}_{\rm{TD}}^{{\rm{eff}}}} \right)$. Thus, by considering Lemma~5-2,~\eqref{C6_Theorem2_der1} becomes
\begin{align}
v_z^{{p},{\rm{T}}}\left( l \right) \ge v_z^{{a},{\rm{T}}}\left( l \right) - \frac{{{{\left( {v_z^{{a},{\rm{T}}}\left( l \right)} \right)}^2}{{\left\| {\bf{h}} \right\|}^2}}}{{v_z^{{a},{\rm{T}}}\left( l \right){{\left\| {\bf{h}} \right\|}^2} + {N_0}}}. \label{C6_Theorem2_der2}
\end{align}
By substituting~\eqref{C6_Theorem2_der2} into~\eqref{C6_state_evolution_A_B}, after some manipulations, we arrive at
\begin{align}
v_z^{{a},{\rm{DD}}}\left( l \right) &= \frac{1}{{\frac{1}{{v_z^{{p},{\rm{T}}}\left( l \right)}} - \frac{1}{{v_z^{{a},{\rm{T}}}\left( l \right)}}}} \notag\\
&\ge \frac{1}{{\frac{1}{{v_z^{{a},{\rm{T}}}\left( l \right) - \frac{{v_z^{{a},{\rm{T}}}\left( l \right){{\left\| {\bf{h}} \right\|}^2}}}{{v_z^{{a},{\rm{T}}}\left( l \right){{\left\| {\bf{h}} \right\|}^2} + {N_0}}}}} - \frac{1}{{v_z^{{a},{\rm{T}}}\left( l \right)}}}}\notag\\
&= \frac{{v_z^{{a},{\rm{T}}}\left( l \right) - \frac{{{{\left( {v_z^{{a},{\rm{T}}}\left( l \right)} \right)}^2}{{\left\| {\bf{h}} \right\|}^2}}}{{v_z^{{a},{\rm{T}}}\left( l \right){{\left\| {\bf{h}} \right\|}^2} + {N_0}}}}}{{\frac{{v_z^{{a},{\rm{T}}}\left( l \right){{\left\| {\bf{h}} \right\|}^2}}}{{v_z^{{a},{\rm{T}}}\left( l \right){{\left\| {\bf{h}} \right\|}^2} + {N_0}}}}} = \frac{{{N_0}}}{{{{\left\| {\bf{h}} \right\|}^2}}}.
\end{align}
This completes the proof of Theorem~5-2.$\hfill\blacksquare$

    \chapter{Performance Analysis of Coded OTFS Systems over High-Mobility Channels}\label{c:literature}

In this chapter, we aim to analyze the error performance of coded OTFS systems over general Rayleigh-fading WSSUS channels~\cite{li2021performance_analysis,Li2020on}. To this end,
we start from the study of conditional PEP \cite{Tarokh1998space,vucetic2003space} for a given channel realization.
In order to obtain an accurate performance analysis, we consider two cases of the OTFS transmission depending on the number of independent resolvable paths of the channel and derive the corresponding conditional PEPs.
Since the exact unconditional PEP is generally intractable~\cite{Lu2000space},
we resort to the application of some proper bounding techniques to study the conditional PEP and derive the unconditional performance upper bounds.
Based on the unconditional performance bound, the impact of channel coding parameters on the performance of OTFS modulation is unveiled. In particular, we find that the squared Euclidean distance between a pair of codewords is the key parameter that determines the coding gain for coded OTFS systems, given the number of independent resolvable paths.
Therefore, the code design criterion is formulated to optimize the coding gain by maximizing the minimum squared Euclidean distance between all codeword pairs.
The main contributions of this chapter can be summarized as follows.
\begin{itemize}
\item We investigate the conditional PEP of OTFS systems for a given channel realization by studying the pairwise Euclidean distance between OTFS codewords. Based on the conditional PEP, we derive the unconditional performance upper bounds for OTFS systems, according to the number of independent resolvable paths. We also show a few important properties of the codeword difference matrix.
    Furthermore, according to the derived bounds, we define the coding gain and diversity gain of OTFS systems. More importantly, we show that the coding gain of OTFS systems depends on the squared Euclidean distance and the number of independent resolvable paths.
\item According to the derived performance bounds, we show that there is a fundamental trade-off between the diversity gain and the coding gain for OTFS systems. In particular, the diversity gain of OTFS systems improves with the number of independent resolvable paths, while the coding gain declines.
\item Based on the derived performance bounds, we propose our code design criterion to optimize the coding gain, which is to maximize the minimum squared Euclidean distance among all codeword pairs.
     In other words, traditional good codes with a large minimum Euclidean distance can be directly applied to OTFS systems.
\item We demonstrate a significant performance improvement achieved by the coded OTFS modulation over the coded OFDM modulation over high-mobility channels by numerical simulations. We also provide numerical results of coded OTFS systems over high-mobility channels with various channel codes, such as classical convolutional codes and state-of-art low-density parity-check (LDPC) codes. Our performance analysis and code design are explicitly verified by these results.
\end{itemize}

\section{Error Performance Analysis}
Without loss of generality, let us focus on the rectangular pulse shaped point-to-point OTFS transmissions with fractional delay and fractional Doppler.
For the ease of derivation, let us write the DD domain system model based on~\eqref{C4_DD_channel_fractional} by
\begin{align}
{\bf{y}}_{\rm DD} = {\bf{\Phi }}_{{{\bm{\omega}} _\tau },{{\bm{\omega}} _\nu }}\left( {\bf{x}}_{\rm DD} \right){\bf{h}} + {\bf{w}} \label{C7_system_model},
\end{align}
where ${{\bm{\omega}} _\tau }$ and ${{\bm{\omega}} _\nu }$ denote the vectors of delay indices and Doppler indices, respectively, i.e., ${{\bm{\omega}} _\tau } = \left[ {{l_1}+{\iota _1},{l_2}+{\iota _2}, \ldots ,{l_P}+{\iota _P}} \right]^{\rm{T}}$, ${{\bm{\omega}} _\nu } = \left[ {{k_1}+{\kappa _1},{k_2}+{\kappa _2}, \ldots ,{k_P}+{\kappa _P}} \right]^{\rm{T}}$, respectively.
In specific, we refer to ${\bf{\Phi }}_{{{\bm{\omega}} _\tau },{{\bm{\omega}} _\nu }}\left( {\bf{x}}_{\rm DD}\right)$ as the
\emph{equivalent codeword matrix} and it is a concatenated matrix of size $MN\times P$ constructed by the column vector ${{\bf{\Xi }}_i}{\bf{x}}_{\rm DD}$ , i.e.,
\begin{equation}
{\bf{\Phi }}_{{{\bm{\omega}} _\tau },{{\bm{\omega}} _\nu }}\left( {\bf{x}_{\rm DD}} \right) = \left[ {{{\bf{\Xi }}_1}{\bf{x}_{\rm DD}}\quad {{\bf{\Xi }}_2}{\bf{x}}_{\rm DD}\quad  \cdots \quad {{\bf{\Xi }}_P}{\bf{x}}_{\rm DD}} \right],
\label{C7_equation_phi}
\end{equation}
and ${{\bf{\Xi }}_i}$ is given by
\begin{equation}
{{\bf{\Xi }}_i} \buildrel \Delta \over = \left( {{{\bf{F}}_N} \otimes {{\bf{I}}_M}} \right){{\bm{\Pi }}^{{l_i}+{\iota _i}}}{{\bm{\Delta}} ^{{k_i}+{\kappa _i}}}\left( {{\bf{F}}_N^{\rm{H}} \otimes {{\bf{I}}_M}} \right), 1 \le i \le P.
\label{C7_equation_Xi}
\end{equation}
In~\eqref{C7_system_model}, $\bf{h}$ is the channel coefficient vector of size $P\times 1$, i.e., ${\bf{h}} = {\left[ {{h_1},{h_2},...,{h_P}} \right]^{\rm{T}}}$, where the elements in ${\bf{h}}$
are assumed to be independent and identically distributed complex Gaussian random variables.
Besides, we assume a uniform power delay and Doppler profile of the channel such that the channel coefficient $h_i$ has mean $\mu $ and variance $1/(2P)$ per real dimension for $1 \le i \le P$ and is independent from the delay and Doppler indices~\cite{molisch2012wireless}.
In particular, we note that if $\mu=0$, $\left| {{h_i}} \right|$ follows the Rayleigh distribution, which will be considered as a special case in our error performance analysis and code design.

In order to investigate the theoretical error performance of the coded OTFS systems, we assume that ideal CSI is available at the receiver, including $\bf{h}$, ${{\bm{\omega}} _\tau }$, and ${{\bm{\omega}} _\nu }$\footnote{Some references regarding the channel estimation for OTFS modulation can be found in~\cite{Raviteja2019embedded}. Meanwhile, recent advances of machine learning based methods can also potentially improve the channel estimation and detection performance of OTFS modulation~\cite{liu2020deep_ABC,liu2020deep_IRS,liu2020deep,liu2019deep}.}.
We note that matrix ${\bf{\Phi }}_{{{\bm{\omega}} _\tau },{{\bm{\omega}} _\nu }}\left( {\bf{x}} \right)$ depends on ${{\bm{\omega}} _\tau }$, ${{\bm{\omega}} _\nu }$, and the transmitted symbol vector ${\bf{x}}_{\rm DD}$. Therefore, for a given channel realization, we define the \emph{conditional Euclidean distance} ${d_{\bf{h},{{\bm{\omega}} _\tau },{{\bm{\omega}} _\nu }}^2\left( {{\bf{x}},{\bf{x'}}} \right)}$ between a pair of codewords ${\bf{x}}$ and ${\bf{x'}}$ (${\bf{x}} \ne {\bf{x'}}$) as
\begin{align}
d_{\bf{h},{{\bm{\omega}} _\tau },{{\bm{\omega}} _\nu }}^2\left( {{\bf{x}},{\bf{x'}}} \right) = &d_{\bf{h},{{\bm{\omega}} _\tau },{{\bm{\omega}} _\nu }}^2\left( {\bf{e}} \right)\notag\\
\buildrel \Delta \over =&{\left\| {{{\bf{\Phi }}_{{{\bm{\omega}} _\tau },{{\bm{\omega}} _\nu }}\left( {\bf{e}} \right)}{\bf{h}}} \right\|^2}={{\bf{h}}^{\rm{H}}}{\bf{\Omega }}_{{{\bm{\omega}} _\tau },{{\bm{\omega}} _\nu }}\left( {\bf{e}} \right){\bf{h}},
\end{align}
where ${\bf{e}} = {\bf{x}} - {\bf{x'}}$ is the corresponding codeword difference (error) sequence and ${\bf{\Omega }}_{{{\bm{\omega}} _\tau },{{\bm{\omega}} _\nu }}\left( {\bf{e}} \right) = {\left( {{\bf{\Phi }}_{{{\bm{\omega}} _\tau },{{\bm{\omega}} _\nu }}\left( {\bf{e}} \right)} \right)^{\rm{H}}}\left( {\bf{\Phi }}_{{{\bm{\omega}} _\tau },{{\bm{\omega}} _\nu }}\left( {\bf{e}} \right)\right)$ is referred to as the \emph{codeword difference matrix}. Without loss of generality and for notational simplicity,
we henceforth drop the subscript of ${\bf{\Omega }}_{{{\bm{\omega}} _\tau },{{\bm{\omega}} _\nu }}\left( {\bf{e}} \right)$ and we now have
\begin{equation}
{\bf{\Omega }}\left( {\bf{e}} \right)\! =\!\! \left[ {\begin{array}{*{20}{c}}
{{{\bf{e}}^{\rm{H}}}{\bf{\Xi }}_1^{\rm{H}}{{\bf{\Xi }}_1}{\bf{e}}}&{{{\bf{e}}^{\rm{H}}}{\bf{\Xi }}_1^{\rm{H}}{{\bf{\Xi }}_2}{\bf{e}}}& \cdots &{{{\bf{e}}^{\rm{H}}}{\bf{\Xi }}_1^{\rm{H}}{{\bf{\Xi }}_P}{\bf{e}}}\\
{{{\bf{e}}^{\rm{H}}}{\bf{\Xi }}_2^{\rm{H}}{{\bf{\Xi }}_1}{\bf{e}}}&{{{\bf{e}}^{\rm{H}}}{\bf{\Xi }}_2^{\rm{H}}{{\bf{\Xi }}_2}{\bf{e}}}&{}& \vdots \\
 \vdots &{}& \ddots & \vdots \\
{{{\bf{e}}^{\rm{H}}}{\bf{\Xi }}_P^{\rm{H}}{{\bf{\Xi }}_1}{\bf{e}}}& \cdots & \cdots &{{{\bf{e}}^{\rm{H}}}{\bf{\Xi }}_P^{\rm{H}}{{\bf{\Xi }}_P}{\bf{e}}}
\end{array}} \right]. \label{C7_Gram}
\end{equation}
The conditional PEP is upper-bounded by \cite{Tarokh1998space,vucetic2003space}
\begin{equation}
\Pr\left( {\left. {{\bf{x}},{\bf{x'}}} \right|{\bf{h}},{{{\bm{\omega}} _\tau }},{{{\bm{\omega}} _\nu }}} \right) \le \exp \left( { - \frac{{{E_s}}}{{4{N_0}}}{d_{\bf{h},{{\bm{\omega}} _\tau },{{\bm{\omega}} _\nu }}^2\left( {{\bf{x}},{\bf{x'}}} \right)}} \right), \label{C7_PEP_derivation1}
\end{equation}
where $E_s$ is the average symbol energy.
Note that the codeword difference matrix ${\bf{\Omega }}\left( {\bf{e}} \right)$ is positive semidefinite Hermitian with a rank $r$, where $r \le P$.
Let us denote by $\left\{ {{{\bf{v}}_1},{{\bf{v}}_2},...,{{\bf{v}}_P}} \right\}$ the eigenvectors of ${\bf{\Omega }}\left( {\bf{e}} \right)$ and
$\left\{ {{\lambda _1},{\lambda _2},...,{\lambda _P}} \right\}$ the corresponding nonnegative real eigenvalues sorted in the descending order,
where ${{\lambda _i}} > 0$ for $1 \le i \le r$ and ${{\lambda _i}} = 0$ for $r+1 \le i \le P$.
Thus,~\eqref{C7_PEP_derivation1} can be further expanded as~\cite{Tarokh1998space,vucetic2003space}
\begin{equation}
\Pr\left( {{\left. {{\bf{x}},{\bf{x'}}} \right|{\bf{h}},{{{\bm{\omega}} _\tau }},{{{\bm{\omega}} _\nu }}}} \right) \le \exp \left( { - \frac{{{E_s}}}{{4{N_0}}}\sum\limits_{i = 1}^r {{\lambda _i}{{\left| {{{\tilde h}_i}} \right|}^2}} } \right), \label{C7_PEP_derivation2}
\end{equation}
where ${{\tilde h}_i} = {\bf{h}}^{\rm T}\cdot{{\bf{v}}_i}$, for $1 \le i \le r$.
It can be shown that $\left\{ {{{\tilde h}_1},{{\tilde h}_2},...,{{\tilde h}_r}} \right\}$ are independent complex Gaussian random variables with mean ${\mu _{{{\tilde h}_i}}}=(\mathbb{E}\left[ {\bf{h}} \right])^{\rm T}\cdot{{\bf{v}}_i}$ and variance $1/(2P)$ per real dimension.
Thus, it is obvious that ${\small| {{{\tilde h}_i}} \small|}$ follows the Rician distribution with a Rician factor ${K_i} = {\left| {{\mu _{{{\tilde h}_i}}}} \right|^2}$ \cite{Tarokh1998space}, and its
PDF is given by
\begin{align}
p\left( {\left| {{{\tilde h}_i}} \right|} \right) = 2P\left| {{{\tilde h}_i}} \right|\exp \left( { - P{{\left| {{{\tilde h}_i}} \right|}^2} - P{K_i}} \right){I_0}\left( {2P\left| {{{\tilde h}_i}} \right|\sqrt {{K_i}} } \right),\label{C7_Rician_PDF}
\end{align}
where ${I_0}(\cdot)$ denotes the Bessel function of the first kind.
In the following, we will target on the analysis of the unconditional PEP. To this end, we aim to calculate the average of~\eqref{C7_PEP_derivation2} over the channel distribution according to~\eqref{C7_Rician_PDF}. More specifically, we will discuss two important cases depending on the number of independent resolvable paths~$P$.

\textbf{Remark 6-1}: It has been defined in the previous works \cite{Raviteja2019effective,Biglieri2019error,Surabhi2019on,Zhiqiang_magzine} that the rank of ${\bf{\Omega }}\left( {\bf{e}} \right)$ is the \textbf{diversity} gain of the OTFS modulation. Specifically, it has been shown in \cite{Surabhi2019on} that the diversity gain of uncoded OTFS modulation systems can be one but the full diversity can be obtained by suitable precoding schemes. Furthermore, \cite{Raviteja2019effective} has shown that the full diversity can be achieved almost surely for the case of $P=2$ when the frame size is sufficiently large, even for uncoded OTFS modulation systems.

\subsection{Error Performance Analysis for Coded OTFS systems}
Notice that $\bf{h}$, ${{\bm{\omega}} _\tau }$, and ${{{\bm{\omega}} _\nu }}$ are independent from each other with the assumptions of WSSUS channel and the uniform power delay profile~\cite{hlawatsch2011wireless}. Therefore, the unconditional PEP can be derived by firstly averaging~\eqref{C7_PEP_derivation2} over ${| {{{\tilde h}_i}} |}$ term by term which results in
\begin{align}
\Pr\left( {\left. {{\bf{x}},{\bf{x'}}} \right|{{{\bm{\omega}} _\tau }},{{{\bm{\omega}} _\nu }}} \right)
 \le \prod\limits_{i = 1}^r {\frac{1}{{1 + \frac{{{E_s}}}{{4{N_0}}}\cdot\frac{{{\lambda _i}}}{{ P }}}}\exp \left( { - \frac{{{K_i}\frac{{{E_s}}}{{4{N_0}}}\cdot\frac{{{\lambda _i}}}{{ P }}}}{{1 + \frac{{{E_s}}}{{4{N_0}}}\cdot\frac{{{\lambda _i}}}{{ P }}}}} \right)}  . \label{C7_PEP_derivation3}
\end{align}
Furthermore, we consider a special case where ${{K_i}}=0$ and ${\small| {{{\tilde h}_i}} \small|}$ follows the Rayleigh distribution, i.e., ${\small| {{{h}_i}} \small|}$ also follows the Rayleigh distribution.
In the case of \textbf{Rayleigh fading},~\eqref{C7_PEP_derivation3} can be further simplified as
\begin{align}
\Pr\left( {\left. {{\bf{x}},{\bf{x'}}} \right|{{{\bm{\omega}} _\tau }},{{{\bm{\omega}} _\nu }}} \right)\le {\left( {\prod\limits_{i = 1}^r {{\lambda _i}/P} } \right)^{ - 1}}{\left( {\frac{{{E_s}}}{{4{N_0}}}} \right)^{ - r}}=\frac{1}{{\prod\limits_{i = 1}^r {{\lambda _i}} }}{\left( {\frac{{{E_s}}}{{4{N_0}P}}} \right)^{ - r}}, \label{C7_New_PEP_derivation1}
\end{align}
It should be noted that~\eqref{C7_New_PEP_derivation1} is consistent with the analysis in \cite{Biglieri2019error}. On the other hand, the PEP in~\eqref{C7_New_PEP_derivation1} depends on the delay and Doppler indices ${\bm{\omega}} _\tau$ and ${\bm{\omega}} _\nu$.
In order to derive the unconditional PEP, we need to find the statistical distribution for the term $\prod\nolimits_{i = 1}^r {{\lambda _i}} $ regarding the delay and Doppler indices.
Unfortunately, such a task is generally intractable~\cite{Lu2000space} and is normally handled by applying the Monte Carlo method without providing any important insight.
Instead of resorting to the Monte Carlo method, we apply proper bounding techniques to evaluate the value of $\prod\nolimits_{i = 1}^r {{\lambda _i}} $ in order to obtain some general results about the unconditional PEP. Specifically, we have the following property.

\textbf{Property 6-1} \emph{(Gram matrix~\cite{Tut_Gram})}:
Let ${\bar{\bf{u}}_i} \buildrel \Delta \over = {{\bf{\Xi }}_i}{\bf{e}}$, for $1\le i\le P$, where ${\bf{\Xi }}_i$ is given by~\eqref{C7_equation_Xi}. Then,
$\left\{ {{\bar{\bf{u}}_1},{\bar{\bf{u}}_2}, \ldots {\bar{\bf{u}}_{P}}} \right\}$ form a list of vectors chosen from the $P$-dimensional complex inner-product subspace $\mathbb{H}^{P}$.
Thus, the codeword difference matrix ${\bf{\Omega }}\left( {\bf{e}} \right)$ is positive semidefinite Hermitian and it is a \emph{Gram matrix} corresponding to the vectors $\left\{ {{\bar{\bf{{u}}}_1},{\bar{\bf{{u}}}_2}, \ldots {\bar{\bf{{u}}}_P}} \right\}$.

Based on the property of ${\bf{\Omega }}\left( {\bf{e}} \right)$, we now introduce four important lemmas, which will be served as the building blocks for our error performance analysis for coded OTFS systems.

\textbf{Lemma 6-1} \emph{(Main diagonal elements of ${\bf{\Omega }}\left( {\bf{e}} \right)$)}:
The main diagonal elements of the codeword difference matrix ${\bf{\Omega }}\left( {\bf{e}} \right)$ are of the same value ${d_{\rm{E}}^2\left( {\bf{e}} \right)}$,
where $d_{\rm{E}}^2\left( {\bf{e}} \right) = {{\bf{e}}^{\rm{H}}}{\bf{e}}$ is the squared Euclidean distance for a pair of codewords ${\bf{x}}$ and ${\bf{x'}}$ corresponding to the error sequence $\bf{e}$.

\emph{Proof}: By considering~\eqref{C7_equation_phi}, the $i$-th diagonal element of ${\bf{\Omega }}\left( {\bf{e}} \right)$ is given by ${{\bf{e}}^{\rm{H}}}{\bf{\Xi }}_i^{\rm{H}}{{\bf{\Xi }}_i}{\bf{e}}$, and it is equal to the inner product of ${{{{\bf{\bar u}}}_i}}$.
Furthermore, we have
\begin{align}
{\bf{\Xi }}_i^{\rm{H}}{{\bf{\Xi }}_i} =& \left( {{{\bf{F}}_N} \otimes {{\bf{I}}_M}} \right){\left( {{{\bm{\Delta}} ^{{k_i}+{\kappa _i}}}} \right)^{\rm{H}}}{\left( {{{\bm{\Pi }}^{{l_i}+{\iota _i}}}} \right)^{\rm{H}}}{{\bm{\Pi }}^{{l_i}+{\iota _i}}}{{\bm{\Delta}} ^{{k_i}+{\kappa _i}}}\left( {{\bf{F}}_N^{\rm{H}} \otimes {{\bf{I}}_M}} \right) \notag\\
=&\left( {{{\bf{F}}_N} \otimes {{\bf{I}}_M}} \right){{\bm{\Delta}} ^{ - {k_i}-{\kappa_i}}}{{\bm{\Pi }}^{ - {l_i}-{\iota _i}}}{{\bm{\Pi }}^{{l_i}+{\iota _i}}}{{\bm{\Delta}} ^{{k_i}+{\kappa _i}}}\left( {{\bf{F}}_N^{\rm{H}} \otimes {{\bf{I}}_M}} \right)
\label{C7_omega_derivation1}\\
=&{{\bf{I}}_{MN}} \label{C7_omega_derivation2},
\end{align}
where~\eqref{C7_omega_derivation1} is due to the properties of ${\bm{\Pi }}$ and $\bm{\Delta}$, respectively, and~\eqref{C7_omega_derivation2} is due to the property of the Kronecker product.
Therefore, the term ${{\bf{e}}^{\rm{H}}}{\bf{\Xi }}_i^{\rm{H}}{{\bf{\Xi }}_i}{\bf{e}}$ is further simplified as ${d_{\rm{E}}^2\left( {\bf{e}} \right)}$.
This completes the proof of Lemma 6-1. \hfill $\blacksquare$

\textbf{Lemma 6-2} \emph{(Trace of ${\bf{\Omega }}\left( {\bf{e}} \right)$)}:
The trace of the codeword difference matrix ${\bf{\Omega }}\left( {\bf{e}} \right)$ is ${Pd_{\rm{E}}^2\left( {\bf{e}} \right)}$. Equivalently, the summation of the first $r$ eigenvalues satisfies $\sum\nolimits_{i = 1}^r {{\lambda _i}}  = Pd_{\rm{E}}^2\left( {\bf{e}} \right)$.

\emph{Proof}: This lemma is a straightforward extension of Lemma 6-2 by noticing that the ${\lambda _i} = 0$ for $r+1 \le i \le P$. \hfill $\blacksquare$

\textbf{Lemma 6-3} \emph{(Lower bound on $\sum\nolimits_{i = 1}^r {1/{\lambda _i}} $)}:
The summation of first $r$ eigenvalue inverses of ${\bf{\Omega }}\left( {\bf{e}} \right)$ is lower-bounded by
\begin{equation}
\sum\limits_{i = 1}^r {\frac{1}{{{\lambda _i}}}}  \ge \frac{{{r^2}}}{{Pd_{\rm{E}}^2\left( {\bf{e}} \right)}}, \label{C7_Inverse_trace}
\end{equation}
where the equality holds if ${\bf{\Omega }}\left( {\bf{e}} \right)$ is a diagonal matrix, i.e., ${\bf{\Omega }}\left( {\bf{e}} \right)=\textrm{diag}\left\{ {d_{\rm{E}}^2{\left( {\bf{e}} \right)}, \ldots ,d_{\rm{E}}^2}{\left( {\bf{e}} \right)} \right\}$.

\emph{Proof}: The proof is given in the Appendix of this chapter.

\textbf{Lemma 6-4} \emph{(Lower bound on $\sum\nolimits_{i = 1}^r {\lambda _i^2} $)}:
The summation of the first $r$ squared eigenvalues of ${{{ {{\bf{\Omega }}\left( {\bf{e}} \right)} }}}$ is lower-bounded by
\begin{equation}
\sum\limits_{i = 1}^r {\lambda _i^2}  \ge \frac{{{P^2}}}{r}{\left( {d_{\rm{E}}^2\left( {\bf{e}} \right)} \right)^2}, \label{C7_Lemma4}
\end{equation}
where the equality holds if ${\bf{\Omega }}\left( {\bf{e}} \right)$ is a diagonal matrix, i.e., ${\bf{\Omega }}\left( {\bf{e}} \right)=\textrm{diag}\left\{ {d_{\rm{E}}^2{\left( {\bf{e}} \right)}, \ldots ,d_{\rm{E}}^2}{\left( {\bf{e}} \right)} \right\}$.

\emph{Proof}: Notice that the first $r$ eigenvalues of ${{\bf{\Omega }}\left( {\bf{e}} \right)}$ are all positive.
Therefore, we apply the Cauchy-Schwarz inequality and the following holds
\begin{equation}
\sum\limits_{i = 1}^r {\lambda _i^2}  \ge \frac{1}{r}{\left( {\sum\limits_{i = 1}^r {{\lambda _i}} } \right)^2} = \frac{{{P^2}}}{r}{\left( {d_{\rm{E}}^2\left( {\bf{e}} \right)} \right)^2},
\end{equation}
where the equality is achieved when the eigenvalues of ${\bf{\Omega }}\left( {\bf{e}} \right)$ are of the same value, e.g., ${\bf{\Omega }}\left( {\bf{e}} \right)=\textrm{diag}\left\{ {d_{\rm{E}}^2{\left( {\bf{e}} \right)}, \ldots ,d_{\rm{E}}^2}{\left( {\bf{e}} \right)} \right\}$. This completes the proof of Lemma 6-4. \hfill $\blacksquare$

The above lemmas show some important properties of the codeword difference matrix ${\bf{\Omega }}\left( {\bf{e}} \right)$. Based on these properties of ${\bf{\Omega }}\left( {\bf{e}} \right)$, we can now consider the following lower bounds of the eigenvalue product.

\textbf{Theorem 6-1} \emph{(Lower bound on $\prod\nolimits_{i = 1}^r {{\lambda _i}} $)}:
The product of the first $r$ eigenvalues of ${\bf{\Omega }}\left( {\bf{e}} \right)$ is lower-bounded by
\begin{equation}
\prod\limits_{i = 1}^r {{\lambda _i}} \ge {\left( {d_{\rm{E}}^2\left( {\bf{e}} \right)} \right)^r}\exp \left( {r - d_{\rm{E}}^2\left( {\bf{e}} \right)\sum\limits_{i = 1}^r {\frac{1}{{{\lambda _i}}}} } \right), \label{C7_determinant_lower_bound}
\end{equation}
where the equality holds if ${\bf{\Omega }}\left( {\bf{e}} \right)$ is a diagonal matrix, i.e., ${\bf{\Omega }}\left( {\bf{e}} \right)=\textrm{diag}\left\{ {d_{\rm{E}}^2{\left( {\bf{e}} \right)}, \ldots ,d_{\rm{E}}^2}{\left( {\bf{e}} \right)} \right\}$.

\emph{Proof}: The proof is given in the Appendix of this chapter.

It should be noted that~\eqref{C7_determinant_lower_bound} still depends on the channel parameters ${\bm{\omega}} _\tau$ and ${\bm{\omega}} _\nu$.
To obtain an unconditional lower bound, we apply an approximation to the lower bound in~\eqref{C7_determinant_lower_bound}, which is summarized in the following Theorem.

\textbf{Theorem 6-2} \emph{(Approximated lower bound on $\prod\nolimits_{i = 1}^r {{\lambda _i}} $)}:
The product of the first $r$ eigenvalues of ${\bf{\Omega }}\left( {\bf{e}} \right)$ can be approximately lower-bounded by
\begin{equation}
\prod\limits_{i = 1}^r {{\lambda _i}} \mathbin{\lower.3ex\hbox{$\buildrel>\over
{\smash{\scriptstyle\sim}\vphantom{_x}}$}} {\left( {d_{\rm{E}}^2\left( {\bf{e}} \right)} \right)^r}, \label{C7_app_determinant_lower_bound}
\end{equation}
where the approximation is exact if ${\bf{\Omega }}\left( {\bf{e}} \right)$ is a diagonal matrix, i.e., ${\bf{\Omega }}\left( {\bf{e}} \right)=\textrm{diag}\left\{ {d_{\rm{E}}^2{\left( {\bf{e}} \right)}, \ldots ,d_{\rm{E}}^2}{\left( {\bf{e}} \right)} \right\}$.

\emph{Proof}: The proof is given in the Appendix of this chapter.

Based on Theorem 6-2, it is not hard to verify that the eigenvalue product can be approximated by the term ${\left( {d_{\rm{E}}^2\left( {\bf{e}} \right)} \right)^r}$, regardless of the specific distributions of the delay and Doppler indices. Meanwhile, the above approximation is quite insightful in the sense that it only relates to the rank of the codeword difference matrix ${\bf{\Omega }}\left( {\bf{e}} \right)$ (this is actually the diversity gain as we will introduce later) and the squared Euclidean distance $d_{\rm{E}}^2\left( {\bf{e}} \right)$, which does not depend on the exact value of the delay and Doppler indices.
In particular, the approximation becomes exact if ${\bf{\Omega }}\left( {\bf{e}} \right)=\textrm{diag}\left\{ {d_{\rm{E}}^2{\left( {\bf{e}} \right)}, \ldots ,d_{\rm{E}}^2}{\left( {\bf{e}} \right)} \right\}$, which can be interpreted as the projections of the error sequence $\bf{e}$ onto each independent resolvable path, i.e., ${\bar{\bf{u}}_i}$, are orthogonal to each other.
According to Theorem 6-2, we approximate the unconditional PEP by
\begin{equation}
\Pr\left( { {{\bf{x}},{\bf{x'}}} } \right)\mathbin{\lower.3ex\hbox{$\buildrel<\over
{\smash{\scriptstyle\sim}\vphantom{_x}}$}}{\left( {\frac{{d_{\rm{E}}^2\left( {\bf{e}} \right)}}{P}} \right)^{ - r}}{\left( {\frac{{{E_s}}}{{4{N_0}}}} \right)^{ - r}}. \label{C7_New_Unconditional_PEP2}
\end{equation}
Based on~\eqref{C7_New_Unconditional_PEP2}, we note that the unconditional PEP for OTFS modulation only depends on ${d_{\rm{E}}^2\left( {\bf{e}} \right)}$, the rank of ${\bf{\Omega }}\left( {\bf{e}} \right)$, and number of independent resolvable paths $P$, and is independent from the specific distribution of delay and Doppler indices.
Our derived result is a generalized framework of the unconditional PEP for small $P$ that is suitable for various OTFS transmission cases, including the case where $r<P$ and the delay and Doppler indices are not uniformly distributed.
In particular, the term $P$ in the denominator can be interpreted as the energy averaging with respect to the number of independent paths, while the term ${\left( {d_{\rm{E}}^2\left( {\bf{e}} \right)} \right)^{ - r}}$
indicates the potential improvement of the error performance introduced by channel coding. Furthermore,
according to~\eqref{C7_New_Unconditional_PEP2}, we refer to the power of the SNR as the \textbf{diversity gain}, which dominates the exponential behaviour of the error performance for OTFS systems against the average SNR. On the other hand, the term ${{d_{\rm{E}}^2\left( {\bf{e}} \right)} \mathord{\left/
 {\vphantom {{d_{\rm{E}}^2\left( {\bf{e}} \right)} r}} \right.
 \kern-\nulldelimiterspace} P}$ is referred to as the \textbf{coding gain}, which characterizes the approximate improvement of coded OTFS systems over the uncoded counterpart with the same diversity gain, i.e., the same exponent $-r$~\cite{Tarokh1998space}.
Considering the diversity property of OTFS modulation discussed in the previous remarks, it is interesting to see from~\eqref{C7_New_Unconditional_PEP2} that there exists a fundamental trade-off between the diversity gain and the coding gain which is formally stated in the following.

\textbf{Corollary 6-1} \emph{(Trade-off between diversity and coding gain)}:
For a given channel code, the diversity gain of OTFS systems improves with the number of independent resolvable paths $P$, while the coding gain declines.

Based on Corollary 6-1, we note that when $P$ (the rank of ${\bf{\Omega }}\left( {\bf{e}} \right)$) is small, the diversity gain is small. In this case, the squared Euclidean distance between codewords
 is crucial for OTFS systems as an optimized code can greatly improve the error performance.
On the other hand, when $P$ (the rank of ${\bf{\Omega }}\left( {\bf{e}} \right)$) is large, there is a large diversity gain. In this case, it is expected that the code design can only offer a limited error performance improvement. However, it should also be noted that the coding gain always improves with the increase of $d_{\rm{E}}^2\left( {\bf{e}} \right)$, regardless of the value of the diversity gain according to~\eqref{C7_New_Unconditional_PEP2}. Therefore, a preliminary guideline for the code design for the OTFS systems is to maximize the minimum value of $d_{\rm{E}}^2\left( {\bf{e}} \right)$ among all pairs of codewords of the code.

To verify the accuracy of the derived unconditional PEP bound, we numerically compare the coding gain and the derived bound corresponding to~\eqref{C7_New_Unconditional_PEP2} and~\eqref{C7_New_PEP_derivation1}.
In particular, recalling~\eqref{C7_New_PEP_derivation1}, after some manipulations, we obtain
\begin{equation}
\Pr\left( {\left. {{\bf{x}},{\bf{x'}}} \right|{{{\bm{\omega}} _\tau }},{{{\bm{\omega}} _\nu }}} \right)\le{\left( {\frac{{{{\left( {\prod\limits_{i = 1}^r {{\lambda _i}} } \right)}^{\frac{{\rm{1}}}{r}}}}}{P}} \right)^{ - r}}{\left( {\frac{{{E_s}}}{{4{N_0}}}} \right)^{ - r}}.
\end{equation}
Hence, we refer to the term ${{{{\left( {\prod\nolimits_{i = 1}^r {{\lambda _i}} } \right)}^{\frac{{\rm{1}}}{r}}}} \mathord{\left/
 {\vphantom {{{{\left( {\prod\nolimits_{i = 1}^r {{\lambda _i}} } \right)}^{\frac{{\rm{1}}}{r}}}} P}} \right.
 \kern-\nulldelimiterspace} P}$ as the \emph{conditional coding gain} of the OTFS systems for given channel parameters ${\bm{\omega}} _\tau$ and ${\bm{\omega}} _\nu$, as the eigenvalues are related to ${{{\bm{\omega}} _\tau }}$ and ${{{\bm{\omega}} _\nu }}$.
Based on the conditional coding gain, we can also obtain the \emph{average coding gain} with respect to various channel parameters ${{{\bm{\omega}} _\tau }},{{{\bm{\omega}} _\nu }}$ and error sequences $\bf{e}$ by means of Monte Carlo simulation.
On the other hand, from the unconditional PEP upper bound~\eqref{C7_New_Unconditional_PEP2}, we call the function $f\left( {d_{\rm{E}}^2\left( {\bf{e}} \right)} \right) = {{d_{\rm{E}}^2\left( {\bf{e}} \right)} \mathord{\left/
 {\vphantom {{d_{\rm{E}}^2\left( {\bf{e}} \right)} P}} \right.
 \kern-\nulldelimiterspace} P}$ the \emph{coding gain bound} of the OTFS systems.
\begin{figure}
\centering
\includegraphics[width=0.7\textwidth]{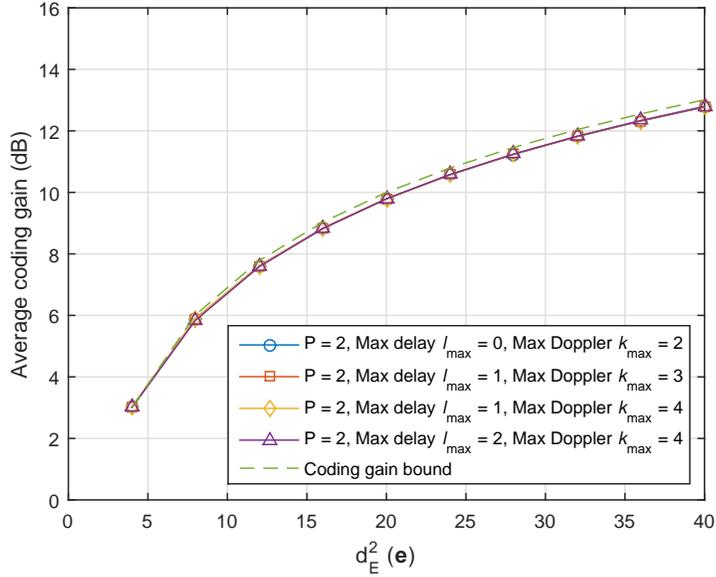}
\caption{Average coding gain for error sequences with $P=2$ and ${d_{\rm{E}}^2\left( {\bf{e}} \right)}$ in terms of various delay and Doppler indices, comparing with the coding gain bound.}
\label{C7_Average_coding_gain_curve1}
\centering
\end{figure}
Now let us compare the average coding gain and the coding gain obtained from the performance bound via simulations.
We numerically average the
conditional coding gains subjected to all error sequences with ${d_{\rm{E}}^2\left( {\bf{e}} \right)}$ and channel parameters ${{{\bm{\omega}} _\tau }},{{{\bm{\omega}} _\nu }}$ to obtain the average coding gain. Without loss of generality, we consider BPSK signals and error sequences with ${d_{\rm{E}}^2\left( {\bf{e}} \right)}$ for OTFS systems with integer delay and Doppler shifts.

Fig.~\ref{C7_Average_coding_gain_curve1} shows the comparison between the average coding gain and the corresponding coding gain bound in decibels for $P=2$ with various maximum delay and Doppler indices,  where the OTFS system with $N=5$ and $M=2$ is considered. As shown in the figure, different values of maximum delay and Doppler indices do not have a strong impact on the average coding gain. Meanwhile, it can be observed in the figure that the average coding gain improves with the increase of the squared Euclidean distance ${d_{\rm{E}}^2\left( {\bf{e}} \right)}$.
Furthermore, the derived coding gain bound shows a close match with the overall average coding gain, especially when ${d_{\rm{E}}^2\left( {\bf{e}} \right)}$ is small.

\begin{figure}
\centering
\includegraphics[width=0.7\textwidth]{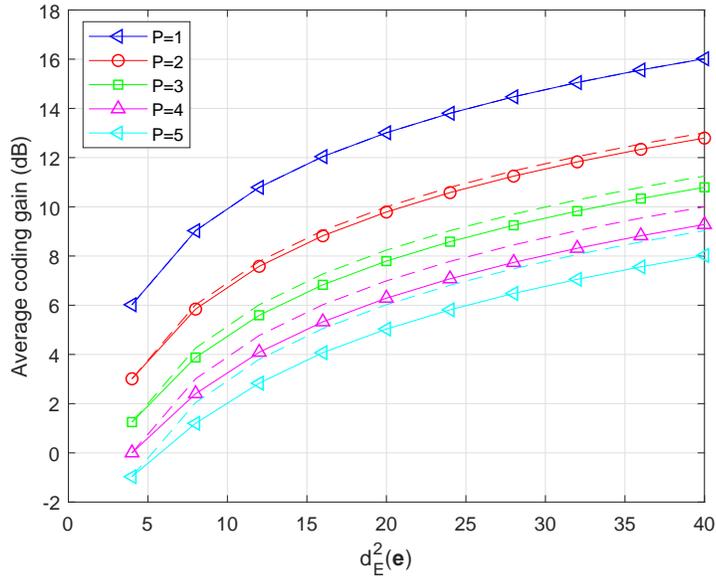}
\caption{Average coding gain for error sequences with ${d_{\rm{E}}^2\left( {\bf{e}} \right)}$ in terms of different numbers of independent resolvable paths, comparing with the coding gain bounds (in dashed lines), where the maximum delay and Doppler indices are set to be $l_{\rm{max}}=2$ and $k_{\rm{max}}=4$, respectively.}
\label{C7_Average_coding_gain_curve2}
\centering
\end{figure}
Fig.~\ref{C7_Average_coding_gain_curve2} illustrates the average coding gains and the corresponding coding gain bounds (in dashed lines) in decibels with respect to different values of $P$, where the OTFS system with $N=5$ and $M=2$ is considered.
As shown in the figure, given ${d_{\rm{E}}^2\left( {\bf{e}} \right)}$, the average coding gain decreases with the increase of the number of paths $P$, which is consistent with Corollary 6-1.
Similar to the previous figure, the coding gain bounds match well with the general behaviour of average coding gains, especially when $P$ is small,
which verifies the correctness of our derivation. On the other hand, we notice that the derived coding gain bound slightly diverge from the average coding gains, when $P$ is large. To verify this observation, we in the following consider the case with a large number of paths and a larger OTFS frame.

\begin{figure}
\centering
\includegraphics[width=0.7\textwidth]{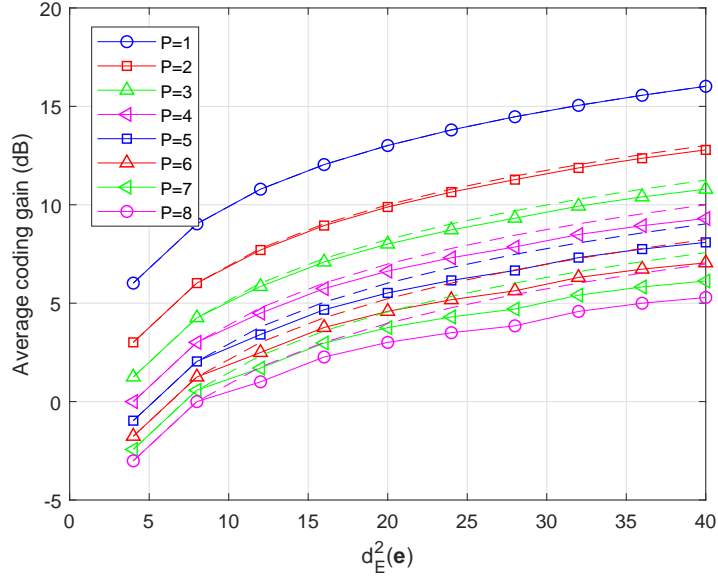}
\caption{Average coding gain for error sequences with ${d_{\rm{E}}^2\left( {\bf{e}} \right)}$ in terms of different numbers of independent resolvable paths, comparing with the coding gain bounds (in dashed lines), where the maximum delay and Doppler indices are set to be $l_{\rm{max}}=2$ and $k_{\rm{max}}=2$, respectively.}
\label{C7_Average_coding_gain_curve3}
\centering
\end{figure}

We consider an OTFS system with $M=32$ and $N=16$ in Fig.~\ref{C7_Average_coding_gain_curve3}, where the maximum delay and Doppler indices are set to be $l_{\rm{max}}=2$ and $k_{\rm{max}}=2$, respectively.
Since numerically emulating all the error sequences with such a frame size is generally intractable in a reasonable time frame even with BPSK mapping, we consider the comparison between the average coding gains and the corresponding coding gain bounds with
given error sequences\footnote{Without loss of generality, the error sequences are of the form ${\bf e}=[2, 0, -2, 2, 0, -2,...,0...0]^{\rm T}$.}.
Similar to the previous figure, we observe that the derived coding gain bounds diverge from the average coding gains for a large number of paths and the gap between them becomes wider with the increase of $P$.
This observation motivates us to derive a more suitable approximation for the coding gain for the case of a large $P$, the details of which will be given in the following subsection.

\subsection{Error Performance Analysis for Large Values of $P$}
When there is a large number of independent resolvable paths of the channel, i.e., the value of $P$ is large, the unconditional PEP can be more accurately bounded by considering the strong law of large number \cite{Yuan2003performance,vucetic2003space}.
In specific, the term $\sum\nolimits_{i = 1}^r {{\lambda _i}{{| {{{\tilde h}_i}} |}^2}} $ in~\eqref{C7_PEP_derivation2} approaches a Gaussian random variable, due to the central limit theorem \cite{papoulis2002probability} and the fact that OTFS modulation can achieve the full diversity with proper designs using precoding techniques~\cite{Surabhi2019on}.

Notice that ${{\tilde h}_i}$ follows the complex Gaussian distribution, with mean ${\mu _{{{\tilde h}_i}}}=(\mathbb{E}\left[ {\bf{h}} \right])^{\rm H}\cdot{{\bf{v}}_i}$ and variance $1/(2P)$ per real dimension.
Therefore, for the ease of derivation, we normalize the variance and rewrite~\eqref{C7_PEP_derivation2} as
\begin{equation}
\Pr\left( {{\left. {{\bf{x}},{\bf{x'}}} \right|{\bf{h}},{{{\bm{\omega}} _\tau }},{{{\bm{\omega}} _\nu }}}} \right) \le \exp \left( { - \frac{{{E_s}}}{{4{N_0}}}\sum\limits_{i = 1}^r {\frac{{{\lambda _i}}}{P}{{\left| {{{\bar h}_i}} \right|}^2}} } \right), \label{C7_New2_PEP_derivation2}
\end{equation}
where ${\bar h_i} = \sqrt P {\tilde h_i}$.
Note that $\left\{ {{{\bar h}_1},{{\bar h}_2},...,{{\bar h}_r}} \right\}$ are independent complex Gaussian random variables with mean ${\mu _{{{\bar h}_i}}}={\sqrt{P}} {\mu _{{{\tilde h}_i}}}$ and variance $1/2$ per real dimension.
Furthermore, it is obvious that $\left| {{{\bar h}_i}} \right|$ follows the Rician distribution with a Rician factor ${{\bar{K}}_i}= {\left| {{\mu _{{{\bar h}_i}}}} \right|^2}$ and a unit variance.
Thus, it can be shown that ${{{\left| {{{\bar h}_i}} \right|}^2}}$ follows a noncentral chi-squared distribution with one DoF and noncentrality parameter $S = {{\bar{K}}_i}$, whose mean and variance are given by~\cite{kay1993fundamentals}
\begin{align}
{\mu _{{{\left| {{{\bar h}_i}} \right|}^2}}} &= 1 + {{\bar{K}}_i}, \label{C7_mean}\\
\sigma _{{{\left| {{{\bar h}_i}} \right|}^2}}^2 &= 2 + 4{{\bar{K}}_i}. \label{C7_variance}
\end{align}
Next, we derive the unconditional PEP by means of Gaussian approximation. To start with, let $\psi  = \sum\limits_{i = 1}^r {{\lambda _i}{{\left| {{{\bar h}_i}} \right|}^2}} $. According to~\eqref{C7_mean} and~\eqref{C7_variance}, we approximate $\psi $ as a Gaussian random variable, whose mean is
${\mu _\psi } = \sum\limits_{i = 1}^r {{\lambda _i}\left( {1 + {{{\bar{K}}_i}}} \right)} $ and variance is $\sigma _\psi ^2 = \sum\limits_{i = 1}^r {\lambda _i^2\left( {2 + 4{{\bar{K}}_i}} \right)} $.
Thus, according to the Gaussian distribution of $\psi $, the conditional PEP in~\eqref{C7_PEP_derivation2} is upper-bounded by
\begin{equation}
\Pr\left( {\left. {{\bf{x}},{\bf{x'}}} \right|{{{\bm{\omega}} _\tau }},{{{\bm{\omega}} _\nu }}} \right)\le \int\limits_{0} ^{ + \infty } {\exp \left( { - \frac{{{E_s}}}{{4{N_0}P}}\psi } \right)p} \left( \psi  \right)d\psi  .\label{C7_PEP_derivation6}
\end{equation}
Considering
\begin{align}
\int\limits_0^{ + \infty } {\exp \left( { - \gamma \psi } \right)p} \left( \psi  \right)d\psi= \exp \left( {\frac{1}{2}{\gamma ^2}\sigma _\psi ^2 - \gamma {\mu _\psi }} \right)Q\left( {\frac{{\gamma \sigma _\psi ^2 - {\mu _\psi }}}{{{\sigma _\psi }}}} \right),\gamma  > 0,
\end{align}
we obtain
\begin{align}
\Pr\left( {\left. {{\bf{x}},{\bf{x'}}} \right|{{{\bm{\omega}} _\tau }},{{{\bm{\omega}} _\nu }}} \right) \le \exp \left( {\frac{1}{2}{{\left( {\frac{{{E_s}}}{{4{N_0}}}} \right)}^2}\cdot\frac{{\sigma _\psi ^2}}{{{P^2}}} - \frac{{{E_s}}}{{4{N_0}}}\cdot\frac{{{\mu _\psi }}}{P}} \right)Q\left( {\frac{{{E_s}}}{{4{N_0}}}\cdot\frac{{{\sigma _\psi }}}{P} - \frac{{{\mu _\psi }}}{{{\sigma _\psi }}}} \right). \label{C7_PEP_derivation7}
\end{align}
Similar to the previous subsection, we consider the special case of Rayleigh fading.

In the case of \textbf{Rayleigh fading}, i.e., $|{{{\bar{h}}_i}}|$ and $|{{h_i}}|$ follow the Rayleigh distribution, we have ${\mu _\psi } = \sum\limits_{i = 1}^r {{\lambda _i}} $ and $\sigma _\psi ^2 = 2 \sum\limits_{i = 1}^r {\lambda _i^2} $.
Therefore, the right hand side of~\eqref{C7_PEP_derivation7} is given by
\begin{align}
\Pr\left( {\left. {{\bf{x}},{\bf{x'}}} \right|{{{\bm{\omega}} _\tau }},{{{\bm{\omega}} _\nu }}} \right)
\le &\exp \left( {{{\left( {\frac{{{E_s}}}{{4{N_0}}}} \right)}^2}\sum\limits_{i = 1}^r {\frac{{\lambda _i^2}}{{{P^2}}}}  - \frac{{{E_s}}}{{4{N_0}}}\sum\limits_{i = 1}^r {\frac{{{\lambda _i}}}{P}} } \right)\notag\\
&Q\left( {\frac{{{E_s}}}{{4{N_0}P}}\sqrt {2\sum\limits_{i = 1}^r {\lambda _i^2} }  - \frac{{\sum\nolimits_{i = 1}^r {{\lambda _i}} }}{{\sqrt {2\sum\nolimits_{i = 1}^r {\lambda _i^2} } }}} \right).
\label{C7_PEP_derivation8}
\end{align}
Furthermore, we consider the Chernoff bound of the Q-function \cite{Yuan2003performance}
\begin{equation}
Q\left( \gamma  \right) \le \exp \left( { - \frac{1}{2}{\gamma ^2}} \right),\gamma  > 0,
\end{equation}
and~\eqref{C7_PEP_derivation8} can be further upper-bounded by
\begin{align}
&\Pr\left( {\left. {{\bf{x}},{\bf{x'}}} \right|{{{\bm{\omega}} _\tau }},{{{\bm{\omega}} _\nu }}} \right) \notag\\
\le &\exp \left( {{{\left( {\frac{{{E_s}}}{{4{N_0}}}} \right)}^2}\sum\limits_{i = 1}^r {\frac{{\lambda _i^2}}{{{P^2}}}}  - \frac{{{E_s}}}{{4{N_0}}}\sum\limits_{i = 1}^r {\frac{{{\lambda _i}}}{P}} } \right)\exp \left( { - {{\left( {\frac{{{E_s}}}{{4{N_0}}}} \right)}^2}\sum\limits_{i = 1}^r {\frac{{\lambda _i^2}}{{{P^2}}}}  - \frac{{{{\left( {\sum\limits_{i = 1}^r {{\lambda _i}} } \right)}^2}}}{{4\sum\limits_{i = 1}^r {\lambda _i^2} }} + \frac{{{E_s}}}{{4{N_0}}}\sum\limits_{i = 1}^r {\frac{{{\lambda _i}}}{P}} } \right) \notag\\
=&\exp \left( { - \frac{{{{\left( {\sum\limits_{i = 1}^r {{\lambda _i}} } \right)}^2}}}{{4\sum\limits_{i = 1}^r {\lambda _i^2} }}} \right),
\label{C7_New_PEP_Large_P_derivation2}
\end{align}
when
\begin{equation}
\frac{{{E_s}}}{{4{N_0}}} \ge \frac{P{\sum\nolimits_{i = 1}^r {{\lambda _i}} }}{{2\sum\nolimits_{i = 1}^r {\lambda _i^2} }}. \label{C7_New_PEP_Large_P_derivation1}
\end{equation}
Based on~\eqref{C7_New_PEP_Large_P_derivation2}, the unconditional PEP can be approximately upper-bounded as shown in the following Theorem.

\textbf{Theorem 6-3} (\emph{Unconditional PEP upper bound for large $P$}):
For a large value of $P$ and a reasonably high SNR, i.e., $\frac{{{E_s}}}{{4{N_0}}} \ge \frac{r}{2{d_{\rm{E}}^2\left( {\bf{e}} \right)}}$, the unconditional PEP of OTFS systems can be approximately upper-bounded by
\begin{equation}
\Pr\left(  {{\bf{x}},{\bf{x'}}} \right) \mathbin{\lower.3ex\hbox{$\buildrel<\over
{\smash{\scriptstyle\sim}\vphantom{_x}}$}} \exp \left( { - \frac{{{E_s}}}{{16{N_0}}}d_{\rm{E}}^2} \left( {\bf{e}} \right)\right) . \label{C7_Large_P_upper_bound}
\end{equation}

\emph{Proof}: The proof is given in the Appendix of this chapter.

It should be noted that, for $r \ge 4$, the approximation in Theorem 3 is sufficiently accurate owing to the strong law of the large number \cite{vucetic2003space,Yuan2003performance}.
On the other hand, the value of $r$ is usually smaller than the squared Euclidean distance $d_{\rm{E}}^2\left( {\bf{e}} \right)$ for practical wireless transmissions with good channel codes\footnote{For example, a popular industry-standard rate-$1/2$ convolutional code with code memory of $6$ has a minimum squared Euclidean distance $d_{\rm{E}}^2 \left( {\bf{e}} \right)=40$ \cite{ryan2009channel}.}.
Therefore, our SNR assumption is reasonable.
Compared with Corollary 6-1, it is not surprising that the unconditional PEP only depends on the squared Euclidean distance $d_{\rm{E}}^2\left( {\bf{e}} \right)$, regardless of the delay and Doppler indices.
Furthermore, we note that~\eqref{C7_Large_P_upper_bound} is of the similar form of the error performance for AWGN channels~\cite{Ventura1997impact}.
This is because the impact of fading is mitigated by a large number of diversity branches and consequently, in other words, the channel with a large number of diversity paths approaches
an AWGN model~\cite{Ventura1997impact}.

Notice that the upper bound in Theorem 6-3 is based on the PEP analysis, which does not directly indicate the average error performance of the coded system~\cite{stefanov2003performance}.
But it can be used to approximate the coding gain for large $P$.
Based on the results from Theorem 6-3, we apply the commonly adopted coding gain approximation for AWGN channels~\cite{ryan2009channel} to evaluate the coding gain of OTFS systems with a large $P$. In specific, we have
\begin{equation}
{\rm{Coding}}\:{\rm{gain}} \simeq 10{\log _{10}}\left( {\frac{{d_{{\rm{c}},\min }^2}}{{d_{{\rm{u}},\min }^2}}} \right) {\rm{ dB}}, \label{C7_coding_gain_large_P}
\end{equation}
where ${d_{{\rm{c}},\min }^2}$ and ${d_{{\rm{u}},\min }^2}$ are the minimum squared Euclidean distances for coded and uncoded OTFS systems, respectively.

In the previous subsections, we have derived the error performance analysis of coded OTFS systems, which is valid for general underspread WSSUS channels including both the integer and fractional delay and Doppler cases.
In the following, we will discuss the design of channel codes according to our error performance analysis.

\subsection{Code Design Issues}
According to the derived analysis, the rule-of-thumb channel code design criterion is discussed in this section.
Without loss of generality, we only consider the Rayleigh fading channel in the following.
We can see from the previous analysis that the code design criterion for the coded OTFS system is to maximize the minimum squared Euclidean distance $d_{\rm{E}}^2\left( {\bf{e}} \right)$.

\textbf{Proposition 6-1} \emph{(The squared Euclidean distance criterion)}:
The channel code should be designed to maximize the minimum squared Euclidean distance among all pairs of possible codewords.


We note that even with the designed code, the error performance of coded OTFS systems may still vary with different channel parameters e.g., ${\bm{\omega}} _\tau$ and ${\bm{\omega}} _\nu$.
This detrimental effect due to channel realizations is widely observed in the system designs for fading channels, such as in~\cite{Lu2000space,Biglieri1998fading}.
In specific, with different channel parameters, the value of the conditional coding gain can be different even with the same error sequence, which may potentially jeopardize the overall error performance of OTFS systems. In order to obtain a more robust performance, it is desirable to apply an interleaver to permute the coded symbols before sending to the constellation mapper or the OTFS modulator in the DD domain.
As pointed out in~\cite{Biglieri1998fading}, such an interleaver can ``whiten'' the transmitted symbols from the information theoretic point of view and the detrimental effect on the error performance due to the channel parameters can thus be alleviated.

To examine our performance analysis of the coded OTFS systems, we perform numerical simulations for OTFS systems over high-mobility channels, the results of which will be shown in the next section.

\section{Numerical Results}
In this section, the error performance of the coded OTFS system with various channel codes is evaluated via numerical simulations by considering perfect CSI at the receiver side.
We consider the BPSK signal for the OTFS system, where the data sequence is firstly encoded and interleaved, and then BPSK mapped.
Without loss of generality, we consider the SPA~\cite{Kschischang2001factor,Yuan2019simple} for OTFS detection, where the details can be found in~\cite{li2021hybrid}.  In order to verify the accuracy of the analytical results, we consider four different convolutional codes (with trellis termination) with different minimum squared Euclidean distance $d_{\rm{E}}^2{\left( {\bf{e}} \right)} $ among all possible codeword pairs. The details of the code parameters are given in Table~\ref{C7_Code_parameters}, including the generator matrix and the memory length. In particular, we also show the smallest value of squared Euclidean distance $d_{\rm{E}}^2\left( {\bf{e}} \right)$ among all possible codeword pairs for each code. In specific, we consider a coded OTFS system with $N=8$ and $M=16$ and correspondingly the codeword length for all considered simulations is $128$ bits unless otherwise specified.
The channel decoder adopts the logarithm domain Bahl-Cocke-Jelinek-Raviv (BCJR) algorithm \cite{Bahl1974optimal}.
Furthermore, we consider the Rayleigh fading case.
If not otherwise specified, we only consider the integer delay and Doppler case and set the maximum delay index as $l_{\max }=3$ and the maximum Doppler index as $k_{\max }=5$,
which is corresponding to a relative speed around $250$ km/h with $4$ GHz carrier frequency and $1.5$ kHz sub-carrier spacing. For each channel realization, we randomly select the
delay and Doppler indices according to the uniform distribution, such that we have $ - {k_{\max }} \le {k_i} \le {k_{\max }}$ and $0 \le {l_i} \le {l_{\max }}$.

\begin{table}[htbp]
\caption{Code Parameters}
\centering
\small
\begin{tabular}{|c|c|c|c|}
\hline
Code structure~&~Generator matrix~&~Memory length~&~Minimum $d_{\rm{E}}^2\left( {\bf{e}} \right)$\\
\hline
A~&~$\left[ 1+D, D \right]$&~1~&~12\\
\hline
B~&~$\left[ 1 + {D^2}, 1 + D + {D^2} \right]$&~2~&~20\\
\hline
C~&~$\left[ {1 + {D^2} + {D^5},1 + D + {D^2} + {D^3} + {D^4} + {D^5}} \right]$~&~5~&~32\\
\hline
D~&~$\left[ {1 \!+\! D \!+\! {D^2} \!+ \!{D^5}\! +\! {D^6},1\! + \!{D^2} \!+\! {D^3}\! +\!{D^4} \!+ \!{D^6}} \right]$&~6~&~40 \\
\hline
\end{tabular}
\label{C7_Code_parameters}
\end{table}

\begin{figure}
\centering
\includegraphics[width=0.7\textwidth]{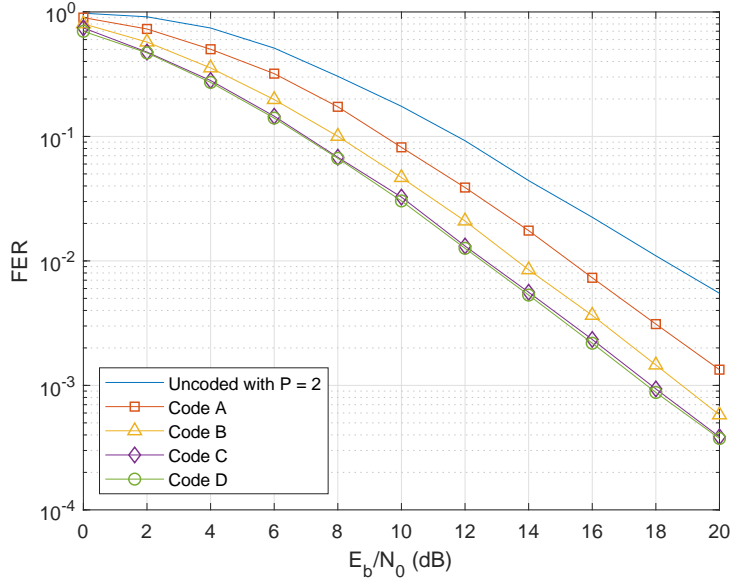}
\caption{FER performance for OTFS modulation with different codes for $P=2$, where the relative user equipment (UE) speed is $250$ km/h.}
\label{C7_Path2}
\centering
\end{figure}
The frame-error-rate (FER) performance of the OTFS systems with $P=2$ is shown in Fig.~\ref{C7_Path2}.
It can be observed that the slope of the FER curve for the uncoded OTFS system is slightly worse than that for coded OTFS systems. This indicates that the uncoded OTFS system with $P=2$ does not guarantee the full diversity for all possible channel realizations, which is consistent with the analysis in \cite{Raviteja2019effective}.
More importantly, this observation also shows that the application of channel coding can improve the diversity gain of OTFS systems in the case where the OTFS modulation fails to achieve the full diversity. 
Moreover, we observe that employing the channel code with a larger minimum squared Euclidean distance $d_{\rm{E}}^2\left( {\bf{e}} \right)$ indeed leads to a larger coding gain. In specific, we observe from the figure that for FER $\approx {10^{ - 2}}$, the required SNRs for codes A, B, C and D, are $15.28$ dB, $13.64$ dB, $12.65$ dB, and $12.54$ dB, respectively. Compared to uncoded OTFS systems, these four coded OTFS systems achieve coding gains roughly $2.99$ dB, $4.63$ dB, $5.62$ dB, and $5.73$ dB, respectively. This observation clearly substantiates the proposed performance analysis and code design criterion in Proposition 6-1.


\begin{figure}
\centering
\includegraphics[width=0.7\textwidth]{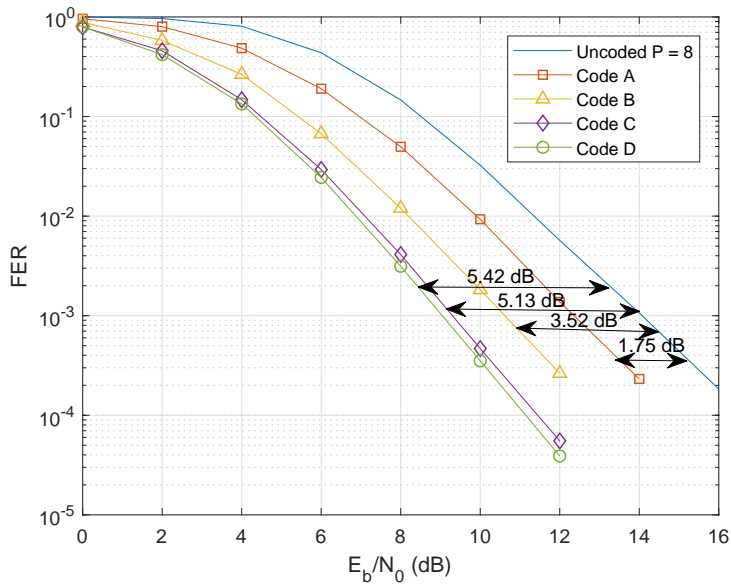}
\caption{FER performance for OTFS modulation with different codes for $P=8$, where the relative UE speed is $250$ km/h.}
\label{C7_Path8}
\centering
\end{figure}
Fig.~\ref{C7_Path8} shows the FER performance of the OTFS systems with $P=8$.
We observe that the channel code with a larger minimum squared Euclidean distance $d_{\rm{E}}^2\left( {\bf{e}} \right)$ enjoys a larger coding gain compared with the uncoded OTFS system, which is consistent with the analysis of Theorem 6-3.
Furthermore, we can calculate the approximated coding gains according to~\eqref{C7_coding_gain_large_P} for the adopted channel codes, which are $1.83$ dB, $4.12$ dB, $6.37$ dB, and $7.42$ dB for code A to D, respectively, while we also obtain the actual coding gains based on the simulation results that are $1.75$ dB, $3.52$ dB, $5.13$ dB, and $5.42$ dB for code A to D, respectively.
These observations verify the accuracy of our analysis for large $P$ in Theorem~6-3 and the coding gain approximation.
Together with the observations from Fig.~\ref{C7_Path2}, we can also conclude that our proposed code design criterion in Proposition 6-1 is universal for general OTFS systems regardless of the channel parameters and the number of paths $P$.

\begin{figure}
\centering
\includegraphics[width=0.7\textwidth]{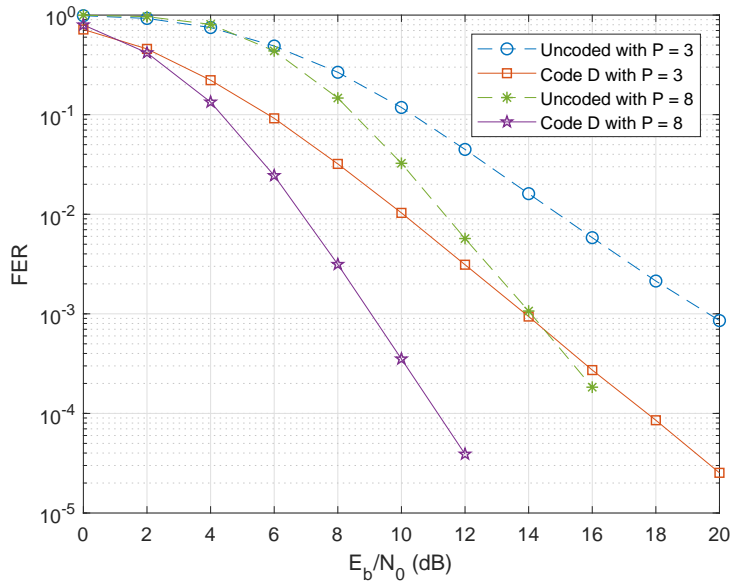}
\caption{FER performance of code D for OTFS modulation with $P=3$ and $P=8$, comparing with that of uncoded OTFS systems, where the relative UE speed is $250$ km/h.}
\label{C7_Coding_vs_Path}
\centering
\end{figure}

Fig.~\ref{C7_Coding_vs_Path} presents the trade-off between the diversity and coding gain. In particular, we consider the FER performance of code D with $P = 3$ and $P = 8$, comparing with that of the corresponding uncoded OTFS systems.
As shown in figure, at FER $\approx {10^{ - 3}}$, the coded OTFS system with $P = 3$ exhibits around $5.7$ dB coding gain compared to that of the uncoded OTFS system with the same FER,
while only around $5.0$ dB coding gain is obtained for the coded OTFS system with $P = 8$. This observation matches the prediction in Corollary 1, which
indicates that the coding gain reduces with the increase of $P$, given the same channel code.

\begin{figure}
\centering
\includegraphics[width=0.7\textwidth]{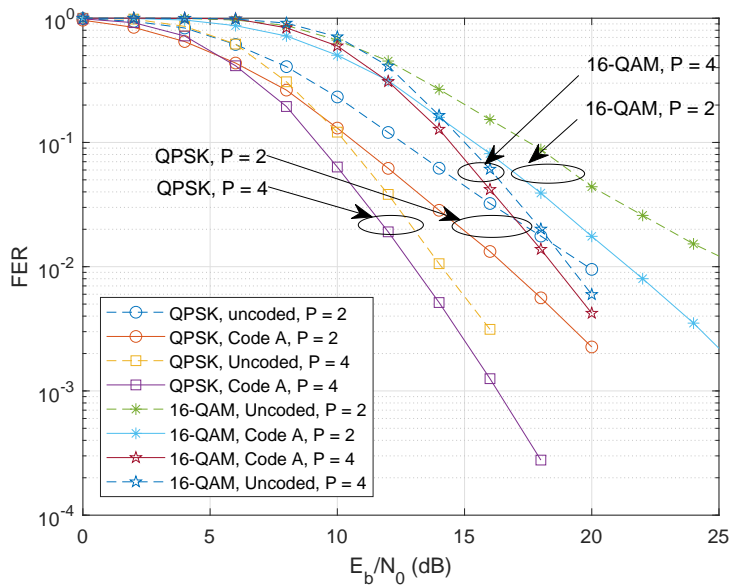}
\caption{FER performance of code A for QPSK/16-QAM mapped OTFS systems with different number of paths and the relative UE speed of $250$ km/h.}
\label{C7_Coding_gain_higher_modulation}
\centering
\end{figure}
Fig.~\ref{C7_Coding_gain_higher_modulation} presents the trade-off between the diversity and coding gain with practical mappings, namely, QPSK and 16-QAM mappings. In particular, we consider the FER performance of code A with $P = 2$ and $P = 4$, comparing with that of the corresponding uncoded OTFS systems. Similar to the previous figure, the trade-off between the diversity and coding gain is clearly demonstrated by the simulation results.  

The FER performance of the OTFS modulation with code D and different number of paths $P$ is illustrated in Fig.~\ref{C7_Coding_gain_higher_modulation}.
It can be observed from the figure that given the same code, the error performance of the coded OTFS systems improves with the increase of number of distinguishable paths. Furthermore, it is obvious that
with the same code, a larger value of $P$ corresponds to a larger diversity advantage as indicated in the figure, which is consistent with our analysis.

\begin{figure}
\centering
\includegraphics[width=0.7\textwidth]{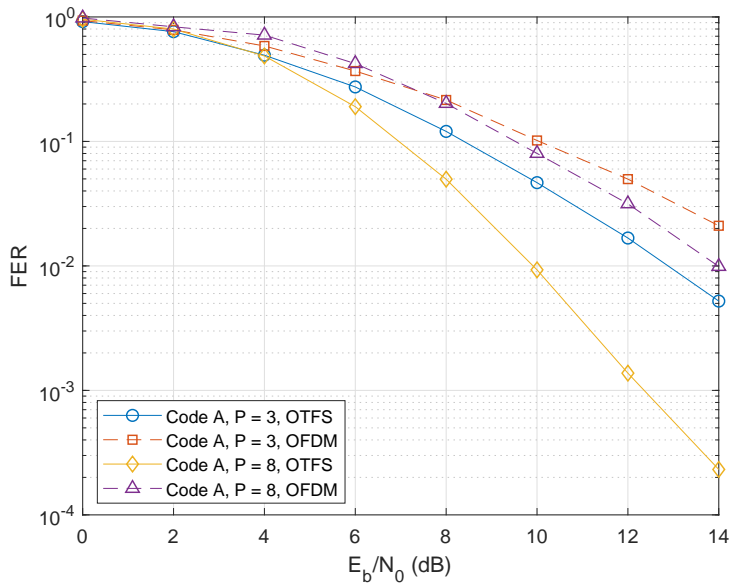}
\caption{FER performance of code A for OTFS modulation with $P=3$ and $P=8$, compared to an OFDM system, where the relative UE speed is $250$ km/h.}
\label{C7_OTFS_vs_OFDM}
\centering
\end{figure}

We compare the FER performance of code A with $P = 3$ and $P = 8$, and that of the corresponding OFDM systems in Fig.~\ref{C7_OTFS_vs_OFDM}. As observed from the figure, the OTFS system enjoy better error performance than that of the OFDM system with the same code, for both $P = 3$ and $P = 8$. Furthermore, the FER curve of the OFDM system shares almost the same slope as that of the OTFS system, for $P = 3$. As for $P = 8$, the achieved diversity gain of the OTFS system is clearly higher than that of the OFDM system.
Note that the diversity gain of coded OFDM systems is determined by the smaller value of the minimum symbol-wise Hamming distance of the code ${\delta _{\rm{H}}}$ and the number of paths $P$\footnote{According to \cite{Diversity_OFDM}, we have $r_{\rm{OFDM}}={\rm{min}}({\delta _{\rm{H}}},P)$, where $r_{\rm{OFDM}}$ is the achievable diversity gain of a coded OFDM system. In specific, we have $r_{\rm{OFDM}}=3$ for both $P=3$ and $P=8$ with Code A.} \cite{Diversity_OFDM}, while OTFS systems can obtain the full diversity almost surely regardless of the employment of channel codes.
Therefore, this observation clearly shows the advantage of the OTFS systems over the OFDM systems.

\begin{figure}
\centering
\includegraphics[width=0.7\textwidth]{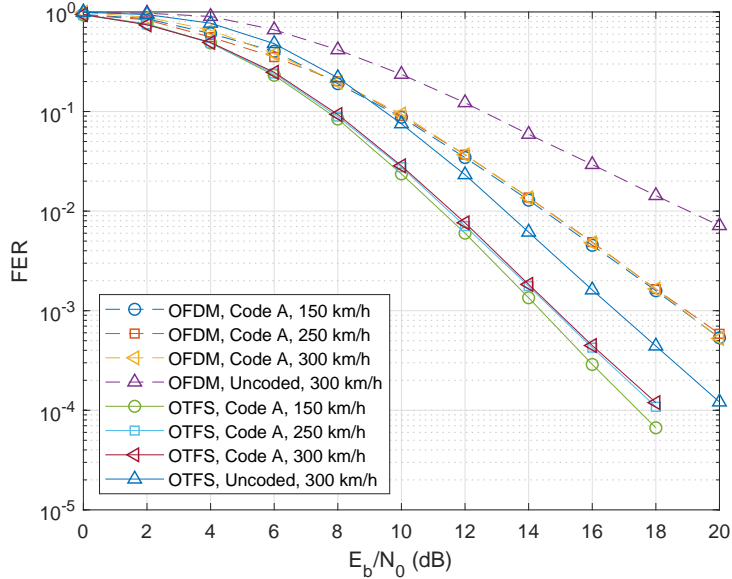}
\caption{FER performance of code A for OTFS modulation and OFDM modulation with $P=4$, where the values of the relative UE speed are $150$ km/h, $250$ km/h, and $300$ km/h, respectively.}
\label{C7_OTFS_vs_OFDM_vs_Speed}
\centering
\end{figure}

The FER performance of both OTFS systems and OFDM systems with various values of the relative UE speed is compared in Fig.~\ref{C7_OTFS_vs_OFDM_vs_Speed}. We consider $l_{\max }=3$ and $k_{\max }=3, 5, 6$, which corresponds to the cases where the relative UE speeds are $150$ km/h, $250$ km/h, and $300$ km/h, respectively.
We apply the near-optimal ML detection~\cite{li2021hybrid} for both the OTFS and OFDM systems to have a fair comparison. We observe that the error performance of both OTFS and OFDM systems with ML detection does not change much with different relative UE speeds, which is consistent with the observations in~\cite{Raviteja2018interference}.
Note that, the ML detection is not practically feasible for conventional OFDM systems due to the high detection complexity. Therefore, frequency domain equalization is usually deployed for OFDM systems~\cite{mostofi2005ici}. In this case, the error performance of OFDM systems will degrade dramatically with the increase of the speed~\cite{mostofi2005ici} due to the severe ICI induced by the Doppler spread.
Furthermore, the FER performance of the OTFS systems outperform that of the OFDM systems, including both coded and uncoded cases with various values of the speed.
Similar to the previous figure, we also observe that the achieved diversity gain of the OTFS system is higher than that of the OFDM system, which agrees with our analysis.

\begin{table*}[htbp]
\caption{Code Parameters for Fig.~\ref{C7_LDPC_Turbo}}
\centering
\begin{tabular}{|c|c|c|c|}
\hline
Code~&~Data length~&~Codeword length~&~Code rate\\
\hline
Convolutional Code D~&~250&~512~&~0.488\\
\hline
5G LDPC~&~256&~512~&~0.5\\
\hline
LTE Turbo~&~250~&~512~&~0.488\\
\hline
\end{tabular}
\label{C7_Code_parameters2}
\end{table*}

\begin{figure}
\centering
\includegraphics[width=0.7\textwidth]{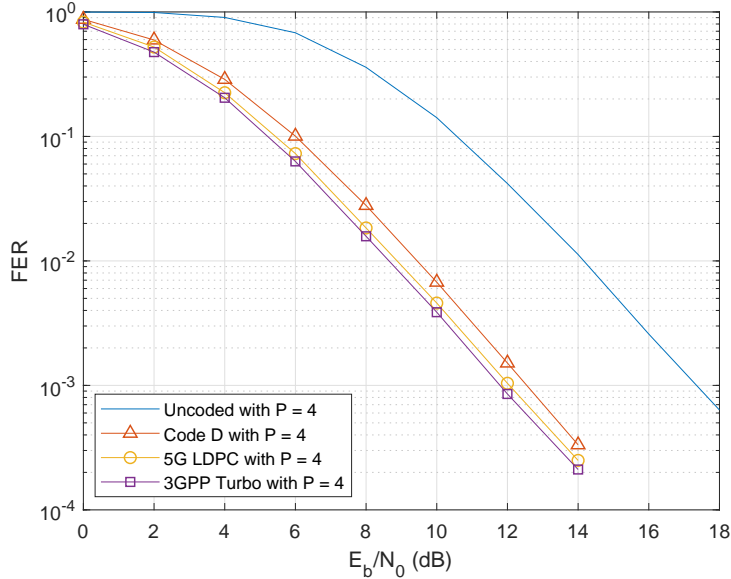}
\caption{FER performance of 5G LDPC code and 3GPP Turbo code for OTFS modulation with $P=4$ and a relative UE speed of $250$ km/h, where the codeword length is $512$ bits.}
\label{C7_LDPC_Turbo}
\centering
\end{figure}

We present the FER results of coded OTFS systems with modern codes in Fig.~\ref{C7_LDPC_Turbo}. We consider an LDPC code from the 5G communication standard \cite{5gChannel1} (referred to as the 5G LDPC code) and the Turbo code from the 3GPP long term evolution (LTE) standard \cite{LTE_Turbo_TS} (referred to as the LTE Turbo code), where code parameters are given in Table~\ref{C7_Code_parameters2} and we have $N=16$ and $M=32$ for OTFS modulation.
As a benchmark, the FER performance of the convolutional code D is also given in Fig.~\ref{C7_LDPC_Turbo}. We observe that, the LTE Turbo code achieves the best error performance compared to the 5G LDPC code and the convolutional code D with $P=4$, although they all share the same diversity gain. More specifically, at FER $\approx {10^{ - 3}}$, 5G LDPC code and LTE Turbo code show around $0.5$ dB and $0.7$ dB SNR gain compared to the convolutional code D. Note that a similar observation of the error performance can be observed over the AWGN channel, where the LTE Turbo code has the best performance while the convolutional code D has the worst performance. Therefore, this observation indicate that codes optimized for AWGN channels can also achieve a good error performance in the OTFS systems, which is consistent with our analysis.

\textbf{Remark 2}:
Based on the above simulation results, we summarize that the slopes of FER curves for both uncoded and coded OTFS systems are generally different. This is because the diversity of uncoded OTFS systems is determined only by channel parameters while the diversity of coded OTFS systems is determined by both the channel and code parameters.
In particular, uncoded OTFS systems cannot guarantee the exploitation of the full diversity in all possible channel coefficients and transmitted symbols realizations, while coded OTFS systems can achieve the full diversity with a careful code design. Note that how to design practical codes to guarantee full diversity is an interesting research work and one initial idea is to ensure the \emph{codeword difference matrix} in~\eqref{C7_Gram} to be a diagonal matrix for any pair of codewords. However, we will explore the details on the code design in our future work due to the space limitation.

\section{Summary of the Chapter}
In this chapter, we studied the performance analysis of coded OTFS systems over high-mobility channels. We first derived the conditional PEP for a given channel realization and then obtained the unconditional PEP by leveraging proper bounding techniques. We discussed two cases of the OTFS transmission according to the number of independent resolvable paths of the channel, where we showed that the coding improvement of OTFS systems depends on the squared Euclidean distance between a pair of codewords. More importantly, we demonstrated the fundamental trade-off between the diversity gain and the coding gain for OTFS systems.
Furthermore, we proposed a code design criterion based on the derived unconditional bound.
The analysis and the code design criterion are verified by numerical simulations with various channel codes.
Our future work will focus on the modern code design for OTFS systems by considering analytical tools such as the density evolution and the extrinsic information transfer (EXIT) chart.
\section{Appendices of the Chapter}
In this section, we provide the related proofs from this chapter.
\subsection{Proof of Lemma 6-3}
Notice that, ${{\bf{\Omega }}\left( {\bf{e}} \right)}$ is positive semidefinite Hermitian. Hence, the eigenvalues $\left\{ {{\lambda _i}} \right\}$ of ${{\bf{\Omega }}\left( {\bf{e}} \right)}$ for $1 \le i \le r$ are all positive.
Considering the arithmetic mean and geometric mean (AM-GM) inequality, we obtain
\begin{equation}
\sum\limits_{i = 1}^r {\frac{1}{{{\lambda _i}}}}   \ge r{\left( {\prod\limits_{i = 1}^r {\frac{1}{{{\lambda _i}}}} } \right)^{\frac{1}{r}}} = \frac{r}{{{{\left( {\prod\limits_{i = 1}^r {{\lambda _i}} } \right)}^{\frac{1}{r}}}}}.
\label{C7_Lm1_1}
\end{equation}
Then, we apply the Cauchy-Schwarz inequality to the denominator of the right hand side of~\eqref{C7_Lm1_1}, which yields
\begin{equation}
\frac{r}{{{{\left( {\prod\limits_{i = 1}^r {{\lambda _i}} } \right)}^{\frac{1}{r}}}}} \ge \frac{{{r^2}}}{{\sum\limits_{i = 1}^r {{\lambda _i}} }}= \frac{r^2}{P{d_{\rm{E}}^2\left( {\bf{e}} \right)}}.
\end{equation}
It is obvious that the equality only holds when the eigenvalues $\left\{ {{\lambda _i}} \right\}$ are of the same value, e.g., the codeword difference matrix ${{\bf{\Omega }}\left( {\bf{e}} \right)}$ is a diagonal matrix, i.e., ${\bf{\Omega }}\left( {\bf{e}} \right)=\textrm{diag}\left\{ {d_{\rm{E}}^2{\left( {\bf{e}} \right)}, \ldots ,d_{\rm{E}}^2}{\left( {\bf{e}} \right)} \right\}$. This completes the proof of Lemma 6-3.\hfill $\blacksquare$

\subsection{Proof of Theorem 6-1}
The product of the first $r$ eigenvalues of ${{\bf{\Omega }}\left( {\bf{e}} \right)}$ can be written as
\begin{equation}
\prod\limits_{i = 1}^r {{\lambda _i}}  = \exp \left( {\ln \left( {\prod\limits_{i = 1}^r {{\lambda _i}} } \right)} \right) = \exp \left( {\sum\limits_{i = 1}^r {\ln \left( {{\lambda _i}} \right)} } \right).
\end{equation}
Let us consider the inequality $\ln \left( \gamma \right) \ge  1 - \frac{1}{\gamma },  \gamma \in \left( {0, + \infty } \right)$,
where the equality only holds when $\gamma=1$.
Therefore, we have
\begin{align}
\prod\limits_{i = 1}^r {{\lambda _i}}  &= \exp \left( {\sum\limits_{i = 1}^r {\ln \left( {{\lambda _i}} \right)} } \right) \notag\\
&= \exp \left( {r\ln \left( {d_{\rm{E}}^2\left( {\bf{e}} \right)} \right) + \sum\limits_{i = 1}^r {\ln \left( {\frac{{{\lambda _i}}}{{d_{\rm{E}}^2\left( {\bf{e}} \right)}}} \right)} } \right)\notag\\
& \ge  \exp \left( {r\ln \left( {d_{\rm{E}}^2\left( {\bf{e}} \right)} \right) + \sum\limits_{i = 1}^r {\left( {1 - \frac{{d_{\rm{E}}^2\left( {\bf{e}} \right)}}{{{\lambda _i}}}} \right)} } \right) \label{C7_inequality}\\
&= {\left( {d_{\rm{E}}^2\left( {\bf{e}} \right)} \right)^r}\exp \left( {r - d_{\rm{E}}^2\left( {\bf{e}} \right)\sum\limits_{i = 1}^r {\frac{1}{{{\lambda _i}}}} } \right).
\end{align}
It is obvious that the equality holds in~\eqref{C7_inequality} only if the first $r$ eigenvalues $\{\lambda _i\}$ of ${{\bf{\Omega }}\left( {\bf{e}} \right)}$ equal to ${d_{\rm{E}}^2\left( {\bf{e}} \right)}$.
Notice that the eigenvalues $\{\lambda _i\}$ of ${{\bf{\Omega }}\left( {\bf{e}} \right)}$ equal to the main diagonal elements when ${{\bf{\Omega }}\left( {\bf{e}} \right)}$ is a diagonal matrix, i.e., ${\bf{\Omega }}\left( {\bf{e}} \right)=\textrm{diag}\left\{ {d_{\rm{E}}^2\left( {\bf{e}} \right), \ldots ,d_{\rm{E}}^2\left( {\bf{e}} \right)} \right\}$, in which case we have $\det \left( {{\bf{\Omega }}\left( {\bf{e}} \right)} \right) = {\left( {d_{\rm{E}}^2}\left( {\bf{e}} \right) \right)^P}$.
This completes the proof of Theorem 6-1.\hfill $\blacksquare$

\subsection{Proof of Theorem 6-2}
Following Theorem 6-1, we note that value of the term $ {r - d_{\rm{E}}^2\left( {\bf{e}} \right)\sum\limits_{i = 1}^r {\frac{1}{{{\lambda _i}}}} } $ is smaller than~1 according to Lemma~6-3. Notice that the value of $f(x)=\exp(-x)$ decays relatively slowly with the increase of $x$, if $x < 1$. This fact motives us to consider the approximation based on Lemma~6-3 as considered in~\cite{Yuan2003performance,vucetic2003space}, such as
\begin{align}
\prod\limits_{i = 1}^r {{\lambda _i}}  &\ge {\left( {d_{\rm{E}}^2\left( {\bf{e}} \right)} \right)^r}\exp \left( {r - d_{\rm{E}}^2\left( {\bf{e}} \right)\sum\limits_{i = 1}^r {\frac{1}{{{\lambda _i}}}} } \right) \notag\\
&\mathbin{\lower.3ex\hbox{$\buildrel>\over
{\smash{\scriptstyle\sim}\vphantom{_x}}$}}{\left( {d_{\rm{E}}^2\left( {\bf{e}} \right)} \right)^r}\exp \left( {r - d_{\rm{E}}^2\left( {\bf{e}} \right){{\left( {\sum\limits_{i = 1}^r {\frac{1}{{{\lambda _i}}}} } \right)}_{\min }}} \right)\notag\\
&={\left( {d_{\rm{E}}^2\left( {\bf{e}} \right)} \right)^r}\exp \left( {r - d_{\rm{E}}^2\left( {\bf{e}} \right)\frac{{{r^2}}}{{Pd_{\rm{E}}^2\left( {\bf{e}} \right)}}} \right)\notag\\
&={\left( {d_{\rm{E}}^2\left( {\bf{e}} \right)} \right)^r}\exp \left( {r - \frac{{{r^2}}}{P}} \right),\notag\\
&\ge {\left( {d_{\rm{E}}^2\left( {\bf{e}} \right)} \right)^r},
\end{align}
where the equality holds if ${\bf{\Omega }}\left( {\bf{e}} \right)$ is a diagonal matrix, i.e., ${\bf{\Omega }}\left( {\bf{e}} \right)=\textrm{diag}\left\{ {d_{\rm{E}}^2{\left( {\bf{e}} \right)}, \ldots ,d_{\rm{E}}^2}{\left( {\bf{e}} \right)} \right\}$.
Mathematically, the above approximation may be loose if ${\bf{\Omega }}\left( {\bf{e}} \right)$ is ill-conditioned. Therefore, we justify the accuracy of our approximation as follows.
A commonly adopted approach in testifying if a matrix is in ill-condition is the $\bf{P}$-condition number~\cite{taylor1978condition}. In specific, we consider the lower bound on the $\bf{P}$-condition number of a Gram matrix~\cite{taylor1978condition}. The ${\bf{P}}$-condition number of ${\bf{\Omega }}\left( {\bf{e}} \right)$ is defined by
\begin{equation}
{\bf{P}}\left( {\bf{\Omega }}\left( {\bf{e}} \right)\right) \buildrel \Delta \over = {\rm Radius}\left( {\bf{\Omega }}\left( {\bf{e}} \right) \right){\rm Radius}\left( {\left( {{\bf{\Omega }}\left( {\bf{e}} \right)} \right)^{ - 1}} \right),
\end{equation}
where ${\rm Radius}\left( {{\bf{\Omega }}\left( {\bf{e}} \right)} \right)$ is the \emph{spectral radius} of ${{\bf{\Omega }}\left( {\bf{e}} \right)}$, i.e., ${\rm Radius}\left( {{\bf{\Omega }}\left( {\bf{e}} \right)} \right)$ equals to the largest eigenvalues of ${{\bf{\Omega }}\left( {\bf{e}} \right)}$.
In particular, the matrix ${{\bf{\Omega }}\left( {\bf{e}} \right)}$ is said to be ill-conditioned if ${\bf{P}}\left({{\bf{\Omega }}\left( {\bf{e}} \right)} \right) $ is large and is to be well-conditioned if ${\bf{P}}\left({{\bf{\Omega }}\left( {\bf{e}} \right)} \right) $ is small.
According to \cite{taylor1978condition}, we have ${\bf{P}}\left( {{\bf{\Omega }}\left( {\bf{e}} \right)} \right)\ge {{{{\left( {{{\left\| {{{\bf{u}}_i}} \right\|}^2}} \right)}_{\max }}} \mathord{\left/
 {\vphantom {{{{\left( {{{\left\| {{{\bf{u}}_i}} \right\|}^2}} \right)}_{\max }}} {{{\left( {{{\left\| {{{\bf{u}}_j}} \right\|}^2}} \right)}_{\min }}}}} \right.
 \kern-\nulldelimiterspace} {{{\left( {{{\left\| {{{\bf{u}}_j}} \right\|}^2}} \right)}_{\min }}}}$ for $1 \le i,j \le P$, which yields
\begin{equation}
{\bf{P}}\left( {{\bf{\Omega }}\left( {\bf{e}} \right)} \right)\ge  {{d_{\rm{E}}^2\left( {\bf{e}} \right)} \mathord{\left/
 {\vphantom {{d_{\rm{E}}^2\left( {\bf{e}} \right)} {d_{\rm{E}}^2\left( {\bf{e}} \right) = 1}}} \right.
 \kern-\nulldelimiterspace} {d_{\rm{E}}^2\left( {\bf{e}} \right) = 1}}.
\end{equation}
We can see that the $\bf{P}$-condition number of ${{\bf{\Omega }}\left( {\bf{e}} \right)}$ always greater than or equal to $1$, which indicates that ${{\bf{\Omega }}\left( {\bf{e}} \right)}$ is generally well-conditioned.
This completes the proof of Theorem 6-2.\hfill $\blacksquare$

\subsection{Proof of Theorem 6-3}
Recalling~\eqref{C7_New_PEP_Large_P_derivation1}, we note that~\eqref{C7_New_PEP_Large_P_derivation2} only holds if
\begin{equation}
\sum\limits_{i = 1}^r {\lambda _{i}^2}  \ge \frac{{P\sum\nolimits_{i = 1}^r {{\lambda _{i}}} }}{{{{{E_s}} \mathord{\left/
 {\vphantom {{{E_s}} {\left( {4{N_0}} \right)}}} \right.
 \kern-\nulldelimiterspace} {\left( {4{N_0}} \right)}}}}=\frac{{{P^2}d_{\rm{E}}^2\left( {\bf{e}} \right)}}{{{{{E_s}} \mathord{\left/
 {\vphantom {{{E_s}} {\left( {4{N_0}} \right)}}} \right.
 \kern-\nulldelimiterspace} {\left( {4{N_0}} \right)}}}}. \label{C7_Th4_derivation1}
\end{equation}
Therefore, we consider the approximation of~\eqref{C7_New_PEP_Large_P_derivation2} as follows
\begin{align}
\Pr\left(  {{\bf{x}},{\bf{x'}}} \right)&\le \exp \left( { - {{{{\left( {\sum\limits_{i = 1}^r {{\lambda _i}} } \right)}^2}} \mathord{\left/
 {\vphantom {{{{\left( {\sum\limits_{i = 1}^r {{\lambda _i}} } \right)}^2}} {\left( {4\sum\limits_{i = 1}^r {\lambda _i^2} } \right)}}} \right.
 \kern-\nulldelimiterspace} {\left( {4\sum\limits_{i = 1}^r {\lambda _i^2} } \right)}}} \right)\notag\\
 & \mathbin{\lower.3ex\hbox{$\buildrel<\over
{\smash{\scriptstyle\sim}\vphantom{_x}}$}} \exp \left( { - {{{{\left( {\sum\limits_{i = 1}^r {{\lambda _i}} } \right)}^2}} \mathord{\left/
 {\vphantom {{{{\left( {\sum\limits_{i = 1}^r {{\lambda _i}} } \right)}^2}} {{{\left( {4\sum\limits_{i = 1}^r {\lambda _i^2} } \right)}_{\min }}}}} \right.
 \kern-\nulldelimiterspace} {{{\left( {4\sum\limits_{i = 1}^r {\lambda _i^2} } \right)}_{\min }}}}} \right) \notag\\
& =\exp \left( { - {{{{\left( {Pd_{\rm{E}}^2\left( {\bf{e}} \right)} \right)}^2}} \mathord{\left/
 {\vphantom {{{{\left( {Pd_{\rm{E}}^2\left( {\bf{e}} \right)} \right)}^2}} {\left( {4\frac{{{P^2}d_{\rm{E}}^2\left( {\bf{e}} \right)}}{{{{{E_s}} \mathord{\left/
 {\vphantom {{{E_s}} {\left( {4{N_0}} \right)}}} \right.
 \kern-\nulldelimiterspace} {\left( {4{N_0}} \right)}}}}} \right)}}} \right.
 \kern-\nulldelimiterspace} {\left( {\frac{{4{P^2}d_{\rm{E}}^2\left( {\bf{e}} \right)}}{{{{{E_s}} \mathord{\left/
 {\vphantom {{{E_s}} {\left( {4{N_0}} \right)}}} \right.
 \kern-\nulldelimiterspace} {\left( {4{N_0}} \right)}}}}} \right)}}} \right)\notag\\
&= \exp \left( { - \frac{{{E_s}}}{{16{N_0}}}d_{\rm{E}}^2\left( {\bf{e}} \right)} \right).
\end{align}
The above approximation is reasonable because the value of $\exp \left( { - {{{{\left( {\sum\limits_{i = 1}^r {{\lambda _i}} } \right)}^2}} \mathord{\left/
 {\vphantom {{{{\left( {\sum\limits_{i = 1}^r {{\lambda _i}} } \right)}^2}} {\left( {4\sum\limits_{i = 1}^r {\lambda _i^2} } \right)}}} \right.
 \kern-\nulldelimiterspace} {\left( {4\sum\limits_{i = 1}^r {\lambda _i^2} } \right)}}} \right)$ changes relatively slowly with the increase of ${4\sum\limits_{i = 1}^r {\lambda _{i}^2} }$~\cite{Yuan2003performance,vucetic2003space}.
On the other hand, the justification of the SNR assumption of~\eqref{C7_New_PEP_Large_P_derivation1} is necessary. According to Lemma 6-4, we have
\begin{equation}
\frac{{P\sum\nolimits_{i = 1}^r {{\lambda _i}} }}{2{\sum\nolimits_{i = 1}^r {\lambda _i^2} }} \le \frac{{{P^2}d_{\rm{E}}^2\left( {\bf{e}} \right)}}{{\frac{{{2P^2}}}{r}{{\left( {d_{\rm{E}}^2\left( {\bf{e}} \right)} \right)}^2}}} = \frac{r}{{2d_{\rm{E}}^2\left( {\bf{e}} \right)}},
\end{equation}
where the equality only holds when the eigenvalues of ${\bf{\Omega }}\left( {\bf{e}} \right)$ share the same value, e.g., ${\bf{\Omega }}\left( {\bf{e}} \right)=\textrm{diag}\left\{ {d_{\rm{E}}^2{\left( {\bf{e}} \right)}, \ldots ,d_{\rm{E}}^2}{\left( {\bf{e}} \right)} \right\}$.
Therefore, we can see that the term $\frac{P{\sum\nolimits_{i = 1}^r {{\lambda _i}} }}{2{\sum\nolimits_{i = 1}^r {\lambda _i^2} }}$ is upper-bounded by $\frac{r}{2{d_{\rm{E}}^2\left( {\bf{e}} \right)}}$.
Hence, the assumption of SNR of~\eqref{C7_New_PEP_Large_P_derivation1} can be further restricted as $\frac{{{E_s}}}{{4{N_0}}} \ge \frac{r}{2{d_{\rm{E}}^2\left( {\bf{e}} \right)}}$.
This completes the proof of Theorem 6-3.\hfill $\blacksquare$

    \chapter{OTFS for Integrated Sensing and Communications}\label{c:literature}
In this chapter, we focus on the design of ISAC by using OTFS waveforms~\cite{li2021novel,Li2022potential}.
Although the applications of OTFS modulation in ISAC transmissions have shown promising performances as mentioned in the literature review, they often rely on sophisticated beamforming schemes~\cite{yuan2021integrated,gaudio2020hybrid,liu2021learning} that are designed according to the CSI at transmitter (CSIT). However, it should be noted that for some practical scenarios, such as co-located radar and communication antennas, channel fading coefficients for communication cannot be directly obtained
from radar sensing. In specific, the strengths of channel fading coefficients for communication usually depend on the path loss and channel
scatterers~\cite{hlawatsch2011wireless}, while the echo strengths for radar sensing also depend on the effective area of the radar receiving antenna and the radar cross section (RCS)~\cite{shrestha2008method}.
Therefore, there is generally a mismatch of the reflection strengths between radar sensing and communication.
In other words, the path with the strongest echo power for radar sensing may not be the strongest path for communication. Consequently, if the beams for communication steer towards the strongest path indicated by radar sensing, the communication performance may degrade dramatically.

Considering the potential mismatch between radar sensing and communication, we propose a novel ISAC transmission framework based on spatially-spread OTFS (SS-OTFS) modulation in this chapter. To facilitate the ISAC design, we introduce the concept of SS-OTFS modulation to further exploit the delay-Doppler-angular (DDA) domain channel characteristics. Compared to conventional MIMO-OTFS modulation, SS-OTFS applies the so-called ``spatial spreading'' and ``spatial de-spreading'' modules at the transmitter and receiver, respectively.  The key novelty of applying those modules is the \textbf{discretization} of the angular domain, which results in simple and insightful input-output relationships for both radar sensing and communication.
The most interesting feature of those relationships is that each antenna (pair) corresponds to a specific angle (pair) according to the \emph{angular resolution}. As such, it is possible to greatly mitigate the multi-path effect, which enables efficient system designs that are based only on estimates of the delays, Dopplers, angles-of-departure (AoDs), and radar reflection coefficients from radar sensing, without the \emph{a priori} knowledge of the fading coefficients of the communication channels.
It is worth emphasizing that most of the existing ISAC work in the literature assume that the communication fading coefficients are available at the BS for system designs~\cite{liu2020joint,liu2018toward}. In contrast, the proposed framework does not require the \emph{a priori} knowledge of communication fading coefficients. In particular, we propose a practical precoding scheme and a power allocation
strategy to improve the communication performance based only on the estimated AoDs, and delay and Doppler shifts from radar sensing at the previous time instant. The main contributions of this chapter can be summarized as follows.
\begin{itemize}
\item We derive both the communication and radar models for SS-OTFS-enabled ISAC transmission. In particular, we show that the interference from spatial multiplexing can be approximately eliminated by spatial spreading and de-spreading with a sufficiently large number of antennas, which results in simple and insightful effective channel matrices for both radar sensing and communication.
\item Based on the radar sensing model, we develop simple beam tracking and AoA estimation algorithms by exploiting the special structure of the effective radar sensing matrix. We show that the transmitted beam width can be easily controlled by the power allocation among the antennas, which is independent from the precoding design. Furthermore, the AoA estimation can be straightforwardly implemented by checking the received power among different antennas, which is due to the discretization of the angular domain. Furthermore, we introduce the power allocation for radar sensing, which is designed to maximize the minimum power of the received echoes.
\item Based on the derived communication model, we analytically unveil the impacts of precoding matrices and power allocation on the PEP. In particular, we show that the PEP is minimized when the \emph{codeword difference matrix} has a diagonal structure and the geometric mean of the allocated power associated to corresponding paths is maximized. Based on this finding, we develop our precoding design by introducing \emph{virtual} delay and Doppler indices to shape the codeword difference matrix, while we show that the equal power allocation can maximize the geometric mean.
\item We notice that radar sensing and communication require different power allocations. Therefore, we discuss the radar sensing and communication performances with respect to different power allocations. Based on our discussions and simulation results, we show that the power allocation should be designed leaning towards radar sensing in practical scenarios. Meanwhile, the effectiveness of the proposed ISAC framework has also been verified by our simulation results.
\end{itemize}

For the sake of clarity, the related notations in this chapter are listed in Table~\ref{C8_LIST_NOTATIONS_Transmitter}, Table~\ref{C8_LIST_NOTATIONS_Channel}, and Table~\ref{C8_LIST_NOTATIONS_Receiver}, corresponding to the ISAC transmitter, ISAC channels, and ISAC receiver, respectively.

\begin{table*}[htbp]
\caption{List of Notations for ISAC Transmitter}
\small
\centering
\begin{tabular}{|l|l|}
\hline
$K$~&~Number of UEs \\
\hline
$N_{\rm BS}$~&~Number of antennas at BS \\
\hline
$M$~&~Number of sub-carriers/delay bins \\
\hline
$N$~&~Number of time slots/Doppler bins \\
\hline
${\Delta f}$~&~Sub-carrier spacing \\
\hline
$T$~&~Time slot duration \\
\hline
${\theta _{{\rm{range}}}}$~&~Beamwidth\\
\hline
${N_{{\rm{range}}}}$~&~Number of antennas to support the beamwidth\\
\hline
${{\mathbb U}_{i,p}} $~&~Antenna set corresponding to the $p$-th path of the $i$-th UE\\
\hline
$\bf X$ and $\bf x$~&~Matrix and vector of the DD domain symbols\\
\hline
$\bf V$ and $\bf v$~&~Matrix and vector of the time-delay (TD) domain symbols\\
\hline
${\bf d}_{n_{\rm t}}$~&~Symbol vector of the $n_{\rm t}$-th antenna after precoding\\
\hline
${\bf d}$~&~Concatenated symbol vector after precoding\\
\hline
${\bf z}_{n_{\rm t}}$~&~Symbol vector of the $n_{\rm t}$-th antenna after power allocation\\
\hline
${\bf z}$~&~Concatenated symbol vector after power allocation\\
\hline
${\bf W}_{n_{\rm t}}$~&~Precoding matrix associated to the $n_{\rm t}$-th antenna\\
\hline
${\bf W}$~&~Concatenated precoding matrix\\
\hline
${\bf W}_{i}^{\rm com}$~&~Set of precoding matrices for the $i$-th UE\\
\hline
${\alpha}_{n_{\rm t}}$~&~Allocated power for the $n_{\rm t}$-th antenna\\
\hline
${\beta}_{i}$~&~Power assigned to the $i$-th UE per antenna\\
\hline
${{\tilde \alpha }_i}$ ~&~Total power for the $i$-th UE\\
\hline
${\alpha}_{\rm total}$~&~Total power budget for transmission\\
\hline
${\bm \alpha}$~&~Power allocation matrix\\
\hline
${\bm \alpha}_{i}^{\rm com}$~&~Set of allocated power values for the $i$-th UE\\
\hline
$\bf S$ and $\bf s$~&~Matrix and vector of the TDS domain symbols\\
\hline
\end{tabular}
\label{C8_LIST_NOTATIONS_Transmitter}
\end{table*}
\begin{table*}[htbp]
\caption{List of Notations for ISAC Channels}
\small
\centering
\begin{tabular}{|l|l|}
\hline
$P$~&~Number of independent resolvable paths \\
\hline
$N_0$ and ${\tilde N}_0$~&~Noise power for communication and radar\\
\hline
$h_{i,p}$ and ${\tilde h}_{i,p}$~&~Communication and radar fading coefficients for the $p$-th path of the $i$-th UE\\
\hline
${\bf{h}}_i^{{\rm{eff}}}$~&~Set of communication fading coefficients for $i$-th UE\\
\hline
${{\varphi _{i,p}}}$~&~AoD for the $p$-th path of the $i$-th UE \\
\hline
$\tau _{i,p}$ and ${\tilde \tau} _{i,p}$~&~Communication and radar delay shifts for $p$-th path of $i$-th UE \\
\hline
$l_{i,p}$ and ${\tilde l} _{i,p}$~&~Communication and radar delay indices for $p$-th path of $i$-th UE \\
\hline
$\nu _{i,p}$ and ${\tilde \nu} _{i,p}$~&~Communication and radar Doppler shifts for $p$-th path of  $i$-th UE \\
\hline
$k_{i,p}$ and ${\tilde k} _{i,p}$~&~Communication and radar Doppler indices for $p$-th path of  $i$-th UE \\
\hline
$\kappa _{i,p}$~&~Fractional communication Doppler index for  $p$-th path of $i$-th UE \\
\hline
${\tilde \kappa} _{i,p}$~&~Fractional radar Doppler index for  $p$-th path of $i$-th UE \\
\hline
${\dot l} _{p}$ and ${\dot k} _{p}$~&~Virtual delay and Doppler indices for the $p$-th path\\
\hline
${\bm{\omega }}_i^\tau$ and ${\bm{\omega }}_i^\nu$~&~Set of communication delay and Doppler indices for the $i$-th UE \\
\hline
${\bf{a}}\left( {{\varphi _{i,p}}} \right)$~&~Steering vector and steering matrix\\
\hline
${\bf{A}}\left( {{\varphi _{i,p}}} \right)$~&~Steering matrix\\
\hline
${\bm{\Pi }}$~&~Permutation matrix (forward cyclic shift) for characterizing delay influence\\
\hline
${\bm{\Delta }}$~&~Diagonal phase shift matrix for characterizing Doppler influence\\
\hline
${\bf{H}}_{{n_{\rm{t}}},i}^{{\rm{TDS}}}$~&~TDS domain communication channel between $n_{\rm t}$-th antenna and $i$-th UE\\
\hline
${\bf{\tilde H}}_{{n_{\rm{t}}},i,{n_{\rm{r}}}}^{{\rm{TDS}}}$~&~TDS domain radar channel for $(n_{\rm t},n_{\rm r})$-th antenna pair and $i$-th UE\\
\hline
${\bf{H}}_{i,p}^{{\rm{TD}}}$ and ${\bf{\tilde H}}_{i,p}^{{\rm{TD}}}$~&~TD domain communication and radar channels for $p$-th path and $i$-th UE\\
\hline
${\bf{h}}_{i,p}^{{\rm{A}}}$~&~Angular domain communication channel for the $p$-th path and the $i$-th UE\\
\hline
${\bf{\tilde H}}_{i,p}^{{\rm{A}}}$~&~Angular domain radar channel for the $p$-th path and the $i$-th UE\\
\hline
${\bf{\tilde H}}_{{\rm{Radar}}}^{{\rm{TDA}}}$~&~TDA domain effective radar sensing matrix\\
\hline
${\bf{ H}}_{i}^{{\rm{DD}}}$~&~Effective DD domain communication channel matrix\\
\hline
\end{tabular}
\label{C8_LIST_NOTATIONS_Channel}
\end{table*}
\begin{table*}[htbp]
\small
\caption{List of Notations for ISAC Receivers}
\centering
\begin{tabular}{|l|l|}
\hline
${\bf r}_{i}$~&~TD domain received symbol vector of the $i$-th UE for communication\\
\hline
${\bf \tilde r}_{n_{\rm r}}$~&~TDS domain received symbol vector of the ${n_{\rm r}}$-th antenna\\
\hline
${\bf r}$~&~Concatenated vector for TD domain received symbols for communication\\
\hline
${\bf \tilde r}$~&~Concatenated vector for TDS domain received symbols for radar\\
\hline
${\bf q}_{i}$~&~TD domain communication noise vector of the $i$-th UE\\
\hline
${\bf \tilde q}_{n_{\rm r}}$~&~TDS domain radar noise vector for the ${n_{\rm r}}$-th antenna\\
\hline
${\bf q}$ and ${\bf \tilde q}$~&~Concatenated TD(S) domain noise vector for communication and radar\\
\hline
${\bm \eta}$ and ${\bm {\tilde \eta}}$~&~DD domain and TDA domain noise vector for communication and radar\\
\hline
${\bf \tilde z}$~&~TDA domain radar received symbol vector\\
\hline
${\bf y}_i$~&~DD domain communication received symbol vector for the $i$-th UE\\
\hline
${\bf{\Phi }}_i^{\omega _i^\tau ,\omega _i^{_\nu },{\bf{W}}_i^{{\rm{com}}}}\left( {\bf{x}} \right)$~&~Equivalent codeword matrix\\
\hline
${\bf{\Omega }}_i\left( {\bf{e}} \right)$~&~Codeword difference matrix\\
\hline
${\bf{\tilde \Omega }}_i\left( {\bf{e}} \right)$~&~Weighted codeword difference matrix\\
\hline
${d_{{\bf{h}}_i^{{\rm{eff}}},{{\bm{\omega}} _\tau },\!{{\bm{\omega}} _\nu },\!{{\bm \alpha} _i^{{\rm{com}}}},\!{\bf{W}}_i^{{\rm{com}}}}^2\!\left( {{\bf{x}},\!{\bf{x'}}} \right)}$~&~Conditional Euclidean distance between $\bf x$ and $\bf x'$\\
\hline
$ {d_{\rm{E}}^2\left( {\bf{e}} \right)}$~&~Euclidean distance of the error sequence $\bf e$\\
\hline
\end{tabular}
\label{C8_LIST_NOTATIONS_Receiver}
\end{table*}

\section{System Model}
\begin{figure}
\centering
\includegraphics[width=0.7\textwidth]{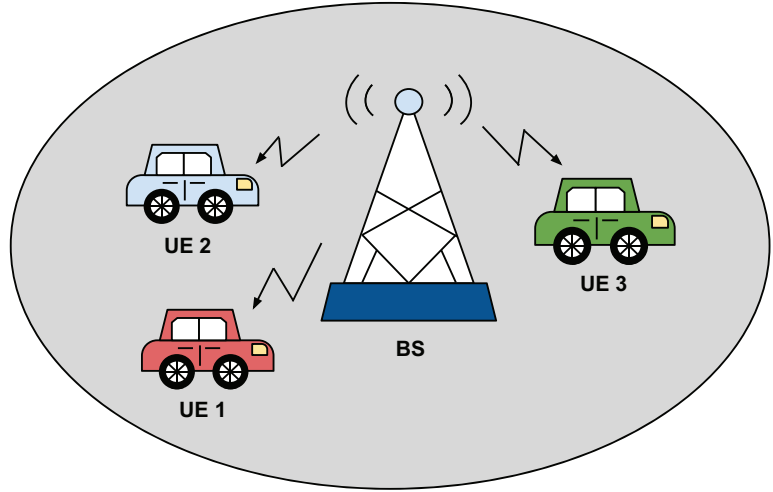}
\caption{A brief diagram of the considered ISAC transmission.}
\label{C8_ISAC}
\centering
\end{figure}

Without loss of generality, let us consider an ISAC system in a mobile network, where one BS broadcasts a common message to $K$ randomly distributed UEs within the service coverage and senses the radar-related information of the UEs based on the received echoes. In particular, we consider a multiple-input single-output (MISO) case, where the BS is equipped $N_{\rm BS}$ antennas while each UE has only one antenna. We assume that the system operates in an open area as shown in Fig.~\ref{C8_ISAC}, where there are $P$ independent resolvable paths between the BS and each UE{\footnote {The existence of resolvable paths indicates that the transmitted electromagnetic wave can be successfully reflected/received by the sensing target/communication receiver, which is crucial for both radar sensing and communication functionalities.}}.

\subsection{Transmitter Structure}
\begin{figure}
\centering
\includegraphics[width=0.7\textwidth]{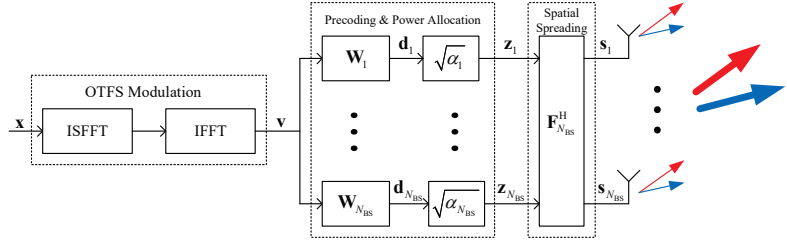}
\caption{The block diagram of the SS-OTFS-enabled ISAC transmitter, where the two arrows represent different beam directions.}
\label{C8_transmitter_model}
\centering
\end{figure}

Without loss of generality, let us consider the SS-OTFS-enabled ISAC transmitter structure as shown in Fig.~\ref{C8_transmitter_model}.

\begin{figure}
\centering
\includegraphics[width=0.7\textwidth]{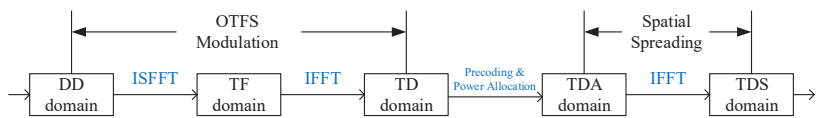}
\caption{The domain transformations the SS-OTFS enabled ISAC transmitter.}
\label{C8_domain_transform}
\centering
\end{figure}

Let ${\bf{X}} \in { {\mathbb A}^{M \times N }}$ be the DD domain transmitted symbol matrix (broadcast information message) of size $M \times N$, where $M$ denotes the number of orthogonal sub-carriers and $N$ denotes the number of time slots, respectively.
Let ${\Delta f}$ and $T$ be the sub-carrier spacing and the time slot duration, respectively.
According to the principle of OTFS modulation~\cite{Hadani2017orthogonal}, the TD domain transmitted symbol matrix ${\bf{V}} \in { {\mathbb A}^{M \times N }}$ of size $M \times N$ can be obtained by performing the ISFFT and IFFT to ${\bf{X}}$.
For simplicity, we consider the vector form representation of OTFS transmissions according to~\cite{Raviteja2019practical}. Let ${\bf{x}} \buildrel \Delta \over= {\rm{vec}}\left( {\bf{X}} \right)$ and  ${\bf{v}} \buildrel \Delta \over= {\rm{vec}}\left( {\bf{V}} \right)$ be the DD domain and the TD domain transmitted symbol vectors of length $MN$, respectively. According to~\cite{Raviteja2019practical,li2021performance_analysis}, we have
\begin{align}
{\bf{v}} \buildrel \Delta \over = {\rm{vec}}\left( {\bf{V}} \right) = {\rm{vec}}\left( {{\bf{F}}_M^{\rm{H}}{{\bf{F}}_M}{\bf{XF}}_N^{\rm{H}}} \right) = \left( {{\bf{F}}_N^{\rm{H}} \otimes {{\bf{I}}_M}} \right){\bf{x}}. \label{C8_TD_transmitted_symbol_vec}
\end{align}
After obtaining the TD domain broadcast message ${\bf{v}}$, the BS multiplexes the message onto each antenna with respect to the precoding matrices{\footnote{It is worth noticing that the conventional precoding matrix for narrow band multiple-input systems is of size $N_{\rm BS}\times N_{\rm BS}$. However, the transmitted signal on each antenna is generally a wideband signal in the considered system. Therefore, we propose to apply precoding to each antenna's transmitted signal in order to combat the multi-path interference, while apply spatial spreading to combat the interference after spatial multiplexing.}} $\left\{ {{{\bf{W}}_1},{{\bf{W}}_2},...,{{\bf{W}}_{{N_{{\rm{BS}}}}}}} \right\}$ and we have
\begin{align}
{{\bf{d}}_{n_{\rm t}}} = {{\bf{W}}_{n_{\rm t}}}{{\bf{v}}}, \quad \forall {n_{\rm{t}}} \in \left\{ {1,2,...,{N_{{\rm{BS}}}}} \right\}, \label{C8_signals_before_PA}
\end{align}
where the precoding matrix ${{\bf{W}}_{n_{\rm t}}}$ is of size $MN \times MN$ and it has a normalized energy with respect to the message length, i.e., ${\left\| {{{\bf{W}}_{{n_{\rm{t}}}}}} \right\|_{\rm{F}}} = MN$, for $1 \le {n_{\rm{t}}} \le {N_{{\rm{BS}}}}$. In particular, we restrict ourselves to only consider the symbol-wise precoding, where each row/column of ${{\bf{W}}_{n_{\rm t}}}$ only has one non-zero element and ${{\bf{W}}_{{n_{\rm{t}}}}}{\bf{W}}_{{n_{\rm{t}}}}^{\rm{H}} = {{\bf{I}}_{MN}}$, for $1 \le {n_{\rm{t}}} \le {N_{{\rm{BS}}}}$.
Based on~\eqref{C8_signals_before_PA}, we allocate power to each antenna's signal, such that the transmitted symbol vector ${\bf z}_{n_{\rm t}}$ after power allocation{\footnote{For the ease of presentation, we henceforth use the term energy and power interchangeably, without raising ambiguities.}} for the $n_{\rm t}$-th antenna is given by
\begin{align}
{{\bf{z}}_{n_{\rm t}}} = \sqrt {{\alpha _{{n_{\rm{t}}}}}} {{\bf{d}}_{n_{\rm t}}}, \label{C8_signals_before_BF}
\end{align}
where $\alpha_{n_{\rm t}}$, for $1 \le {n_{\rm{t}}} \le {N_{{\rm{BS}}}}$, is the allocated power for the ${n_{\rm{t}}}$-th antenna, and $\sum\nolimits_{{N_{\rm{t}}} = 1}^{{N_{{\rm{BS}}}}} {{\alpha _{{n_{\rm{t}}}}}}  = {\alpha _{{\rm{total}}}}$ with ${\alpha _{{\rm{total}}}}$ being the total transmit power budget.
Let us define the transmitted symbol vector before and after power allocation by ${\bf{d}} \buildrel \Delta \over = {\left[ {{\bf{d}}_1^{\rm{H}},{\bf{d}}_2^{\rm{H}},...,{\bf{d}}_{{N_{{\rm{BS}}}}}^{\rm{H}}} \right]^{\rm{H}}}$ and ${\bf{z}} \buildrel \Delta \over = {\left[ {{\bf{z}}_1^{\rm{H}},{\bf{z}}_2^{\rm{H}},...,{\bf{z}}_{{N_{{\rm{BS}}}}}^{\rm{H}}} \right]^{\rm{H}}}$, respectively.
Then, it can be shown that
\begin{align}
{\bf{z}} = \left( {{\bm \alpha}  \otimes {{\bf{I}}_{MN}}} \right){\bf{d}}, \label{C8_power_allocation_vec_form}
\end{align}
where ${\bm \alpha}\buildrel \Delta \over = {\rm diag}\left\{ {\sqrt{\alpha _1},\sqrt{\alpha _2},...,\sqrt{\alpha _{{N_{{\rm{BS}}}}}}} \right\}$ is the diagonal power allocation matrix of size $N_{\rm BS}\times N_{\rm BS}$.
Let us rearrange the transmitted symbol vectors for different antennas into a matrix $\bf Z$ of size $MN \times N_{\rm BS}$ based on ${\bf{z}} = {\rm{vec}}\left( {\bf{Z}} \right)$.
Then, we apply $N_{\rm BS}$-point IFFT, i.e., ${\bf F}_{N_{\rm BS}}^{\rm H}$, to the symbols among different antennas for spatial spreading, yielding
\begin{align}
{\bf{S}} = {\bf{ZF}}_{{N_{{\rm{BS}}}}}^{\rm{H}}, \label{C8_signals_after_BF}
\end{align}
where ${\bf{S}}$ is the time-delay-spatial (TDS) domain transmitted symbol matrix of size $MN \times N_{\rm BS}$.
Assuming that a rectangular pulse is applied as the transmitter shaping pulse, it can be shown that the TDS domain transmitted signal for the $n_{\rm t}$-th antenna can be fully characterized by the $n_{\rm t}$-th column of ${\bf{S}}$~\cite{Raviteja2019practical}. Denote by ${\bf s}_{n_{\rm t}}$ the $n_{\rm t}$-th column of ${\bf{S}}$, and we have ${\bf{s}} = {\rm{vec}}\left( {\bf{S}} \right)$, where ${\bf{s}} \buildrel \Delta \over = {\left[ {{\bf{s}}_1^{\rm{H}},{\bf{s}}_2^{\rm{H}},...,{\bf{s}}_{{N_{{\rm{BS}}}}}^{\rm{H}}} \right]^{\rm{H}}}$.
By combining~\eqref{C8_TD_transmitted_symbol_vec},~\eqref{C8_power_allocation_vec_form}, and~\eqref{C8_signals_after_BF}, and considering the property of Kronecker product, we have
\begin{align}
{\bf{s}} &= \left( {{\bf{F}}_{{N_{{\rm{BS}}}}}^{\rm{H}} \otimes {{\bf{I}}_{MN}}} \right){\bf{z}} = \left( {{\bf{F}}_{{N_{{\rm{BS}}}}}^{\rm{H}} \otimes {{\bf{I}}_{MN}}} \right)\left( {{\bm{\alpha }} \otimes {{\bf{I}}_{MN}}} \right){\bf{Wv}}\notag\\
 &= \left( {{\left( {{\bf{F}}_{{N_{{\rm{BS}}}}}^{\rm{H}}{\bm \alpha} } \right)} \otimes {{\bf{I}}_{MN}}} \right){\bf{W}}\left( {{\bf{F}}_N^{\rm{H}} \otimes {{\bf{I}}_M}} \right){\bf{x}},
\label{C8_signals_after_BF_vec}
\end{align}
where ${\bf{W}} \buildrel \Delta \over = {\left[ {{\bf{W}}_1^{\rm{H}},{\bf{W}}_2^{\rm{H}},...,{\bf{W}}_{{N_{{\rm{BS}}}}}^{\rm{H}}} \right]^{\rm{H}}}$ is the concatenated precoding matrix of size $N_{\rm BS}MN \times MN$. For a better understanding, we provide a diagram in Fig.~\ref{C8_domain_transform}, characterizing the domain transformations for the transmitter step by step.
As shown in Fig.~\ref{C8_domain_transform}, the OTFS modulation transforms the broadcast message from the DD domain to the TF domain and then to the TD domain. The precoding and power allocation are performed at the time-delay-angular (TDA) domain after repeating the signals onto each antenna. After that, the IFFT converts the signals from the TDA domain to the  TDS domain for signal transmission.
\subsection{Communication Model}
Without loss of generality, let us assume that there is $P$ independent resolvable paths between the BS and each UE{\footnote{In practice, the number of paths $P$ could be estimated by the DD domain channel estimation algorithms. Some related discussions could be found in~\cite{wei2020transmitter,Wei2021off}.}}.
According to the far-field assumption~\cite{tse2005fundamentals} and the DD domain channel characteristics~\cite{hlawatsch2011wireless},
the communication channel with respect to the antenna index $n_{\rm t}$, for $1 \le n_{\rm t} \le N_{\rm BS}$, and the UE index $i$, for $1 \le i \le K$, can be modeled by
\begin{align}
h\left( {{n_{\rm{t}}},i,\tau ,\nu } \right) = \sum\limits_{p = 1}^P {{h_{i,p}}\exp \left( {j\pi \left( {{n_{\rm{t}}} - 1} \right)\sin {\varphi _{i,p}}} \right)}\delta \left( {\tau  - {\tau _{i,p}}} \right)\delta \left( {\nu  - {\nu _{i,p}}} \right), \label{C8_com_DD_response}
\end{align}
where we assume that the distance between the adjacent antennas is half of the signal wavelength. In~\eqref{C8_com_DD_response}, $h_{i,p} \in {\mathbb C}$, ${{\varphi _{i,p}}}$, $\tau _{i,p}$, and $\nu _{i,p}$ are the communication fading coefficient, AoD, delay shift, and Doppler shift corresponding to the $p$-th path of the $i$-th UE, respectively.
We assume that the communication fading coefficient $h_{i,p}$ follows the uniform power delay and power Doppler profiles, such that
the $h_{i,p}$ has zero mean and variance $1/(2P)$ per real dimension for $1 \le p \le P$, and is independent from the delay and Doppler indices~\cite{molisch2012wireless}.

Assuming that a rectangular pulse is applied as the matched-filtering pulse for each UE, then with a reduced CP structure (introduced in Chapter 3)~\cite{Raviteja2019practical}, the TDS domain channel response based on~\eqref{C8_com_DD_response} can be equivalently represented by its matrix form~\cite{Raviteja2019practical}, i.e.,
\begin{align}
{\bf{H}}_{{n_{\rm{t}}},i}^{{\rm{TDS}}}
 \buildrel \Delta \over = &\sum\limits_{p = 1}^P {{h_{i,p}}} \exp \left( {j\pi \left( {{n_{\rm{t}}} - 1} \right)\sin {\varphi _{i,p}}} \right){{\bf{\Pi }}^{{l_{i,p}}}}{{\bf{\Delta }}^{{k_i} + {\kappa _{i,p}}}}, \forall i,p . \label{C8_com_TDS_matrix}
\end{align}
Specifically, we denote by $l_{i,p}$ and $k_{i,p}$ the indices of delay and Doppler associated with the $p$-th path of the $i$-th UE, respectively, where we have
\begin{equation}
{\tau _{i,p}} = \frac{{l_{i,p}}}{{M\Delta f}},\quad {\rm and }\quad
{\nu _{i,p}} = \frac{{{k_{i,p}} + {\kappa _{i,p}}}}{{NT}},
\label{C8_com_DD_resolution}
\end{equation}
respectively. Similar to the descriptions in Chapter~3, we have $0 \le l_{i,p} \le M-1$, and $0 \le k_{i,p} \le N-1$, respectively~\cite{Raviteja2019practical}.
Note that the term $- {1 \mathord{\left/
 {\vphantom {1 2}} \right.
 \kern-\nulldelimiterspace} 2} \le {\kappa _{i,p}} \le {1 \mathord{\left/
 {\vphantom {1 2}} \right.
 \kern-\nulldelimiterspace} 2}$ denotes the fractional Doppler which corresponds to the fractional shift from the nearest Doppler grid \cite{Raviteja2018interference}. On the other hand, since the typical value of the sampling time ${1 \mathord{\left/
 {\vphantom {1 {M\Delta f}}} \right.
 \kern-\nulldelimiterspace} {M\Delta f}}$ in the delay domain is usually sufficiently small, the impact of
fractional delays in typical wide-band systems can be neglected~\cite{tse2005fundamentals}.
With~\eqref{C8_com_TDS_matrix}, the TD domain{\footnote{Since the UE only has one antenna, the spatial/angular domain features at the receiver side, e.g., receiver steering vector, are disappeared. Therefore, we use the term TD instead of TDS/TDA for the relevant descriptions for the communication receiver in sequel. }} received symbol vector for the $i$-th UE is written by
\begin{equation}
{{\bf{r}}_i} = \sum\limits_{{n_{\rm{t}}} = 1}^{{N_{{\rm{BS}}}}} {{\bf{H}}_{{n_{\rm{t}}},i}^{{\rm{TDS}}}{{\bf{s}}_{{n_{\rm{t}}}}} + {{\bf{q }}_i}},  \label{C8_com_TDS_sum_r}
 \end{equation}
where ${{\bf{q }}_i}$ denotes the AWGN samples with one-sided power spectral density (PSD) $N_0$.
Equivalently, by separating the angular features in~\eqref{C8_com_TDS_matrix},~\eqref{C8_com_TDS_sum_r} can be rearranged as
\begin{align}
{{\bf{r}}_i} = \sum\limits_{p = 1}^{P} \left( {{{\bf{a}}^{\rm{T}}}\left( {{\varphi _{i,p}}} \right) \otimes {\bf{H}}_{i,p}^{{\rm{TD}}}} \right){\bf{s}} + {{\bf{q }}_i},
\end{align}
where ${\bf{a}}\left( {{\varphi _{i,p}}} \right)$ is the transmit steering vector given by{\footnote{For the sake of a better clarification of the advantages of the proposed scheme, we consider the energy normalization of the underlying channel in the following contents of this chapter. In practice, the normalization term in~\eqref{C8_steering_vector} should not exist, yielding a multiplexing gain related to the number of antennas $N_{\rm BS}$.}}
\begin{align}
&{\bf{a}}\left( {{\varphi _{i,p}}} \right) \buildrel \Delta \over = \frac{1}{{\sqrt {{N_{{\rm{BS}}}}} }}{\left[ {1,\exp \left( {j\pi \sin {\varphi _{i,p}}} \right),...,\exp \left( {j\pi \left( {{N_{{\rm{BS}}}} - 1} \right)\sin {\varphi _{i,p}}} \right)} \right]^{\rm{T}}},
\label{C8_steering_vector}
\end{align}
and
${{\bf{H}}_{i,p}^{{\rm{TD}}}}$ is defined as the TD domain equivalent communication channel matrix for the $p$-th path of the $i$-th UE, given by
\begin{align}
{\bf{H}}_{i,p}^{{\rm{TD}}} \buildrel \Delta \over =  {{h_{i,p}}} {{\bf{\Pi }}^{{l_{i,p}}}}{{\bf{\Delta }}^{{k_{i,p}} + {\kappa _{i,p}}}} . \label{C8_com_TD_P_matrix}
\end{align}
Then, according to the connections between the TD domain to the DD domain~\cite{Raviteja2019practical}, the DD domain received signal for the $i$-th UE is given by
\begin{align}
{{\bf{y}}_i} = \left( {{{\bf{F}}_N} \otimes {{\bf{I}}_M}} \right)\sum\limits_{p = 1}^P {\left( {{{\bf{a}}^{\rm{T}}}\left( {{\varphi _{i,p}}} \right) \otimes {\bf{H}}_{i,p}^{{\rm{TD}}}} \right)} {\bf{s}} + {{\bm{\eta }}_i}, \label{C8_com_DD_y-der1}
\end{align}
where ${{\bm{\eta }}_i}  \buildrel \Delta \over =  \left( {{{\bf{F}}_N} \otimes {{\bf{I}}_M}} \right){{\bf{q}}_i}$ is the equivalent AWGN noise vector in the DD domain.
Finally, by substituting~\eqref{C8_signals_after_BF_vec} into~\eqref{C8_com_DD_y-der1}, we obtain
\begin{align}
{{\bf{y}}_i} =& \left( {{{\bf{F}}_N} \otimes {{\bf{I}}_M}} \right)\sum\limits_{p = 1}^P {\left( {{{\bf{a}}^{\rm{T}}}\left( {{\varphi _{i,p}}} \right) \otimes {\bf{H}}_{i,p}^{{\rm{TD}}}} \right)} \left( {{\left( {{\bf{F}}_{{N_{{\rm{BS}}}}}^{\rm{H}}{\bm \alpha} } \right)} \otimes {{\bf{I}}_{MN}}} \right){\bf{W}}\left( {{\bf{F}}_N^{\rm{H}} \otimes {{\bf{I}}_M}} \right){\bf{x}} + {{\bm{\eta }}_i} \notag\\
=&\left( {{{\bf{F}}_N} \otimes {{\bf{I}}_M}} \right)\sum\limits_{p = 1}^P {\left( {\left( {{{\bf{a}}^{\rm{T}}}\left( {{\varphi _{i,p}}} \right){\bf{F}}_{{N_{{\rm{BS}}}}}^{\rm{H}}{\bm \alpha}} \right) \otimes {\bf{H}}_{i,p}^{{\rm{TD}}}} \right)}{\bf{W}}\left( {{\bf{F}}_N^{\rm{H}} \otimes {{\bf{I}}_M}} \right){\bf{x}} + {{\bm{\eta }}_i}. \label{C8_com_DD_y-der2}
\end{align}

\subsection{Radar Model}
Similar to the communication model, we consider the radar channel response with respect to the transmit antenna index $n_{\rm t}$, for $1 \le n_{\rm t} \le N_{\rm BS}$, the UE index $i$, for $1 \le i \le K$, the receive antenna index $n_{\rm r}$, for $1 \le n_{\rm r} \le N_{\rm BS}$, which is modeled by
\begin{align}
&\tilde h\left( {{n_{\rm{t}}},i,{n_{\rm{r}}},\tilde \tau ,\tilde \nu } \right) \notag\\
=& \sum\limits_{p = 1}^P {{{\tilde h}_{i,p}}\exp \left( {j\pi \left( {{n_{\rm{t}}} - 1} \right)\sin {\varphi _{i,p}}} \right)}\exp \left( {j\pi \left( {{n_{\rm{r}}} - 1} \right)\sin {\varphi _{i,p}}} \right)\!\delta \left( {\tau  \!-\! {{\tilde \tau }_{i,p}}} \right)\!\delta \left( {\nu  \!-\! {{\tilde \nu }_{i,p}}} \right) , \label{C8_radar_DD_response}
\end{align}
where we assume that the receive and transmit antennas are co-located such that AoDs and AoAs are of the same values~\cite{liu2020joint}.
In~\eqref{C8_radar_DD_response}, ${\tilde h}_{i,p}\in {\mathbb C}$, ${\tilde \tau} _{i,p}$, and ${\tilde \nu} _{i,p}$ are the radar reflection coefficient, round-trip delay shift, and round-trip Doppler shift associated to the $p$-th path of the $i$-th UE, respectively, where the round-trip delay and Doppler shifts satisfy ${{{\tilde \tau }_{i,p}}}=2{{{ \tau }_{i,p}}}$ and ${{{\tilde \nu }_{i,p}}}=2{{{ \nu }_{i,p}}}$~\cite{liu2020joint,yuan2021integrated}.
In practice, the radar reflection coefficient ${\tilde h}_{i,p}$ relates to the distance between the $i$-th UE and the BS with respect to the $p$-th path, the effective area of the radar receiving antenna, the RCS, the wavelength, and the transmit and receive antenna gains~\cite{shrestha2008method}.
Similar to the communication model, we assume that a rectangular pulse is applied as the filtering pulse at the receiver, then with the reduced CP structure~\cite{Raviteja2019practical}, the TDS domain equivalent radar sensing matrix is given by~\cite{Raviteja2019practical}
\begin{align}
{\bf{\tilde H}}_{{n_{\rm{t}}},i,{n_{\rm{r}}}}^{{\rm{TDS}}} \buildrel \Delta \over =& \sum\limits_{p = 1}^P {{{\tilde h}_{i,p}}} \exp \left( {j\pi \left( {{n_{\rm{t}}} - 1} \right)\sin {\varphi _{i,p}}} \right)\exp \left( {j\pi \left( {{n_{\rm{r}}} - 1} \right)\sin {\varphi _{i,p}}} \right){{\bf{\Pi }}^{{{\tilde l}_{i,p}}}}{{\bf{\Delta }}^{{{\tilde k}_i} \!+\! {{\tilde \kappa }_{i,p}}}}, \label{C8_radar_TDS_matrix}
\end{align}
where ${{{\tilde l}_{i,p}}}=2{{{ l}_{i,p}}}$ and ${{\tilde k}_i} + {{\tilde \kappa }_{i,p}} = 2\left( {{k_i} + {\kappa _{i,p}}} \right)$.
Let ${{\tilde N}_0}$ denote the one-sided PSD for the radar noise, which takes into account of both the AWGN noise power from the channel and the interference power from the transmit signals after interference cancellation~\cite{liu2020joint}. Then, similar to~\eqref{C8_com_TDS_sum_r}, the radar received TDS domain symbol vector for the $n_{\rm r}$-th antenna is written by
\begin{equation}
{{{\bf{\tilde r}}}_{{n_{\rm{r}}}}} = \sum\limits_{i = 1}^K {\sum\limits_{{n_{\rm{t}}} = 1}^{{N_{{\rm{BS}}}}} {{\bf{\tilde H}}_{{n_{\rm{t}}},i,{n_{\rm{r}}}}^{{\rm{TDS}}}{{\bf{s}}_{{n_{\rm{t}}}}} + {{{\bf{\tilde q}}}_{n_{\rm r}}}} }  ,  \label{C8_radar_TDS_sum_r}
 \end{equation}
where ${{{\bf{\tilde q}}}_{n_{\rm r}}} $ denotes the TDS domain radar AWGN vector.
By stacking the TDS domain received symbols from each antenna, we have
\begin{align}
{\bf{\tilde r}} &= {\left[ {{\bf{\tilde r}}_1^{\rm{H}},{\bf{\tilde r}}_2^{\rm{H}},...,{\bf{\tilde r}}_{{N_{{\rm{BS}}}}}^{\rm{H}}} \right]^{\rm{H}}}= \sum\limits_{i = 1}^K {\sum\limits_{p = 1}^P {\left( {{\bf{A}}\left( {{\varphi _{i,p}}} \right) \otimes {\bf{\tilde H}}_{i,p}^{{\rm{TD}}}} \right)} } {\bf{s}} + {\bf{\tilde q}}, \label{C8_radar_TDS_vec_r_der1}
\end{align}
where ${\bf{\tilde q}}  \buildrel \Delta \over =  {\left[ {{\bf{\tilde q}}_1^{\rm{H}},{\bf{\tilde q}}_2^{\rm{H}},...,{\bf{\tilde q}}_{{N_{{\rm{BS}}}}}^{\rm{H}}} \right]^{\rm{H}}}$ is the equivalent TDS domain radar noise,
\begin{align}
{\bf{A}}\left( {{\varphi _{i,p}}} \right) \buildrel \Delta \over = {\bf{a}}\left( {{\varphi _{i,p}}} \right){{\bf{a}}^{\rm{T}}}\left( {{\varphi _{i,p}}} \right), \forall i, p, \label{C8_radar_steering_mtx}
\end{align}
is the steering matrix associated with the $p$-th path of the $i$-th UE, and
\begin{align}
{\bf{\tilde H}}_{i,p}^{{\rm{TD}}} \buildrel \Delta \over = {{{\tilde h}_{i,p}}} {{\bf{\Pi }}^{{{\tilde l}_{i,p}}}}{{\bf{\Delta }}^{{{\tilde k}_i} + {{\tilde \kappa }_{i,p}}}},\label{C8_radar_TD_P_matrix}
\end{align}
is the TD domain equivalent radar channel matrix for the $p$-th path of the $i$-th UE, respectively.
Without loss of generality, we consider the radar sensing in the TDA domain. Therefore, we apply the spatial de-spreading to the TDS domain radar received symbol vector ${\bf{\tilde r}}$, yielding
\begin{align}
{\bf{\tilde z}} =& \left( {{{\bf{F}}_{{N_{{\rm{BS}}}}}} \otimes {{\bf{I}}_{MN}}} \right){\bf{\tilde r}}\notag\\
 =& \left( {{{\bf{F}}_{{N_{{\rm{BS}}}}}} \otimes {{\bf{I}}_{MN}}} \right)\sum\limits_{i = 1}^K {\sum\limits_{p = 1}^P \left( {{\bf{A}}\left( {{\varphi _{i,p}}} \right) \otimes {\bf{\tilde H}}_{i,p}^{{\rm{TD}}}} \right)}{\left( {{\bf{F}}_{{N_{{\rm{BS}}}}}^{\rm{H}} \otimes {{\bf{I}}_{MN}}} \right){\bf{z}}}  + {\bm{\tilde \eta}} \notag\\
 =& \sum\limits_{i = 1}^K {\sum\limits_{p = 1}^P {\left( {\left( {{{\bf{F}}_{{N_{{\rm{BS}}}}}}{\bf{A}}\left( {{\varphi _{i,p}}} \right){\bf{F}}_{{N_{{\rm{BS}}}}}^{\rm{H}}{\bm{\alpha }}} \right) \otimes {\bf{\tilde H}}_{i,p}^{{\rm{TD}}}} \right){\bf{d}}}  + {\bm{\tilde \eta}} }  , \label{C8_radar_TDA_after_de_spreading}
\end{align}
where ${\bm{\tilde \eta}}$ is the TDA domain equivalent radar AWGN vector with one-sided PSD ${{\tilde N}_0}$.

\subsection{Model Simplifications with Spatial Spreading and De-spreading}
To characterize the effect of spatial spreading and de-spreading, we are interested in the structure of the equivalent angular domain channel vector/matrix for both communication and radar.
For the $p$-th path of the $i$-th UE, let us define the equivalent angular domain channel vector for the communication channel by
\begin{align}
{\bf{h}}_{i,p}^{\rm{A}} \buildrel \Delta \over = {{\bf{a}}^{\rm{T}}}\left( {{\varphi _{i,p}}} \right){\bf{F}}_{{N_{{\rm{BS}}}}}^{\rm{H}}{\bm \alpha}, \label{C8_com_H_A}
\end{align}
and the equivalent angular domain channel matrix for the radar channel by
\begin{align}
{\bf{\tilde H}}_{i,p}^{\rm{A}} \buildrel \Delta \over = {{\bf{F}}_{{N_{{\rm{BS}}}}}}{\bf{A}}\left( {{\varphi _{i,p}}} \right){\bf{F}}_{{N_{{\rm{BS}}}}}^{\rm{H}}{\bm \alpha},\label{C8_radar_H_A}
\end{align}
respectively.
Based on~\eqref{C8_com_H_A} and~\eqref{C8_radar_H_A}, we can derive the elements in ${\bf{h}}_{i,p}^{\rm{A}}$ and ${\bf{\tilde H}}_{i,p}^{\rm{A}}$ after some manipulations. The corresponding results are summarized in Lemma~7-1.

\textbf{Lemma 7-1} (\emph{Angular Domain Channel}):
Given $1 \le k \le N_{\rm BS}$, and $1 \le l \le N_{\rm BS}$, the elements in the equivalent angular domain communication and radar channels can be expressed by
\begin{align}
h_{i,p}^{\rm{A}}\left[ l \right] = \frac{{\sqrt {{\alpha _l}} }}{{{N_{{\rm{BS}}}}}}\sum\limits_{{n_{\rm{t}}} = 0}^{{N_{{\rm{BS}}}} - 1} {\exp \left( {j{n_{\rm{t}}}\pi \left( {\sin {\varphi _{i,p}} + \frac{{2\left( {l - 1} \right)}}{{{N_{{\rm{BS}}}}}}} \right)} \right)} , \label{C8_com_H_A_value}
\end{align}
and
\begin{align}
\tilde H_{i,p}^{\rm{A}}\left[ {k,l} \right] = &\frac{{\sqrt {{\alpha _l}} }}{{{{\left( {{N_{{\rm{BS}}}}} \right)}^2}}}\left( {\sum\limits_{{n_{\rm{r}}} = 0}^{{N_{{\rm{BS}}}} - 1} {\exp \left( {j\pi {n_{\rm{r}}}\left( {\sin {\varphi _{i,p}} - \frac{{2\left( {k - 1} \right)}}{{{N_{{\rm{BS}}}}}}} \right)} \right)} } \right)\notag\\
&\quad\quad \quad\!\!\quad\left( {\sum\limits_{{n_{\rm{t}}} = 0}^{{N_{{\rm{BS}}}} - 1} {\exp \left( {j\pi {n_{\rm{t}}}\left( {\sin {\varphi _{i,p}} + \frac{{2\left( {l - 1} \right)}}{{{N_{{\rm{BS}}}}}}} \right)} \right)} } \right), \label{C8_radar_H_A_value}
\end{align}
respectively.

\emph{Proof}: The proof is given in the Appendix of this chapter.

Based on~\eqref{C8_com_H_A_value} and~\eqref{C8_radar_H_A_value}, we notice that if the value of $\sin {\varphi _{i,p}}$ is integer multiples of $2/N_{\rm BS}$, the values of ${{h}}_{i,p}^{\rm{A}}\left[ l \right]$ and ${{\tilde H}}_{i,p}^{\rm{A}}\left[ k,l \right]$ will be either $\sqrt {\alpha_l}$ or zero.
Therefore, let us define the \textbf{angular resolution} by $2/N_{\rm BS}$.
According to the angular resolution, we further define the \textbf{transmit angular index} by ${a_{i,p}} \buildrel \Delta \over = {\left[ {{N_{{\rm{BS}}}} - \frac{{\sin \left( {{\varphi _{i,p}}} \right){N_{{\rm{BS}}}}}}{2}} \right]_{{N_{{\rm{BS}}}}}}+1$ and the \textbf{receive angular index} by ${{\tilde a}_{i,p}} \buildrel \Delta \over = {\left[ {{N_{{\rm{BS}}}} + \frac{{\sin \left( {{\varphi _{i,p}}} \right){N_{{\rm{BS}}}}}}{2}} \right]_{{N_{{\rm{BS}}}}}}+1$.
Based on these, we have the following corollary.

\textbf{Corollary 7-1} (\emph{Asymptotical Orthogonality}):
With a sufficiently large number of antennas at the BS, the angular index ${a_{i,p}} $ is of an integer value. In this case,
we have ${{h}}_{i,p}^{\rm{A}}\left[ l \right]=0$, for $ l \ne {{a_{i,p}}}$, while ${{h}}_{i,p}^{\rm{A}}\left[ l \right]=\sqrt{\alpha_l}$, for $ l = {{a_{i,p}}}$. Furthermore, we have ${{\tilde H}}_{i,p}^{\rm{A}}\left[ k,l \right]=0$, for $ k \ne {\tilde a}_{i,p} , l \ne {{a_{i,p}}}$, while ${{\tilde H}}_{i,p}^{\rm{A}}\left[ k,l \right]=\sqrt{\alpha_l}$, for $ k ={{\tilde a}_{i,p}} , l = {a_{i,p}}$.

\emph{Proof}: The proof is given in the Appendix of this chapter.

\begin{figure}
\centering
\includegraphics[width=0.7\textwidth]{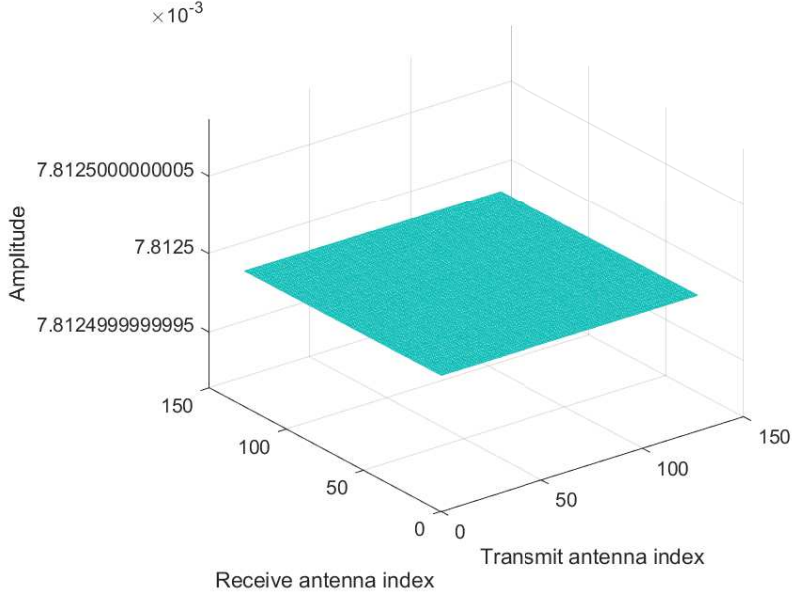}
\caption{Absolute values of ${\bf{\tilde H}}_{i,p}^{\rm{A}}$ without spreading and de-spreading.}
\label{C8_withoutSS}
\centering
\end{figure}

\begin{figure}
\centering
\includegraphics[width=0.7\textwidth]{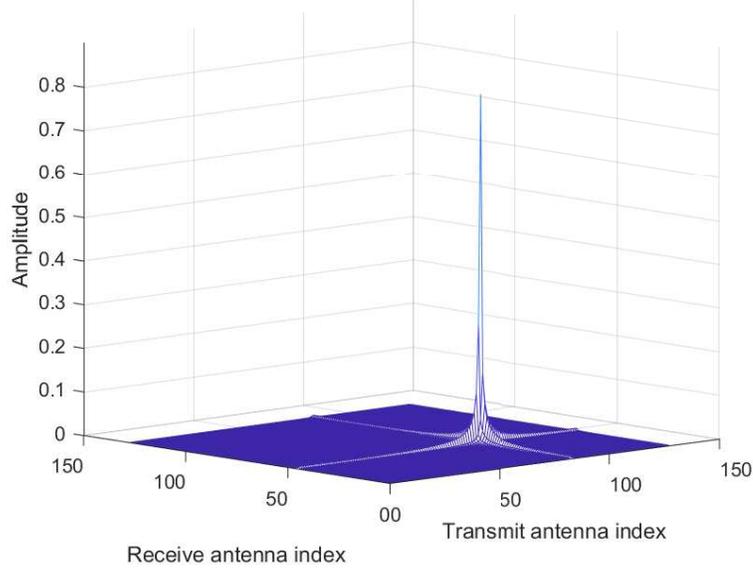}
\caption{Absolute values of ${\bf{\tilde H}}_{i,p}^{\rm{A}}$ with spreading and de-spreading.}
\label{C8_withSS}
\centering
\end{figure}


As indicated from Lemma 7-1, we notice that the interference for both the communication and radar channels due to the multipath and multiuser transmissions can be approximately eliminated by spatial spreading and de-spreading, given a sufficient number of antennas.
For a better understanding, we provide an example of the equivalent angular domain radar channel matrix with $\varphi _{i,p}=\pi/4$ and $N_{\rm BS}=128$ in Fig.~\ref{C8_withoutSS} and Fig.~\ref{C8_withSS}. As indicated by the figure, the channel matrix with spatial spreading and de-spreading is much more sparse, i.e., only the grids around the $84$-th transmitted antenna and the $46$-th received antenna are of values, which is consistent with our analysis.
In what follows, we assume that $N_{\rm BS}$ is sufficiently large such that the angular indices are of integer values{\footnote{Given the accuracy of AoD and AoA estimates for practical systems, this assumption is valid for practical BS setups of ISAC transmissions~\cite{liu2020joint,yuan2021integrated}. }}.
Particularly, notice that when $N_{\rm BS}$ is sufficiently large, the angular resolution is sufficiently high. Hence, we further assume that all the paths can be fully separated by its angular domain features, i.e., $ {{a_{i,p}} \ne {a_{i',p'}}}$, for $i \ne i'$ or $p \ne p'$.
Consequently, both the communication and radar models in~\eqref{C8_com_DD_y-der2} and~\eqref{C8_radar_TDA_after_de_spreading} can be further simplified, yielding
\begin{align}
{{\bf{y}}_i} = \sum\limits_{p = 1}^P {\sqrt{\alpha_{a_{i,p}}}}{\left( {{{\bf{F}}_N} \otimes {{\bf{I}}_M}} \right){\bf{H}}_{i,p}^{{\rm{TD}}}{{\bf{W}}_{{a_{i,p}}}}\left( {{\bf{F}}_N^{\rm{H}} \otimes {{\bf{I}}_M}} \right){\bf{x}}}+ {{\bm{\eta }}_i} ,\label{C8_com_model}
\end{align}
and
\begin{align}
{\bf{\tilde z}} = {\bf{\tilde H}}_{{\rm{Radar}}}^{{\rm{TDA}}}{\bf{z}} + {\bm{\tilde \eta}}  ,\label{C8_radar_model}
\end{align}
respectively, where ${\bf{\tilde H}}_{{\rm{Radar}}}^{{\rm{TDA}}}$ is the TDA domain effective radar sensing matrix.
In particular, it can be shown that ${\bf{\tilde H}}_{{\rm{Radar}}}^{{\rm{TDA}}}$
is a block matrix of size $N_{\rm BS}MN \times N_{\rm BS}MN$, whose $({\tilde a}_{i,p}, a_{i,p})$-th sub-block is given by ${\sqrt{\alpha_{a_{i,p}}}}{\bf{\tilde H}}_{i,p}^{{\rm{TD}}}$ and the rest sub-blocks are given by ${{\bf{0}}_{MN}}$.

\textbf{Remark 7-1}:
It is of importance to discuss the connections and differences between the spatial spreading/de-spreading and conventional beamforming. In conventional beamforming, the beams are formed delicately according to the exact AoDs or AoAs. In contrast, the proposed spatial spreading/de-spreading can be viewed as a special type of beamforming, where the beams are formed according to a pre-defined angular grid according to the DFT/inverse DFT matrices, as implied by~\eqref{C8_com_H_A} and~\eqref{C8_radar_H_A}. Thanks to the perfect orthogonality among the rows/columns of the the DFT/inverse DFT matrices, there is no inter-beam interference when applying spatial spreading/de-spreading, which is different from the conventional beamforming. However, there could be a potential beam alignment issue for spatial spreading/de-spreading, when the AoDs/AoAs associated to the paths are not perfectly aligned with the pre-defined angular grid. In this case, there may be a power reduction due to the misalignment. Note that the misalignment penalty could be mitigated by introducing a phase offset before the spatial spreading/de-spreading according to the \emph{a priori} knowledge of AoDs and AoAs, i.e., rotate the whole pre-defined angular grid with respect to a certain angle, such that the beams are more likely to be aligned with the AoDs or AoAs associated to the paths. Some recently proposed methods based on machine learning could also be useful for this problem~\cite{liu2021learning,liu2022Predictive}.

\textbf{Remark 7-2}: As indicated by the above discussion, the proposed spatial spreading and de-spreading leads to the discretization of angular features, which simplifies the input-output relationship for both communication and radar channels. In particular, the angular domain discretization enables direct interactions between the transmitted signal on a specific antenna and the channel distortion associated to a specific path.
Therefore, it is suitable for ISAC transmissions{\footnote{It is worth mentioning that the proposed framework can naturally be extended to the general case of multi-user MIMO by introducing the \emph{indicator matrices} that assign the data stream for each UE to the corresponding antenna before spatial spreading. Some related details could be found in~\cite{Li2022potential}.}}.

\section{Radar Sensing Designs based on SS-OTFS Modulation}
In this section, we are focusing on the radar sensing designs based on the SS-OTFS modulation. According to the angular discretization enabled by spatial spreading and de-spreading, we aim to design practical beam tracking and AoA estimation algorithms based on the previous AoA estimates and radar reflection coefficients. It should be noted that, from the radar point of view, not all the targets are of interest, because they may introduce clutter interference that can potentially undermine the sensing performance. However, some of the clutter might come from scatterers that can significantly contribute to the total received power for communication transmissions. Therefore, it is still necessary to estimate the parameters corresponding all the scatterers
for ISAC systems~\cite{liu2020joint}.

\subsection{Beam Tracking}
Let us focus on the radar channel model given in~\eqref{C8_radar_model}. In particular, with a sufficiently large number of antennas, the effective radar sensing matrix ${\bf{\tilde H}}_{{\rm{Radar}}}^{{\rm{TDA}}}$ is a sparse block matrix as discussed in Section~7.1. Thanks to the sparsity enabled by the SS-OTFS transmission, the angular domain is sufficiently discretized. Therefore, the signals can be transmitted towards the desired directions by simply assigning power to the corresponding antennas without sophisticated precoding designs.

For the case of beam tracking, the BS knows the previous AoA estimates associated to the different paths of the UEs from the previous time instant.
Therefore, a common approach for beam tracking is to send relatively wider beams towards the AoAs estimated from the previous time instant.
To achieve this, we only need to allocate power on the corresponding antennas. Denote by ${\theta _{{\rm{range}}}}$ the desired beam width. According to the angular resolution, we need to allocate powers on ${N_{{\rm{range}}}} \buildrel \Delta \over = \frac{{{\theta _{{\rm{range}}}}}}{2}{N_{{\rm{BS}}}}$ antennas. In specific, let ${{\mathbb U}_{i,p}}$ denote the antenna set corresponding to those antennas for the $p$-th paths of the $i$-th UE and it is given by
\begin{align}
{{\mathbb U}_{i,p}}\!  \buildrel \Delta \over = & \left\{ {{{\left[ {{N_{{a_{i,p}}}} \!\!- \!1 \!- \!\frac{{{N_{{\rm{range}}}}}}{2}} \right]}_{{N_{{\rm{BS}}}}}}\!\!\! \!\!+ \!1,{\left[ {{N_{{a_{i,p}}}} \!\!\!- \!\frac{{{N_{{\rm{range}}}}}}{2}} \right]_{{N_{{\rm{BS}}}}}}\!\!\! \!\!+ \!1,} \right. ...,\left. {{{\left[ {{N_{{a_{i,p}}}}\!\! - \!1 + \frac{{{N_{{\rm{range}}}}}}{2}} \right]}_{{N_{{\rm{BS}}}}}}\!\!\! \!\!+ \!1} \right\},
\label{C8_antenna_set}
\end{align}
where we assume that ${N_{{\rm{range}}}}$ is an even number.
Given the high angular resolution, we further assume that the AoAs are sufficiently separated, such that ${{\mathbb U}_{i,p}} \cap {{\mathbb U}_{i',p'}} = \varnothing , \forall i \ne i',p \ne p'$.
To ensure the reliability of radar sensing and communication, we propose to send the precoded signals with equal powers towards all the antennas in ${{\mathbb U}_{i,p}}$, such that the ISAC signal will be guaranteed to be reflected/received by the corresponding UE.

\subsection{AoA Estimation}
Now let us focus on the AoA estimation.
According to the radar model in~\eqref{C8_radar_model}, we notice that the TDA domain equivalent radar sensing matrix is a sparse block matrix, where only the sub-blocks related to the angular indices ${\tilde a}_{i,p}$ are of non-zero values.
This observation motivates us to consider a correlation-based AoA estimation method. According to~\eqref{C8_radar_model}, the covariance matrix of the TDA domain radar received symbol vector ${\bf{\tilde z}}$ can be calculated by
${{\bf{R}}_{{\bf{\tilde z}}}} \buildrel \Delta \over = {\mathbb E}\left[ {{\bf{\tilde z}}{{{\bf{\tilde z}}}^{\rm{H}}}} \right] = {\bf{\tilde H}}_{{\rm{Radar}}}^{{\rm{TDA}}}{\left( {{\bf{\tilde H}}_{{\rm{Radar}}}^{{\rm{TDA}}}} \right)^{\rm{H}}} + {{\tilde N}_0}{{\bf{I}}_{{N_{{\rm{BS}}}}MN}}$.
Furthermore, since ${\bf{\tilde H}}_{{\rm{Radar}}}^{{\rm{TDA}}}$ is a block matrix, ${{\bf{R}}_{{\bf{\tilde z}}}}$ is also a block matrix with $N_{\rm BS}$ row groups and $N_{\rm BS}$ column groups. In specific, it can be shown that the $n_{\rm r}$-th diagonal sub-block of ${{\bf{R}}_{{\bf{\tilde z}}}}$
is given by
\begin{align}
{{\bf{R}}_{{\bf{\tilde z}}}}\left[ {{n_{\rm{r}}},{n_{\rm{r}}}} \right] =\left\{ \begin{array}{l}
{\alpha _{{a_{i,p}}}}\!{\bf{\tilde H}}_{i,p}^{{\rm{TD}}}{{\bf{W}}_{{a_{i,p}}}}\!\!{\bf{W}}_{{a_{i,p}}}^{\rm{H}}{\left( {{\bf{\tilde H}}_{i,p}^{{\rm{TD}}}} \right)^{\rm{H}}}\!\! + \!{{\tilde N}_0}{{\bf{I}}_{MN}},\quad\;\; \!\!{n_{\rm{r}}} = {{\tilde a}_{i,p}}.\\
{{\tilde N}_0}{{\bf{I}}_{MN}},\!\!\quad\quad\quad\quad\quad\quad\quad\quad\quad\quad\quad\quad\quad\quad\quad{n_{\rm{r}}} \ne {{\tilde a}_{i,p}} ,
\end{array} \right. \label{C8_AoA_der1}
\end{align}
where we assume that ${\bf{x}}{{\bf{x}}^{\rm{H}}} = {{\bf{I}}_{MN}}$.
Furthermore, since ${{\bf{W}}_{{{ a}_{i,p}}}}{\bf{W}}_{{{ a}_{i,p}}}^{\rm{H}} = {{\bf{I}}_{MN}}$ according to our definition,~\eqref{C8_AoA_der1} can be further simplified by
\begin{align}
{{\bf{R}}_{{\bf{\tilde z}}}}\left[ {{n_{\rm{r}}},{n_{\rm{r}}}} \right] = \left\{ \begin{array}{l}
\left( {{\alpha _{{a_{i,p}}}}{{\left| {{{\tilde h}_{i,p}}} \right|}^2} + {{\tilde N}_0}} \right){{\bf{I}}_{MN}},\!\quad{n_{\rm{r}}} = {{\tilde a}_{i,p}},\\
{{\tilde N}_0}{{\bf{I}}_{MN}},\quad\quad\quad\quad\quad\quad\quad\quad\;\;{n_{\rm{r}}} \ne {{\tilde a}_{i,p}}.
\end{array} \right.\label{C8_AoA_der2}
\end{align}
As implied by~\eqref{C8_AoA_der2}, it can be shown that the diagonal sub-blocks of ${{\bf{R}}_{{\bf{\tilde z}}}}$ are also of diagonal structure. Furthermore, by noticing that the value of ${\alpha _{{a_{i,p}}}}{{\left| {{{\tilde h}_{i,p}}} \right|}^2}$ can be much larger than ${{\tilde N}_0}$ in the high SNR regime, we propose to calculate the traces of the diagonal sub-blocks of ${{\bf{R}}_{{\bf{\tilde z}}}}$ in order to find ${{\tilde a}_{i,p}}$. In specific, the AoA estimation can be carried out by calculating the traces of $N_{\rm BS}$ diagonal sub-blocks first and then finding the corresponding indices associated to the $KP$ largest of the trace values and finally calculating the values of the AoDs based on the indices and the angular resolution as discussed in Section~7.1.
An important issue at this point is how to associate the estimated AoAs with the paths of different UEs. Unfortunately, we do not have enough space to discuss this issue. We refer the interested readers to~\cite{liu2020joint} for more information.

\subsection{Power Allocation for Radar Sensing}
As indicated by~\eqref{C8_AoA_der2}, the power allocation will affect estimation performance. In particular, the value of ${{\left| {{{\tilde h}_{i,p}}} \right|}^2}$ can be largely different for different paths. Therefore, it is important to assign different powers to the related antennas according to the value of ${{\left| {{{\tilde h}_{i,p}}} \right|}^2}$.
Let us assume that the value of ${{\left| {{{\tilde h}_{i,p}}} \right|}^2}$ of the current time instant is accurately derived based on the corresponding estimates from the previous time instant. As the above AoA estimation is based on calculation of the correlation matrix, we aim to maximize the minimum effective radar SNR among all the paths, such that the radar can provide a relatively accurate sensing performance for each path.
Notice that ${\tilde N}_0$ is assumed to be equal to each UE. Therefore, to maximize the minimum radar SNR is equivalent to maximize the received power ${{\alpha _{{a_{i,p}}}}{{\left| {{{\tilde h}_{i,p}}} \right|}^2}}$ associated to each path.
Therefore, our power allocation is designed as follows
\begin{align}
\mathop {\max }\limits_{{\alpha _{{a_{1,1}}}},...,{\alpha _{{a_{K,P}}}}}&\quad {\left( {{\alpha _{{a_{i,p}}}}{{\left| {{{\tilde h}_{i,p}}} \right|}^2}} \right)_{\min }} \label{C8_maximum_power}\\
{\rm{s}}{\rm{.t}}{\rm{.}}&\quad\left( {\sum\limits_{i = 1}^K {\sum\limits_{p = 1}^P {{\alpha _{{a_{i,p}}}}} } } \right)\left( {{N_{{\rm{range}}}} + 1} \right) = {\alpha _{{\rm{total}}}}, \notag\\ &\quad\quad{{a_{i,p}}} \ge 0. \label{C8_sum_power_constraint}
\end{align}
It can be shown that the maximum of~\eqref{C8_maximum_power} is achieved when all the received power are of the same value, i.e.,
${\alpha _{{a_{1,1}}}}{\left| {{{\tilde h}_{1,1}}} \right|^2} = {\alpha _{{a_{1,2}}}}{\left| {{{\tilde h}_{1,2}}} \right|^2} = ... = {\alpha _{{a_{K,P}}}}{\left| {{{\tilde h}_{K,P}}} \right|^2}$.
Therefore, according to~\eqref{C8_sum_power_constraint}, the power allocation should satisfy
\begin{align}
{\alpha _{{a_{i,p}}}} = {{\left( {\frac{{{\alpha _{{\rm{total}}}}}}{{{N_{{\rm{range}}}} + 1}}\frac{1}{{{{\left| {{{\tilde h}_{i,p}}} \right|}^2}}}} \right)} \mathord{\left/
 {\vphantom {{\left( {\frac{{{\alpha _{{\rm{total}}}}}}{{{N_{{\rm{range}}}} + 1}}\frac{1}{{{{\left| {{{\tilde h}_{i,p}}} \right|}^2}}}} \right)} {\left( {\sum\limits_{j = 1}^K {\sum\limits_{p' = 1}^P {\frac{1}{{{{\left| {{{\tilde h}_{j,p'}}} \right|}^2}}}} } } \right)}}} \right.
 \kern-\nulldelimiterspace} {\left( {\sum\limits_{j = 1}^K {\sum\limits_{p' = 1}^P {\frac{1}{{{{\left| {{{\tilde h}_{j,p'}}} \right|}^2}}}} } } \right)}}.
\end{align}

\textbf{Remark 7-3}: Intuitively speaking, the proposed power allocation aims to assign a larger power towards to the direction where the power of the radar echo is small. This is contradict to the conventional water-filling principle from the communication point of view, where more power should be assigned towards the direction with a larger channel gain in order to improve the achievable rate. However, since the communication fading coefficients cannot be directly obtained based on the radar sensing estimates, it may be difficult for the BS to design a suitable power allocation that can provide a good trade-off for both radar sensing and communication performances. Consequently, the best we can do is to adapt the proposed power allocation with respect to the statistical distribution of the communication fading coefficients. The related issues will be discussed in detail in Section~7.3.3.

\section{Sensing-Assisted Communication Design}
In this section, we will develop communication schemes based on the estimated parameters from radar sensing. We notice that there is a deterministic relationship between the round-trip delay and Doppler from radar sensing and the delay and Doppler for communication. Furthermore, given the AoA estimates from radar, we can also determine the corresponding transmit angular indices ${a_{i,p}}$. Unfortunately, there is no direct relationship between the communication fading coefficients and radar reflection coefficients. Therefore, our design criterion is to minimize the PEP for communication with respect to the \emph{a priori} AoA, delay and Doppler estimates that are obtained from radar sensing.
According to the quasi-static property of the DD domain channel response~\cite{Zhiqiang_magzine,Li2020on,hlawatsch2011wireless}, we assume that the minor changes of the related parameters from the previous time instant are
well-compensated~\cite{liu2020joint}, including the delay shifts, and Doppler shifts associated to the corresponding paths.
On the other hand, we note that the AoAs at the current time instant may be different from the estimates from the previous time instant, due to the high angular resolution and mobility. However, as will be explained in detail later, our proposed design can be easily combined with the beam tracking scheme introduced in the previous section. Therefore, we first assume that the AoA estimates are accurate in this section for the ease of derivation.

\subsection{Pair-wise Error Probability Analysis}
Recalling~\eqref{C8_com_model}, we define the effective DD domain communication channel matrix for the $i$-th UE by
\begin{align}
{\bf{H}}_i^{{\rm{DD}}} \buildrel \Delta \over = &\sum\limits_{p = 1}^P {{\sqrt{\alpha_{a_{i,p}}}}{h_{i,p}}} \left( {{{\bf{F}}_N} \otimes {{\bf{I}}_M}} \right){{\bf{\Pi }}^{{l_{i,p}}}}{{\bf{\Delta }}^{{k_{i,p}} + {\kappa _{i,p}}}}{{\bf{W}}_{{a_{i,p}}}}\left( {{\bf{F}}_N^{\rm{H}} \otimes {{\bf{I}}_M}} \right).
\label{C8_H_DD_i}
\end{align}
Then,~\eqref{C8_com_model} can be rewritten by
\begin{align}
{{\bf{y}}_i} = {\bf{H}}_i^{{\rm{DD}}}{\bf{x}} + {{\bm \eta}_i}.
\label{C8_new_com_model}
\end{align}
In what follows, we will study the PEP performance with the ML detection\footnote{We note that the channel estimation and channel equalization for SS-OTFS can be carried out in a similar way to the conventional OTFS. Some related discussions can be found in~\cite{Zhiqiang_magzine,wei2020transmitter,li2021cross}. } in order to facilitate our precoding design.
For the ease of presentation, let us define the following vectors and matrices for related parameters. We define the
effective fading coefficients by ${\bf{h}}_i^{{\rm{eff}}} \buildrel \Delta \over = {\left[ { {h_{i,1}}, {h_{i,2}},..., {h_{i,P}}} \right]^{\rm{T}}}$, the delay shifts by
${\bm{\omega }}_i^\tau  \buildrel \Delta \over = {\left[ {{l_{i,1}},{l_{i,2}},...,{l_{i,P}}} \right]^{\rm{T}}}$, the Doppler shifts by
${\bm{\omega }}_i^\nu  \buildrel \Delta \over = {\left[ {{k_{i,1}} + {\kappa _{i,1}},{k_{i,2}} + {\kappa _{i,2}},...,{k_{i,P}} + {\kappa _{i,P}}} \right]^{\rm{T}}}$, the precoding matrices by ${\bf{W}}_i^{{\rm{com}}} \buildrel \Delta \over = {\left[ {{\bf{W}}_{{a_{i,1}}}^{\rm{H}},{\bf{W}}_{{a_{i,2}}}^{\rm{H}},...,{\bf{W}}_{{a_{i,P}}}^{\rm{H}}} \right]^{\rm{H}}}$, and the allocated power for the $i$-th UE by
${\bm \alpha} _i^{{\rm{com}}} \buildrel \Delta \over = {\rm{diag}}\left\{ {\sqrt {{\alpha _{{a_{i,1}}}}} ,\sqrt {{\alpha _{{a_{i,2}}}}} ,...,\sqrt {{\alpha _{{a_{i,P}}}}} } \right\}$, respectively.
According to \cite{Raviteja2019effective} and \cite{li2021performance_analysis},~\eqref{C8_new_com_model} can be rewritten by
\begin{align}
{{\bf{y}}_i} = {\bf{\Phi }}_i^{\omega _i^\tau ,\omega _i^{_\nu },{\bf{W}}_i^{{\rm{com}}}}\left( {\bf{x}} \right){\bm \alpha} _i^{{\rm{com}}}{\bf{h}}_i^{{\rm{eff}}} + {{\bm \eta} _i} \label{C8_new_com_model2},
\end{align}
where ${\bf{\Phi }}_i^{\omega _i^\tau ,\omega _i^{_\nu },{\bf{W}}_i^{{\rm{com}}}}\left( {\bf{x}} \right)$ is referred to as the \emph{equivalent codeword matrix} and it is a concatenated matrix of size $MN\times P$, constructed by the column vector ${{\bf{\Xi }}_{i,p}}{\bf{x}}$ , i.e.,
\begin{equation}
{\bf{\Phi }}_i^{\omega _i^\tau ,\omega _i^{_\nu },{\bf{W}}_i^{{\rm{com}}}}\left( {\bf{x}} \right) = \left[ {{{\bf{\Xi }}_{i,1}}{\bf{x}}\quad {{\bf{\Xi }}_{i,2}}{\bf{x}}\quad  \cdots \quad {{\bf{\Xi }}_{i,P}}{\bf{x}}} \right],
\label{C8_equation_phi}
\end{equation}
and ${{\bf{\Xi }}_{i,p}}$ is given by
\begin{equation}
{{\bf{\Xi }}_{i,p}} \buildrel \Delta \over = \left( {{{\bf{F}}_N} \otimes {{\bf{I}}_M}} \right){{\bm{\Pi }}^{{l_{i,p}}}}{{\bm{\Delta}} ^{{k_{i,p}}+{\kappa _{i,p}}}}{{\bf{W}}_{{a_{i,p}}}}\left( {{\bf{F}}_N^{\rm{H}} \otimes {{\bf{I}}_M}} \right).
\label{C8_equation_Xi}
\end{equation}
To analyze the PEP performance for communication, we start from the study of the conditional PEP based on~\eqref{C8_new_com_model2}. In particular, let us define  the \emph{conditional Euclidean distance} ${d_{{\bf{h}}_i^{{\rm{eff}}},{{\bm{\omega}} _\tau },{{\bm{\omega}} _\nu },{{\bm \alpha} _i^{{\rm{com}}}},{\bf{W}}_i^{{\rm{com}}}}^2\left( {{\bf{x}},{\bf{x'}}} \right)}$ between a pair of codewords ${\bf{x}}$ and ${\bf{x'}}$ (${\bf{x}} \ne {\bf{x'}}$) by
\begin{align}
d_{{\bf{h}}_i^{{\rm{eff}}},{\bm \omega} _i^\tau ,{\bm \omega}_i^{_\nu },{\bm \alpha} _i^{{\rm{com}}},{\bf{W}}_i^{{\rm{com}}}}^2\left( {{\bf{x}},{\bf{x'}}} \right)
=& d_{{\bf{h}}_i^{{\rm{eff}}},{\bm \omega} _i^\tau ,{\bm \omega}_i^{_\nu },{\bm \alpha} _i^{{\rm{com}}},{\bf{W}}_i^{{\rm{com}}}}^2\left( {\bf{e}} \right)\notag\\
\buildrel \Delta \over =&{\left\| {{\bf{\Phi }}_i^{{\bm \omega} _i^\tau ,{\bm \omega} _i^{_\nu },{\bf{W}}_i^{{\rm{com}}}}\left( {\bf{e}} \right){\bm \alpha} _i^{{\rm{com}}}{{\bf{h}}_i^{{\rm{eff}}}}} \right\|^2}\notag\\
=&{\left( {{\bf{h}}_i^{{\rm{eff}}}} \right)^{\rm{H}}} {\left({{\bm \alpha} _i^{{\rm{com}}}}\right)^{\rm H}}{\bf{\Omega }}_i^{_{{\bm \omega} _i^\tau ,{\bm \omega} _i^{_\nu },{\bf{W}}_i^{{\rm{com}}}}}\left( {\bf{e}} \right){{\bm \alpha} _i^{{\rm{com}}}}{\bf{h}}_i^{{\rm{eff}}}\notag\\
=&{\left( {{\bf{h}}_i^{{\rm{eff}}}} \right)^{\rm{H}}}{\bf{\tilde \Omega }}_i^{{\bm \omega }_i^\tau ,{\bm \omega }_i^\nu,{\bf{W}}_i^{{\rm{com}}} }\left( {\bf{e}} \right){\bf{h}}_i^{{\rm{eff}}}, \label{C8_too_long}
\end{align}
where ${\bf{e}} = {\bf{x}} - {\bf{x'}}$ is the corresponding codeword difference (error) sequence.
In particular, we refer to ${\bf{\Omega }}_i^{_{{\bm \omega} _i^\tau ,{\bm \omega} _i^{_\nu },{\bf{W}}_i^{{\rm{com}}}}}\left( {\bf{e}} \right) \buildrel \Delta \over = {\left( {{\bf{\Phi }}_i^{{\bm \omega} _i^\tau ,{\bm \omega} _i^{_\nu },{\bf{W}}_i^{{\rm{com}}}}\left( {\bf{e}} \right)} \right)^{\rm{H}}}{\bf{\Phi }}_i^{{\bm \omega} _i^\tau ,{\bm \omega} _i^{_\nu },{\bf{W}}_i^{{\rm{com}}}}\left( {\bf{e}} \right)
$ in~\eqref{C8_too_long} as the \emph{codeword difference matrix}, while to ${\bf{\tilde \Omega }}_i^{_{{\bm \omega} _i^\tau ,{\bm \omega} _i^{_\nu },{\bf{W}}_i^{{\rm{com}}}}}\left( {\bf{e}} \right) \buildrel \Delta \over = {\left({{\bm \alpha} _i^{{\rm{com}}}}\right)^{\rm H}}{\bf{\Omega }}_i^{_{{\bm \omega} _i^\tau ,{\bm \omega} _i^{_\nu },{\bf{W}}_i^{{\rm{com}}}}}\left( {\bf{e}} \right){{\bm \alpha} _i^{{\rm{com}}}}$ as the \emph{weighted codeword difference matrix}.
For notational simplicity, we henceforth drop the superscript of ${\bf{\Omega }}_i^{_{{\bm \omega} _i^\tau ,{\bm \omega} _i^{_\nu },{\bf{W}}_i^{{\rm{com}}}}}\left( {\bf{e}} \right)$ and ${\bf{\tilde \Omega }}_i^{_{{\bm \omega} _i^\tau ,{\bm \omega} _i^{_\nu },{\bf{W}}_i^{{\rm{com}}}}}\left( {\bf{e}} \right)$, and we have
\begin{align}
{\bf{\Omega }}_i\left( {\bf{e}} \right)= \left[ {\begin{array}{*{20}{c}}
{{{\bf{e}}^{\rm{H}}}{\bf{\Xi }}_{i,1}^{\rm{H}}{{\bf{\Xi }}_{i,1}}{\bf{e}}}&{{{\bf{e}}^{\rm{H}}}{\bf{\Xi }}_{i,1}^{\rm{H}}{{\bf{\Xi }}_{i,2}}{\bf{e}}}& \cdots &{{{\bf{e}}^{\rm{H}}}{\bf{\Xi }}_{i,1}^{\rm{H}}{{\bf{\Xi }}_{i,P}}{\bf{e}}}\\
{{{\bf{e}}^{\rm{H}}}{\bf{\Xi }}_{i,2}^{\rm{H}}{{\bf{\Xi }}_{i,1}}{\bf{e}}}&{{{\bf{e}}^{\rm{H}}}{\bf{\Xi }}_{i,2}^{\rm{H}}{{\bf{\Xi }}_{i,2}}{\bf{e}}}&{}& \vdots \\
 \vdots &{}& \ddots & \vdots \\
{{{\bf{e}}^{\rm{H}}}{\bf{\Xi }}_{i,P}^{\rm{H}}{{\bf{\Xi }}_{i,P}}{\bf{e}}}& \cdots & \cdots &{{{\bf{e}}^{\rm{H}}}{\bf{\Xi }}_{i,P}^{\rm{H}}{{\bf{\Xi }}_{i,P}}{\bf{e}}}
\end{array}} \right]. \label{C8_Gram}
\end{align}
According to~\cite{li2021performance_analysis}, the conditional PEP is upper-bounded by
\begin{align}
\Pr\left( {\left. {{\bf{x}},{\bf{x'}}} \right|{{\bf{h}}_i^{{\rm{eff}}}},{{\bm{\omega }}_i^\tau},{{\bm{\omega }}_i^\nu}},{\bf{W}}_i^{{\rm{com}}},{\bm \alpha} _i^{{\rm{com}}} \right)
\le& \exp \left( { - \frac{{{1}}}{{4{N_0}}}{d_{{\bf{h}}_i^{{\rm{eff}}},{\bm \omega} _i^\tau ,{\bm \omega}_i^{_\nu },{\bm \alpha} _i^{{\rm{com}}},{\bf{W}}_i^{{\rm{com}}}}^2\left( {{\bf{x}},{\bf{x'}}} \right)}} \right)\notag\\
=& \exp \left( { - \frac{{{1}}}{{4{N_0}}}} {\left( {{\bf{h}}_i^{{\rm{eff}}}} \right)^{\rm{H}}}{\bf{\tilde \Omega }}_i\left( {\bf{e}} \right){\bf{h}}_i^{{\rm{eff}}}\right) .\label{C8_PEP_derivation1}
\end{align}
To further simplify~\eqref{C8_PEP_derivation1}, let us focus on the structures of both ${\bf{\tilde \Omega }}_i\left( {\bf{e}} \right)$ and ${\bf{\Omega }}_i\left( {\bf{e}} \right)$.
We observe that both ${\bf{\tilde \Omega }}_i\left( {\bf{e}} \right)$ and ${\bf{\Omega }}_i\left( {\bf{e}} \right)$ are positive-semidefinite Hermitian matrices by their definitions. Furthermore, with a proper design of power allocation, i.e., ${{\bm \alpha} _i^{{\rm{com}}}}$ is of full-rank, both ${\bf{\tilde \Omega }}_i\left( {\bf{e}} \right)$ and ${\bf{\Omega }}_i\left( {\bf{e}} \right)$ share the same rank.
Based on this observation, we consider the eigenvalue decomposition to further our derivation.
Let $r_i$ denote the rank of both ${\bf{\tilde \Omega }}_i\left( {\bf{e}} \right)$ and ${\bf{\Omega }}_i\left( {\bf{e}} \right)$, where $r_i \le P$.
Furthermore, let us denote by $\left\{ {{{\bf{u}}_{i,1}},{{\bf{u}}_{i,2}},...,{{\bf{u}}_{i,P}}} \right\}$ the eigenvectors of ${\bf{\tilde \Omega }}_i\left( {\bf{e}} \right)$ and
$\left\{ {{\lambda _i}\left[ 1 \right],{\lambda _i}\left[ 2 \right],...,{\lambda _i}\left[ P \right]} \right\}$ the corresponding nonnegative real eigenvalues sorted in the descending order,
where ${{\lambda _i\left[ j \right]}} > 0$ for $1 \le j \le r_i$ and ${{\lambda _i\left[ j \right]}} = 0$ for $r_i+1 \le j \le P$.
Then,~\eqref{C8_PEP_derivation1} can be further expanded by~\cite{Tarokh1998space}
\begin{align}
\Pr\left( {\left. {{\bf{x}},{\bf{x'}}} \right|{{\bf{h}}_i^{{\rm{eff}}}},{{\bm{\omega }}_i^\tau},{{\bm{\omega }}_i^\nu}},{\bf{W}}_i^{{\rm{com}}},{\bm \alpha} _i^{{\rm{com}}} \right)\le  \exp \left( { - \frac{1}{{4{N_0}}}\sum\limits_{j = 1}^{{r_i}} {{\lambda _i}\left[ j \right]} {{\left| {\bar h_i^{{\rm{eff}}}\left[ j \right]} \right|}^2}} \right), \label{C8_PEP_derivation2}
\end{align}
where $\bar h_i^{{\rm{eff}}}\left[ j \right] = {{\bf{u}}_{i,j}}{\bf{h}}_i^{{\rm{eff}}}$, for $1 \le j \le r_i$.
Note that the exact values of the elements in ${\bf{h}}_i^{\rm{eff}}$ are unknown to the BS. Therefore, we need to consider the distributions of those elements in order to further our derivation.
It can be shown that $\left\{ {\bar h_i^{{\rm{eff}}}\left[ 1 \right],\bar h_i^{{\rm{eff}}}\left[ 2 \right],...,\bar h_i^{{\rm{eff}}}\left[ r_i \right]} \right\}$ are independent complex Gaussian random variables with zero mean and variance $1/(2P)$ per real dimension.
Consequently, ${\small| \bar h_i^{{\rm{eff}}}\left[ j \right] \small|}$ follows the Rayleigh distribution~\cite{Tarokh1998space}, whose
PDF is given by ${f_{{\rm{PDF}}}}\left( x \right) = 2Px\exp \left( { - P{{x}^2}} \right)$.
With the uniform power delay and Doppler profile, we can get rid of the influence of effective fading coefficients in~\eqref{C8_PEP_derivation2} by averaging ${\small| \bar h_i^{{\rm{eff}}}\left[ j \right] \small|}$ term by term, yielding
\begin{align}
\Pr\left( {\left. {{\bf{x}},{\bf{x'}}} \right|{{\bm{\omega }}_i^\tau},{{\bm{\omega }}_i^\nu}},{\bf{W}}_i^{{\rm{com}}},{\bm \alpha} _i^{{\rm{com}}} \right)
\le \prod\limits_{j = 1}^{{r_i}} {\frac{1}{{1 + \frac{{{\lambda _i[j]}}}{{4{N_0}P}}}}}  \le \frac{1}{{\prod\limits_{j = 1}^{{r_i}} {{\lambda _i[j]}} }}{\left( {\frac{1}{{4{N_0}P}}} \right)^{ - {r_i}}}. \label{C8_PEP_derivation3}
\end{align}
As indicated by~\eqref{C8_PEP_derivation3}, the PEP decreases exponentially with an order of ${r_i}$ with the reduction of the noise PSD. In fact, this exponent is the \textbf{diversity gain} of the transmission~\cite{Tarokh1998space,vucetic2003space,Yuan2003performance}.

In order to enable reliable transmissions, we aim to minimize the upper-bound in~\eqref{C8_PEP_derivation3} by designing suitable precoding matrices and power allocation. To facilitate our design, let we first assume that there is a set of precoding matrices ${\bf{ W}}_i^{{\rm{com}}}$ and a specific power allocation ${\bm { \alpha} _i^{{\rm{com}}}}$ that can minimize the PEP upper-bound in~\eqref{C8_PEP_derivation3} for any given $\bf e$ with respect to all possible delay and Doppler shifts, i.e., ${\bm{\omega }}_i^\tau$ and ${\bm{\omega }}_i^\nu$.
Then, we will develop practical precoding designs and power allocation in the next subsection, such that this lowest PEP upper-bound is approachable.
Notice that both ${\bf{\Omega }}_i\left( {\bf{e}} \right)$ and ${\bf{\tilde \Omega }}_i\left( {\bf{e}} \right)$ are \emph{Gram} matrices~\cite{Tut_Gram} of size $P \times P$ and thus the maximum value of the ranks of both ${\bf{\Omega }}_i\left( {\bf{e}} \right)$ and ${\bf{\Omega }}_i\left( {\bf{e}} \right)$ is $P$. In particular, when ${\bf{\Omega }}_i\left( {\bf{e}} \right)$ is of full-rank,~\eqref{C8_PEP_derivation3} can be further simplified by
\begin{align}
\Pr\left( {\left. {{\bf{x}},{\bf{x'}}} \right|{{\bm{\omega }}_i^\tau},{{\bm{\omega }}_i^\nu}},{\bf{W}}_i^{{\rm{com}}},{\bm \alpha} _i^{{\rm{com}}} \right)
 \le &\frac{1}{{\prod\limits_{j = 1}^{{P}} {{\lambda _i[j]}} }}{\left( {\frac{1}{{4{N_0}P}}} \right)^{ - {P}}}= \frac{1}{{\det \left( {{{{\bf{\tilde \Omega }}}_i\left( {\bf{e}} \right)}} \right)}}{\left( {\frac{1}{{4{N_0}P}}} \right)^{ - P}}\notag\\
=& \frac{1}{{\det \left( {{{\bf{\Omega }}_i\left( {\bf{e}} \right)}} \right)}}{\left( {\frac{{{{\left( {\prod\limits_{p = 1}^P {{\alpha _{{a_{i,p}}}}} } \right)}^{\frac{1}{P}}}}}{{4{N_0}P}}} \right)^{ - P}}. \label{C8_PEP_derivation4}
\end{align}
Based on~\eqref{C8_PEP_derivation4}, we notice that the determinant of ${\bf{\Omega }}_i\left( {\bf{e}} \right)$ is related to the delay shifts ${{\bm{\omega }}_i^\tau}$, the Doppler shifts ${{\bm{\omega }}_i^\nu}$, and the precoding matrices ${\bf{W}}_i^{{\rm{com}}}$, but it is independent from the power allocation ${\bm \alpha} _i^{{\rm{com}}}$. This fact indicates that the we can analyze the influence from the precoding matrices and power allocation on the error performance separately.
In particular,~\eqref{C8_PEP_derivation4} indicates that in the case of full-rank, the precoding scheme should be designed to maximize the determinant of the codeword difference matrix ${\bf{\Omega }}_i\left( {\bf{e}} \right)$.
To maximize the determinant of ${\bf{\Omega }}_i\left( {\bf{e}} \right)$, let us consider the following theorem.

\textbf{Theorem 7-1} (\emph{Upper-bound on the determinant of ${\bf{\Omega }}_i\left( {\bf{e}} \right)$}):
The determinant of the codeword difference matrix ${\bf{\Omega }}_i\left( {\bf{e}} \right)$ can be upper-bounded by
\begin{equation}
\det \left( {{{\bf{\Omega }}_i}\left( {\bf{e}} \right)} \right) \le {\left( {d_{\rm{E}}^2\left( {\bf{e}} \right)} \right)^P}, \label{C8_determinant_upper_bound1}
\end{equation}
where $ {d_{\rm{E}}^2\left( {\bf{e}} \right)}$ denotes the Euclidean distance of the error sequence $\bf e$, i.e.,
$ {d_{\rm{E}}^2\left( {\bf{e}} \right)}  \buildrel \Delta \over = {\bf e}^{\rm H}{\bf e}$. Furthermore, a sufficient condition for achieving the equality is ${\bf{\Omega }}_i\left( {\bf{e}} \right)$ being a diagonal matrix.

\emph{Proof}: The proof is given in the Appendix of this chapter.

In fact, Theorem~7-1 indicates that the PEP upper-bound can be minimized if the received signals from different paths are orthogonal to each other~\cite{Li2020on,li2021performance_analysis}.
Notice that Theorem~7-1 provides an upper-bound of the determinant that is independent from the delay and Doppler shifts. Therefore, in the best case scenario, i.e., orthogonal paths, by substituting~\eqref{C8_determinant_upper_bound1} into~\eqref{C8_PEP_derivation4}, we arrive at
\begin{align}
\Pr\left( \left. {{\bf{x}},{\bf{x'}}} \right|{\bf{W}}_i^{{\rm{com}}},{\bm \alpha} _i^{{\rm{com}}} \right)\le \frac{1}{{\left( {d_{\rm{E}}^2\left( {\bf{e}} \right)} \right)^P}}{\left( {\frac{{{{\left( {\prod\limits_{p = 1}^P {{\alpha _{{a_{i,p}}}}} } \right)}^{\frac{1}{P}}}}}{{4{N_0}P}}} \right)^{ - P}}
={\left( {\frac{{d_{\rm{E}}^2\left( {\bf{e}} \right)}}{P}} \right)^{ - P}}{\left( {\frac{{{{\left( {\prod\limits_{p = 1}^P {{\alpha _{{a_{i,p}}}}} } \right)}^{\frac{1}{P}}}} }{{4{N_0}}}} \right)^{ - P}}.\label{C8_PEP_derivation5}
\end{align}
According to~\eqref{C8_PEP_derivation5}, we refer to the term ${d_{\rm{E}}^2\left( {\bf{e}} \right)}/P$ as the \textbf{maximum coding gain} of the underlying transmission~\cite{Li2020on,li2021performance_analysis}, which indicates how much can the error performance be possibly improved by varying $\bf e$, for all possible values of the delay and Doppler shifts.
In the next subsection, we will design suitable precoding matrices based on~\eqref{C8_PEP_derivation5}.

\subsection{Precoding Design}
We notice that the delay and Doppler indices among different paths will affect the PEP performance.
Motivated by this observation, we propose our precoding design by considering the concepts of \textbf{virtual delay index} and \textbf{virtual Doppler index}, whose definitions are given as follows.

\textbf{Definition 7-1} (\emph{Virtual Delay and Doppler Indices}):
The virtual delay and Doppler indices are defined by $0 \le {{{\dot l}_p}} \le M-1$ and $0 \le {{{\dot k}_p}} \le N-1$, for $1 \le p \le P$, where ${{{\dot l}_p}}$ and ${{{\dot k}_p}}$ are of integer values, for $1 \le p, p' \le P$.

Recalling the discussions in previous subsections, we note that our precoding design is to shape the codeword difference matrix ${\bf{\Omega }}_i\left( {\bf{e}} \right)$, such that it can be a diagonal matrix, for any possible ${\bm{\omega }}_i^\tau$, ${\bm{\omega }}_i^\nu$, and $\bf e$.
By observing the structure of~\eqref{C8_Gram} and according to Theorem~7-1, we notice that the aforementioned design criterion for precoding matrices is satisfied if
\begin{align}
{\bf e}^{\rm H}{\bf{\Xi }}_{_{i,p}}^{\rm{H}}{{\bf{\Xi }}_{i,p'}} {\bf e}= 0, \label{C8_precoding_CR1}
\end{align}
and
\begin{align}
{\bf{W}}_{{a_{i,p}}}^{\rm{H}}{{\bf{W}}_{{a_{i,p}}}} = {{\bf{I}}_{MN}}, \label{C8_precoding_CR2}
\end{align}
for any possible $\bf e$, and any $ 1 \le p,p' \le P$ and $p' \ne p$. Corresponding to both~\eqref{C8_precoding_CR1} and~\eqref{C8_precoding_CR2}, the following lemma shows an interesting fact of the precoding design problem.

\textbf{Lemma 7-3} (\emph{Determinant Dilemma}):
The precoding matrices cannot satisfy ${\bf{W}}_{{a_{i,p}}}^{\rm{H}}{{\bf{W}}_{{a_{i,p}}}} = {{\bf{I}}_{MN}}$ and ${\bf{W}}_{{a_{i,p}}}^{\rm{H}}{{\bf{W}}_{{a_{i,p'}}}} = {{\bf{0}}_{MN}}$ at the same time, for $ 1 \le p,p' \le P$ and $p' \ne p$.

\emph{Proof}: The proof is given in the Appendix of this chapter.

As indicated by Lemma~7-3, we note that an explicit design algorithm of the precoding matrices satisfying the above criteria is not realizable.
Therefore, we consider a relaxation for the precoding design, where ${\bf{\Omega }}_i\left( {\bf{e}} \right)$ is a diagonally-dominant matrix~\cite{horn2012matrix} instead of a strict diagonal matrix.
To achieve this, we require that both the virtual delay and Doppler indices are different for different paths, i.e.,
${{{\dot l}_p}} \ne {{{\dot l}_{p'}}}$ and ${{{\dot k}_p}} \ne {{{\dot k}_{p'}}}$ for any $ 1 \le p,p' \le P$ and $p' \ne p$.
Therefore, for given \emph{a priori} AoA, delay and Doppler estimates from radar sensing, the proposed precoding matrices are of the form
\begin{align}
{{\bf{W}}_{{a_{i,p}}}} \buildrel \Delta \over = {{\bf{\Delta }}^{ - {{\hat k}_{i,p}} - {{\hat \kappa }_{i,p}}}}{{\bf{\Pi }}^{ - {{\hat l}_{i,p}}}}{{\bf{\Pi }}^{{{\dot l}_p}}}{{\bf{\Delta }}^{{{\dot k}_p}}}, \label{C8_precoding_design}
\end{align}
where ${{\hat l}_{i,p}} $ and ${{{\hat k}_{i,p}} + {{\hat \kappa }_{i,p}}}$ are the estimated delay and Doppler indices, while ${{{\dot l}_p}}$ and ${{{\dot k}_p}}$ are the \emph{virtual} delay and Doppler indices with different values for different paths.

The rationale of the proposed precoding design is to improve the orthogonality devised by OTFS transmissions based on the nature of delay and Doppler shifts of the channel. Note that the delay and Doppler shifts are two physical parameters that determines the distortion characteristics of the resolvable path for transmission and different resolvable paths cannot share the same delay and Doppler shifts according to the definition~\cite{tse2005fundamentals,hlawatsch2011wireless}.
However, it is possible that different resolvable paths share the same delay or Doppler shift and in this case the natural orthogonality among different paths may be undermined. In the following proposition, we prove that the proposed precoding scheme can improve the orthogonality in the case where the delay or Doppler shifts associated to different paths are of the same values.

\textbf{Proposition 7-1} (\emph{Diagonal Dominance}):
Let ${\varepsilon}$ denote the event where at least two paths share the same delay or Doppler indices. Then, it can be shown that
\begin{align}
\Pr \left( {\left. {\left| {{{\bf{e}}^{\rm{H}}}{{\bf{\Xi }}_{i,p}}{{\bf{\Xi }}_{i,p}}{\bf{e}}} \right| \ge \sum\limits_{\scriptstyle {p'} = 1\hfill\atop
\scriptstyle {p'} \ne p\hfill}^P {\left| {{{\bf{e}}^{\rm{H}}}{{\bf{\Xi }}_{i,p}}{{\bf{\Xi }}_{i,p'}}{\bf{e}}} \right|} } \right|{{\bf{W}}_{{a_{i,p}}}},\varepsilon } \right)
 \ge \Pr \left( {\left. {\left| {{{\bf{e}}^{\rm{H}}}{{\bf{\Xi }}_{i,p}}{{\bf{\Xi }}_{i,p}}{\bf{e}}} \right| \ge \sum\limits_{\scriptstyle {p'} = 1\hfill\atop
\scriptstyle {p'} \ne p\hfill}^P {\left| {{{\bf{e}}^{\rm{H}}}{{\bf{\Xi }}_{i,p}}{{\bf{\Xi }}_{i,p'}}{\bf{e}}} \right|} } \right|\varepsilon } \right) .\label{C8_Prop1_add}
\end{align}

\emph{Proof}: The proof is given in the Appendix of this chapter.

\begin{figure}
\centering
\includegraphics[width=0.7\textwidth]{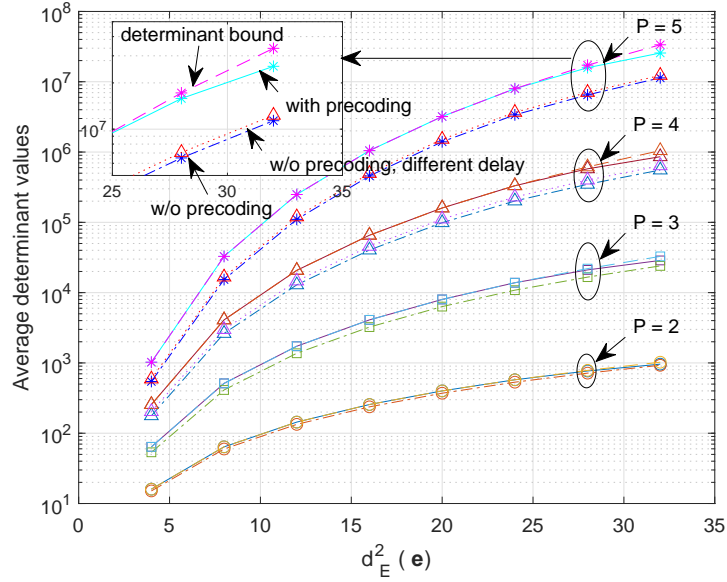}
\caption{Average determinant values of ${{\bf{\Omega }}_i\left( {\bf{e}} \right)}$ with (in dash-dotted lines) and without precoding (in solid lines for random delay indices and in dotted lines for different delay indices), comparing with the determinant bound in~\eqref{C8_determinant_upper_bound1} (in dashed lines), where $M=8,N=8,l_{\rm max}=2$, and $k_{\rm max}=2$, respectively. }
\label{C8_Determinant_evaluation}
\centering
\end{figure}

The definition of diagonally-dominant matrix suggests that the amplitude of the diagonal elements is larger than the sum of the amplitudes of the other elements in the same row. According to Proposition~7-1, we can infer that the proposed precoding can improve the probability of the codeword difference matrix ${\bf{\Omega }}_i\left( {\bf{e}} \right)$ having a diagonally-dominant structure.
Generally speaking, diagonally-dominant matrix are well-structured, whose determinant value is close to the corresponding diagonal matrix~\cite{horn2012matrix}.
To further qualify the effectiveness of the proposed precoding, we compare the average determinant of ${{\bf{\Omega }}_i\left( {\bf{e}} \right)}$ with and without precoding by numerical simulations.
Without loss of generality, we consider $M=8,N=8$ in Fig.~\ref{C8_Determinant_evaluation}, where the maximum delay and Doppler indices are set to be $l_{\rm{max}}=2$ and $k_{\rm{max}}=2$, respectively.
Since numerically emulating all the error sequences with such a frame size is generally intractable in a reasonable time frame even with BPSK mapping, we consider the comparison between the average determinant values of ${{\bf{\Omega }}_i\left( {\bf{e}} \right)}$ with and without precoding
for given error sequences{\footnote{We use the same error sequence as in~\cite{li2021performance_analysis}, i.e., ${\bf e}=[2, 0, -2, 2, 0, -2,...,0...0]^{\rm T}$.}}~\cite{Li2020on,li2021performance_analysis}, with respect to all possible channel realizations{\footnote{Without loss of generality, we require the absolute value of the difference between any two Doppler indices no smaller than $0.2$.}}.
For a better illustration, we also plot the determinant upper-bound in~\eqref{C8_determinant_upper_bound1} for comparison.
As indicated by the figure, the proposed precoding can indeed increase the determinant value of ${{\bf{\Omega }}_i\left( {\bf{e}} \right)}$ compared to the case without precoding, where the determinant value with precoding aligns well with the upper-bound, especially for small values of $ {d_{\rm{E}}^2\left( {\bf{e}} \right)}$. In particular, we observe that the improvement becomes more obvious with more resolvable paths, which indicates that the proposed precoding is more helpful for communication transmissions in rich scattering scenarios.
On the other hand, it has reported in the literature that the fractional Doppler may potentially degrade the error performance~\cite{wei2020transmitter}. Therefore, we have also plotted the curves of the average determinant values corresponding to the case of different delay indices and fractional Doppler indices in Fig.~\ref{C8_Determinant_evaluation}, for both $P=4$ and $P=5$. As indicated by the figure, the determinant values increase slightly compared to the non-precoded case (random delay and fractional Doppler indices), but still shows a noticeable gap compared to the precoded case. This observation agrees with the previous conclusions in~\cite{wei2020transmitter}.

\textbf{Remark 7-4}: It should also be noted that the effective DD domain channel matrix for the case of fractional Doppler shifts is generally dense and complex~\cite{Zhiqiang_magzine}. Consequently, the detection complexity required for fractional Doppler case is quite high~\cite{Raviteja2018interference}. Therefore, our proposed precoding design can also reduce the detection complexity at the UE side.

\textbf{Remark 7-5}: We briefly discuss the extension of our communication design in the case of inaccurate AoA estimates.
According to the proposed beam tracking scheme, it is suggested to allocate power to all the antennas in ${{\mathbb U}_{i,p}}$. Hence, given that the minor changes of the related parameters from the previous time instant are
well-compensated~\cite{liu2020joint}, we can apply the proposed precoding design to all the antennas in ${{\mathbb U}_{i,p}}$, and evenly assign the power among related antennas. We note that this will inevitably lead to
an SNR reduction for communication. However, this reduction is small if the beam width is small.

\subsection{Power Allocation for Communication}
According to~\eqref{C8_PEP_derivation5}, the power allocation problem for communication is formulated by
\begin{align}
\mathop {\rm minimize }\limits_{{\alpha _{{a_{i,1}}}},...,{\alpha _{{a_{i,P}}}}} &{\left( {\frac{{d_{\rm{E}}^2\left( {\bf{e}} \right)}}{P}} \right)^{ - P}}{\left( {\frac{{{{\left( {\prod\limits_{p = 1}^P {{\alpha _{{a_{i,p}}}}} } \right)}^{\frac{1}{P}}}}}{{4{N_0}}}} \right)^{ - P}} \\
{\rm{s}}{\rm{.t}}{\rm{.}}\quad&\sum\limits_{p = 1}^P {{\alpha _{{a_{i,p}}}}}  = {{\tilde \alpha }_i}\quad {\rm and }\quad {{a_{i,p}}} \ge 0,
\end{align}
where ${{\tilde \alpha }_i}$ denotes the allocated power for the $i$-th UE.
As indicated by the above optimization problem, the power allocation should be designed to maximize the power product ${ {\prod\nolimits_{p = 1}^P {{\alpha _{{a_{i,p}}}}} } }$, i.e., the geometric mean of the allocated power associated to the corresponding paths in order to minimize the PEP. According to the
arithmetic mean-geometric mean (AM-GM) inequality, it is not hard to show that with a total power constraint ${{\tilde \alpha }_i}$, the optimal power allocation for minimizing PEP is the equal power allocation.
This is not unexpected, because the communication fading coefficients are of the same distribution and therefore the power allocation should not provide any bias to any path.

\textbf{Remark 7-6}: It is interesting to see that both radar sensing and communication require different power allocations.
Therefore, we in the following briefly discuss how to adapt those two allocations in practical systems.
Let us first consider the total power constraint for the $i$-th UE for the communication transmission, i.e., $\sum\nolimits_{p = 1}^P {{\alpha _{{a_{i,p}}}}} $. According to the AM-GM inequality, we have $\prod\nolimits_{p = 1}^P {{\alpha _{{a_{i,p}}}}}  \le {\left( {\frac{{\sum\nolimits_{p = 1}^P {{\alpha _{{a_{i,p}}}}} }}{P}} \right)^P}$. Notice that there is an exponent $P$ on the right hand side. Therefore, the communication performance degradation due to unfit power allocation may become more severe with a larger number of resolvable paths. However, it should be noted that the communication fading coefficients are not known at the BS. Therefore, the random nature of the communication channel may also mitigate the performance degradation induced by the undesirable power allocation.
On the other hand, since the proposed AoA estimation algorithm in Section~7.2.2 is based on the principle of matched-filtering, where the received power will be the key factor determining the estimation performance. Different from the communication counterpart, the radar reflection coefficients are assumed to be known at the BS. Therefore, a suitable power allocation that is designed specifically for each signal transmission can largely improve the sensing performance.
Therefore, in practical ISAC scenarios, radar sensing performance should be the priority for power allocation designs.

It should be noticed that a precise analysis on the relationship of the power allocation between radar sensing and communication performances requires detailed statistical models for both the radar reflection coefficients and communication fading coefficients, which is also related to system settings, e.g., the frame size. Since the major focus of this chapter is to propose the ISAC transmission framework based on SS-OTFS modulation, we leave this interesting issue for our future work.

\section{Numerical Results}
We demonstrate the numerical results for the proposed ISAC transmissions in this section, where the SPA detection~\cite{li2021hybrid} for OTFS equalization is adopted at the UE side.
In specific, we set $N_{\rm BS}=128$, $M=32$, $N=16$, $\tau_{\max}=10$, and $\nu_{\max}=6$, respectively, unless specified otherwise, where the transmitted signals are BPSK  modulated.
To evaluate the communication performance, we define the average symbol SNR by $\frac{{{E_s}}}{{{N_0}}} \buildrel \Delta \over = \frac{\beta_i}{{{N_0}}}$, where $\beta_i$ denotes the average power assigned to each antenna for the $i$-th UE, i.e., ${\beta _i} = \sum\nolimits_{p = 1}^P {{\alpha _{{a_{i,p}}}}} /P$.
Meanwhile, we define the radar SNR by the ratio between the total power $\alpha_{\rm total}$ and the radar noise PSD ${\tilde N}_0$.
Without loss of generality, we assume that the reflection coefficient coefficients follow a uniform complex Gaussian distribution, i.e., ${{\tilde h}_{i,p}} \sim {\cal CN}\left( {0,1} \right)$.

\subsection{Beam Tracking Performance}
\begin{figure}
\centering
\includegraphics[width=0.5\textwidth]{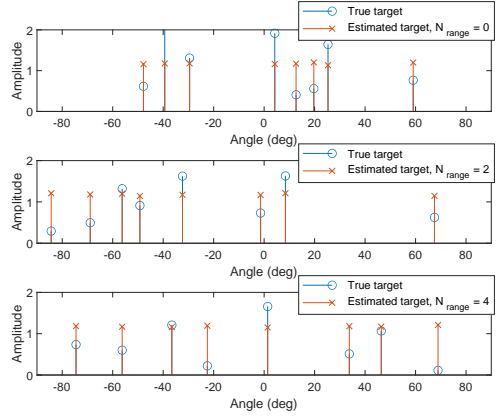}
\caption{AoA estimation performance with various beam widths for $K=4$ UEs, where each UE has $P=2$ paths.}
\label{C8_AoA}
\centering
\end{figure}

We present the AoA estimation performance for the proposed ISAC transmission with various beam widths in Fig.~\ref{C8_AoA}, where we assume that there are in total $K=4$ UEs and each UE has $P=2$ paths. In specific, we consider a radar SNR at $5$ dB in Fig.~\ref{C8_AoA}, where the proposed power allocation in Section~7.2.3 is applied and beam widths are controlled by the value of $N_{\rm range}$ as discussed in Section~7.2.1. In particular, the amplitude in the figure represents the normalized trace of~\eqref{C8_AoA_der2} with respect to the number of transmitted symbols.
As can be observed from the figure, the proposed beam tracking and AoA estimation can provide an accurate estimation performance for the considered scenario with various beam widths.
Note that the received echo power is reduced with the increase of the beam widths. In specific, it can be shown the average received power with $N_{\rm range}=4$ is only $1/5$ of that with $N_{\rm range}=0$, since we evenly assign the transmit power among all $N_{\rm range}+1$ antennas. Therefore, this observation also indicates that the proposed power allocation is suitable for the considered radar sensing issues.

\subsection{Precoding Performance}
\begin{figure}
\centering
\includegraphics[width=0.7\textwidth]{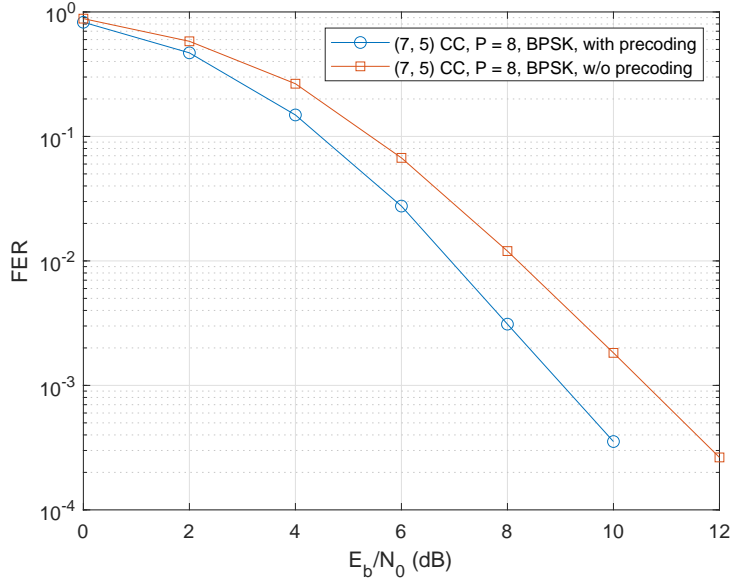}
\caption{FER comparisons between the proposed ISAC transmissions with and without precoding, where (7, 5) convolutionally coded BPSK signals are considered with $P = 8$.}
\label{C8_Path8_coded}
\centering
\end{figure}

We verify the effectiveness of the proposed precoding scheme in Fig.~\ref{C8_Path8_coded}, where the FER performance for a specific UE with only integer delay and Doppler indices is illustrated. In particular, we consider the coded BPSK signals with $P=8$ for transmission. Due to the high detection/decoding complexity, we consider a smaller frame size for simulation, where we have $N=8$, $M=16$.
Without loss of generality, we apply the terminated (7, 5) convolutional code as the channel code, and the virtual delay and Doppler indices are randomly generated for precoding.
Meanwhile, we consider the equal power allocation in this example.
As can be observed from the figure, with the same channel coding and power allocation, the transmission with precoding has a roughly $1.7$ dB gain in terms of average bit SNR compared to the transmission without precoding at FER $4 \approx  \times 10^{-4}$.
Furthermore, we also notice that the FER slope for precoded transmission is steeper than the transmission without precoding. This observation indicates that the proposed precoding can also improve the diversity gain, which is due to the fact that the codeword difference matrix is more likely to have full-rank when different delay and Doppler indices are of different values.

\begin{figure}
\centering
\includegraphics[width=0.7\textwidth]{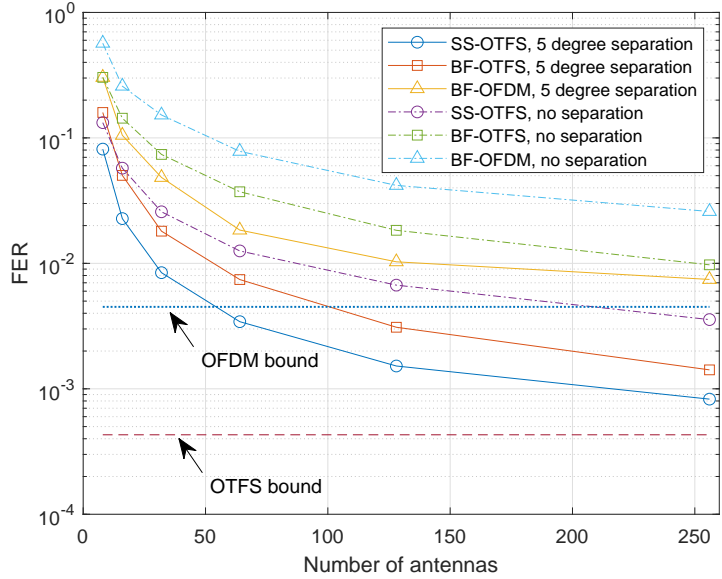}
\caption{FER performance comparisons between SS-OTFS, BF-OTFS, and BF-OFDM for the first UE with $P=4$. }
\centering
\label{C8_New_compare_SS_OTFS_OTFS_OFDM}
\end{figure}

We demonstrate the FER performances of SS-OTFS{\footnote{According to Remark 7-1, we introduce a phase offset according to the AoD associated to the first path of the first UE before spatial spreading.}} for the first UE with $P=4$ with respect to different numbers of antennas in Fig.~\ref{C8_New_compare_SS_OTFS_OTFS_OFDM}, where a rate-half (3, 2) convolutional code is used for ISAC transmission. To show the superior performance of SS-OTFS for ISAC transmissions, we also provide the FER performances for the beamforming counterparts, including the beamforming-based OTFS (BF-OTFS) and the beamforming-based OFDM (BF-OFDM), where the proposed precoding are applied for both SS-OTFS and BF-OTFS. In particular, we do not assume perfect beam separation by using spatial spreading, as mentioned in our system model. Instead, we take into account of the interferences from different beams and study the potential performance improvements with increased number of antennas, while the perfect beam separation cases for both OTFS and OFDM are served as the corresponding error performance lower-bounds for references.
Furthermore, two ISAC transmission cases are considered for a fair comparison. In the first case, the AoDs associated to different paths are randomly generated from $\left[ { - {{90}^ \circ },{{90}^ \circ }} \right]$ with a uniform distribution, while in the second case, the AoDs are uniformly generated at random from $\left[ { - {{90}^ \circ },{{90}^ \circ }} \right]$ with a uniform distribution but we require an at least ${{5^ \circ }}$ separation between any two paths, which is more reasonable in practical ISAC transmissions. As can be observed from Fig.~\ref{C8_New_compare_SS_OTFS_OTFS_OFDM}, SS-OTFS outperforms BF-OTFS and BF-OFDM counterparts for both the two ISAC cases with the same number of antennas. Also, the FER performances for SS-OTFS, BF-OTFS and BF-OFDM improve with more antennas and the AoD separation. This is not unexpected, because the shaped beams become narrower with more antennas and the AoD separation can effectively mitigate the level of crosstalk between different beams in multipath transmissions. Moreover, we notice that the FER performances for SS-OTFS approach the error performance lower-bound, i.e., perfect beam separation case, with an increased number of antennas, which implies that SS-OTFS can effectively mitigate the interferences from different beams.
\subsection{Radar Sensing and Communication Performances with Power Allocations}
As discussed in Remark 7-6, both radar sensing and communication requires different power allocations. Therefore, we evaluate the performances of radar sensing and communication with different power allocations in this subsection. In particular, we consider both the equal power allocation (designed for communication, referred to as ``communication power allocation" in the figures) and the power allocation (designed for radar sensing, referred to as ``radar power allocation" in the figures) proposed in Section~7.2.3. Without loss of generality, we set $N_{\rm range}=0$.

\begin{figure}
\centering
\includegraphics[width=0.7\textwidth]{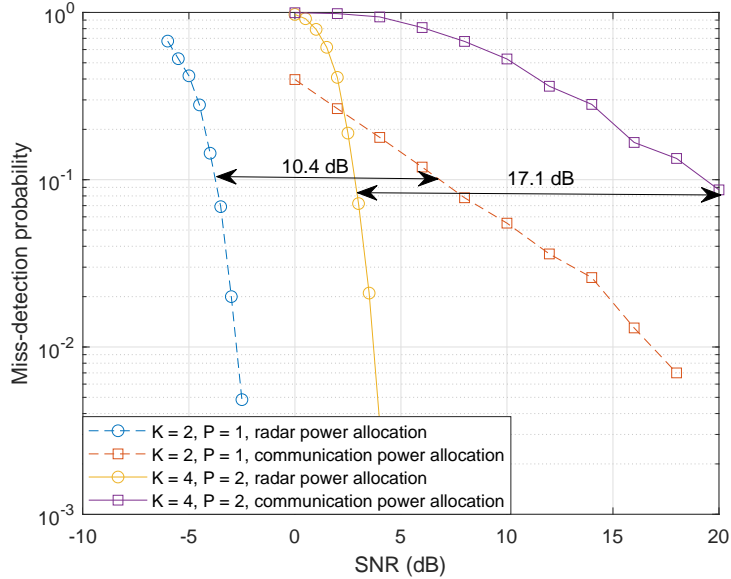}
\caption{Miss-detection probability for radar sensing with radar/communication power allocation for different number of UEs and paths.}
\label{C8_Misdetection}
\centering
\end{figure}

To demonstrate the advantages of the proposed ISAC transmission, we consider a strict performance metric for radar sensing. As the angular domain is discretized, we are interested in the ``miss-detection probability" which is defined by the ratio between the times when the radar does not accurately detect the receive antenna indices ${\tilde a}_{i,p}$ for $1 \le i \le K$ and $1\le p\le P$ and the total number of ISAC transmissions. Without loss of generality, we consider two cases for radar sensing in Fig.~\ref{C8_Misdetection}, where the number of UEs are $K=4$ and $K=2$, and the number of paths are $P=2$ and $P=1$, respectively. As indicated by the figure, suitable power allocation can provide significant performance improvements for radar sensing, especially when the number of targets is large. This indicates that the proposed power allocation is indeed suitable for radar sensing.

\begin{figure}
\centering
\includegraphics[width=0.7\textwidth]{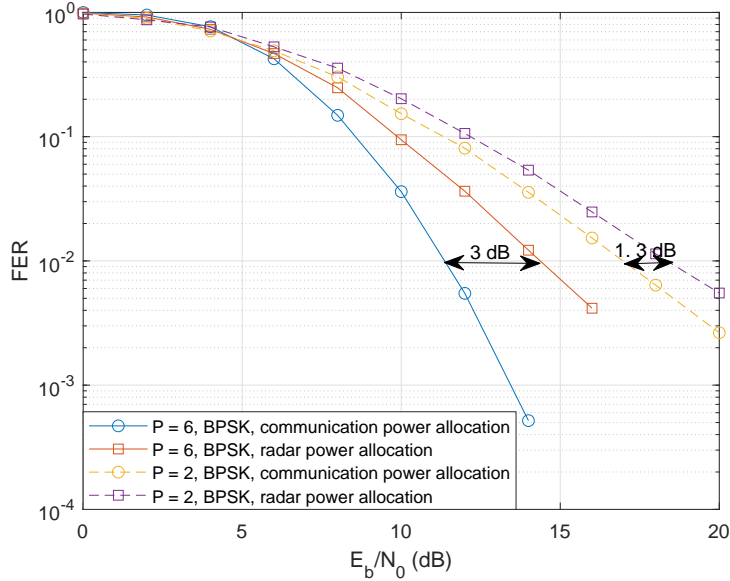}
\caption{Comparisons between the FER performances for a specific UE with communication power allocation and with radar power allocation.}
\label{C8_FER_PA}
\centering
\end{figure}
We show the FER performances for a specific UE with different power allocations for communications in Fig.~\ref{C8_FER_PA}, where the proposed precoding scheme is applied. As shown in the figure, power allocation designed for communication can provide a better error performance for communication, and the performance improvement becomes larger with more paths, which are consistent with our discussions in Section~7.3.3. On the other hand, we also notice that with the power allocation designed for radar sensing, the communication performance degrades. However, this performance degradation is relatively small compared to the performance improvement for radar sensing. Therefore, it is desirable to design the power allocation with a priority for improving the radar sensing performance.

\section{Summary of the Chapter}
In this chapter, we proposed a novel framework for ISAC transmissions based on the SS-OTFS modulation by considering the mismatch of the reflection strengths between the radar sensing and communication. We first derived the
channel models for both radar sensing and communication, which are then simplified based on the properties of SS-OTFS modulation.
Based on the radar model, we proposed simple beam tracking, AoA estimation algorithms, and power allocation for radar sensing. Furthermore, we carried out a detailed analysis on the
PEP for communication, where we showed that the PEP can be minimized with equal power allocation if the received signals from different paths are orthogonal to each other.
Based on this conclusion, we proposed a symbol-wise precoding design to improve this orthogonality by introducing virtual delay and Doppler indices. We also noticed that radar sensing and communication require different power allocations.
To facilitate the power allocation design for practical ISAC systems, we briefly discussed the radar sensing and communication performances with different power allocations and concluded
that the power allocation should be designed leaning towards radar sensing.
Finally, the effectiveness of the proposed framework is verified by numerical results.

\section{Appendices of the Chapter}
In this section, we provide the related proofs from this chapter.
\subsection{Proof of Lemma 7-1}
The proof of this lemma is based on the calculations of~\eqref{C8_com_H_A} and~\eqref{C8_radar_H_A}. Let us start with the calculation for the equivalent angular domain channel vector for the communication channel. For $1 \le l \le N_{\rm BS}$, it can be shown that the $l$-th column of ${\bf{F}}_{{N_{{\rm{BS}}}}}^{\rm{H}}$ is given by
\begin{align}
\left[ {\frac{1}{{\sqrt {{N_{{\rm{BS}}}}} }},\frac{1}{{\sqrt {{N_{{\rm{BS}}}}} }}\exp \left( {j2\pi \frac{1}{{{N_{{\rm{BS}}}}}}} \right),...,} \right.
{\left. {\frac{1}{{\sqrt {{N_{{\rm{BS}}}}} }}\exp \left( {j2\pi \frac{{{N_{{\rm{BS}}}} - 1}}{{{N_{{\rm{BS}}}}}}} \right)} \right]^{\rm{T}}}.
\end{align}
Furthermore, notice that $\bm \alpha$ is a diagonal matrix, whose $l$-th diagonal element is $\alpha_l$. Thus, according to the rule of matrix multiplication, we have
\begin{align}
h_{i,p}^{\rm{A}}\left[ l \right] = \frac{{\sqrt {{\alpha _l}} }}{{{N_{{\rm{BS}}}}}}\sum\limits_{{n_{\rm{t}}} = 0}^{{N_{{\rm{BS}}}} - 1} {\exp \left( {j{n_{\rm{t}}}\pi \left( {\sin {\varphi _{i,p}} + \frac{{2\left( {l - 1} \right)}}{{{N_{{\rm{BS}}}}}}} \right)} \right)} .
\end{align}
The radar channel elements can be calculated in a similar way to the above calculation for communication channel elements, according to the rule of matrix multiplication. This completes the proof of Lemma~7-1.\hfill $\blacksquare$
\subsection{Proof of Corollary 7-1}
Let $\sin {\varphi _{i,p}} = q\frac{2}{{{N_{{\rm{BS}}}}}}$, where $q$ is an arbitrary integer number. Then,~\eqref{C8_com_H_A} becomes
\begin{align}
h_{i,p}^{\rm{A}}\left[ l \right] = \frac{{\sqrt {{\alpha _l}} }}{{{N_{{\rm{BS}}}}}}\sum\limits_{{n_{\rm{t}}} = 0}^{{N_{{\rm{BS}}}} - 1} {\exp \left( {j2\pi \frac{{{n_{\rm{t}}}}}{{{N_{{\rm{BS}}}}}}\left( {l + q - 1} \right)} \right)}. \label{C8_Corollary_der1}
\end{align}
Notice that
\begin{align}
\sum\limits_{{n_{\rm{t}}} = 0}^{{N_{{\rm{BS}}}} - 1} {\exp \left( {j2\pi \frac{{{n_{\rm{t}}}}}{{{N_{{\rm{BS}}}}}}l} \right)}  = \left\{ \begin{array}{l}
{N_{{\rm{BS}}}},\quad\quad l = 0,\\
0,\quad\quad\quad\; l \ne 0.
\end{array} \right.\notag
\end{align}
Therefore, it can be shown that~\eqref{C8_Corollary_der1} only has non-zero values if $l={a_{i,p}}={\left[ {{N_{{\rm{BS}}}} - \frac{{\sin \left( {{\varphi _{i,p}}} \right){N_{{\rm{BS}}}}}}{2}} \right]_{{N_{{\rm{BS}}}}}} + 1$, where the modulo operation is due to the fact that $\sin {\varphi _{i,p}}$ could take negative values while the antenna index has to be positive. Similarly,~\eqref{C8_radar_H_A} can be expressed by
\begin{align}
\tilde H_{i,p}^{\rm{A}}\left[ {k,l} \right] = &\frac{{\sqrt {{\alpha _l}} }}{{{{\left( {{N_{{\rm{BS}}}}} \right)}^2}}}\left( {\sum\limits_{{n_{\rm{r}}} = 0}^{{N_{{\rm{BS}}}} - 1}\! {\exp \left( {j2\pi \frac{{{n_{\rm{r}}}}}{{{N_{{\rm{BS}}}}}}\left( {q - k + 1} \right)} \right)} } \right)\left( {\sum\limits_{{n_{\rm{t}}} = 0}^{{N_{{\rm{BS}}}} - 1} \!{\exp \left( {j2\pi \frac{{{n_{\rm{t}}}}}{{{N_{{\rm{BS}}}}}}\left( {l + q - 1} \right)} \right)} } \right). \label{C8_Corollary_der2}
\end{align}
Thus,~\eqref{C8_Corollary_der2} only has non-zero values if $l={a_{i,p}}$ and $k = {{\tilde a}_{i,p}} = {\left[ {{N_{{\rm{BS}}}} + \frac{{\sin \left( {{\varphi _{i,p}}} \right){N_{{\rm{BS}}}}}}{2}} \right]_{{N_{{\rm{BS}}}}}} + 1$. On the other hand, the values of $h_{i,p}^{\rm{A}}\left[ l \right]$ and $\tilde H_{i,p}^{\rm{A}}\left[ {k,l} \right]$ for $l={a_{i,p}}$ and $k = {{\tilde a}_{i,p}}$ can be easily calculated based on~\eqref{C8_Corollary_der1} and~\eqref{C8_Corollary_der2}. This completes the proof of Corollary~7-1. \hfill $\blacksquare$
\subsection{Proof of Theorem 7-1}
According to~\eqref{C8_Gram}, we notice that the codeword difference matrix ${{\bf{\Omega }}_i\left( {\bf{e}} \right)}$ is a Gram matrix corresponding to the vectors $\left\{ {{{\bf{{u}}}_{i,1}},{{\bf{{u}}}_{i,2}}, \ldots {{\bf{{u}}}_{i,P}}} \right\}$, where ${{\bf{u}}_{i,j}} \buildrel \Delta \over = {{\bf{\Xi }}_{i,j}}{\bf{e}}$.
In particular, the determinant of the Gram matrix ${{\bf{\Omega }}\left( {\bf{e}} \right)}$, i.e., the Gram determinant, is equal to the square of the $P$-dimensional volume of the parallelotope constructed on $\left\{ {{{\bf{{u}}}_{i,1}},{{\bf{{u}}}_{i,2}}, \ldots {{\bf{{u}}}_{i,P}}} \right\}$.
Let us refer to ${\rm{GD}}\left(\left\{ {{{\bf{{u}}}_{i,1}},{{\bf{{u}}}_{i,2}}, \ldots {{\bf{{u}}}_{i,P}}} \right\} \right)$ as the Gram determinant of ${{\bf{\Omega }}_i\left( {\bf{e}} \right)}$. It can be shown that the Gram determinant can be calculated recursively, such as~\cite{Tut_Gram}
\begin{align}
{\rm{GD}}\left( \left\{ {{{\bf{{u}}}_{i,1}},{{\bf{{u}}}_{i,2}}, \ldots {{\bf{{u}}}_{i,j}}} \right\} \right)= {\rm{GD}}\left( \left\{ {{{\bf{{u}}}_{i,1}},{{\bf{{u}}}_{i,2}}, \ldots {{\bf{{u}}}_{i,j-1}}} \right\} \right){\left\| {{{{\bf{\tilde u}}}_{i,j}}} \right\|^2},     \label{C8_Gram_derivation1}
\end{align}
where the term ${{{{\bf{\tilde u}}}_{i,j}}}$ denotes the \emph{orthogonal projection} of ${{{{\bf{u}}}_{i,j}}}$ onto the \emph{orthogonal complement} of ${\rm{span}}\left( {{{{\bf{ u}}}_{i,1}},{{{\bf{ u}}}_{i,2}}, \ldots {{{\bf{ u}}}_{i,j - 1}}} \right)$.
Considering the property of orthogonal projection, we have
\begin{align}
{\rm{GD}}\left(\left\{ {{{\bf{{u}}}_{i,1}},{{\bf{{u}}}_{i,2}}, \ldots {{\bf{{u}}}_{i,j}}} \right\} \right)=& {\rm{GD}}\left( \left\{ {{{\bf{{u}}}_{i,1}},{{\bf{{u}}}_{i,2}}, \ldots {{\bf{{u}}}_{i,j-1}}} \right\} \right){\left\| {{{{\bf{\tilde u}}}_{i,j}}} \right\|^2} \notag\\
\le& {\rm{GD}}\left( {\left\{ {{{\bf{{u}}}_{i,1}},{{\bf{{u}}}_{i,2}}, \ldots {{\bf{{u}}}_{i,j-1}}} \right\}} \right){\left\| {{{{\bf{u}}}_{i,j}}} \right\|^2},\label{C8_Gram_derivation2}
\end{align}
where the equality holds if ${{{\bf{u}}_{i,j}}}$ is orthogonal to ${{{\bf{u}}_{i,j'}}}$, for $1\le j' < j$, i.e., ${\left( {{{\bf{\Xi }}_{i,j'}}{\bf{e}}} \right)^{\rm{H}}}\left( {{{\bf{\Xi }}_{i,j}}{\bf{e}}} \right) = 0$, for $1\le j' < j$.
Hence, by considering~\eqref{C8_Gram_derivation2}, the Gram determinant can be upper-bounded by
\begin{align}
\det \left( {{{\bf{\Omega }}_i}\left( {\bf{e}} \right)} \right)
=& {\rm{GD}}\left( {\left\{ {{{\bf{u}}_{i,1}},{{\bf{u}}_{i,2}}, \ldots {{\bf{u}}_{i,P - 1}}} \right\}} \right){\left\| {{{{\bf{\tilde u}}}_{i,P}}} \right\|^2} \notag\\
\le& {\rm{GD}}\left( {\left\{ {{{\bf{u}}_{i,1}},{{\bf{u}}_{i,2}}, \ldots {{\bf{u}}_{i,P - 1}}} \right\}} \right){\left\| {{{\bf{u}}_{i,P}}} \right\|^2} \notag\\
\le& \prod\limits_{j = 1}^P {{{\left\| {{{{\bf{ u}}}_{i,j}}} \right\|}^2}}\notag\\
=&\prod\limits_{j = 1}^{P} {{\bf{e}}^{\rm{H}}}\left( {{{\bf{F}}_N} \otimes {{\bf{I}}_M}} \right){\bf{W}}_{{a_{i,j}}}^{\rm{H}}{{\bf{W}}_{{a_{i,j}}}}\left( {{\bf{F}}_N^{\rm{H}} \otimes {{\bf{I}}_M}} \right){\bf{e}}.\label{C8_Gram_derivation3}
\end{align}
By noticing that ${\bf{W}}_{{a_{i,j}}}^{\rm{H}}{{\bf{W}}_{{a_{i,j}}}}= {\bf I}_{MN}$,~\eqref{C8_Gram_derivation3} can be further derived as~\eqref{C8_determinant_upper_bound1}.
This completes the proof of Theorem~7-1.
\subsection{Proof of Lemma 7-3}
Assuming that ${\bf{W}}_{{a_{i,p}}}^{\rm{H}}{{\bf{W}}_{{a_{i,p'}}}} = {{\bf{0}}_{MN}}$, for $ 1 \le p,p' \le P$ and $p' \ne p$. Then, it is obvious that there exists an index $p$, where $1 \le p \le P$, such that ${\bf{W}}_{{a_{i,p}}}$ has a zero determinant, which is contradict to ${\bf{W}}_{{a_{i,p}}}^{\rm{H}}{{\bf{W}}_{{a_{i,p}}}} = {{\bf{I}}_{MN}}$.
This completes the proof of Lemma~7-3.
\subsection{Proof of Proposition 7-1}
Let ${\varepsilon '}$ denote the event that different paths do have the same delay or Doppler indices. To prove~\eqref{C8_Prop1_add}, we only need to prove that
\begin{align}
&\Pr \left( {\left. {\left| {{{\bf{e}}^{\rm{H}}}{{\bf{\Xi }}_{i,p}}{{\bf{\Xi }}_{i,p}}{\bf{e}}} \right| \ge \sum\limits_{\scriptstyle {p'} = 1\hfill\atop
\scriptstyle{p'} \ne p\hfill}^P {\left| {{{\bf{e}}^{\rm{H}}}{{\bf{\Xi }}_{i,p}}{{\bf{\Xi }}_{i,p'}}{\bf{e}}} \right|} } \right|\varepsilon '} \right) \notag\\
\ge& \Pr \left( {\left. {\left| {{{\bf{e}}^{\rm{H}}}{{\bf{\Xi }}_{i,p}}{{\bf{\Xi }}_{i,p}}{\bf{e}}} \right| \ge \sum\limits_{\scriptstyle{p'} = 1\hfill\atop
\scriptstyle{p'} \ne p\hfill}^P {\left| {{{\bf{e}}^{\rm{H}}}{{\bf{\Xi }}_{i,p}}{{\bf{\Xi }}_{i,p'}}{\bf{e}}} \right|} } \right|\varepsilon } \right) .\label{C8_Prop1_add_der1}
\end{align}
To this end, let us focus on the $(p,p')$-th element of ${{\bf{\Omega }}_i}\left( {\bf{e}} \right)$, and it is rewritten by
\begin{align}
{{\bf{e}}^{\rm{H}}}{\bf{\Xi }}_{i,p}^{\rm{H}}{{\bf{\Xi }}_{i,p'}}{\bf{e}} =& {{\bf{e}}^{\rm{H}}}\!\left( {{{\bf{F}}_N} \otimes {{\bf{I}}_M}} \right)\!{{\bm \Delta}\! ^{ - {{ k}_p}-{\kappa _p}}}{{\bm \Pi} ^{ - {{ l}_p}}}{{\bm \Pi} ^{{{ l}_{p'}}}}\!{{\bm \Delta} ^{{{ k}_{p'}+{\kappa _{p'}}}}}\!\left( {{\bf{F}}_N^{\rm{H}} \otimes {{\bf{I}}_M}} \right)\!{\bf{e}}\notag\\
 = &{{{\bf{\tilde e}}}^{\rm{H}}}{{\bm\Delta} ^{ - {{ k}_p}-{\kappa _{p}}}}{{\bm\Pi} ^{{{ l}_{p'}} - {{ l}_p}}}{{\bm\Delta} ^{{{ k}_{p'}+{\kappa _{p'}}}}}{\bf{\tilde e}},
\label{C8_Pro1_der1}
\end{align}
where ${\bf{\tilde e}} \buildrel \Delta \over = \left( {{\bf{F}}_N^{\rm{H}} \otimes {{\bf{I}}_M}} \right){\bf{e}}$ is the TDA domain error sequence.
Furthermore, let $n' = {\left[ {n - 1 - \left( {{l_{p'}} - {l_p}} \right)} \right]_{MN}} + 1$,~\eqref{C8_Pro1_der1} can be further derived by
\begin{align}
{{\bf{e}}^{\rm{H}}}{\bf{\Xi }}_{i,p}^{\rm{H}}{{\bf{\Xi }}_{i,p'}}{\bf{e}}=\sum\limits_{n = 1}^{MN} {{e^{j\frac{{2\pi }}{{MN}}\left( {\left( {n' - 1} \right)\left( {{k_{p'}} + {\kappa _{p'}}} \right) - \left( {n - 1} \right)\left( {{k_p} + {\kappa _p}} \right)} \right)}}{{\tilde e}^*}\left[ n \right]\tilde e\left[ {n'} \right]}. \label{C8_Pro1_der2}
\end{align}
In particular, it can be shown that for $p=p'$,~\eqref{C8_Pro1_der2} becomes ${{d_{\rm{E}}^2\left( {\bf{e}} \right)}}$, which is independent with the delay and Doppler distributions.
Thus, we notice that~\eqref{C8_Prop1_add_der1} can be verified if the absolute value of ${{\bf{e}}^{\rm{H}}}{\bf{\Xi }}_{i,p}^{\rm{H}}{{\bf{\Xi }}_{i,p'}}{\bf{e}}, \forall p \ne p'$ becomes smaller when different paths share the same delay/Doppler indices.
However, the values of the elements in ${\bf{e}}$ vary with respect to different transmitted symbol vectors and detection outputs, which are usually studied statically for the related analysis~\cite{li2021cross}.
Note that each TDA domain OTFS symbol is a superposition of $N$ DD domain OTFS symbols with specific phase rotations according to the spreading effect of IFFT. Thus, in practical systems with a sufficiently large $N$, the elements in ${\bf{\tilde e}} $ behave like i.i.d. Gaussian variable due to the law of large numbers~\cite{li2021cross}.
In the case where $l_p=l_{p'}$, i.e., different paths share the same delay index,~\eqref{C8_Pro1_der2} can be simplified as
\begin{align}
{{\bf{e}}^{\rm{H}}}{\bf{\Xi }}_{i,p}^{\rm{H}}{{\bf{\Xi }}_{i,p'}}{\bf{e}}= {{{\bf{\tilde e}}}^{\rm{H}}}{{\bm\Delta} ^{ {{ k}_{p'}+{\kappa _{p'}}}- {{ k}_p-{\kappa _{p}}}}}{\bf{\tilde e}}=\sum\limits_{n = 1}^{MN} {{e^{j\frac{{2\pi }}{{MN}}\left( {n - 1} \right)\left( {{k_{p'}} + {\kappa _{p'}} - {k_p} - {\kappa _p}} \right)}}{{\left| {\tilde e\left[ n \right]} \right|}^2}} .
\label{C8_Pro1_der3}
\end{align}
Comparing~\eqref{C8_Pro1_der2} and~\eqref{C8_Pro1_der3}, we observe that the absolute value of~\eqref{C8_Pro1_der3} is larger than that of~\eqref{C8_Pro1_der2} with a sufficiently large $MN$, because the expectation of the term ${{{\tilde e}^*}\left[ n \right]\tilde e\left[ {n'} \right]}$ is zero based on the i.i.d. assumption~\cite{li2021cross}, while the term ${{\left| {\tilde e\left[ n \right]} \right|}^2}$ is strictly non-negative.
On the other hand, when ${{k_p} + {\kappa _p}}={{k_{p'}} + {\kappa _{p'}}}$, i.e., different paths share the same Doppler index,~\eqref{C8_Pro1_der2} can be simplified as
\begin{align}
{{\bf{e}}^{\rm{H}}}{\bf{\Xi }}_{i,p}^{\rm{H}}{{\bf{\Xi }}_{i,p'}}{\bf{e}}= \sum\limits_{n = 1}^{MN} {{e^{j\frac{{2\pi }}{{MN}}\left( {n' - n} \right)\left( {{k_p} + {\kappa _p}} \right)}}{{\tilde e}^*}\left[ n \right]\tilde e\left[ {n'} \right]} \approx {e^{j\frac{{2\pi }}{{MN}}\left( {{l_p} - {l_{p'}}} \right)\left( {{k_p} + {\kappa _p}} \right)}}\sum\limits_{n = 1}^{MN} {{{\tilde e}^*}\left[ n \right]\tilde e\left[ {n'} \right]}  ,
\label{C8_Pro1_der4}
\end{align}
whose absolute value is given by $\left| {\sum\limits_{n = 1}^{MN} {{{\tilde e}^*}\left[ n \right]\tilde e\left[ {n'} \right]} } \right|$. Therefore, it can be shown that the absolute value of~\eqref{C8_Pro1_der2} is no larger than the absolute value of~\eqref{C8_Pro1_der4}, due to $\left| {\sum\limits_{n = 1}^{MN} {a_n b_n} } \right| \le \sum\limits_{n = 1}^{MN} {\left| a_n \right|\left| b_n \right|} $, for any two arbitrary complex vectors $\bf a$ and $\bf b$.
Based on the above discussions, we have shown that the absolute value of ${{\bf{e}}^{\rm{H}}}{\bf{\Xi }}_{i,p}^{\rm{H}}{{\bf{\Xi }}_{i,p'}}{\bf{e}}, \forall p \ne p'$ generally becomes smaller when different paths share the same delay/Doppler indices, while the value of ${{\bf{e}}^{\rm{H}}}{\bf{\Xi }}_{i,p}^{\rm{H}}{{\bf{\Xi }}_{i,p'}}{\bf{e}}$ remains constant with different delay and Doppler distributions. Thus,~\eqref{C8_Prop1_add_der1} holds and this completes the proof of Proposition~7-1.

    \chapter{Conclusions and Future Work}\label{c:literature}
In this thesis, we discussed the related subjects of OTFS modulation. Specifically, we have studied the topics of wireless channels, OTFS concepts, OTFS detection, OTFS performance analysis, and the application of OTFS in ISAC transmissions.

In Chapter~1, we reviewed the motivation and needs for the application of OTFS modulation. We also provided a literature review for OTFS modulation. In specific, the literature review covered the related studies of OTFS modulation including the concept and implementation, channel estimation, signal detection, and applications.
The main structure of this thesis was highlighted and the main research contributions were summarized.

In Chapter~2, the properties wireless channels were reviewed and the related discussions on the channel responses in different domains were presented. We also adopted several examples to describe the characteristics of the channel responses in different domains.

In Chapter~3, the concept of OTFS modulation was derived by using ZT and DZT. In particular, we offered DD domain interpretations of key components of modulation, such as pulse shaping and matched-filtering. Furthermore, the input-output relationship for rectangular pulse-shaped OTFS modulation was highlighted.

In Chapter~4, the hybrid-MAP-PIC detection for OTFS modulation was studied. We presented the derivations of the hybrid-MAP-PIC detection according to the SPA framework and also discussed its error performance and parameter selection. Our numerical results implied that the hybrid-MAP-PIC detection outperforms the conventional message passing algorithm and approaches the optimal detection performance with a reduced complexity.

In Chapter~5, we introduced the cross domain iterative detection for OTFS modulation. The main novelty of this detection is that the extrinsic information is passed between different domains via the corresponding unitary transformation. Error performance analysis of this detection was conducted based on the state evolution. Our numerical results have demonstrated the superior performance of the cross domain iterative detection over the conventional OTFS detection approaches.

In Chapter~6, we studied the error performance of coded OTFS systems based on the PEP analysis. In particular, our derivations have shown that there is a trade-off between the diversity gain and the coding gain, i.e., the diversity gain of OTFS systems improves with the number of resolvable paths, while the coding gain declines. Rule-of-thumb code design guidelines were also provided according to the trade-off. Numerical results agreed with our analysis and clearly demonstrated the trade-off between diversity gain and coding gain.

In Chapter~7, the application of OTFS modulation in ISAC transmissions was discussed. To facilitate the ISAC designs, we introduced the concept of spatial-spreading, which leads to simple and insightful input-output relationships for both communications and radar sensing. A framework of ISAC transmission using OTFS waveform was then investigated and low-complexity radar estimation and communication precoding schemes were proposed. Finally, we applied numerical approaches to verify the effectiveness of the considered scheme.

Although we have provided some related discussions of OTFS modulation, it should be noted that the related development of OTFS modulation is still at an early stage and some important issues of OTFS modulation still require to prompt attention. In the following, we list several related future works for OTFS modulation.

\section{Future Work}

\subsection{Exploiting DD Domain Channel Characteristics}
The application of OTFS modulation gives a direct way of exploiting the DD domain channel characteristics.
However, several appealing DD domain channel characteristics have not yet been fully exploited. For example, the commonly adopted channel model for OTFS modulation assumes that the channel geometry remains unchanged for a period of time. However, this assumption may not also hold in practical scenarios, especially for complex communication applications, such as high-speed train communication, where practical issues, such as path live-or-die, need to be considered.
Another example may be the case of channel inference based on DD domain channel reciprocity. This problem is of practical importance for a cellular system, where the uplink and downlink channels may share the same channel geometry over a period of time that is longer than the conventional coherence time. However, how to exploit the DD domain channel reciprocity still needs investigation.
As such, there is still a challenge for related OTFS designs to fully copilot the DD domain channel characteristics, and this challenge could be solved by conducting extensive real-world channel measurements for various wireless channels.

\subsection{Fundamental Limits for DD Domain Communications}
Although the DD domain communication has shown appealing advantages over the conventional TF domain counterpart, its fundamental limits are still unclear in the literature.
Despite the earlier works on the achievable rate analysis for OTFS modulation, the information-theoretical understanding of OTFS modulation still needs further investigations. For example, the DD domain has the abilities to fully localize a pulse without violating the Heisenberg's uncertainty principle~\cite{Hadani2017orthogonal,Hadani2018OTFS_long}. But how much information we can gain from the DD domain localization is still unclear.
Furthermore, the capacity scaling law for MIMO-OTFS has not been fully studied in the literature, despite the fact that it is crucial for many practical applications.
On top of that, security and privacy performances are also important aspects of future wireless networks. However, the related studies on security and privacy performances for DD domain communications still require further investigations.

\subsection{DD Domain Communication Designs}
Although most of the related works on OTFS modulation have been focused on the system design, there are still many important aspects of DD domain communication designs that need to be studied.
For instance, the long latency may be an issue for OTFS transmission. In particular, the DD domain received symbols can only be obtained once all the information symbols in the TF domain are received, which is generally time-consuming compared to the conventional OFDM. Therefore, it is important to design OTFS receivers with a low latency. Another issue could be the code design for OTFS. We have studied the error performance for coded OTFS systems in Chapter~7, but the detailed design criteria still need to be investigated. In specific, how to design good channel codes for OTFS systems, such as LDPC codes~\cite{Yihuan2021design,Yihuan2019LDPC,KP_TVT_2022,XiaoweiTCOM}, coupled codes~\cite{Zhang2019tail_biting,Shuangyang2018self,PIC2020}, Bose-Chaudhuri-Hocquenghem (BCH) codes~\cite{Bryan2020deep}, could be an important issue for practical application of OTFS modulation.
Furthermore, the combination of OTFS with emerging new technologies~\cite{ZhiqiangNOMAOMAGAIN,wei2018multiNOMA,MyAsynchronizedNOMA,hu2021robust,Yuanxin2022TWCnew,liu2020location,liu2019maximum,liu2016optimal,liu2016blind,liu2014blind,SunZhuoUAV,ZhiqiangUAV,WeiBeamWidthControl,WeiProceeding,QMin_IT,QMin_noSIC,QMin_lattice_NOMA,QMin_lattice} is also of importance to exploit.
On the other hand, machine learning has shown great potential to facilitate the design of wireless systems~\cite{Bryan2021novel,Bryan2019deep_FTN,liu2019deep,liu2020deep_ABC,liu2020deep_IRS}, and some preliminary works on machine learning-based OTFS designs appeared in~\cite{Yosef20212D,Yosef_MIMOOTFS_ICC}. However, there are still many aspects of OTFS modulation that could be improved with the assistance of machine learning.


    \clearpage


    \clearpage

    \backmatter

    \pagestyle{noHeading}
    \bibliographystyle{IEEEtran}
    \bibliography{references}

\end{document}